\documentclass[10pt]{article}
\pdfoutput=1
\usepackage[english]{babel}
\usepackage{latexsym,amsmath,amssymb,amsbsy,amstext,amscd,amsfonts,amsthm}
\usepackage{graphics,graphicx}
\usepackage{color}
\usepackage{tabularx}
\usepackage{multirow}
\usepackage{booktabs}
\usepackage{float}
\usepackage{epstopdf}
\theoremstyle{definition}
\newtheorem{remark}{Remark}
\usepackage{array}
\date{\today}

\title{Universal hydrofracturing algorithm for shear-thinning fluids: particle velocity based simulation}

\author{Monika Perkowska$^{(1,2)}$, Michal Wrobel$^{(1,3)}$, Gennady Mishuris$^{(1)}$
\\
{\it $^{(1)}$ Department of Mathematics,
Aberystwyth University, }
\\ {\it Ceredigion SY23 3BZ, Wales, UK}
\\{\it $^{(2)}$\!EUROTECH Sp. z o.o.,}
\\ {\it Wojska Polskiego 3, 39-300, Mielec, Poland}
\\{\it $^{(3)}$\!EnginSoft TRENTO,}
\\ {\it Via della Stazione 27, fraz. Mattarello, 38123 Trento, Italy}}

\begin{document}

\maketitle

\begin{abstract}

A universal particle velocity based algorithm for simulating hydraulic fractures, previously proposed  for Newtonian fluids, is extended to the class of shear-thinning fluids. The scheme is not limited to any particular elasticity operator or crack propagation  regime. The computations are based on two dependent variables: the crack opening and the reduced particle velocity.  The application of the latter facilitates utilization of the local condition of Stefan type (speed equation) to trace the fracture front.  The condition is given in a general explicit form which relates the crack propagation speed (and the crack length) to the solution tip asymptotics. The utilization of a modular structure, and the adaptive character of its basic blocks, result in a flexible numerical scheme. The computational accuracy of the proposed algorithm is validated against a number of analytical benchmark solutions.

\end{abstract}

\section{Introduction}

The phenomenon of a hydraulically propelled fracture that propagates in a solid material is present in many natural processes and in technology \cite{board_1992,pine_cun_1985,rubin_1995,tsai_rice_2010}. The multiphysical character of the problem makes mathematical and numerical modeling of hydraulic fractures (HFs) an extremely challenging task. The main computational difficulties stem from: i) strong non-linearities resulting from interaction between the solid and fluid phases, ii) singularities in the physical fields, iii) moving boundaries, iv) degeneration of the governing equations at the singular points of the domain, v) leak-off to the rock formation, vi) pronounced multiscale effects, vii) complex geometry, and others.

The first simplified mathematical models of hydraulic fractures were introduced in the 1940s and 1950s \cite{Crittedon,Harrison,Howard,Hubbert,Sneddon_Elliot}. These works, together with some later studies, evolved into basic 1D models: i) the PKN model \cite{nord_1972,perkins_kern_1961}, ii) the KGD model \cite{geertsma_1969,khr_zhelt_1955},  iii) the radial or penny shaped model \cite{sned_1946}. These models were used for decades in practical applications to design the hydrofracturing treatments. However, a continuous increase in both the number and size of fracking installations necessitated further development of the mathematical models describing the process, which resulted in the introduction of the so-called pseudo 3D models \cite{mack_warp_2000}, planar 3D models \cite{Advani,Clifton,Vandamme} and recent attempts to develop fully 3D models \cite{ad_et_al_2007,Damjanac_2013,Wangen,Yamamoto_2004}. A comprehensive review of the topic can be found in \cite{ad_et_al_2007}.

Simultaneously, the studies of the basic solution features pertaining to the underlying physics of the process have been carried out. Special attention has been devoted to the near-tip region of the fracture. The asymptotic behaviour of the solution for various hydraulic fracture models and crack propagation regimes has been identified in a number of works -- see e.g. \cite{Det_Gar_2003,Linkov_umbrella,MKP,MKP_2007a} and references therein. The importance of the multiscale character of the problem has been fully recognized. It is now well understood that the global behaviour of the hydraulic fracture is primarily controlled by its near-tip region \cite{Gar_Det_Ad}. As a result of the moving boundary of the domain, the governing equations and boundary conditions degenerate at the crack tip, which makes tracing the fracture front an extremely difficult task \cite{Peirce_2014}. Moreover, the solution tip asymptotics essentially depend on the fracture propagation regime. Here, four basic modes have been identified \cite{Gar_Det_Ad,Hu_2010}: i) leak-off dominated, ii) storage dominated, iii) toughness dominated, iv) viscosity dominated.  All these factors clearly indicate that a proper understanding of the solution structure and the correct application of the tip asymptotics in the computational schemes are crucial for the accuracy, credibility and robustness of computations. In the recent paper by Lecampion et al. \cite{Lecampion_Brisbane} it was shown that the algorithms which utilize the appropriate multiscale hydraulic fracture tip asymptote perform much better than those that do not apply it. The advantages of employing enhanced asymptotic representation in the numerical algorithm have been demonstrated in \cite{gar_2006,Hu_2010,Kovalyshen}  (special tip element), \cite{ad_et_al_2007,Chen_2013,Gordeliy_2013,Gordeliy_2013a,Gordeliy_2015,Lecampion_2009,Weber_2013} (BEM and XFEM formulations).

Another crucial element affecting the credibility and efficiency when numerically modeling hydraulic fractures is the mechanism of fracture front tracing. As in any other problem involving a moving boundary, a proper Stefan type condition should be imposed. However,  its direct utilization in HF is a numerically challenging task \cite{Det_Peirce_2014}. Alternatively, in the case of 1D problems (PKN, KGD and radial models), the fluid balance equation can be used  to trace the crack tip position. The non-local nature of such a condition has a clear advantage from the computational point of view, in comparison with any local counterpart.  When properly employed in the numerical scheme, it is equivalent to the Stefan condition and allows one to obtain accurate results for a reasonable computational cost. As for the Stefan type boundary condition, its applicability and importance in hydraulic fracture problems was pointed out in \cite{kemp_1990}. Two decades later its relevance was rediscovered in \cite{linkov_2011}, where the condition itself was called the \emph{speed equation}.

The situation is quite different when one analyzes 2D or 3D models of hydraulic fractures. Thus, it is of vital importance to utilize a method based on the local condition to establish the fracture front location. To this end an implicit level set method was developed in \cite{Peirce_2008}. The technique was successfully implemented in \cite{Peirce_2014}. Recently, an explicit level set scheme was proposed in \cite{Linkov_umbrella}. It combines the information in the solution tip asymptotics with the speed equation.

Such an approach for the 1D PKN model has already been presented in \cite{kus_mis_wr_2013,mis_wr_lin_2012,wr_mis_2013} while its generalization, introducing explicit relations between crack tip asymptotics and crack propagation speed for all classical 1D models, can be found in \cite{wr_mis_2015}. It was shown that, when properly combined with other numerical devices, the algorithm represents an accurate and universal tool, which allows one to circumvent the main problems found in classical approaches used to establish the crack tip position. In this paper, we will consequently use the methodology developed in \cite{wr_mis_2015}.

Appropriate  modeling of the interaction between solid and fluid phases necessitates a thorough investigation into the underlying physical mechanisms of fluid flow inside the fracture. One of the most important problems here is a proper description of, and accounting for, the rheological properties of the fracturing fluid. Depending on the environment in which the fracking process is carried out, various types of fluids are in use. In particular, in low permeability reservoirs (small leak-off to the rock formation) one usually employs water with some additives. In such a case the fluid can be  modeled as a Newtonian one. Most of the fracturing fluids utilized in the conventional reservoirs exhibit non-Newtonian properties that can be described by a power law \cite{Cameron_1989,Eskin_2009,Lakhtychkin_2012,lavrov_2013,lavrov_2013a,Valko_1995}:

\begin{equation}\label{power_law_fluid}
\tau=M\dot{\gamma}^n.
\end{equation}
Here $\tau$ denotes the shear stress, $\dot{\gamma}$ stands for the shear strain rate, $n$ is the "fluid behaviour index" and $M$ the "consistency index". The range of $0<n < 1$ refers to the so-called shear-thinning fluids, while $n>1$ pertains to the shear-thickening case. For $n=1$, one has the Newtonian fluid with the consistency index equal to the dynamic viscosity: $M=\mu$. A special case when $n=0$ pertains to the perfectly plastic fluid, where $M=\sigma_Y$ is the yield stress.

However, as pointed out in \cite{Shah_1995}, such a simple two-parameter model cannot correctly describe the real fluid behavior at low and high shear rates.  In such cases more advanced approximations, like the Carreau model \cite{Carreau_1972,Darby_2001}, are appropriate. Unfortunately, a closed-form solution for the Poiseuille type flow in a channel with parallel walls cannot be obtained with the Carreau law. Thus, the lubrication theory is no longer in use here. In order to circumvent this deficiency, the so-called truncated power law was recently proposed in \cite{Lavrov_runk_pl}. This four-parameter approximation uses cut-off viscosities at low and high shear  rates, with the power law dependence in the intermediate range. It provides a satisfactory imitation of the Carreau model and simultaneously allows one to obtain the closed-form solution for the flow between parallel plates. A three-parameter rheological model for the Hershel-Buckley fluid type \cite{Herschel_1926} was used in \cite{Sun_2014}, where the power law \eqref{power_law_fluid} was combined with  a lower cut-off when the shear stress becomes equal to the value of the yield stress. A closed-form solution for the Poiseuille type flow in the channel was embedded in the PKN model with the Carter leak-off law.

Nevertheless, as far as the simulation of hydraulic fractures for non-Newtonian fluids is concerned, most of the results available in literature refer to the pure power law dependence of the form \eqref{power_law_fluid}.

An analytical solution of the PKN model for shear-thinning fluids was obtained in \cite{linkov_2013}. The author reformulated the problem in terms of new dependent variables, using the power of the crack opening (a modified opening) and the fluid velocity. A self-similar solution in the form of  an infinite power series was given for the case of a  predefined leak-off. Particular solutions for an impermeable rock (zero leak-off) for two limiting values of the fluid behaviour index, $n=0$ and $n=1$ (perfectly plastic and Newtonian fluid), were later used in \cite{linkov_2014a} to draw conclusions concerning the equivalence of different shear-thinning fluids with respect to their hydrofracturing action. In \cite{fareo_mason_2013}, the authors analyzed the PKN problem for the zero leak-off case. They introduced two simple analytical solutions: for constant fluid velocity along the fracture length and for constant crack volume in time. A numerical algorithm based on  reducing the problem to a dynamic system of the first order was proposed. A number of numerical results for shear-thinning and shear-thickening fluids were presented.

A penny-shaped fracture induced by a power-law fluid was analyzed in \cite{Advani_1987}. The governing equations of the problem were derived and explicit time-dependent particular solutions were proposed for both Newtonian and shear-thinning fluids. Parametric sensitivity studies were conducted to estimate the influence of both the fluid viscosity and the leak-off (Carter law) on the fracture evolution.

The KGD model in viscosity dominated regime for impermeable rock was investigated in \cite{ad_det_2002}. The authors reduced the problem to a self-similar formulation and delivered semi-analytical solutions, in the form of power series, for both the Newtonian and a number of shear-thinning fluids. In \cite{golovin_2015} the KGD model for a viscosity dominated regime was analyzed in the case of permeable rock. A finite element  based algorithm with no explicit fracture front tracing mechanism was proposed to find the solution. The credibility of computations was verified against the reference solutions from \cite{ad_det_2002}. The KGD model for a finite material toughness was investigated in \cite{gar_2006}. The author proved that for shear-thinning fluids the crack propagation regime evolves in time from toughness dominated to viscosity dominated. Corresponding approximate self-similar solutions for these limiting cases were delivered in the form of power series. A numerical algorithm for the transient solution was also proposed.

In this paper we analyze the problem of a 1D hydraulic fracture driven by a non-Newtonian power law fluid whose rheology is described by equation \eqref{power_law_fluid}. Two classical hydrofracturing models, the PKN and the KGD (for the viscosity dominated and toughness dominated regimes), are under consideration. A unified numerical algorithm, being a direct extension of the one introduced for Newtonian fluids in \cite{wr_mis_2015}, is proposed. The basic assumptions of the numerical scheme are: i) a proper choice of independent and dependent variables (reduced particle velocity), ii) a mechanism of fracture front tracing based on the Stefan condition, utilizing a universal explicit relation between the crack propagation speed and the coefficients in the asymptotic expansion of the crack opening, iii) proper regularization techniques, iv) an improved temporal approximation, v) modular algorithm architecture.

Some of the aforementioned elements were already presented in \cite{kus_mis_wr_2013,mis_wr_lin_2012,wr_mis_2013}, where one can find a broad analysis on the application of proper independent and dependent variables and regularization techniques. The theoretical background of the mechanism of fracture front tracing together with a detailed description of the techniques of its application was given in \cite{wr_mis_2015}. Adaptation of the numerical scheme from \cite{wr_mis_2015} to the class of shear-thinning fluids was primarily related to the different form of the Poiseuille equation and necessitated: i) updating the asymptotic expansions, and the interrelations between them, for respective dependent variables, ii) modifying the regularization techniques, iii) development of new analytical benchmark solutions utilized to verify the algorithm. The accuracy and efficiency of computations, investigated against dedicated analytical benchmarks, demonstrate the ultimate performance of the proposed numerical scheme.

The structure of the paper is as follows. In Section \ref{sec:problem_formulation}, a general formulation of the problem is given. After subsequent normalization the complete system of equations is reformulated in terms of a new pair of dependent variables: the crack opening and the reduced particle velocity. The mechanism of fracture front tracing is introduced. The employed methodology is based on proper utilization of the solution tip asymptotics.
 Section \ref{sec:self_sim_formulation} is devoted to the time-independent (self-similar) problem formulation. The self-similar variant of the universal algorithm is introduced and thoroughly tested for the considered hydraulic fracture models. The performance of the algorithm is verified against analytical benchmarks presented in Appendix A, as well as reference solutions found in literature. A new approximation for the classical problem of the KGD model in viscosity dominated regime is delivered.
 In Section \ref{sec:transient} the transient version of the algorithm is presented. Again, its performance is validated in a number of tests. The main peculiarities of the scheme are shown. The remaining part of the paper contains discussion of the results and final conclusions. Additional material is given in the appendices.

\section{Problem formulation}
\label{sec:problem_formulation}

Let us consider a rectilinear symmetrical crack of length $2l$ situated in the plane $x\in [-l,l]$. The crack is fully filled by a non-Newtonian liquid injected at the middle point ($x=0$) at a known rate $q_0(t)$. As a result, the crack front ($x=\pm l$) moves and the crack length, $l=l(t)$, is a function of time. Below we present a classic set of governing equations for the 1D formulation of the problem, which can be found in e.g. \cite{ad_det_2002,gar_2006,linkov_2013} for various hydraulic fracture (HF) models. As usual, due to symmetry of the problem, we restrict our analysis to one of the symmetrical parts of the crack $x \in [0, l(t)]$.

The mass conservation principle is expressed by the continuity equation:
\begin{equation}\label{continuity}
\frac{\partial w}{\partial t}+\frac{\partial q}{\partial x}+q_l=0,\quad t > t_0,\quad 0< x < l(t),
\end{equation}
while fluid flow inside the fracture is described by the Poiseuille equation:
\begin{equation}\label{poiseuille}
q^n = - \frac{1}{M'} w^{2n+1} \frac{\partial p}{\partial x}.
\end{equation}

In the above equations $w=w(t,x)$ is the crack opening, $q=q(t,x)$ stands for the fluid flow rate, $p=p(t,x)$ ($p=p_{f}-\sigma_0$, $\sigma_0$ -
confining stress) refers to the net fluid pressure. The constant $M'$ in the Poiseuille equation is a modified fluid consistency index $M'= 2^{n+1}(2n+1)^n/n^n M$. The function $q_l=q_l(t,x)$ on the right hand side of \eqref{continuity} is the volumetric fluid loss to the rock formation in the direction perpendicular to the crack surfaces per unit length of the fracture. In general, this function is itself a part of the solution, for which an additional diffusion problem in the surrounding rock needs to be solved. Such a task goes beyond the scope of our analysis and thus we assume that the leak-off function is predefined.

Fluid equations \eqref{continuity} -- \eqref{poiseuille} are supplemented by the equation describing rock deformation under applied hydraulic pressure. Here we use the following symbolic relation:
\begin{equation}\label{elasticity}
p(t,x)={\cal A}w(t,x), \quad 0<x<l(t),
\end{equation}
where the operator ${\cal A}$ refers to the chosen HF model (related to the fracture geometry).
In this paper we consider two hydraulic fracture models:
\begin{itemize}
\item{the PKN model \cite{linkov_2013}}
\begin{equation}\label{elasticity_PKN}
{\cal A}_1w=k_1w,
\end{equation}
\item{the KGD model \cite{ad_det_2002,gar_2006}}
\begin{equation}
\label{elasticity_KGD}
{\cal A}_2w=\frac{k_2}{2\pi}\int_{0}^{l(t)}\frac{\partial w(t,s)}{\partial s}\frac{s}{x^2-s^2}ds.
\end{equation}
\end{itemize}
For the classical PKN model, the constant $k_1$ can be found in \cite{nord_1972} for an elliptical crack of height $h$, where $E$ and $\nu$ are the elasticity modulus and Poisson's ratio respectively. The multiplier $k_2$ in the KGD model follows, for example, from \cite{mus_1992,sned_low_1969}:
\begin{equation}\label{k1_k2}
k_1=\frac{2E}{\pi h(1-\nu)},\quad k_2=\frac{E}{(1-\nu^2)}.
\end{equation}
The inverse operators for \eqref{elasticity_PKN} -- \eqref{elasticity_KGD} are:
\begin{equation}\label{inverse_PKN}
{\cal A}_1^{-1} p= \frac{1}{k_1}p,
\end{equation}
\begin{equation}
\label{inverse_KGD}
{\cal A}_2^{-1} p=\frac{4}{\pi k_2}\int_0^{l(t)} p(t,s) \ln \left|\frac{\sqrt{l^2(t)-x^2}+\sqrt{l^2(t)-s^2}}{\sqrt{l^2(t)-x^2}-\sqrt{l^2(t)-s^2}}\right|ds.
\end{equation}
The form of the operators \eqref{elasticity_KGD} and \eqref{inverse_KGD} implies that:
\begin{equation}
\label{bc_p} \frac{\partial w}{\partial x}(t,0)=0.
\end{equation}
The foregoing equations are supplemented by the influx boundary condition at the crack mouth ($x=0$):
\begin{equation}\label{bc_0} q
(t,0)=q_0(t),
\end{equation}
and two boundary conditions at the crack tip:
\begin{equation}\label{bc_1}
w(t,l(t))=0, \quad q(t,l(t))=0.
\end{equation}
In this paper we assume non-zero initial conditions:
\begin{equation}
\label{ic}
l(0)=l_\diamond,\quad w(0,x)=w_\diamond(x),\quad x\in(0,l_\diamond).
\end{equation}
Note that \eqref{ic} in its general form accounts for a preexisting hydraulic fracture.

In the toughness dominated KGD model the fracture evolution is analyzed in the framework of Linear Elastic Fracture Mechanics (LEFM), and the following propagation condition holds:
\begin{equation}
\label{K_I_criterion}
K_{I}(t)=K_{IC},
\end{equation}
where $K_{IC}$ is the material toughness \cite{rice_1968}, and $K_{I}$ is the stress intensity factor (SIF) defined as
\cite{mus_1992}:
\begin{equation}
\label{K_I}
K_{I}= 2\sqrt{\frac{l(t)}{\pi}} \int_0^{l(t)} \frac{p(t,s)ds}{\sqrt{l^2(t)-s^2}}.
\end{equation}
As a result the crack opening asymptote yields \cite{mus_1992}:
\begin{equation}
\label{K_1}
w(t,x)\sim\frac{8(1-\nu^2)}{\sqrt{2\pi}}\frac{K_{I}}{E}\sqrt{l(t)-x},\quad x\to l(t).
\end{equation}

Finally, the global fluid balance equation, called also the solvability condition \cite{ad_et_al_2007}, can be written as:
\begin{equation}\label{global_balance}
\int_0^{l(t)}[w(t,x)-w_\diamond(x)]dx-\int_{0}^tq_0(t)dt+\int_0^{l(t)}\int_{0}^tq_l(t,x)dtdx=0.
\end{equation}

The above equations constitute the classical model of hydraulic fracture \cite{ad_et_al_2007,gar_2006,linkov_2012}.

In our analysis we shall use another dependent variable -- the particle velocity, $v$. It describes the average speed of fluid flow through the fracture cross section and can be defined as:
\begin{equation}\label{particle_velocity}
v(t,x)=\frac{q(t,x)}{w(t,x)}=\left(-\frac{1}{M'} w^{n+1} \frac{\partial p}{\partial x}\right)^{1/n}.
\end{equation}

The advantages of applying this variable were shown in \cite{wr_mis_2015}. Moreover, its introduction is related to the mechanism of fracture front tracing utilized throughout this paper. Namely, provided that there is neither lag between fluid front and crack tip nor an invasive zone ahead of the fracture (the fluid front coincides with the fracture tip), the crack propagation speed equals the value of particle velocity at the crack tip (see discussion in \cite{Det_Peirce_2014,linkov_2012,wr_mis_2015}). This allows us to formulate the relation for fracture front tracing as an evolution of the Stefan condition \cite{kemp_1990,stefan_1889}, called also the {\it{speed equation}} \cite{linkov_2011}:
\begin{equation}\label{SE}
\frac{dl}{dt}=v(t,l(t))=\left(-\frac{1}{M'} w^{n+1} \frac{\partial p}{\partial x}\right)^{1/n}\bigg|_{x=l(t)}.
\end{equation}
The technical implementation of condition \eqref{SE} and advantages of such an approach are extensively discussed in \cite{wr_mis_2015}.

\begin{remark}
In general, \eqref{SE} should be modified when the leak-off is described by the Carter law \cite{Det_Peirce_2014}.
\end{remark}

\subsection{Normalization}
\label{basic}

Let us normalize the problem by introducing the following dimensionless variables:
\[
\tilde x=\frac{x}{l(t)}, \quad \tilde t = \frac{t}{t_r^{1/n}},\quad \tilde w(\tilde t,\tilde x)=\frac{w(t,x)}{l_*}, \quad L(\tilde
t)=\frac{l(t)}{l_*},\quad\tilde q_l(\tilde t,\tilde x)= \frac{t_r^{1/n}}{l_*} q_l( t, x),
\]
\begin{equation}\label{norm_V}
 \quad
\tilde q(\tilde t,\tilde x)=
\frac{t_r^{1/n}}{l_*^2}q(t,x),\quad \tilde p(\tilde t, \tilde x)=\frac{1}{k_m}p(t,x),\quad \tilde v(\tilde t,\tilde x)=\frac{t_r^{1/n}}{l_*}v(t,x),
\quad t_r=\frac{M'}{k_m},
\end{equation}
where $\tilde x\in[0,1]$, $n$ is the fluid behaviour index defined above \eqref{power_law_fluid}, and we have that $m=1-r$, with $r$ being the homogeneity order of the operator ${\cal A}_m$. The parameter $m$ has value 1 or 2 for the PKN and KGD model respectively.
This type of normalization has already been used in \cite{wr_mis_2015}. It is not attributed to any particular influx regime, elasticity operator or asymptotic behaviour of the solution.

Using the transformations \eqref{norm_V} and expressing the fluid flow rate in terms of the crack aperture and the particle velocity, one can rewrite the continuity equation \eqref{continuity} as:
\begin{equation}\label{cont_norm}
\frac{\partial \tilde w}{\partial \tilde t}-\frac{L'}{L}\tilde x \frac{\partial \tilde w}{\partial \tilde x}+\frac{1}{L}\frac{\partial (\tilde w \tilde v)}{\partial \tilde x}+\tilde q_l=0.
\end{equation}
The normalized particle velocity \eqref{particle_velocity} yields:
\begin{equation}\label{v_norm}
\tilde v=\frac{\tilde q}{\tilde w}=\left(-\frac{1}{L}\tilde w^{n+1} \frac{\partial \tilde p}{\partial x}\right)^{1/n}.
\end{equation}
The notation for the elasticity equation \eqref{elasticity} is:
\begin{equation}\label{elas_norm}
{\tilde p=}{\cal \tilde A} \tilde w,
\end{equation}
with operators now defined as:
\begin{equation}\label{elas_norm_PKN}
{\cal \tilde A}_1 \tilde w= \tilde w,
\end{equation}
\begin{equation}\label{elas_norm_KGD}
{\cal \tilde A}_2 \tilde w= \frac{1}{2 \pi} \frac{1}{L(\tilde t)} \int_0^1\frac{\partial \tilde w(\tilde t,\eta)}{\partial \eta} \frac{\eta}{\tilde x^2-\eta^2}d\eta,
\end{equation}
while their inverses are:
\begin{equation}\label{inv_norm_PKN}
\tilde w={\cal \tilde A}_1^{-1} \tilde p=\tilde p,
\end{equation}
\begin{equation}\label{inv_norm_KGD}
\tilde w={\cal \tilde A}_2^{-1} \tilde p=\frac{4}{\pi}L(\tilde t)\int_0^1 \tilde p(t,\eta)\ln\left|\frac{\sqrt{1-\tilde x^2}+\sqrt{1-\eta^2}}{\sqrt{1-\tilde x^2}-\sqrt{1-\eta^2}}\right|d\eta.
\end{equation}

Following \cite{wr_mis_2015}, we shall use an alternative form of \eqref{inv_norm_KGD} defined as:
\begin{equation}
\label{inv_norm_KGD_2}
\tilde w={\cal \tilde A}_2^{-1} \tilde p=-\frac{4}{\pi}L(\tilde t)\int_0^1 \frac{\partial \tilde p(t,\eta)}{\partial \eta}K(\eta,x)d\eta +\frac{4}{\sqrt{\pi}} \sqrt{L(\tilde t)}\tilde K_I\sqrt{1-\tilde x^2},
\end{equation}
where the kernel $K(\eta,x)$ is:
\begin{equation}\label{K}
K(\eta,x)=(\eta-x)\ln \left | \frac{\sqrt{1-x^2}+\sqrt{1-\eta ^2}}{\sqrt{1-x^2}-\sqrt{1-\eta ^2}}\right |+x \ln \left (\frac{1+\eta x+\sqrt{1-x^2}\sqrt{1-\eta^2}}
{1+\eta x-\sqrt{1-x^2}\sqrt{1-\eta^2}}\right).
\end{equation}
$\tilde K_{I}=K_{Ic}(1-\nu^2)/(E\sqrt{l_*})$ stands for the dimensionless toughness. Note that the fracture propagation condition \eqref{K_I_criterion} is already included in the formulation.
Consequently, the asymptotic estimate \eqref{K_1} can be rewritten as:
\begin{equation}\label{K_1_norm}
\tilde w (\tilde t, \tilde x)=\frac{8}{\sqrt{2\pi}}\tilde K_{I} \sqrt{L(\tilde t)}\sqrt{1-\tilde x},\quad \tilde x \to 1.
\end{equation}

From definition \eqref{K_I} one can determine the dimensionless toughness as:
\begin{equation}\label{K_1_norm2}
\tilde{K_{I}}= 2\sqrt{\frac{L(\tilde t)}{\pi}} \int_0^1 \frac{\tilde p(\tilde t,s)ds}{\sqrt{1-s^2}}.
\end{equation}

The normalized global fluid balance equation gives:
\begin{equation}\label{global_balance_norm}
\int_0^1\left[L(\tilde t)\tilde w(\tilde t, \tilde x)-L(0)w_*(\tilde x)\right]d \tilde x-\int_0^{\tilde t}\tilde q_0(\tilde t)d\tilde t+\int_0^{\tilde t}L(\tilde t)\int_0^1 \tilde q_l(\tilde t,\tilde x)d \tilde x d\tilde t=0,
\end{equation}
while the speed equation \eqref{SE} converts to:
\begin{equation}\label{v_0}
\tilde v_0(t)=\frac{dL}{dt}=\left[-\frac{1}{L} \tilde w^{n+1}\frac{\partial \tilde p}{\partial \tilde x}\right]^{1/n}_{\tilde x=1} < \infty.
\end{equation}
Here, the value of $\tilde v_0(t)$ is bounded and equal to the crack propagation speed: $\tilde v_0(\tilde t)=\tilde v(\tilde t,1)$. This assumption is valid if the crack opening and the leak-off functions behave asymptotically as:
\begin{equation}
\label{ql_asymp}
\tilde w=\tilde w_0 (\tilde t)(1-\tilde x)^{\alpha_0}+o\left((1-\tilde x)^{\alpha_0}\right),\quad \tilde q_l=\tilde q_{l0}(\tilde t)(1-\tilde x)^\eta+o\left((1-\tilde x)^\eta\right), \quad \tilde x \to 1,
\end{equation}
where $1+\eta>\alpha_0$ (see discussion in \cite{wr_mis_2015}). Note also that relation \eqref{global_balance_norm} assumes that the process of the fracture growth is monotonic ($\tilde v_0(\tilde t)=L'(\tilde t)>0$).

As previously, for the KGD model, the fracture symmetry condition holds:
\begin{equation}
\label{dw_0}
\frac{\partial \tilde w}{\partial \tilde x}(\tilde t,0)=0.
\end{equation}
The boundary conditions \eqref{bc_0} -- \eqref{bc_1} assume the form:
\begin{equation}\label{bc0_norm}
\tilde w(\tilde t,0)\tilde v(\tilde t,0)=\tilde q_0(\tilde t),\quad
\tilde w (\tilde t,1)=0.
\end{equation}
Note that the equivalent of condition \eqref{bc_1}$_2$, when represented in terms of $\tilde w$ and $\tilde v$, and accounting for condition \eqref{v_0}, is satisfied automatically.

The initial conditions are now as follows:
\begin{equation}\label{ic_norm}
L(0)=\frac{l_{\diamond}}{l_*}, \quad \tilde w(0,\tilde x)=\tilde w_*(\tilde x)\equiv\frac{1}{l_*}w_\diamond(l_\diamond\tilde x), \quad \tilde x \in [0,1].
\end{equation}

For simplicity, from now on we omit the "$\sim$" symbol.

\subsection{Tip asymptotics and crack propagation speed}

Recently, a universal algorithm for numerical simulation of hydraulic fractures has been proposed \cite{wr_mis_2015}. One of its main elements is the mechanism of fracture front tracing based on the Stefan condition. It also extensively utilizes the crack tip asymptotics. In this paper we are going to extend this approach to the case of shear-thinning fluids. To this end, below we provide the basic information on the asymptotic behaviour of the solution at the fracture front.

As was shown in \cite{ad_det_2002,gar_2006,linkov_2013}, the crack aperture in the vicinity of the fracture tip can be expressed for some $\delta>0$ as:
\begin{equation}\label{w_asym}
w(t,x)=w_0(t)(1-x^m)^{\alpha_0}+w_1(t)(1-x^m)^{\alpha_1}+w_2(t)(1-x^m)^{\alpha_2}+O\Big((1-x^m)^{\alpha_2+\delta}\Big),\quad x \to 1,
\end{equation}
where powers $\alpha_i$ for respective HF models are given in Table \ref{T1} for the case of impermeable rock ($q_l=0$). Note that for the KGD model in toughness dominated regime in the case of perfectly plastic fluid ($n=0$),
the second term of the crack tip asymptotics additionally contains a logarithmic multiplier \cite{gar_2006}.

\begin{table}[h]

\vspace{3mm}
\renewcommand{\arraystretch}{1.5}
\begin{center}
\begin{tabular}{|c|c|c|c|c|c|c|c|c|c|c|}
  \hline
    HF model & $m$ &  $\alpha_0 $ &$\alpha_1$& $\alpha_2$ &$\beta_1$& $\beta_2$ &$\zeta_0$ &$\zeta_1$ &$\zeta_2$\\
  \hline\hline
  PKN & 1  &$\frac{1}{n+2}$&$\frac{n+3}{n+2}$ &$\frac{2n+5}{n+2}$&1 &2 &1 &2 &3 \\
  \hline
  KGD (viscosity dominated)&2 &$\frac{2}{n+2} $&$\frac{n+4}{n+2}$& $\frac{2n+6}{n+2}$& 1&$\frac{2n+2}{n+2}$&1 &$\frac{4}{n+2}$&2 \\
  \hline
  KGD (toughness dominated) & 2   & $\frac{1}{2}$ & $\frac{3-n}{2}$ & $\frac{5-2n}{2}$ &$\frac{2-n}{2}$&1&$\frac{2-n}{2}$&1&min$(2-n,\frac{3+n}{2})$ \\
  \hline
\end{tabular}
\end{center}

\vspace{-2mm}

\caption{The values of basic constants involved in the asymptotic expansions of $w$, $v$ and $\phi$ for $0<n<1$. }
\label{T1}
\end{table}
As a consequence, the particle velocity asymptotics yield:
\begin{equation}\label{v_asym}
v(t,x)=v_0(t)+v_1(t)(1-x^m)^{\beta_1}+O\Big((1-x^m)^{\beta_2}\Big),\quad x \to 1.
\end{equation}
The values of $\beta_i$ are collected in Table \ref{T1}.

The value of $v_0(t)$ should be greater than zero to make the crack move forward. In the cases under consideration $v_0(t)$ is defined by no more than the two leading terms of asymptotic expansion for the crack opening.

Let us adopt the following symbolic representation:
\begin{equation}
\label{C_gen}
\lim_{x\to1}w^{n+1}\frac{\partial}{\partial x}{\cal A}w=-C_{{\cal A}}\frac{{\cal L}(w)}{L^{m-1}}<0,
\end{equation}
where ${\cal L}(w)$ is a functional on the fracture aperture $w$, related to the form of elasticity operator, while $C_{{\cal A}}$ is a known constant. Thus, the formula
\begin{equation}\label{v_0_univ}
v_0=\left[\frac{1}{L^m} C_{{\cal A}}{\cal L}(w)\right]^{1/n}
\end{equation}
is a universal one (valid for both analyzed elasticity operators and all crack propagation regimes) and constitutes a relation between the crack propagation speed and the multiplier(s) of the leading term(s) of the crack opening tip asymptotics.

The values of $C_{{\cal A}}$ and ${\cal L}(w)$ for respective HF models are:
\begin{itemize}
\item{PKN model}
\begin{equation}\label{LC_PKN}
C_{{\cal A}}=\frac{1}{n+2},\quad {\cal L}(w)=w_0^{n+2},
\end{equation}
\item{KGD model - viscosity dominated regime}
\begin{equation}\label{LC_KGD_fluid}
C_{{\cal A}}= \frac{2n}{(n+2)^2} \cot\frac{n\pi}{n+2},\quad {\cal L}(w)=w_0^{n+2},
\end{equation}
\item{KGD model - toughness dominated regime}
\begin{equation}\label{LC_KGD_toughness}
C_{{\cal A}}=\frac{(3-n)(1-n)}{4}\tan\frac{n\pi }{2}, \quad {\cal L}(w)=w_0^{n+1} w_1.
\end{equation}
\end{itemize}

Taking into account that $v_0(t)=L'(t)$, equation \eqref{v_0_univ} can be directly integrated to compute the crack length:
\begin{equation}\label{L_int}
L(t)=\left[L^{\frac{m}{n}+1}(0)+\left(\frac{m}{n}+1\right)C_{\cal A}^{\frac{1}{n}}\int_0^t {\cal L}^{\frac{1}{n}}(w) d\tau \right]^\frac{n}{m+n}.
\end{equation}
Note that, for the toughness dominated regime of the KGD model, the elasticity relation \eqref{inv_norm_KGD_2} (and asymptotic estimate \eqref{K_1_norm}) imply that:
\begin{equation}\label{w_0_KI}
w_0(t)=\frac{4}{\sqrt{\pi}}K_I\sqrt{L(t)}.
\end{equation}
This, together with relation \eqref{LC_KGD_toughness}, suggests that in the two limiting cases: small toughness ($K_I\to0$) and large toughness ($K_I\to\infty$), the coefficients $w_0$ and $w_1$ should assume extremely different values, i.e. when one of them tends to zero the other magnifies.

Combining \eqref{v_0_univ} and \eqref{v_0} with \eqref{w_0_KI},
after some algebra and integration, gives the following relation, which is valid for any real $\kappa$:
\begin{equation}\label{v_0_univ_2}
L(t)=\left(L^\zeta(0)+\zeta C_{\cal A}^\frac{1}{n} \left(\frac{\pi}{16K^2_I}\right)^\kappa \int_0^tw_0^{\frac{n+1}{n}+2\kappa}(\tau)w_1^\frac{1}{n}(\tau)d\tau\right)^\frac{1}{\zeta},
\end{equation}
where $\zeta=\kappa+2/n+1$. For $\kappa=0$, the latter coincides with \eqref{L_int} (here $m=2$). Taking $\kappa=-(n+1)/(2n)$ one can eliminate $w_0$
from the formula to have
\begin{equation}
\label{L_tough}
L(t)=\left[L^{\frac{3+n}{2n}}(0)+\frac{3+n}{2n}C_{\cal{A}}^{1/n}\left(\frac{4K_I}{\sqrt{\pi}}\right)^{\frac{n+1}{n}}\int_0^t w_1^{1/n}(\tau)d\tau\right]^\frac{2n}{3+n}.
\end{equation}
Such a formulation is conducive to computing the so-called small toughness case.
However, bearing in mind that the value of the functional ${\cal L}(w)$ is directly evaluated during the computations, relation \eqref{L_int} is probably the best choice, and will consequently be used in our algorithm.

\subsection{Reformulation of the problem in terms of reduced particle velocity}

Following \cite{wr_mis_2015}, let us introduce a new dependent variable called the {\it reduced particle velocity} ${\phi}(t,x)$:
\begin{equation}\label{phi}
{\phi}(t,x)=v(t,x)-xv_0(t).
\end{equation}
The advantages of its utilization have been shown in the recalled paper. The asymptotics of $\phi$ can be described qualitatively as (compare Table \ref{T1}):
\begin{equation}
\label{fi_asym}
\phi(t,x)=\phi_0(t)(1-x^m)^{\zeta_0}+\phi_1(t)(1-x^m)^{\zeta_1}+O\Big((1-x^m)^{\zeta_1+\delta}\Big),\quad x \to 1.
\end{equation}

In numerical implementation $\phi$ possesses all the advantages of the particle velocity function. Moreover, an unknown value of the crack propagation speed, $v_0$, now can be used as an additional
regularization parameter in the continuity equation.

By combining \eqref{phi} with \eqref{v_norm} and substituting the result into the continuity equation \eqref{cont_norm} one obtains a modified governing PDE:
\begin{equation}
\label{phi_cont}
\frac{\partial w}{\partial t}+\frac{1}{L}\frac{\partial }{\partial x}(w \phi)+\frac{L'}{L}w+q_l=0.
\end{equation}
The boundary condition \eqref{bc0_norm}$_1$ can now be replaced by:
\begin{equation}\label{q0_phi}
w(t,0)\phi(t,0)=q_0(t),
\end{equation}
while the tip condition \eqref{bc0_norm}$_2$ and the speed equation \eqref{v_0_univ} yield:
\begin{equation}\label{phi_tip}
w(t,1)=0,\quad \phi(t,1)=0.
\end{equation}
Note that the following solvability condition for equation \eqref{phi_cont} should be satisfied:
\begin{equation}\label{phi_solv}
\int_0^1\left(\frac{\partial w}{\partial t}(t,x)+q_l(t,x)\right)dx+\frac{v_0(t)}{L(t)}\int_0^1w(t,x)dx=\frac{1}{L(t)}q_0(t),
\end{equation}
and, since $w>0$ for any $x\in[0,1)$, it allows one to uniquely define the value of the crack velocity, $v_0$, at every time step.
As a result, equation \eqref{phi_cont} always has a unique solution with respect to the reduced particle velocity $\phi$.

Following \eqref{v_norm}, expressing the pressure derivative in terms of the reduced velocity function gives:
\begin{equation}\label{p_prim}
\frac{\partial p}{\partial x}=-\frac{L}{w^{n+1}}\left(\phi+xv_0\right)^n.
\end{equation}
In this way, the basic system of equations is now built on two dependent variables: the crack opening, $w(t,x)$, and the reduced particle velocity, $\phi(t,x)$, with an additional relation for the pressure derivative \eqref{p_prim}. However, the latter will later be combined with the respective transformed form of the elasticity relation \eqref{inv_norm_PKN} -- \eqref{inv_norm_KGD} to eliminate $p'_x$. We will be looking for a solution of the lubrication equation \eqref{phi_cont}, under boundary conditions \eqref{q0_phi}, \eqref{phi_tip} and \eqref{dw_0} for the KGD model, initial conditions together with the speed equation \eqref{v_0} and resulting expressions for the crack length. Additionally, the fluid balance condition \eqref{phi_solv} and the respective elasticity relation, \eqref{elas_norm_PKN} or \eqref{elas_norm_KGD}, together with their inverses, \eqref{inv_norm_PKN} or \eqref{inv_norm_KGD}, are to be satisfied.

\section{Self-similar formulation of the problem}
\label{sec:self_sim_formulation}

Let us assume that a solution to the problem described in the previous section can be expressed in the following manner:
\begin{equation}\label{self_sim}
w(t,x)=\psi(t) \hat w(x),\quad p(t,x)=\frac{\psi(t)}{L^{m-1}(t)}\hat p(x),
\end{equation}
\[
q(t,x)=\frac{\psi^{2+\frac{2}{n}}(t)}{L^{\frac{m}{n}}}\hat q(x),  \quad
v(t,x)=\frac{\psi^{1+\frac{2}{n}}(t)}{L^{\frac{m}{n}}}\hat v(x),  \quad
\phi(t,x)=\frac{\psi^{1+\frac{2}{n}}(t)}{L^{\frac{m}{n}}}\hat \phi(x),
\]
where $\psi(t)=\psi(t,\gamma)$ is a smooth continuous function of time. Such a separation of variables enables one to reduce the problem to the time-independent form in the case when $\psi(t)$ is a power law or exponential function of time. The spatial components of the solution will henceforth be marked by 'hat'-symbol. The resulting time-independent formulation is called the self-similar problem.

\subsection{Representation of the solution}

As a consequence of \eqref{self_sim}, the qualitative asymptotic behaviour of the pertinent spatial functions remains the same as their time-dependent counterparts (i.e. $\hat w(x)$ complies with \eqref{w_asym}, $\hat v$ with \eqref{v_asym} and $\hat \phi$ with \eqref{fi_asym}). The elasticity equation \eqref{elas_norm} is converted to:
\begin{equation}\label{p_ss}
\hat p(x)=\hat {\cal A} \hat w(x),
\end{equation}
where the self-similar elasticity operator yields respectively:
\begin{equation}\label{p_ss_2}
\hat {\cal A}_1 \hat w(x)=\hat p(x),\quad \hat {\cal A}_2 \hat w(x)=\frac{1}{2\pi}\int_0^1 \frac{d \hat w }{d \eta} \frac{\eta}{x^2-\eta^2}d\eta.
\end{equation}
The inversions of \eqref{p_ss_2} produce:
\begin{equation}\label{p_ss_inv_PKN}
\hat w(x)=\hat p(x),
\end{equation}
\begin{equation}\label{p_ss_inv_KGD}
\hat w=-\frac{4}{\pi}\int_0^1 \frac{d \hat p }{ds}K(s,x)ds+\frac{4}{\sqrt{\pi}} \hat K_I \sqrt{1-x^2},
\end{equation}
for the PKN and KGD models, respectively. In general, \eqref{p_ss_inv_KGD} holds provided that the normalized stress intensity factor is a function of time defined as:
\begin{equation}
\label{K_Is}
K_I(t)=\hat K_I \frac{\psi (t)}{\sqrt{L(t)}},
\end{equation}
where:
\begin{equation}\label{KI_ss_p}
\hat K_I=\frac{2}{\sqrt{\pi}}\int_0^1\frac{\hat p(s)}{\sqrt{1-s^2}}ds.
\end{equation}
However, as shall be specified later, there is a combination of the self-similar parameters which allows one to reduce the problem to the time-independent version for constant $K_I$.
The self-similar particle velocity is:
\begin{equation}
\label{v_ss}
\hat v(x)=\left[-\hat w^{n+1}(x) \frac{d \hat p}{dx} \right]^{1/n},
\end{equation}
with the crack propagation speed computed as:
\begin{equation}
\label{v0_sep}
\hat v_0=\Big(C_{\cal A}{\cal L}(\hat w)\Big)^{\frac{1}{n}}.
\end{equation}
The reduced particle velocity is expressed in the same manner as in \eqref{phi}:
\begin{equation}
\label{fi_sep}
\hat \phi(x)=\hat v(x)-x \hat v_0.
\end{equation}
The pressure derivative expressed through $\hat \phi$ yields:
\begin{equation}\label{p_prim_ss}
\frac{d \hat p}{d x}=-\frac{1}{\hat w^{n+1}}\left(\hat \phi+x \hat v_0\right)^n.
\end{equation}
Under these assumptions equation \eqref{phi_cont} can be transformed into the form:
\begin{equation}\label{ODE_gen}
\frac{1}{\hat v_0} \frac{d}{dx}(\hat w \hat \phi)=-\rho_1 \hat w- \rho_2 \hat q_l,
\end{equation}
where $\rho_1$ and $\rho_2$ for two cases of $\psi(t)$ are given in Table \ref{T22}, and:
\begin{equation}
q_l(t,x)=\frac{1}{\gamma}\psi'(t)\hat q_l(x).
\end{equation}
The system has to be equipped with the following boundary conditions
\begin{equation}\label{exp_ss_BC}
\hat w(1)=0,\quad \hat \phi(1)=0,\quad \hat w(0) \hat \phi(0)=\hat q_0,
\end{equation}
with an additional one for the KGD model:
\begin{equation}\label{sym_ss}
\frac{d \hat w}{dx}\big | _{x=0}=0.
\end{equation}
The fluid balance condition in its self-similar form yields:
\begin{equation}\label{balance_ss}
\frac{1}{\hat v_0}\hat q_0-\rho_1\int_0^1 \hat w dx-\rho_2\int_0^1 \hat q_l dx=0.
\end{equation}

\begin{table}[h]
\begin{center}
\renewcommand{\arraystretch}{2}

\begin{tabular}{|c|c|c|c|}
  \hline
    Type of the self-similar law & $\rho_1$ & $\rho_2$ & $L(t)$\\
  \hline\hline
  $\psi(t)=e^{\gamma t}$ &$\frac{m+n}{2+n}+1$ & $\frac{m+n}{(2+n)\gamma}$ &  $\left(\frac{(m+n)\hat v_0}{(n+2)\gamma}\right)^\frac{n}{m+n} e^\frac{(n+2)\gamma t}{m+n}$\\
  \hline
  $\psi(t)=(a+t)^{\gamma}$  & $ \frac{(m+n)\gamma}{n+\gamma(2+n)}+1 $& $ \frac{m+n}{n+\gamma(n+2)}$ & $\left(\frac{(m+n)\hat v_0}{n+\gamma(2+n)}\right)^\frac{n}{m+n}(a+t)^\frac{n+\gamma(2+n)}{m+n}$\\

  \hline
\end{tabular}
\end{center}
\caption{Values of the auxiliary parameters, $\rho_1$, $\rho_2$ and the crack length, $L(t)$, in the self-similar formulations reflecting different time-dependent behaviour. Here, $0\le n \le 1$, while $m$ and $v_0$ are defined in Table \ref{T1} and \eqref{v0_sep}, respectively.}
\label{T22}
\end{table}

The foregoing reduction of the problem to the self-similar version is possible for two types of the time-dependent functions:
\begin{equation}
\label{psi_def}
\psi (t)=e^{\gamma t}, \quad \psi (t)=(a+t)^\gamma,
\end{equation}
where $a$ and $\gamma$ are some non-negative constants. Values of the respective coefficients in ODE \eqref{ODE_gen}, together with the pertinent expressions for the crack length for both functions $\psi$, are collected in Table \ref{T22}. Note that the reformulation of the problem to the self-similar form for constant non-zero $K_I$ is only feasible for \eqref{psi_def}$_2$ with $\gamma=n/(2+n)$.

\subsection{Description of the algorithm for self-similar solution}

The proposed unified algorithm, based on the reduced particle velocity and the speed equation, has a common feature for the case $n>0$. Namely, equation \eqref{v0_sep} (and its time-dependent equivalent \eqref{v_0_univ}) can be directly used to compute the particle velocity. This is no longer true for $n=0$, as the equation degenerates. Thus the case with $n=0$ is a special one and the respective algorithm is built separately, although it remains very similar to the basic version. The other limiting case, $n=1$, has been completely analysed in \cite{wr_mis_2015} using the same approach as adopted in this paper. In fact the only difference, in comparison with the general case ($0<n<1$), is that the pressure in the toughness dominated regime of the KGD model behaves differently near the crack tip for Newtonian fluid (logarithmic singularity). However, as can be easily seen from \eqref{LC_KGD_toughness}, the speed equation is still valid and has the same form based on the coefficients of the asymptotics for the crack opening. Taking all this into account, we built a universal algorithm to compute the solution for values of the fluid behaviour index in the range $0.01<n<0.99$. We checked that the difference between the limiting cases ($n=0$ and $n=1$) and the end of the aforementioned $n$-interval differs negligibly (that is between $n=0.01$ and $n=0$, as well as $n=0.99$ and $n=1$). Nevertheless, while presenting graphical results for the whole domain $0\le n \le 1$, using our numerical and semi-analytical examples, we incorporate the accumulated knowledge about the limiting cases as well. All computations were carried out using MATLAB environment on conventional laptop.

\subsubsection{Viscous fluid ($0<n<1$)}

Below we collect a basic set of equations, used later to formulate the universal algorithm for numerically simulating  hydraulic fractures. The numerical scheme utilizes:
\begin{itemize}
\item the equation defining the reduced velocity following from \eqref{ODE_gen}:
\begin{equation}\label{ss_1}
\hat \phi(x)=\frac{\hat v_0}{\hat w(x)}\int_x^1\Big(\rho_1 \hat w(\xi)+ \rho_2 \hat q_l(\xi)\Big)d\xi,
\end{equation}
where $\hat v_0$ is given in \eqref{v0_sep},
 \item  the equation allowing computation of the crack opening, which follows from \eqref{p_prim_ss} and the respective inverse
elasticity operator \eqref{p_ss_inv_PKN}, \eqref{p_ss_inv_KGD}, written symbolically as:
\begin{equation}\label{ss_2}
\hat w (x)={\cal B}\left[\left(\hat\phi(x)+x \hat v_0 \right)^n\right],
\end{equation}
\item the boundary conditions \eqref{exp_ss_BC}, and additionally \eqref{sym_ss} for the KGD model,

\item the solvability condition \eqref{balance_ss} which follows immediately from \eqref{ss_1} and \eqref{exp_ss_BC}$_3$.
\end{itemize}

For the PKN model operator ${\cal B}$ in \eqref{ss_2} assumes one of the following alternative forms:
\begin{equation}\label{p_prim_1aaa}
{\cal B}[\nu(x)]=\int_x^1\frac{\nu(\eta)}{\hat w^{n+1}(\eta)}d\eta,
\end{equation}
or
\begin{equation}\label{w_B1_PKNaa}
{\cal B}[\nu(x)]=\left[(n+2)\int_x^1 \nu(\eta) d\eta\right]^{1/(n+2)}.
\end{equation}
For the KGD variant the operator is directly defined by relation \eqref{p_ss_inv_KGD} which is a self-similar equivalent of \eqref{inv_norm_KGD_2}:
\begin{equation}\label{w_to_findaaa}
{\cal B}[\nu(x)]= \frac{4}{\pi}\int_0^1 \frac{\nu(\eta)}{\hat w^{n+1}(\eta)}
K(\eta,x) d\eta+\frac{4}{\sqrt{\pi}}\hat K_I\sqrt{1- x^2},
\end{equation}
where the kernel $K(\eta,x)$ is defined in \eqref{K} and $\hat K_I=0$ for the viscosity dominated regime.

Note that the general form of the speed equation \eqref{v_0} and the additional boundary condition \eqref{dw_0} for the KGD model (\eqref{sym_ss} in self-similar formulation) are satisfied by the system \eqref{ss_1} -- \eqref{w_to_findaaa} automatically.

\begin{remark}\label{3PD}
In our computations for the PKN model we will always use relation \eqref{p_prim_1aaa} for the following two reasons. First of all, when dealing with the P3D model (see \cite{linkov_mis_2013,mack_warp_2000}), this is the only form of the equation which is available. But more importantly numerical tests revealed that, although formula  \eqref{w_B1_PKNaa} produces more accurate results than \eqref{p_prim_1aaa} in the self-similar formulation, no such advantage exists in the transient regime. In this case, the error introduced by the temporal derivative approximation is appreciably higher than that resulting from the computation of the operator \eqref{ss_2}, regardless of its form. This feature of the algorithm was also identified in \cite{wr_mis_2015}. It is noteworthy that due to this computational strategy, the accuracy estimation given in Section \ref{sec:ss_alg} always describes the upper bound of the error, i.e. the accuracy can be improved by using relation \eqref{w_B1_PKNaa}.
\end{remark}

\subsubsection{Special case: perfectly plastic fluid ($n=0$)}
\label{n_0_section}

Note that for the perfectly plastic fluid ($n=0$) the Poiseulle equation degenerates and formula \eqref{v_ss} becomes:
\begin{equation}\label{n0_special_case}
-\hat w \frac{d\hat p}{d x}=1,
\end{equation}
which does not allow us to compute the particle velocity. Thus, this case requires special treatment.

In fact, for the PKN model, one can obtain an analytical solution to the problem provided that the leak-off function is predefined and independent of the solution itself.
Indeed, combining \eqref{n0_special_case} and \eqref{p_ss_inv_PKN} one has:
\begin{equation}\label{w_PKN_n0_ss}
\hat w(x)=\sqrt{2} (1-x)^{\frac{1}{2}},
\end{equation}
which, after substitution into the continuity equation \eqref{ODE_gen}, allows us to derive a formula for the reduced particle velocity:
\begin{equation}\label{v_PKN_n0_ss}
\hat \phi(x)=\hat{v}_0\left[1-x+\frac{ \int_x^1 \hat{q_l} ds}{2\sqrt{2}\gamma (1-x)^{\frac{1}{2}}}\right].
\end{equation}
In the case of the KGD model, relation \eqref{p_ss_inv_KGD} transforms to a nonlinear integral equation:
\begin{equation}\label{w_operator_n0}
\hat w(x)= \frac{4}{\pi}\int_0^1 \frac{1}{\hat w(\eta)}
K(\eta,x) d\eta+\frac{4}{\sqrt{\pi}}\hat K_I\sqrt{1- x^2},
\end{equation}
while the reduced particle velocity is computed from \eqref{ss_1}.

Finally, the self-similar crack propagation speed for both the PKN and KGD models has to be computed from the balance equation \eqref{balance_ss}:
 \begin{equation}\label{v0_n0_ss}
\hat v_0=\frac{\hat{q_0}}{\rho_1\int_0^1\hat{w}dx+\rho_2\int_0^1\hat{q_l}dx}.
\end{equation}

\subsection{Numerical realization of the algorithm for the self-similar solution}
\label{sec:num_alg_self_sim}

The solution to the self-similar problem formulated above is sought in the framework of the {\it universal algorithm}, first introduced in \cite{wr_mis_2015}. Even though the algorithm was designed to work for Newtonian fluids, in this paper we prove that it is efficient in the case of non-Newtonian fluids as well.

The adaptation of the numerical scheme addresses new computational challenges related to the rheological properties of fluid: i) the order of non-linearity of Poiseulle equation \eqref{poiseuille}, ii)
the qualitative behaviour of the crack-tip asymptotics dependent on the fluid behaviour index $n$, iii) the degeneration of the Poiseulle equation for perfectly plastic fluid ($n=0$).

The algorithm consists of the following iterative steps:
\begin{itemize}
\item
In the first stage, some initial approximation of $\hat w=\hat w^{(j-1)}$ is taken\footnote{ Here the superscript $j-1$ ($j=1$ at the first step) refers to the number of the consequent iteration.} in a form satisfying the main constraints for the crack opening (boundary conditions and asymptotic behaviour). Equations \eqref{balance_ss} and \eqref{v0_sep} are utilized to compute $\hat v_0^{(j)}$, which is then substituted into \eqref{ss_1}, enabling one to integrate the latter and obtain the reduced particle velocity $\hat \phi^{(i)}$.

As the governing ODE \eqref{ODE_gen} degenerates at the crack tip, we use the so-called $\varepsilon$-regularization technique (see e.g. \cite{kus_mis_wr_2013,linkov_2011} for details), which means that the integration is carried out over the truncated spatial interval $x\in[0,1-\varepsilon]$, where $\varepsilon$ is a small parameter. The boundary condition \eqref{exp_ss_BC}$_2$ is replaced by an approximate one resulting from the asymptotics \eqref{fi_asym}, specified at $x=1-\varepsilon$. The regularized boundary condition is introduced in the form:
\begin{equation}\label{fi_reg}
\hat \phi_N=s_1(\zeta_0,\zeta_1)\hat \phi_{N-1}+s_2(\zeta_0,\zeta_1)\hat \phi_{N-2},
\end{equation}
where subscripts of $\hat \phi$ refer to the indices of nodal points of the spatial mesh (containing $N$ nodes). The values of multipliers $s_{1(2)}(\zeta_0,\zeta_1)$ depend on the particular asymptotic behaviour of function $\hat \phi$ (see \eqref{fi_asym} and Table \ref{T1}).

Here $\zeta_0$ and $\zeta_1$ are known exponents of the first two asymptotic terms of the function $\hat \phi$, and as such do not change during the iteration process. As a result, functions $\hat \phi^{(j)}$ and $\hat v_0^{(j)}$ computed at this stage satisfy, together with predefined $\hat w^{(j-1)}$, the: i) fluid balance equation \eqref{balance_ss}, ii) continuity equation \eqref{ODE_gen}, iii) regularized boundary condition for $\hat \phi$ \eqref{fi_reg} (equivalent to \eqref{exp_ss_BC}$_2$), iv) influx boundary condition \eqref{exp_ss_BC}$_3$ -- indirectly, through the fluid balance equation.

\item
At the second stage of each iterative loop, the values of $\hat \phi^{(j)}$ and $\hat v_0^{(j)}$ obtained in the previous step are utilized to compute the next approximation of $\hat w^{(j)}$ using \eqref{p_prim_1aaa} -- \eqref{w_to_findaaa}.

While computing the respective integrals in \eqref{p_prim_1aaa} -- \eqref{w_to_findaaa}, it is crucial to preserve the appropriate asymptotic behaviour of the integrands, resulting from \eqref{w_asym} and \eqref{fi_asym}. Hence, $\hat w^{(j)}$ computed at this stage satisfies the respective elasticity relation and boundary condition \eqref{exp_ss_BC}$_1$ through the regularized boundary condition:
\begin{equation}\label{w_reg}
\hat w_N=s_1(\alpha_0,\alpha_1)\hat w_{N-1}+s_2(\alpha_0,\alpha_1)\hat w_{N-2}.
\end{equation}
\item
The aforementioned two stages of the iterative loop are repeated until all components of the solution $\hat v_0$, $\hat \phi$ and $\hat w$ have converged to within prescribed tolerances.
\end{itemize}

\begin{remark}\label{delta_computations}
Similarly to that found in \cite{wr_mis_2013}, the performance of the algorithm improves significantly when, instead of the dependent variables $\hat w$ and $\hat \phi$, one uses the difference between them and their leading asymptotic terms:
\begin{equation}
\Delta \hat w=\hat w-\hat w_0(1-x^m)^{\alpha_0},\quad \Delta \hat \phi=\hat \phi -\hat \phi_0(1-x^m)^{\zeta_0}.
\end{equation}
Then the leading terms in the asymptotics of the left and right-hand sides of equations \eqref{ss_1} and \eqref{p_prim_1aaa} -- \eqref{w_to_findaaa} are canceled analytically and only the values of $\Delta \hat w$ and $\Delta \hat \phi$ are iterated. Moreover, while searching for the regularization parameter $\hat v_0$, we take into account its relationship with the respective coefficients in the solutions asymptotic expansion. This in turn produces a nonlinear equation which is solved using the Newton - Raphson method. Finally, the qualitative asymptotic behaviour of the new dependent variables $\Delta \hat w$ and $\Delta \hat \phi$ is known in advance, and the respective exponents should be appropriately adopted in the tip asymptotics. Interestingly, sensitivity analysis conducted by the authors proved that, even if the employed exponent values are inaccurate, the algorithm remains efficient and stable, however the accuracy level deteriorates by one or two orders of magnitude.
\end{remark}

\begin{remark}\label{n0_computations}
In the case of a perfectly plastic fluid ($n=0$) a modified computational strategy has to be enforced. For the PKN model it is sufficient to directly employ relations \eqref{w_PKN_n0_ss} -- \eqref{v_PKN_n0_ss}, where integration of the leak-off function is carried out numerically. For the KGD model, the crack propagation speed $\hat v_0$ is computed from \eqref{v0_n0_ss}, while the crack opening calculation utilizes \eqref{w_operator_n0} with $\hat w^{(j-1)}$ on the right hand side.
\end{remark}

\subsection{Analysis of the algorithm performance for the self-similar formulation}
\label{sec:ss_alg}

Let us now analyze the performance of the algorithm presented above. In the following three subsections we investigate the accuracy of computations carried out using the proposed numerical scheme. To this end a set of analytical benchmark solutions described in Appendix A is employed.

\subsubsection{Analysis of the algorithm - PKN model}
\label{sec:PKN_ss_1}

For the PKN model, we use a benchmark solution in the form \eqref{w_rep} for three base functions of the type \eqref{h_PKN}. All resulting quantities can be obtained using the methods described in Appendix A. Note that this benchmark solution assumes a predefined non-zero leak-off function. All results presented in this subsection were computed for $\gamma=\frac{1}{3+2n}$, which, in the applied scaling, corresponds to a constant injection flux rate.

In this subsection we use various error measures in order to demonstrate the algorithms peculiarities and to define an optimal set of accuracy parameters, used later on to investigate the performance of the proposed numerical scheme. We will consider both the average and maximal relative errors of the following: the crack opening, $\delta \hat w$, the particle velocity, $\delta \hat v$, and the
reduced particle velocity, $\delta \hat \phi$.
The relative errors over $x$ are defined by the following standard form:
\begin{equation}
\label{del_av}
\delta \varpi_{av}=\sqrt{\int_0^1(\varpi-\varpi_N)^2dx\cdot\left(\int_0^1\varpi^2 dx\right)^{-1}},\quad \delta \varpi_{max}=\max_{0\leq x<1}\frac{|\varpi(x)-\varpi_N(x)|}{|\varpi(x)|},
\end{equation}
where $\varpi$ stands for the benchmark solution, while $\varpi_N$ refers to the respective numerical result.

\begin{figure}[h!]
		\hspace{-6mm}
		\includegraphics [scale=0.33]{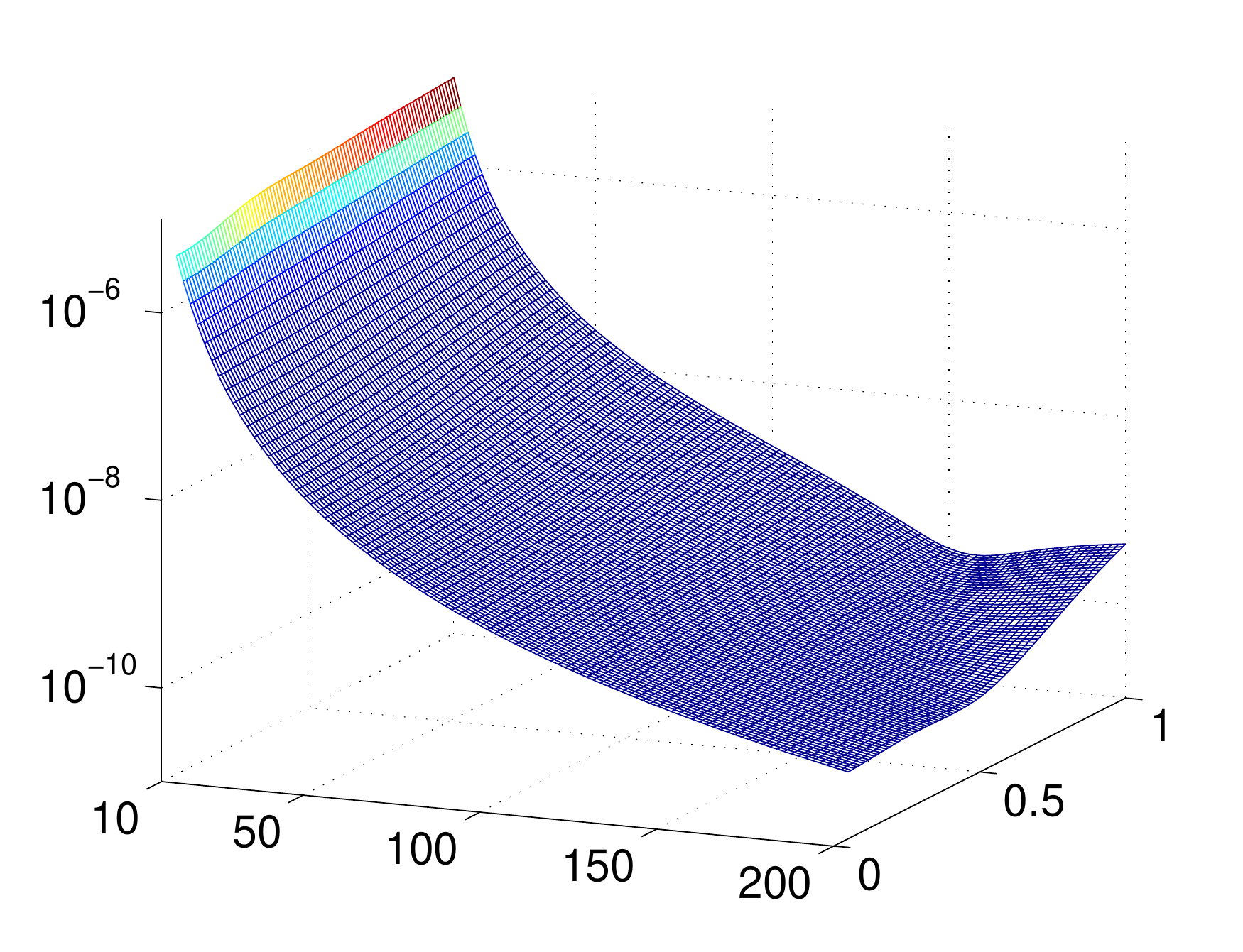}
		\put(-115,0){$N$}
		\put(-33,5){$n$}
    \put(-50,100){$\delta \hat w_{av}$}
    \hspace{-6mm}
    \includegraphics [scale=0.33]{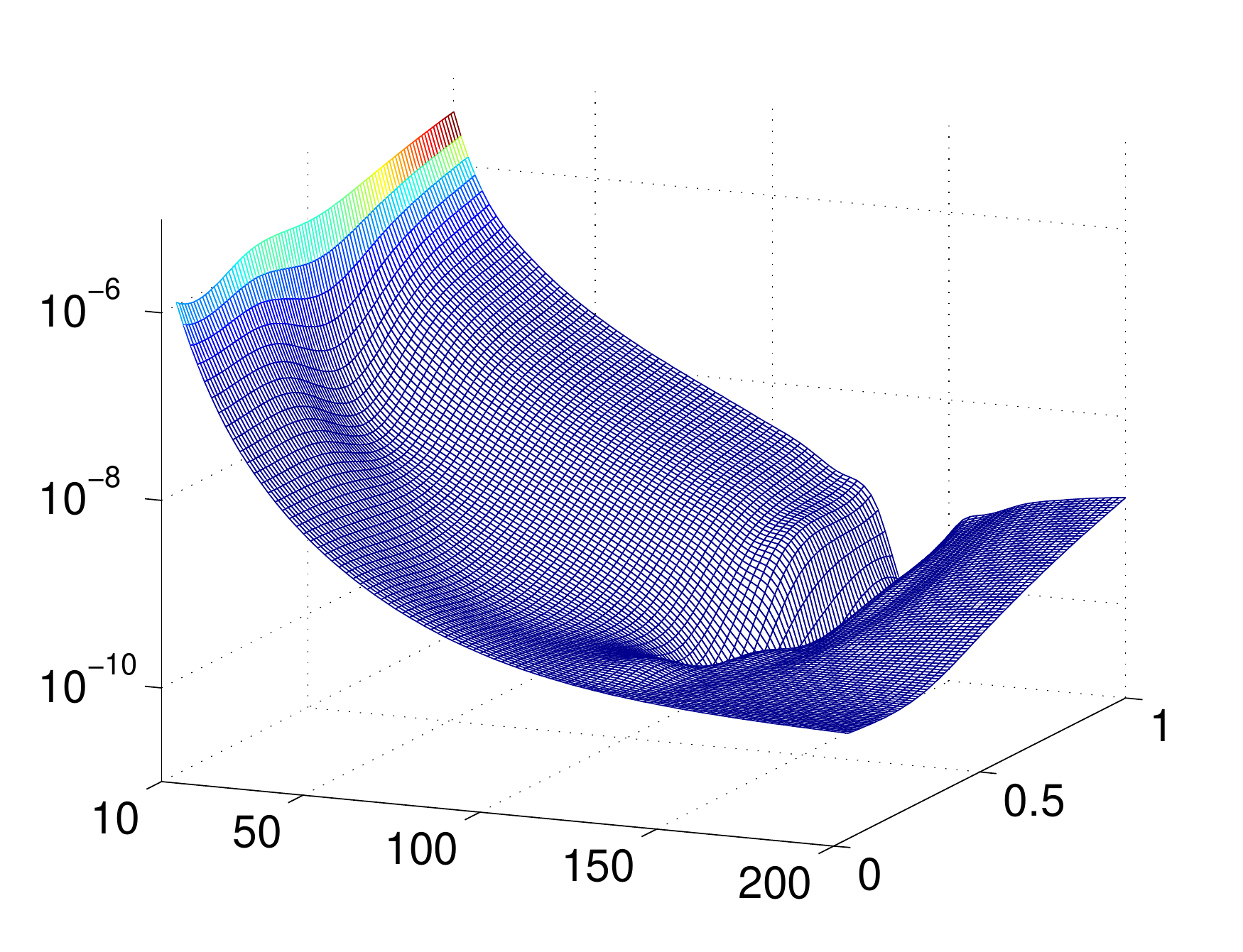}
		\put(-115,0){$N$}
		\put(-33,5){$n$}
    \put(-50,100){$\delta \hat v_{av}$}
		\hspace{-6mm}
    \includegraphics [scale=0.33]{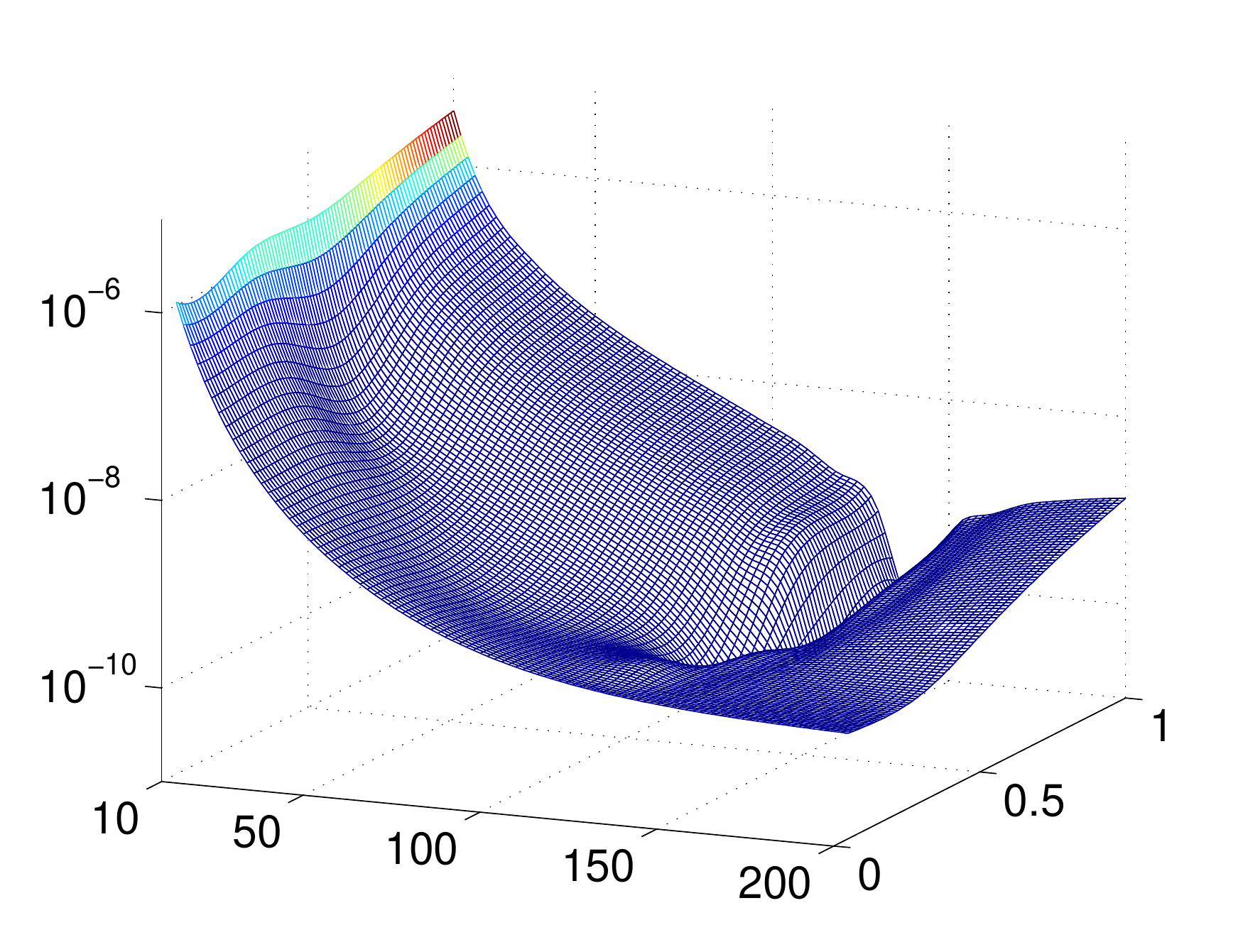}
  	\put(-115,0){$N$}
		\put(-33,5){$n$}
    \put(-50,100){$\delta \hat \phi_{av}$}
    \caption{PKN model -- average relative error in the self-similar solution for different numbers of nodal points, $N$, and values of the fluid behaviour index, $n$: a) the crack opening $\hat w$, b) the particle velocity $\hat v$, c) the reduced particle velocity $\hat \phi$. }

\label{av_rel_PKN}
\end{figure}

In Fig. \ref{av_rel_PKN} the average error of the basic variables, as functions of the mesh density, $N$, and the fluid behaviour index, $n$, are depicted. The errors were estimated for a number of nodal points varying from 10 to 200, where the spatial mesh density was increased at both ends of the interval, in a manner described in \cite{wr_mis_2013}. The smallest computed value of $n$ was moved away from zero (to 0.01), as for the PKN model the solution for the perfectly plastic fluid is found in the framework of a simplified scheme -- see Remark \ref{n0_computations} (it can, in fact, be derived analytically for the considered benchmark).

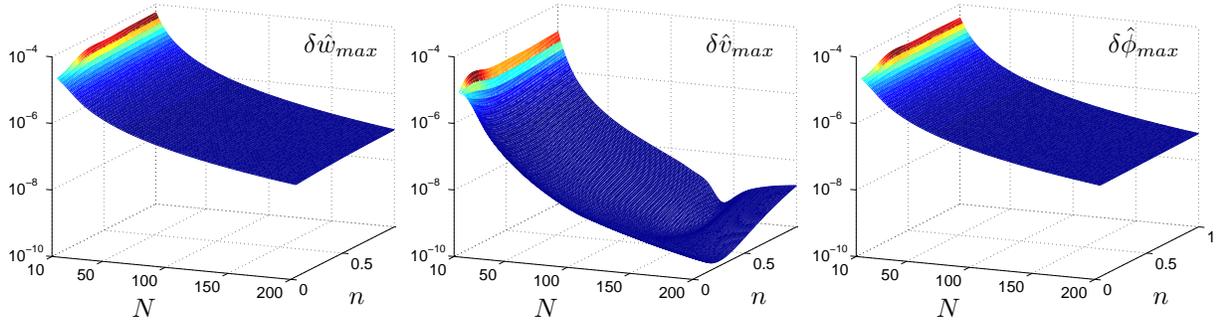
\begin{figure}[h!]
		\hspace{-6mm}
		\includegraphics [scale=0.33]{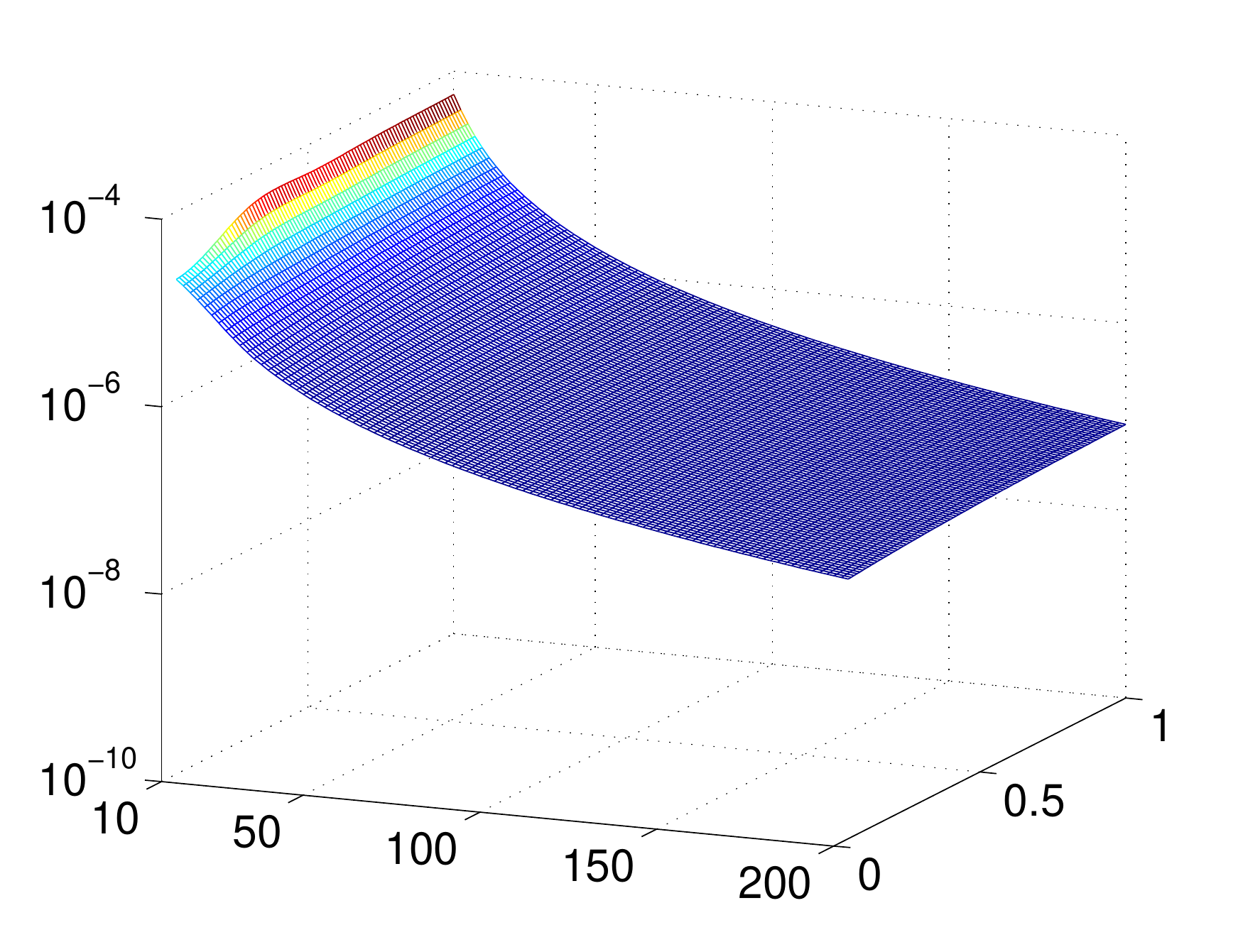}
		\put(-115,0){$N$}
		\put(-33,5){$n$}
    \put(-50,100){$\delta \hat w_{max}$}
		\hspace{-6mm}
    \includegraphics [scale=0.33]{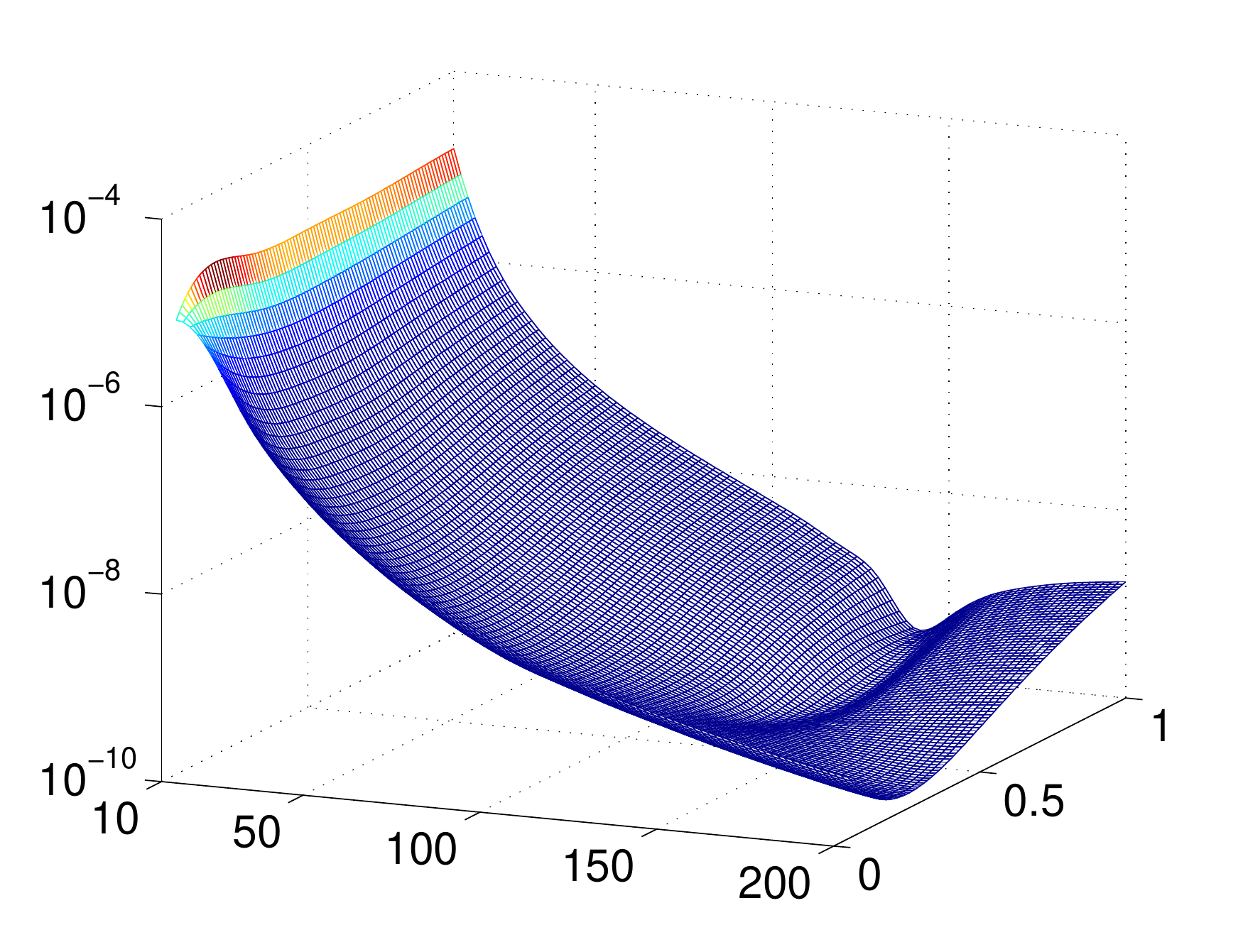}
		\put(-115,0){$N$}
		\put(-33,5){$n$}
    \put(-50,100){$\delta \hat v_{max}$}
		\hspace{-6mm}
    \includegraphics [scale=0.33]{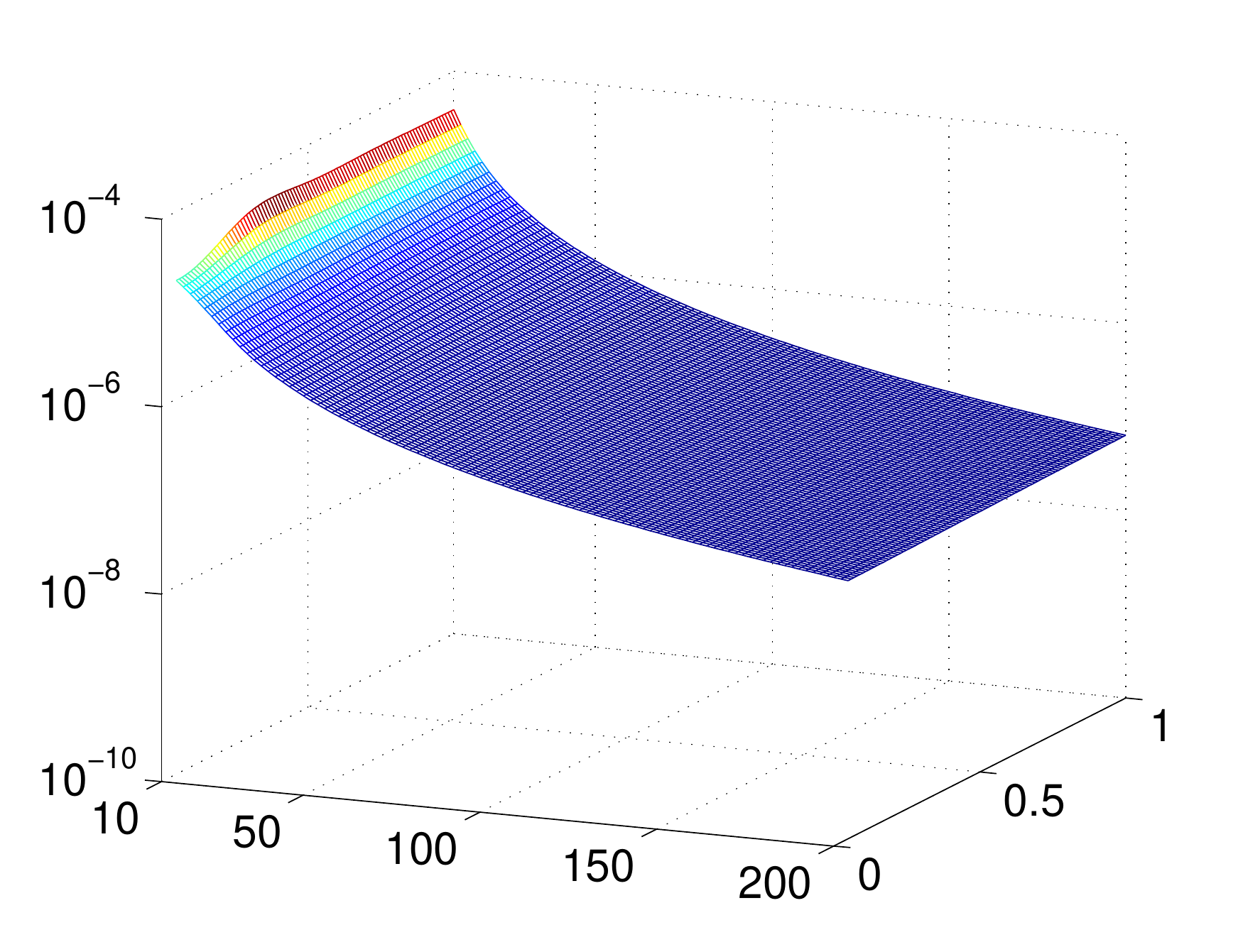}
		\put(-115,0){$N$}
		\put(-33,5){$n$}
		\put(-50,100){$\delta \hat \phi_{max}$}
    \caption{PKN model -- maximal relative error in the self-similar solution for different numbers of nodal points, $N$, and values of the fluid behaviour index, $n$: a) the crack opening $\hat w$, b) the particle velocity $\hat v$, c) the reduced particle velocity $\hat \phi$. }

\label{max_rel_PKN}
\end{figure}

From Fig. \ref{av_rel_PKN} it follows that the average computational error is hardly sensitive to the value of fluid behaviour index, $n$. Indeed, it is only for the largest number of nodal points that a discernible difference in accuracy for different $n$ can be observed. Simultaneously, for any fixed $n$, a constant improvement in the solution accuracy when increasing the number of nodal points $N$ can be observed over almost the whole range of $N$. Some irregularities (local minima) in the error distribution can be observed for the particle velocity and the reduced particle velocity for $N$ around 150. For a fluid behaviour index greater than approximately 0.5 when taking $N>150$ the error is no longer reduced. No noticeable difference between the error distributions for $\hat v$ and $\hat \phi$ is observed.

In Fig. \ref{max_rel_PKN} we depict the distributions of the maximal relative errors for $\hat w $, $\hat v$ and $\hat \phi$ as functions of $N$ and $n$. It shows that the general trends in accuracy (low sensitivity to the value of $n$, error reduction with growing $N$) are similar to those reported above. However, the level of the respective errors for the crack opening and the reduced particle velocity are up to two orders of magnitude higher than they were for the average error. This trend does not hold for the particle velocity, where the maximal error is of a similar order to that for the average formulation.

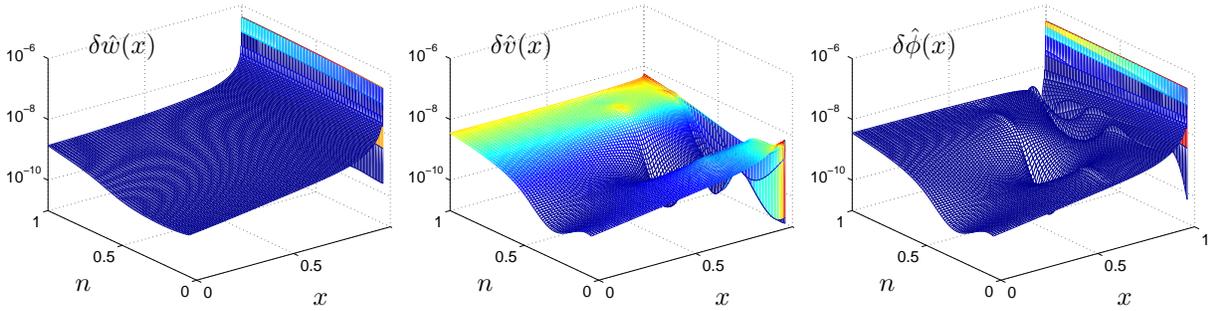
\begin{figure}[h!]
		\hspace{-6mm}
		\includegraphics [scale=0.33]{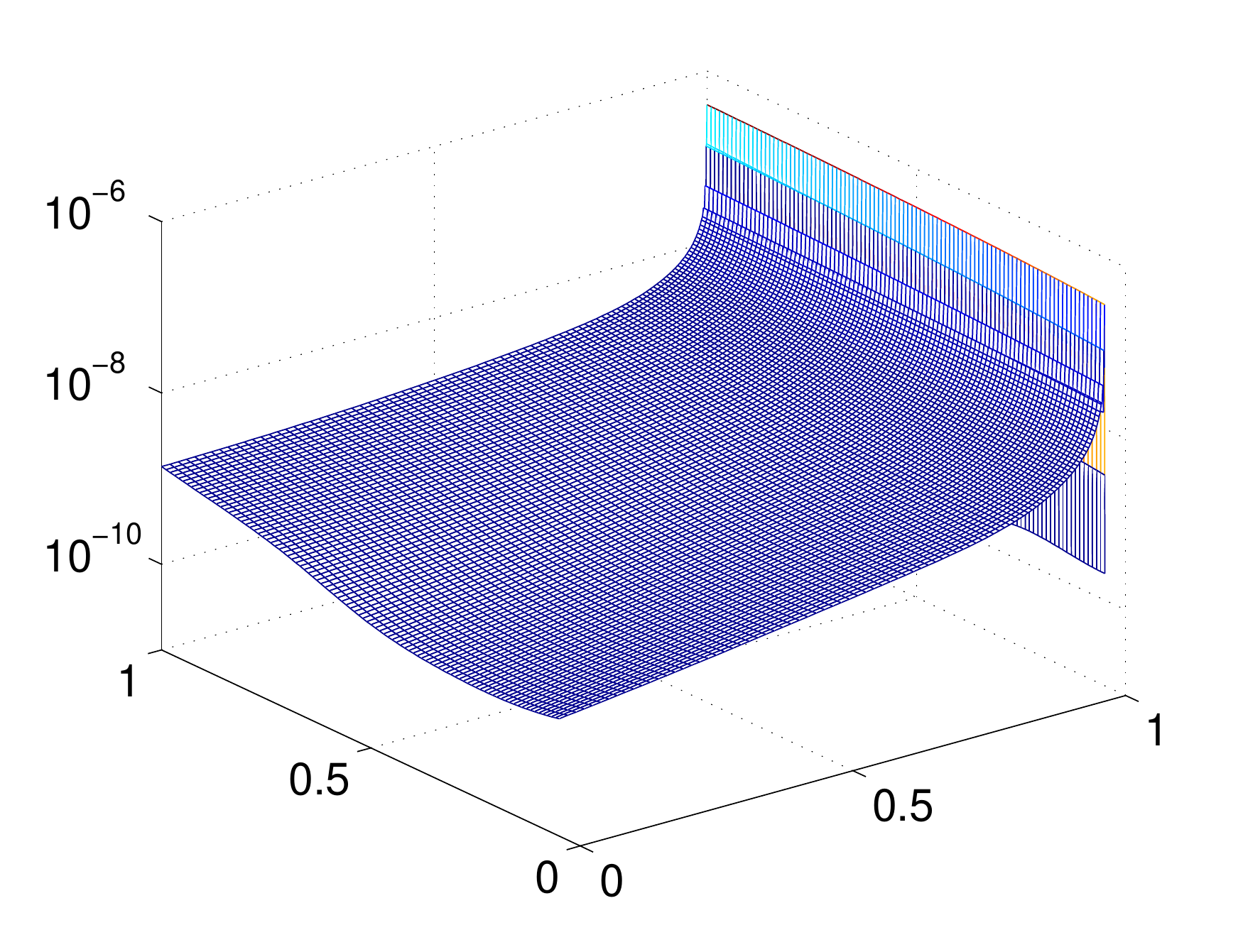}
		\put(-135,10){$n$}
		\put(-45,5){$x$}
\put(-130,100){$\delta \hat w(x)$}
		\hspace{-6mm}
    \includegraphics [scale=0.33]{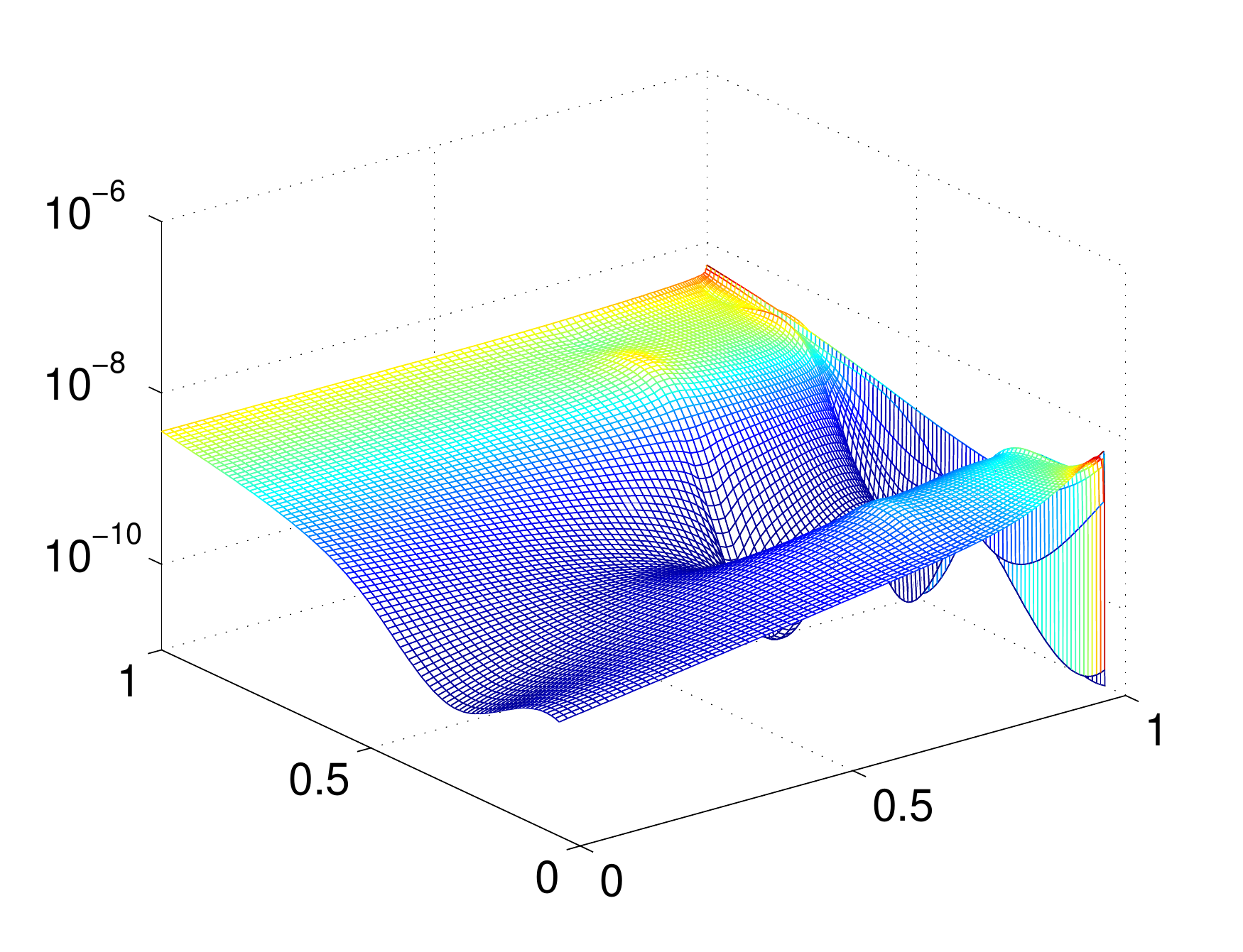}
		\put(-135,10){$n$}
		\put(-45,5){$x$}
\put(-130,100){$\delta \hat v(x)$}
		\hspace{-6mm}
    \includegraphics [scale=0.33]{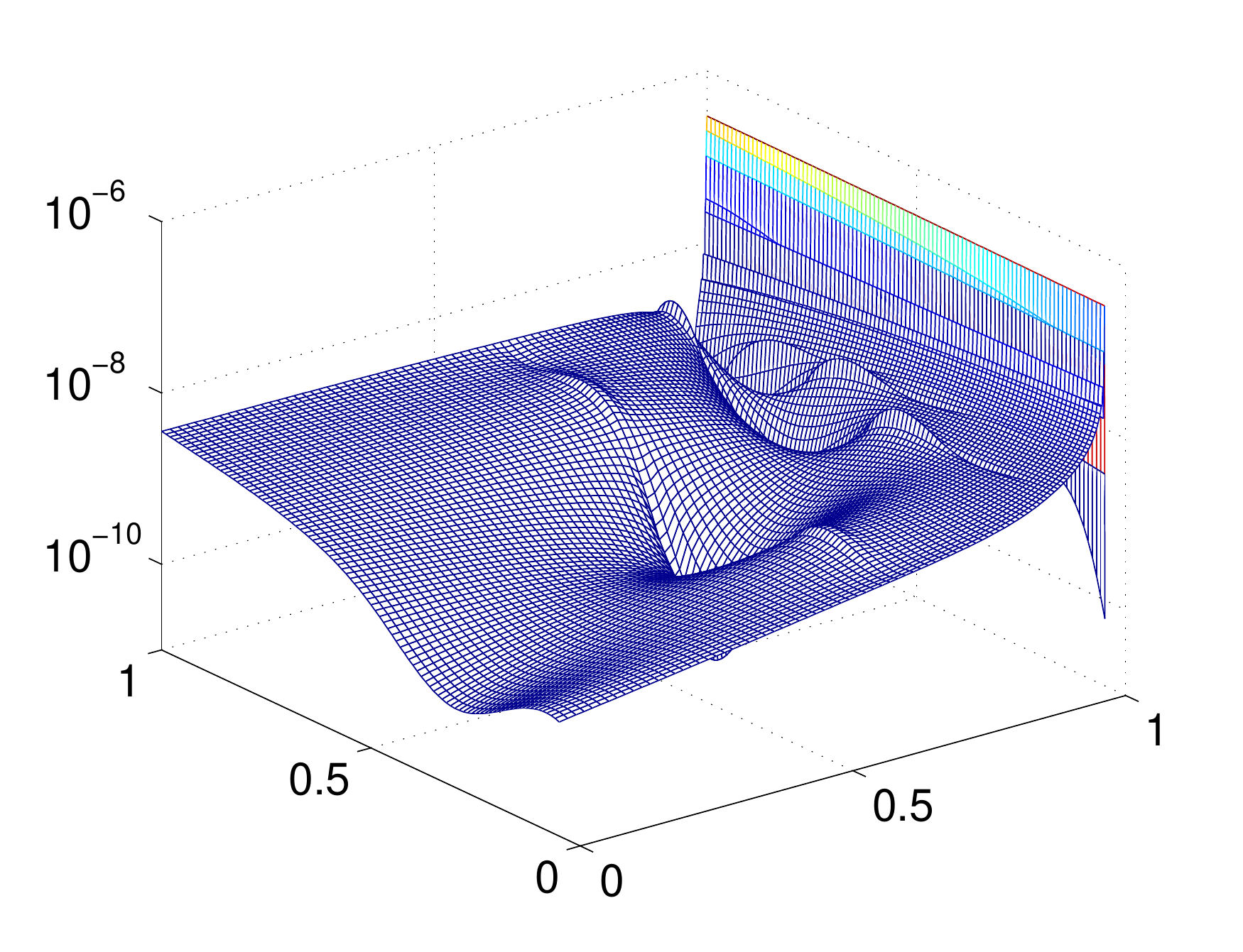}
		\put(-135,10){$n$}
		\put(-45,5){$x$}
		\put(-130,100){$\delta \hat \phi(x)$}
    \caption{PKN model -- distribution of the relative error for the self-similar solution: a) the crack opening $\hat w$, b) the particle velocity $\hat v$, c) the reduced particle velocity $\hat \phi$. The mesh is composed of $N=100$ nodal points. }

\label{rel_error_PKN}
\end{figure}

In order to explain the difference between the average and maximal errors let us analyze the error distribution over $x$ for a given number of nodal points, $N=100$. In Fig. \ref{rel_error_PKN} we present the relative errors of $\hat w$, $\hat \phi$ and $\hat v$ as functions of $x$ and $n$. The error distribution of the crack aperture and the reduced particle velocity is rather non-uniform, with a distinct increase in the near-tip region. In the case of the particle velocity (Fig. \ref{rel_error_PKN}b) one does not observe the accuracy deterioration at the crack tip, which is a result of a different asymptotic behaviour. Indeed, the spike growth of $\delta \hat w(x)$ and $\delta \hat \phi(x)$ is a direct consequence of the fact that both variables vanish at the crack tip (compare with \eqref{exp_ss_BC}$_1$ and \eqref{exp_ss_BC}$_2$) while the particle velocity tends to a non-zero value $\hat v(1)=\hat v_0\ne0$. Moreover, the graph of $\delta \hat v$ indicates the quality of computations for the crack propagation speed, $\hat v_0$, itself. The reduced particle velocity, $\hat \phi$, and particle velocity, $\hat v$, exhibit some tendency towards error reduction at the crack inlet for lower values of $n$, however this does not result in a decrease of the maximal error, which is similar for the whole range of $n$. Thus, the relative error magnification at the crack tip has little influence on the average error values, as the accuracy over most of the $x$ interval is up to \emph{three orders of magnitude better}.

As a consequence of the foregoing analysis, from now on we will present as graphical information only the following measures of the solution accuracy: i) the average relative error over $x$ of the crack opening and the particle velocity; ii) the relative error of the crack opening and the particle velocity for a fixed value of $N$. The first group of parameters is very useful when analyzing the general trends in accuracy for various $N$ and $n$. Moreover, it has been shown that the average error for $\hat v$ is practically the same as that for $\hat \phi$, which in light of the straightforward physical interpretation of the former makes it more convenient for use in accuracy analysis. The second group of parameters (point-wise relative errors) provide an illustration of the error distributions local peculiarities, while simultaneously depicting the solutions maximal errors.

Before we move on to the KGD models, we would like to note that the values of computational parameters such as the regularization parameter $\varepsilon=10^{-6}$ and mesh density parameters defining the distribution of points on the spatial interval $(0,1)$ were kept identical in all computations. It is clear from the results presented above that, by optimizing these parameters for each specific value of the fluid behaviour index, $n$, and number of nodal points, $N$, one can reduce the error further, and there is an apparent potential for greater accuracy improvement with $N$ growing beyond 200. However, the level of computational accuracy reached already is at least a few orders of magnitude better than any reported in the literature (see results in Section \ref{sec:comparison}).

\subsubsection{Analysis of the algorithm - KGD model in viscosity dominated regime}
\label{sec:KGD_f_ss_1}

Let us now perform accuracy analysis for the viscosity dominated regime of the KGD model. The benchmark solution is based on representation \eqref{w_rep} for three test functions of type \eqref{h_KGD_fluid}. All resulting quantities can be obtained in the manner described in Appendix A. As for the PKN model, this benchmark solution assumes a predefined non-zero leak-off function. All results presented in this section were computed for $\gamma=\frac{1}{2+n}$, which corresponds to a constant influx rate.

We begin by investigating the relationship between the accuracy of computations, mesh density (number of nodal points $N$) and the fluid behaviour index, $n$. For different numbers of nodal points $N$ (ranging from 30 to 200) the computations were performed for $n$ continuously changing between 0 and 1. Again, the mesh density was increased at both ends of the spatial interval. The lower limit of $N$ was set to 30 instead of 10, as proper numerical computation of the inverse elasticity operator \eqref{p_ss_inv_KGD} necessitates finer meshing than its equivalent in the PKN model (identity operator). It should be emphasized that it is still possible to obtain very accurate results even for $N<30$, however this requires modifying the nodal spacing at the ends of the interval for each value of $n$.

\begin{figure}[h!]
\begin{center}
		\hspace{5mm}
    \includegraphics [scale=0.4]{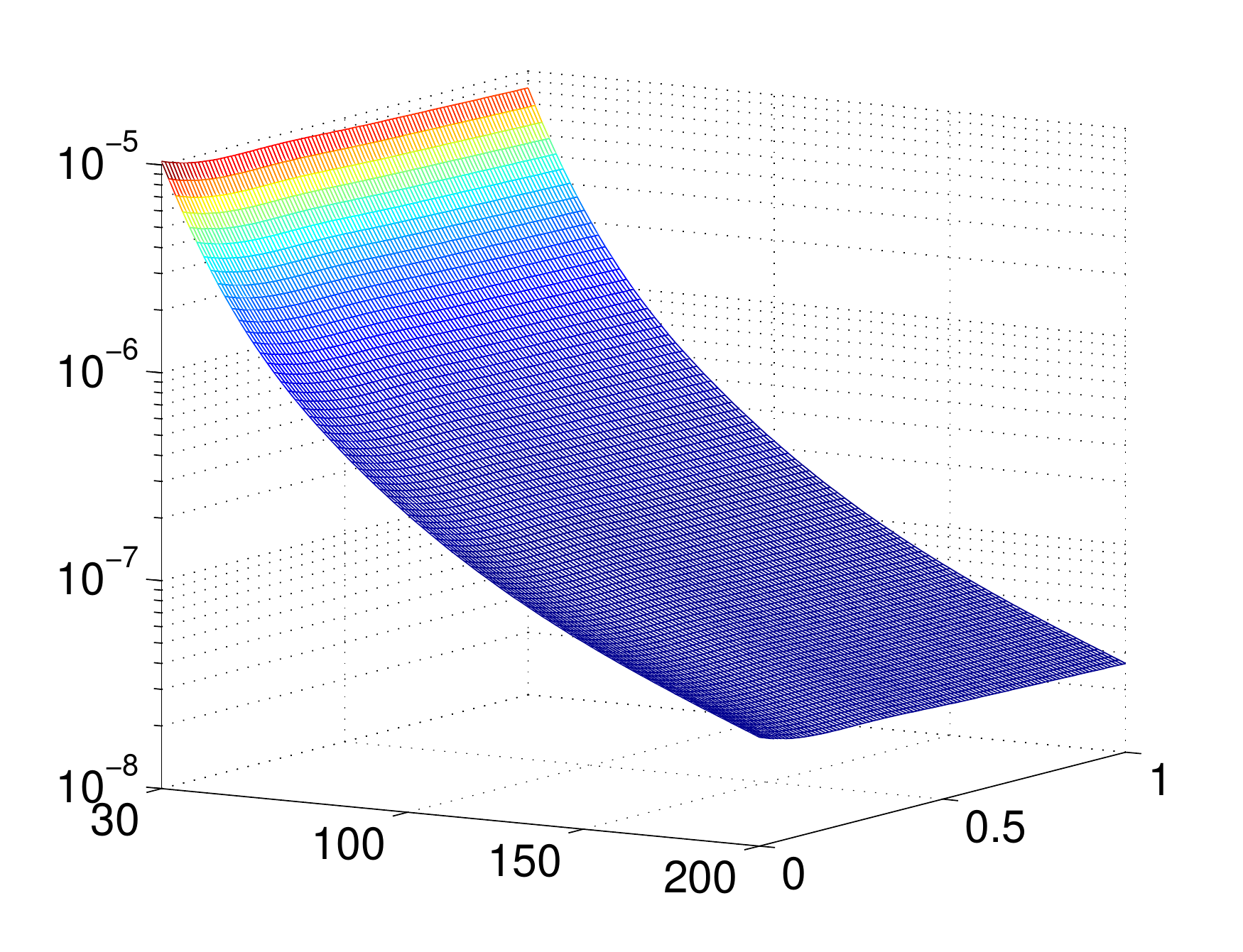}
		\put(-150,0){$N$}
		\put(-35,5){$n$}
    \put(-210,80){$\delta \hat w_{av}$}
    \put(-230,140){$\textbf{a)}$}
    \hspace{6mm}
    \includegraphics [scale=0.4]{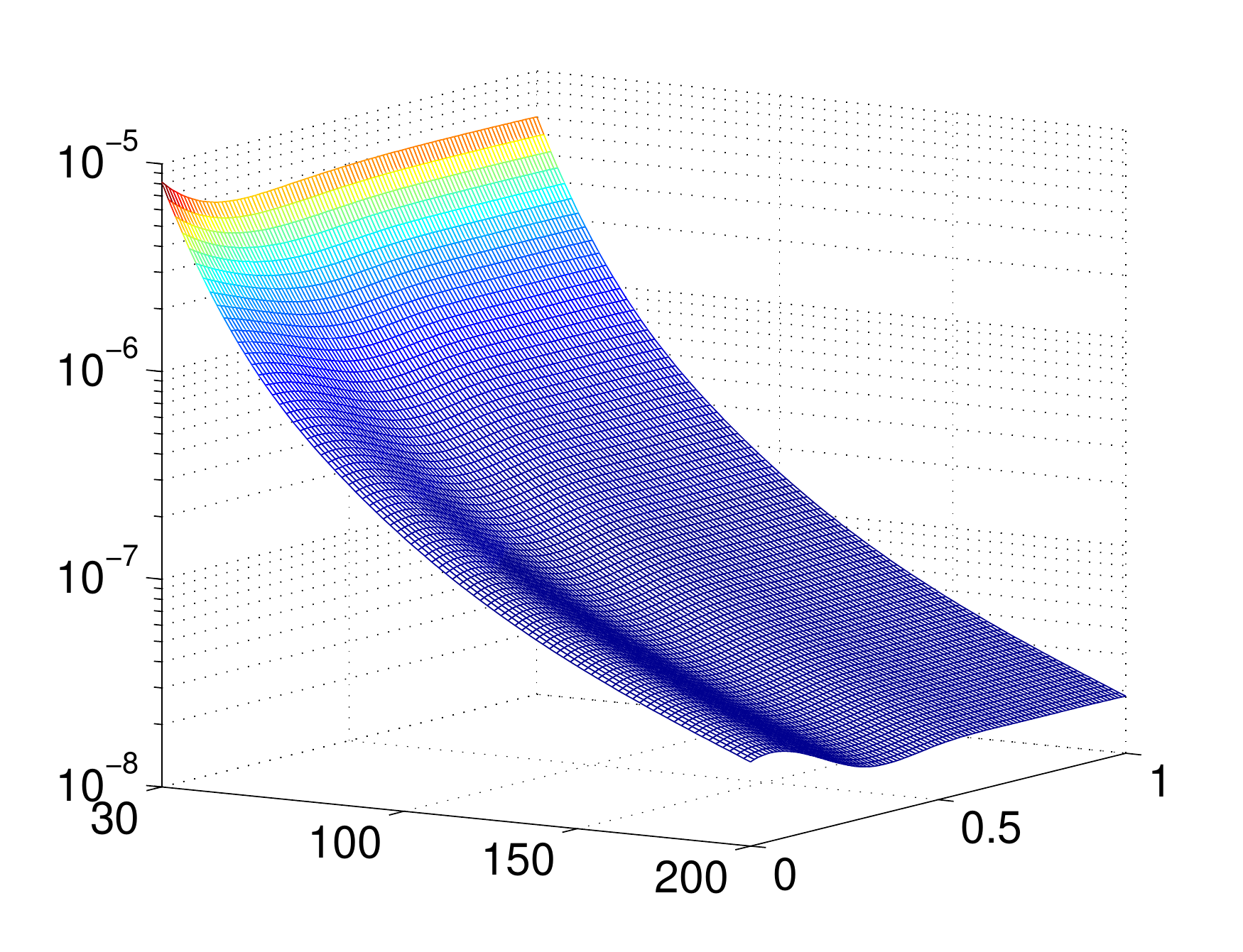}
		\put(-150,0){$N$}
		\put(-35,5){$n$}
		\put(-210,80){$\delta \hat v_{av}$}
    \put(-230,140){$\textbf{b)}$}
		\end{center}
    \caption{KGD model (viscosity dominated regime) -- average relative error for the self-similar solution with different numbers of nodal points, $N$, and values of the fluid behaviour index, $n$: a) the crack opening $\hat w$, b) the particle velocity $\hat v$. }

\label{av_rel_KGD_f}
\end{figure}
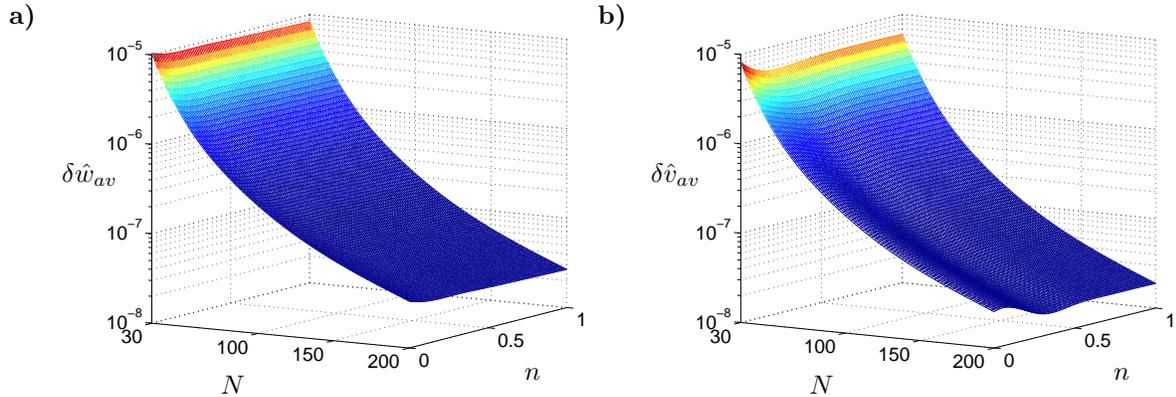

Similarly to that shown in Fig. \ref{av_rel_PKN} for the PKN model, the numerical simulations performance in the KGD viscosity dominated case is demonstrated using the average relative error over $x$ for the crack opening, $\delta \hat w_{av}$, and the particle velocity, $\delta \hat v_{av}$, with the results presented in Fig. \ref{av_rel_KGD_f}. It shows that the level of both errors is the same, however they are distinctly greater (up to two orders of magnitude for large $N$) than was the case for the PKN model. In the analyzed range of $N$, increasing the mesh density leads to a monotonous reduction in the error of computations. However, for $N<200$ the stabilization level is still not achieved. Low sensitivity of the results to the value of $n$ is observed. Only for coarse meshing does the solution exhibits some susceptibility to this parameter. The trends reported for $\delta \hat \phi_{av}$ (not depicted here) were the same as those for $\delta \hat v_{av}$.

\begin{figure}[h!]
		\hspace{-6mm}
		\includegraphics [scale=0.33]{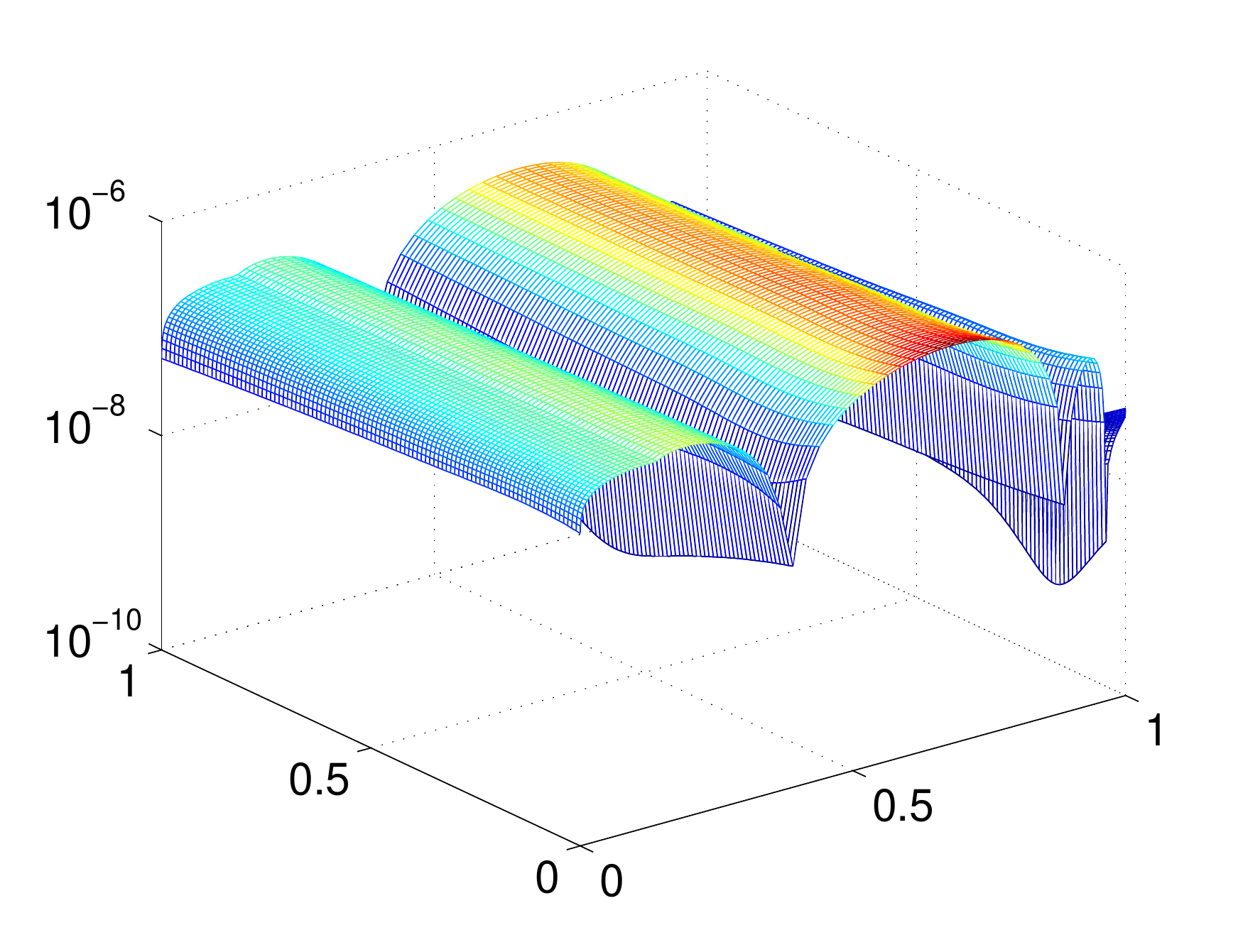}
		\put(-135,10){$n$}
		\put(-45,5){$x$}
\put(-130,110){$\delta \hat w(x)$}
		\hspace{-6mm}
    \includegraphics [scale=0.33]{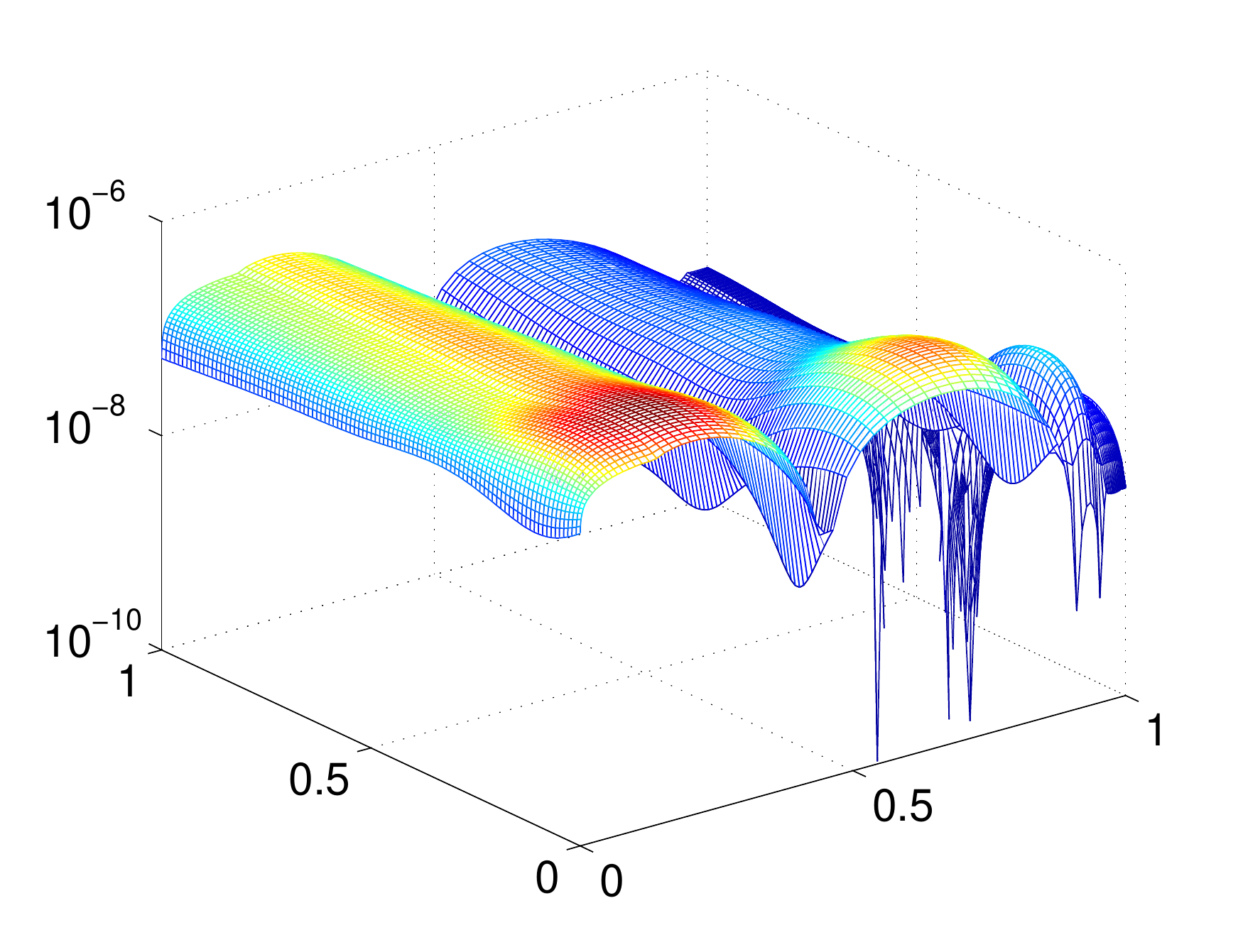}
		\put(-135,10){$n$}
		\put(-45,5){$x$}
\put(-130,110){$\delta \hat v(x)$}
		\hspace{-6mm}
    \includegraphics [scale=0.33]{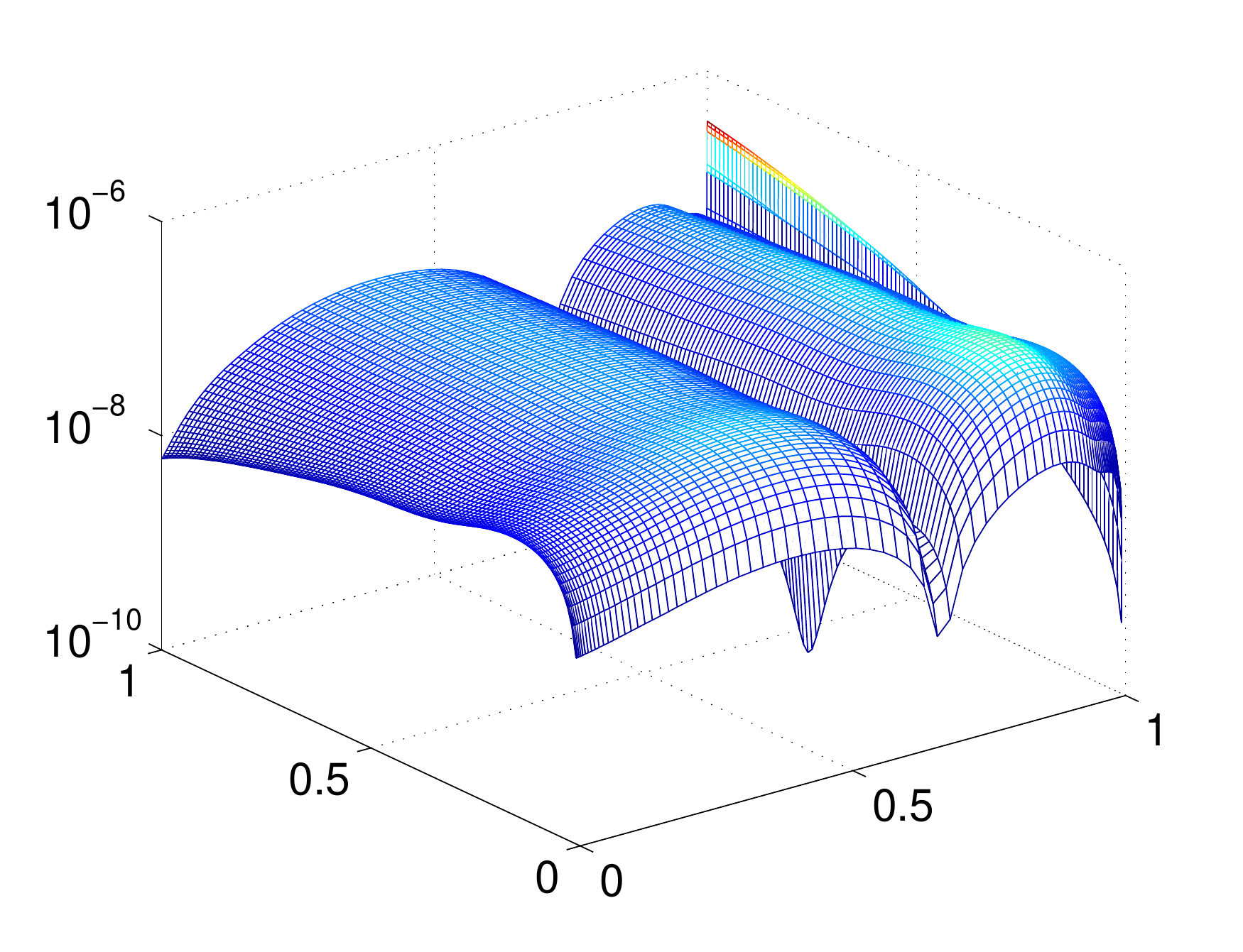}
		\put(-135,10){$n$}
		\put(-45,5){$x$}
	\put(-140,110){$\Delta \hat p(x)$}
    \caption{KGD model (viscosity dominated regime) -- error distribution for the self-similar solution: a) relative error of the crack opening $\delta \hat w$, b) relative error of the particle velocity $\delta \hat v$, c) absolute error of the net fluid pressure $\Delta \hat p$. Mesh is composed of $N=100$ nodes. }
\label{rel_error_KGD_f_1}
\end{figure}
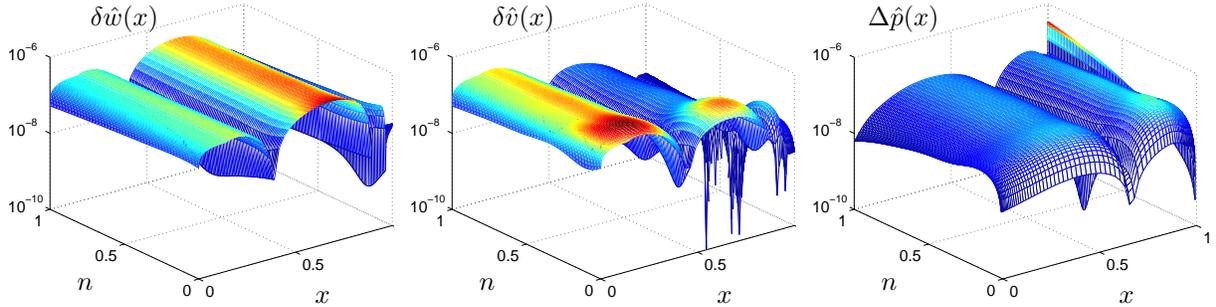

As previously for the PKN model, here we also present the spatial distribution of the solution error for $N=100$. The analyzed measures of error are: the relative error of the crack opening, $\delta \hat w$, and the relative error of the particle velocity, $\delta \hat v$. Additionally, the absolute error of the net fluid pressure, $\Delta \hat p$, is introduced, as later on we are going to compare our results for the pressure with others that are available in literature. Note that in the case of the PKN model, the error of the net fluid pressure was solely defined by the error of the crack opening. We do not show the pressures relative error, as the function $\hat p$ assumes a zero value at some point in the interval. All computations were carried out for $N=100$ nodal points and non uniform spatial meshing (the same for all $n$). The respective results are presented in Fig. \ref{rel_error_KGD_f_1}.

First and foremost, regardless of the error measure, the same accuracy level is retained over the whole interval of $n$. This proves that the proposed numerical scheme is highly stable. The relative error of the crack opening does not exhibit a magnification at the crack tip, which was evident in the PKN model. The maximal value of $\delta \hat w$ remains similar to that in the PKN case (compare Fig. \ref{rel_error_PKN}a), but its location has moved away from the crack tip. In general, the error level over most of the spatial interval (except for the near-tip region) has increased in comparison with the PKN model by at least an order of magnitude for the crack width and the particle velocity. This is not a surprise as the KGD problem is more complicated than the PKN model. Surprisingly, the accuracy of the net fluid pressure is of the same level even near the fracture front, where $\hat p$ exhibits singular behaviour.

\subsubsection{Analysis of the algorithm - KGD model in toughness dominated regime}
\label{sec:KGD_t_ss_1}

The benchmark solution, utilized here for accuracy analysis, is based on representation \eqref{w_rep} for five base functions of type \eqref{h_KGD_fluid} and \eqref{w_0_toughness}. The applied power $\gamma=\frac{n}{2+n}$ corresponds to a constant injection flux rate and simultaneously allows the reduction of the time-dependent problem to its self-similar version for a constant (independent of time) value of the normalized stress intensity factor. The self-similar SIF for this benchmark was $\hat K_I\approx 1$ and depended on $n$.

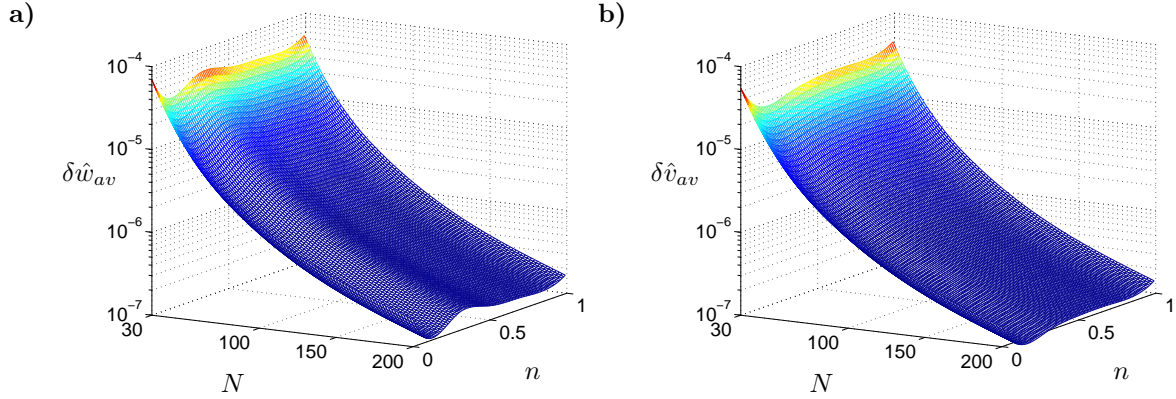
\begin{figure}[h!]
\begin{center}
		\hspace{5mm}
    \includegraphics [scale=0.4]{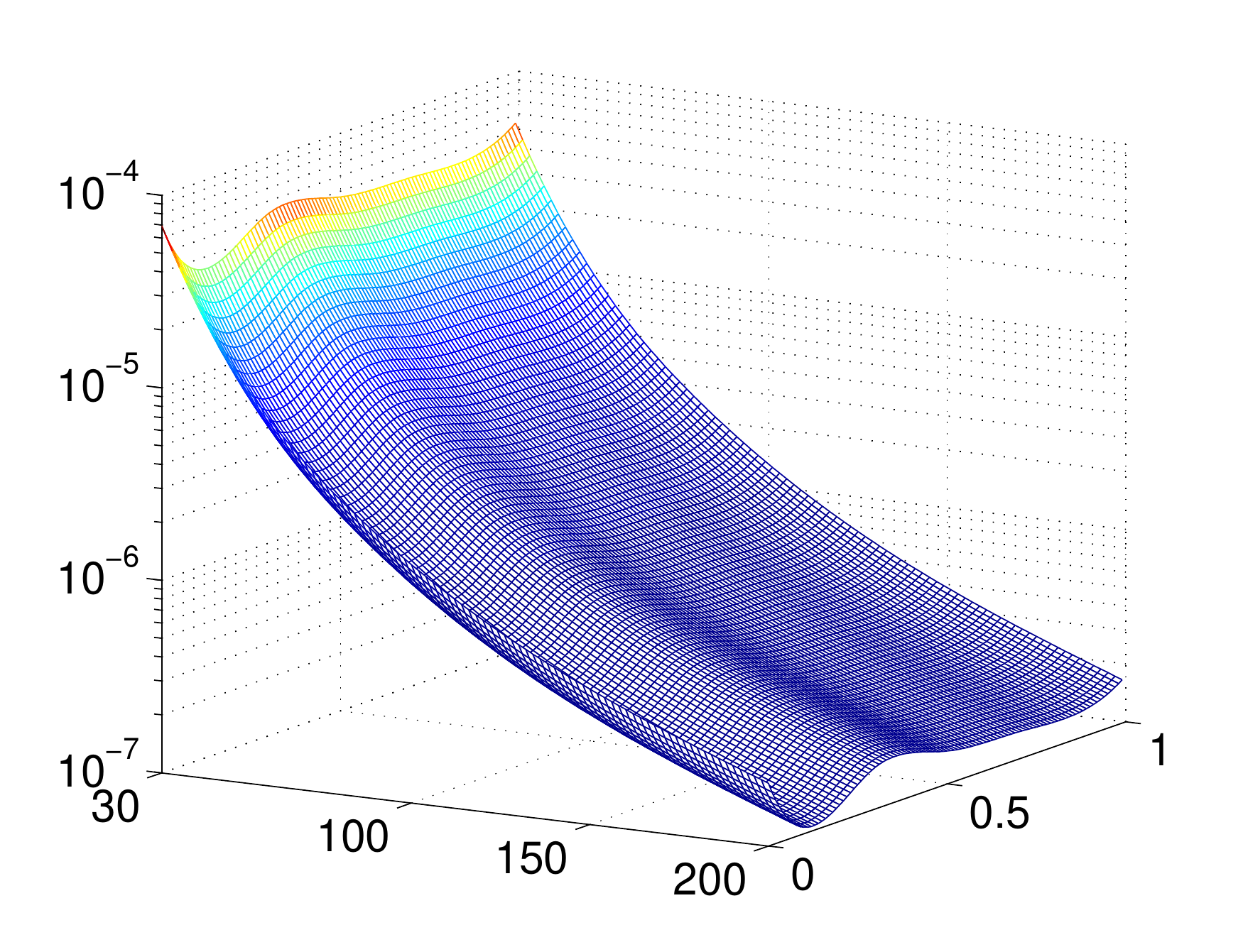}
		\put(-150,0){$N$}
		\put(-35,5){$n$}
    \put(-210,80){$\delta \hat w_{av}$}
    \put(-230,140){$\textbf{a)}$}
    \hspace{6mm}
    \includegraphics [scale=0.4]{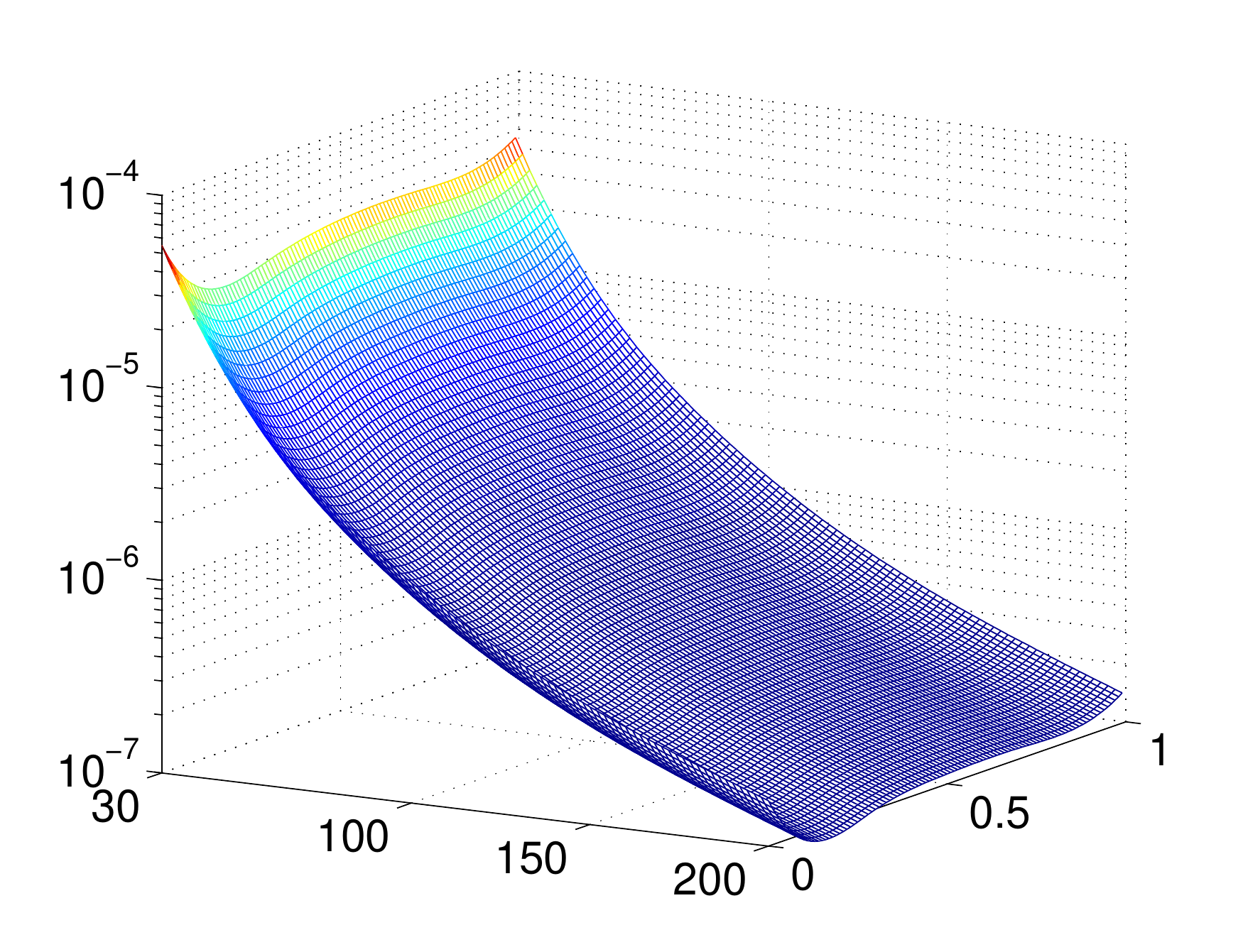}
		\put(-150,0){$N$}
		\put(-35,5){$n$}
		\put(-210,80){$\delta \hat v_{av}$}
    \put(-230,140){$\textbf{b)}$}
		\end{center}

    \caption{KGD model (toughness dominated regime) -- average relative error for the self-similar solution for different numbers of nodal points, $N$, and values of the fluid behaviour index, $n$: a) the crack opening $\hat w$, b) the particle velocity $\hat v$. }

\label{av_rel_KGD_t}
\end{figure}

Let us start by analyzing the average relative errors, similar to that conducted previously for the PKN model and the KGD model in viscosity dominated regime. In Fig. \ref{av_rel_KGD_t} we depict the average error of the crack opening and the particle velocity for various $N$ ($30\leq N \leq 200$), with $n$ in the range of interest. As previously, the adaptive spatial mesh (the same refined meshing at the ends of the interval for all values of $n$) was used.

The results show that again the accuracy of computations is not sensitive to the value of the fluid behaviour index $n$. The error distributions for both analyzed dependent variables are very similar. The error level is slightly higher than it was for the viscosity dominated regime of the KGD model (compare Fig. \ref{av_rel_KGD_f}). However, it is sufficient to take as few as 50 nodal points to have the average relative error of order $10^{-5}$. In the analyzed range of $N$ an increase in the mesh density produces a constant error reduction. No saturation level is achieved for $N<200$.

\begin{figure}[h!]
		\hspace{-2mm}
		\includegraphics [scale=0.30]{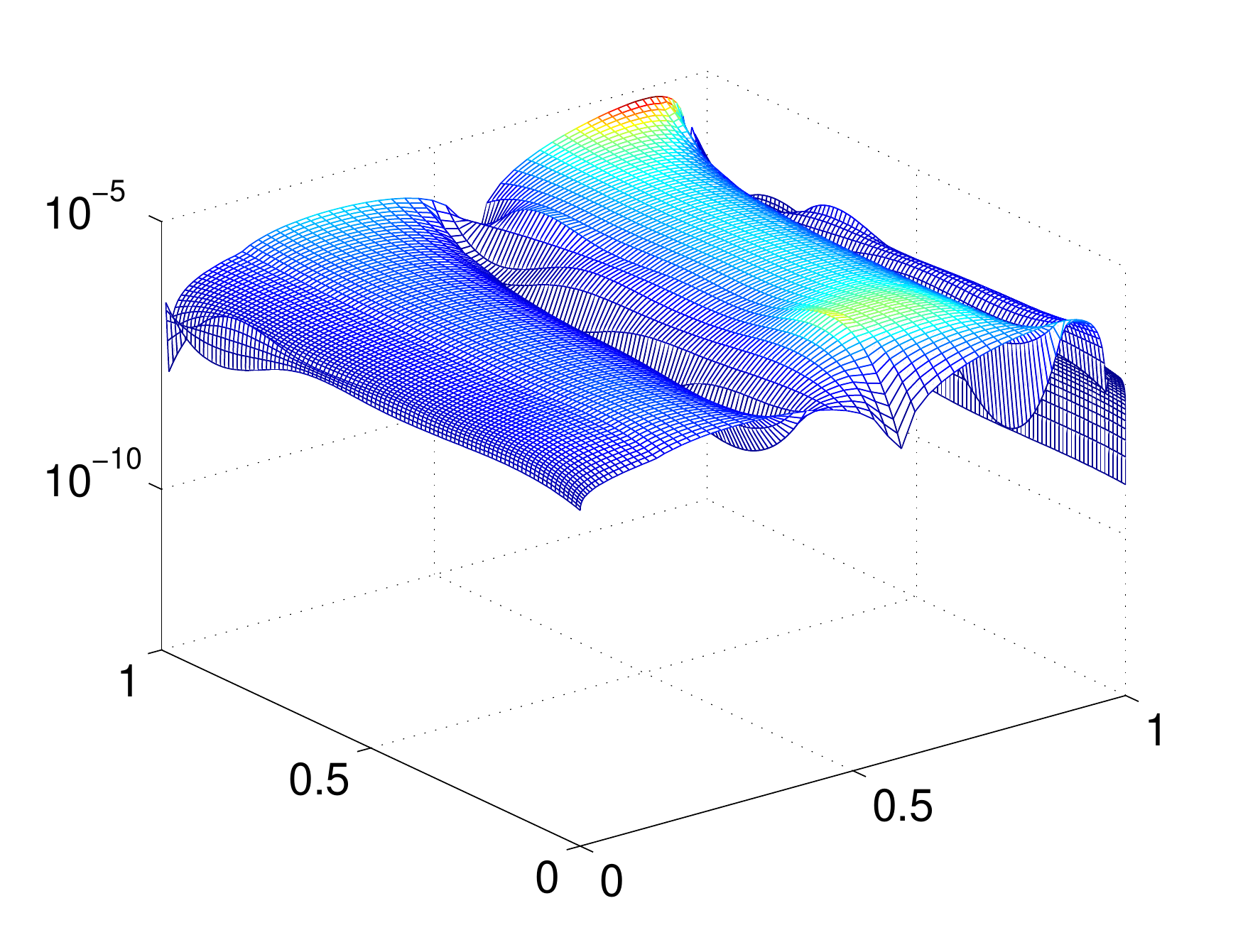}
		\put(-130,2){$n$}
		\put(-45,5){$x$}
\put(-130,110){$\delta \hat w(x)$}
		\hspace{-2mm}
    \includegraphics [scale=0.30]{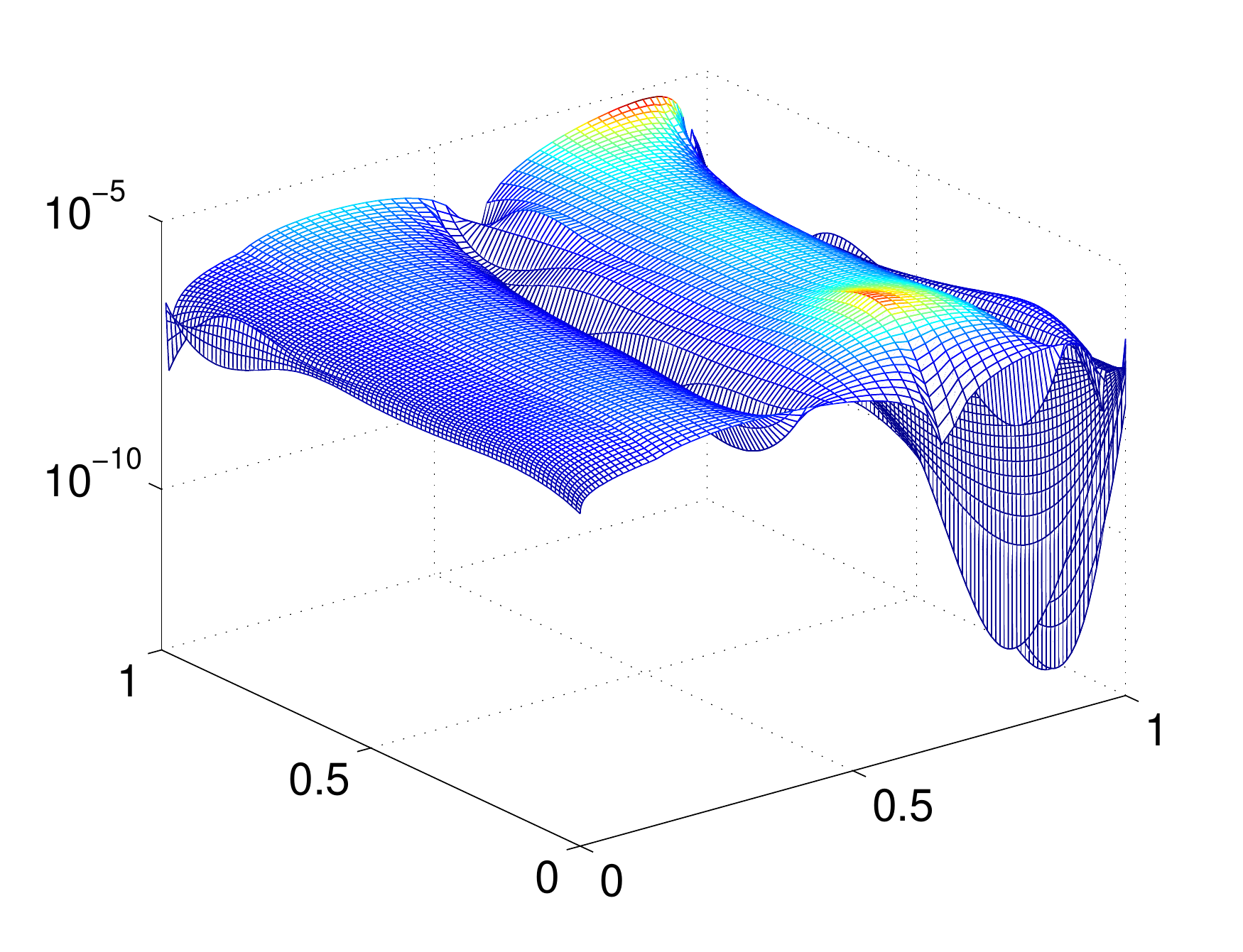}
		\put(-130,2){$n$}
		\put(-45,5){$x$}
\put(-130,110){$\delta \hat v(x)$}
		\hspace{-2mm}
    \includegraphics [scale=0.33]{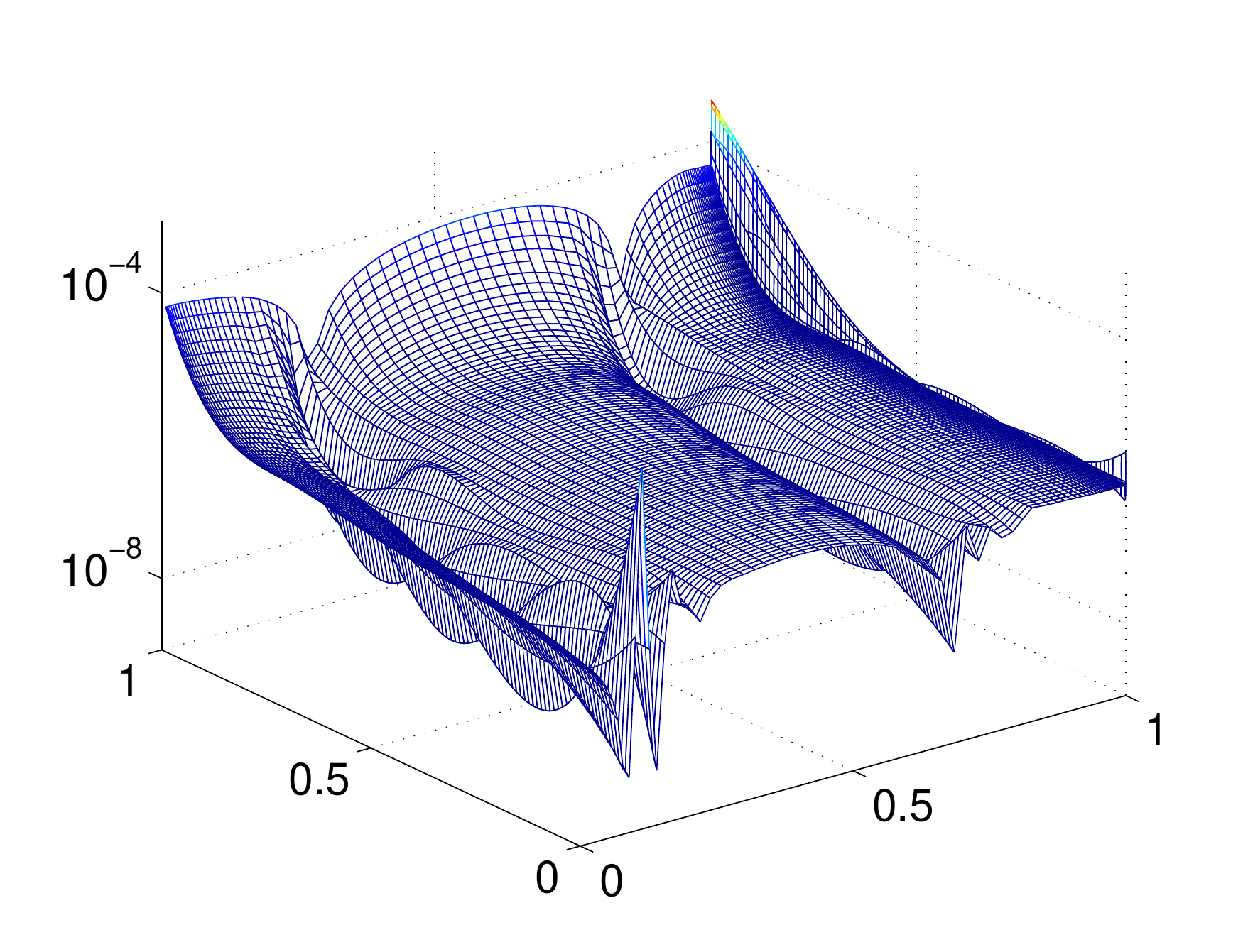}
		\put(-130,2){$n$}
		\put(-45,5){$x$}
	\put(-140,110){$\Delta \hat p(x)$}
    \caption{KGD model (toughness driven regime) -- error distribution for the self-similar solution: a) relative error of the crack opening $\delta \hat w$, b) relative error of the particle velocity $\delta \hat v$, c) absolute error of the net fluid pressure $\Delta \hat p$. The mesh is composed of $N=100$ nodes. }
\label{rel_error_KGD_t_1}
\end{figure}
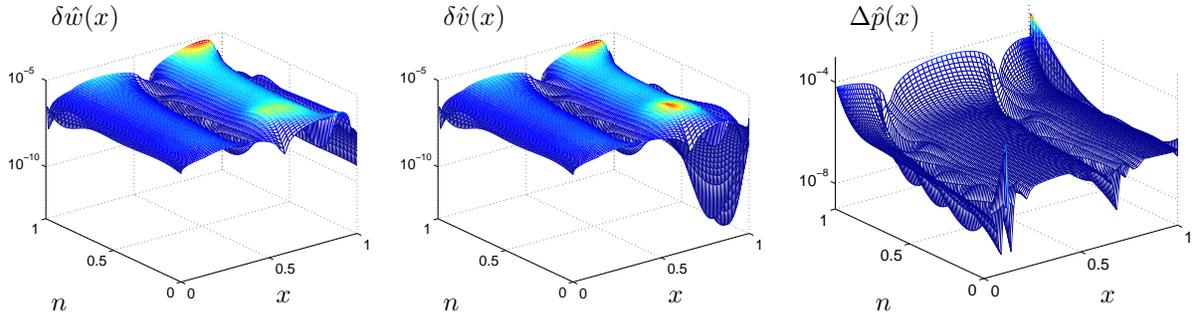

In the next stage of our analysis we investigate the spatial distribution of relative errors. In Fig. \ref{rel_error_KGD_t_1} we present the relative error for $\hat w$ and $\hat v$ and the absolute error of $\hat p$ as functions of $x$ and $n$. In general, the accuracy for the crack opening and the particle velocity is about half an order worse than it was for the viscosity dominated regime and does not change with $n$. No error magnification at the crack tip is observed. Also, the error of the net fluid pressure proves to be very stable for various $n$. A small increase in $\Delta \hat p$ can be seen at the crack tip for growing $n$, however considering the tip singularity of $\hat p$ this phenomenon does not detract from the high quality of the results.

\subsubsection{Summary of the presented results}
\label{sec:Sammary_1}

In order to present a brief summary of the foregoing accuracy analysis let us collect representative values of the computational error for the main components of the self-similar solution: the crack opening, $\hat w$, and the particle velocity, $\hat v$. The results are given in Table \ref{T33}. For each hydraulic fracture model and accuracy parameter, the presented figures should be considered as averaged over $n$. In some cases the difference between extreme values for different $n$ was negligible, in others there was a distinct discrepancy but they still remained within the same order of accuracy. The quality of computations for the fluid flow rate can be easily reconstructed from the presented data (bearing in mind that $\hat q=\hat w \hat v$). One can also estimate the accuracy of the net fluid pressure for each HF model under consideration from Fig. \ref{rel_error_PKN}a, Fig. \ref{rel_error_KGD_f_1}c and Fig. \ref{rel_error_KGD_t_1}c.
All computations were carried out using a mesh composed of $N=100$ nodal points, with increased density at the ends of the interval.

To enrich the presented investigation even further, we have also conducted additional tests for different values of the self-similar SIF, $\hat K_I=100,1,0.1,0.01$.  To this end, the analytical benchmark solutions from Appendix A have been used for different magnitudes of parameters $\lambda_j$. We wish to underline that, even for the lower values of $\hat K_I$, the problem should not necessarily be viewed as the so-called 'small toughness' regime. It results from the presence of a non-zero leak-off, which in the considered class of benchmark solutions is properly combined with the fracture opening and the particle velocity to balance the continuity equation. Thus it can alleviate, to some degree, the computational difficulties created by this extreme regime of crack propagation. Indeed, the results presented in Table \ref{T33} seem to confirm this prediction, as only a limited deterioration in the accuracy is observed for the two lowest values of $\hat K_I$. The problem of small toughness will be analyzed in the next subsection.

\begin{table}[h]
\begin{center}
\renewcommand{\arraystretch}{1.5}
\begin{tabular}{|c|c|c|c|c|c|c|}
\hline
  \multirow{2}{*}{Variable}  &\multicolumn{1}{c|}{ \multirow{2}{*}{PKN}}&\multicolumn{1}{c|}{{KGD }}&\multicolumn{4}{c|}{KGD toughness dominated}\\
\cline{4-7}
 &&viscosity dominated &$\hat K_I=100$&$\hat K_I=1$&$\hat K_I=0.1$&$\hat K_I=0.01$\\
\hline\hline
  $\delta \hat w_{av}$ &$10^{-9}$ & $3\cdot 10^{-7}$ &  $5\cdot 10^{-7}$&$10^{-6}$& $6\cdot 10^{-6}$& $10^{-5}$\\
  \hline
  $\delta \hat v_{av}$  & $10^{-9}$ & $10^{-7}$ & $10^{-7}$& $5\cdot 10^{-7}$& $10^{-6}$& $10^{-6}$ \\
  \hline
  $\delta \hat w_{max}$  & $4\cdot 10^{-7}$ & $4\cdot 10^{-7}$ & $5\cdot 10^{-7}$& $3\cdot 10^{-6}$& $10^{-6}$& $4\cdot 10^{-6}$ \\
  \hline
	  $\delta \hat v_{max}$  & $10^{-8}$ & $2\cdot 10^{-7}$ & $10^{-6}$& $5\cdot 10^{-6}$& $10^{-5}$& $10^{-5}$ \\
  \hline
\end{tabular}
\end{center}
\caption{The level of computational error for the self-similar crack opening, $\hat w$ and the particle velocity, $\hat v$, computed by the universal algorithm for different HF models, based on analytical benchmarks from Appendix A. For all computations the non-uniform mesh, distributed over the integral $(0,1)$, consisted of $N=100$ points and was more dense near the endpoints. Meshing parameters, and all other technical parameters, remained identical irrespective of the chosen model.}
\label{T33}
\end{table}

\subsubsection{KGD model: small toughness case}
\label{sec:fluid_toughness}

It is well known that the small toughness case constitutes an extremely challenging computational problem due to the immense reduction of the process zone \cite{Lecampion_Brisbane}. It necessitates special measures to stabilize the computations and preserve the solution accuracy, which results in an appreciable increase in the computational cost (in comparison with a moderate toughness case, the solution time was 100 times higher). In our test we shall utilize a reference solution for the viscosity dominated regime of the KGD model, namely a numerical solution obtained for $\hat q_l=0$ and $\hat q_0=1$. It is obvious that if for zero leak-off one imposes the same influx value in the toughness dominated regime, then for sufficiently small $\hat K_I$ the solution should approach that of the viscosity dominated regime.

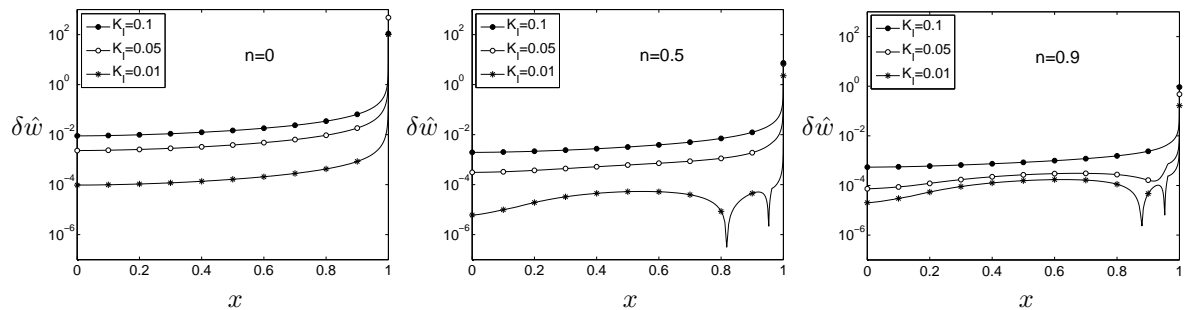
\begin{figure}[h!]
\begin{center}
		\includegraphics [scale=0.30]{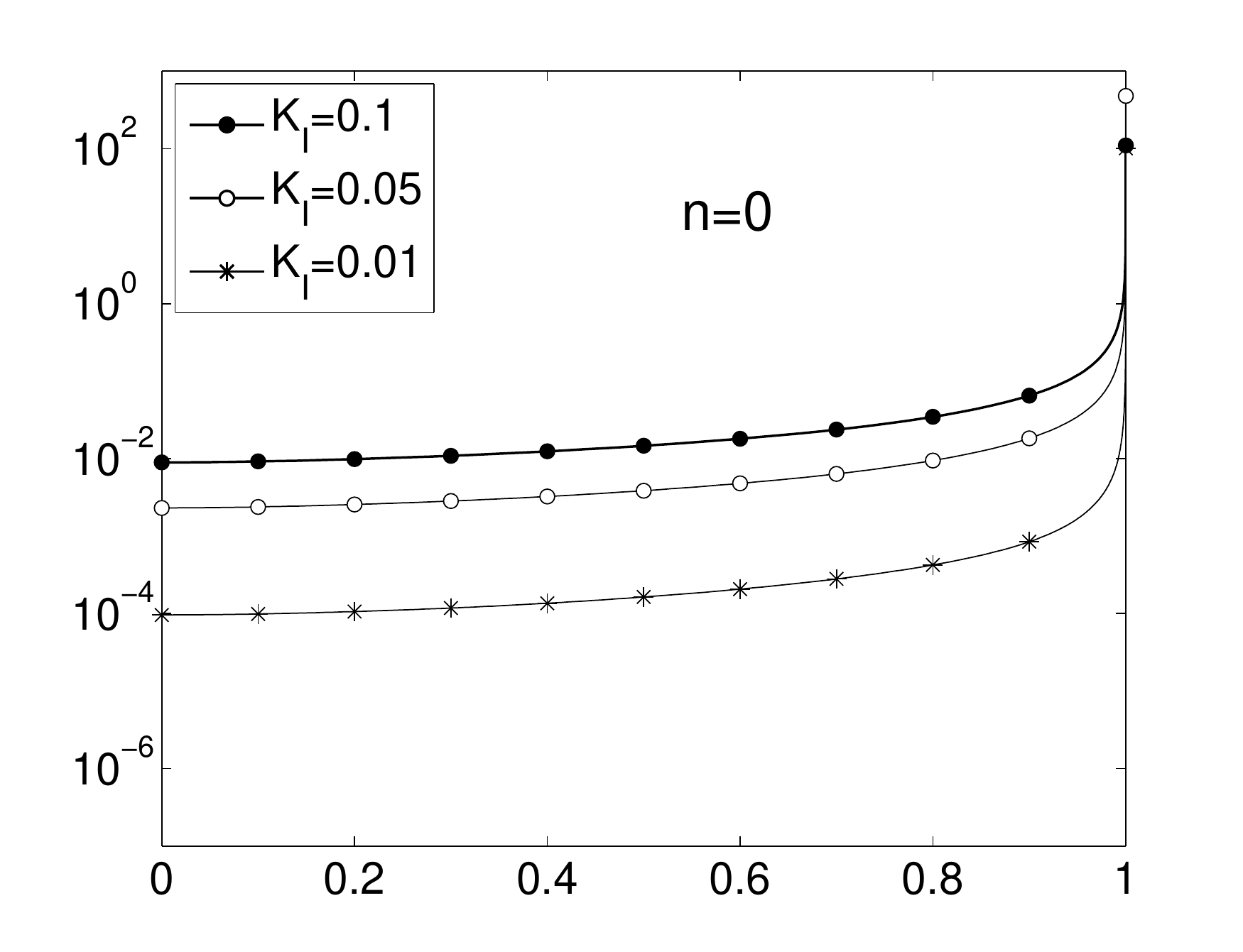}
		\put(-157,60){$\delta \hat w$}
		\put(-75,-5){$x$}
    \hspace{-2mm}
    \includegraphics [scale=0.30]{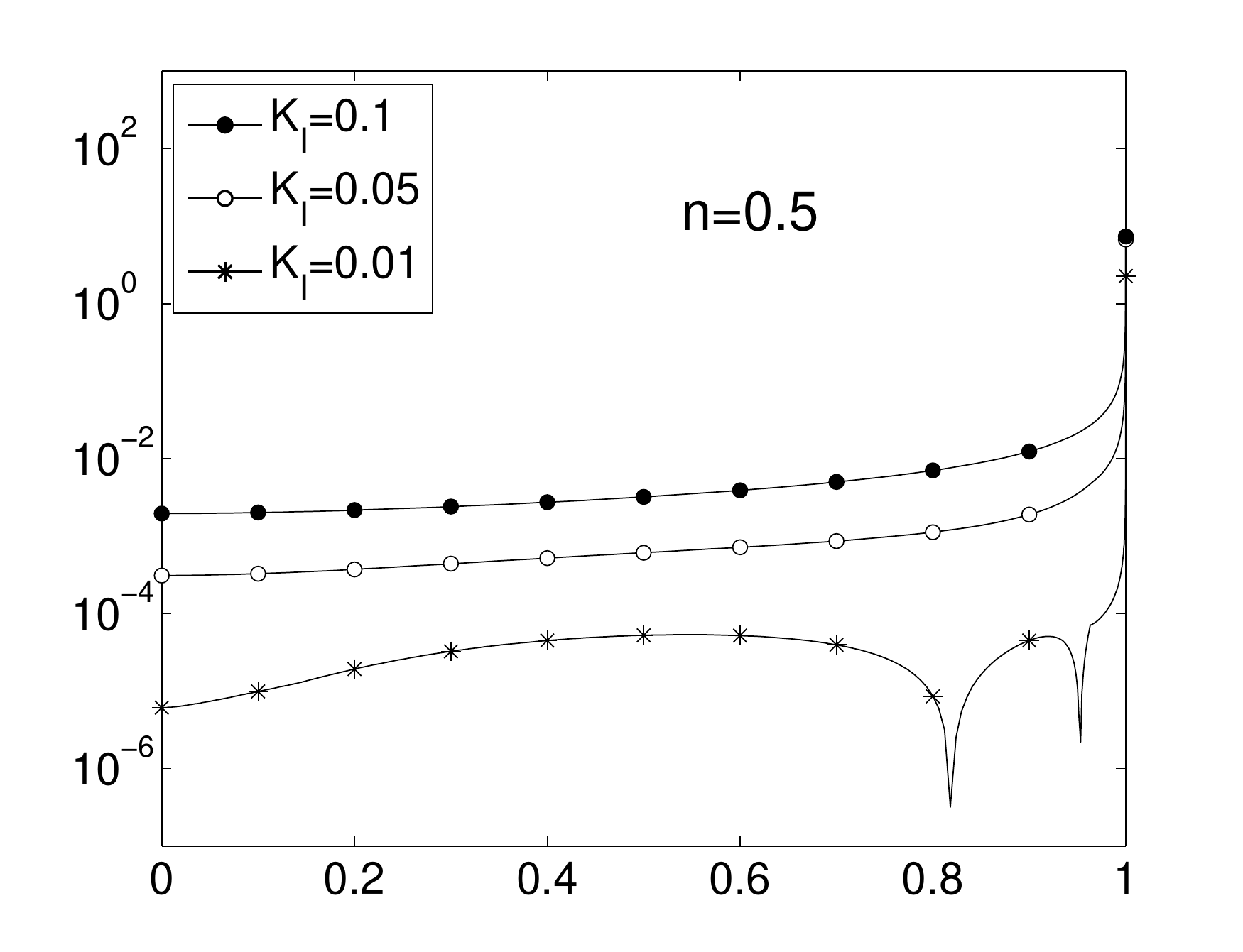}
 		\put(-157,60){$\delta \hat w$}
		\put(-75,-5){$x$}
    \hspace{-2mm}
    \includegraphics [scale=0.30]{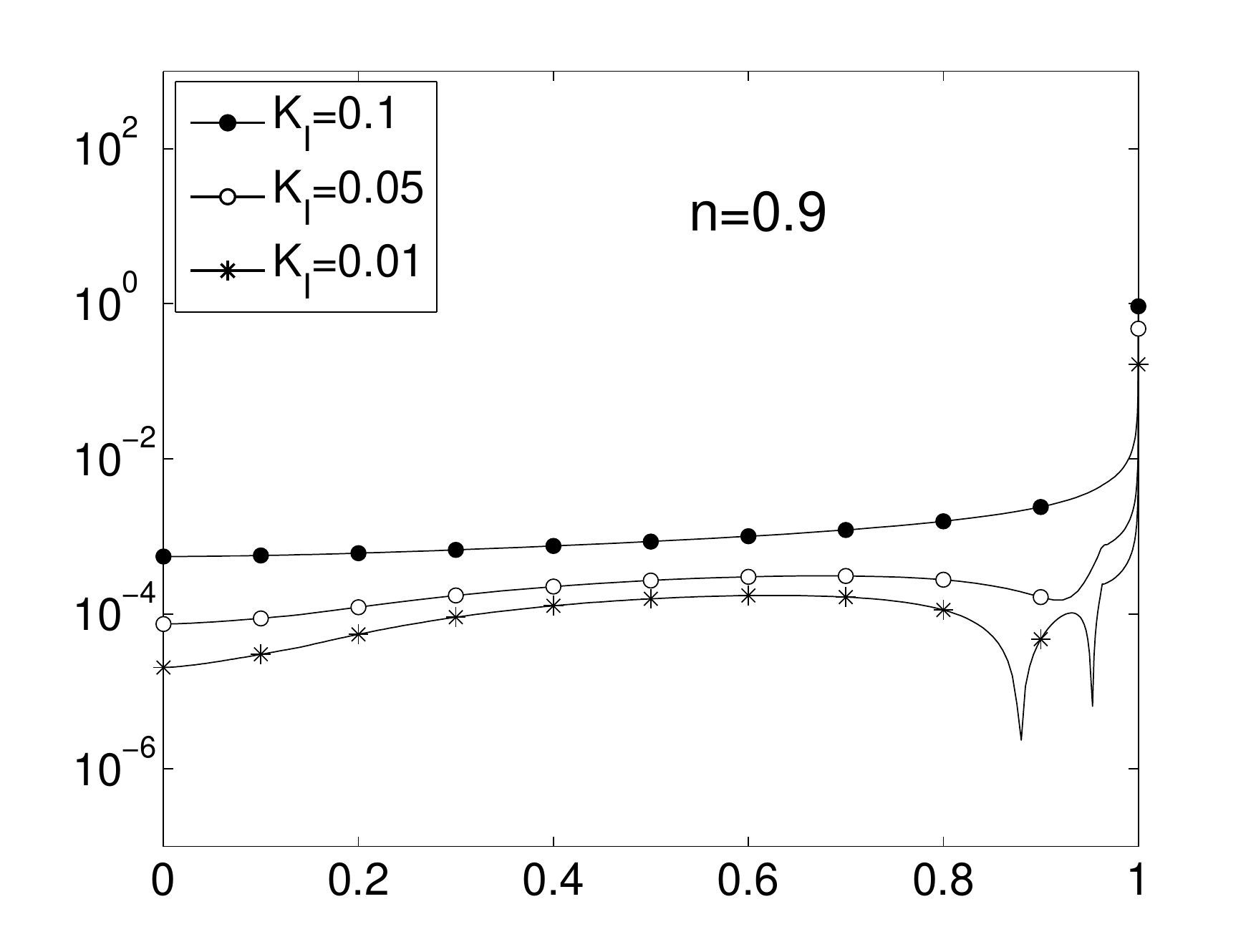}
		\put(-157,60){$\delta \hat w$}
  	\put(-75,-5){$x$}
\end{center}
    \caption{Crack opening: the relative deviation of the numerical solution for the small toughness KGD model from the KGD viscosity dominated reference solution for different values of the self-similar stress intensity factor, $\hat K_I$. The mesh is composed of $N=100$ nodes. For each $\hat K_I$ appropriate value of the regularization parameter $\varepsilon$ was taken.}
\label{small_toughness}
\end{figure}

The strategy for our test is as follows. For three values of the fluid behaviour index $n=\{0,0.5,0.9\}$ we conduct  computations while steadily decreasing the self-similar stress intensity factor, $\hat K_I$. At every step we investigate the relative deviation of the crack opening in the toughness dominated regime ($\hat w_T$) from the viscosity dominated reference solution ($\hat w_V$): $\delta \hat w=\left |(\hat w_T-\hat w_V)/{\hat w_V}\right |$. The results for $\hat K_I=\{0.1,0.05,0.01\}$ are depicted in Fig. \ref{small_toughness}.

It shows that for a fixed $\hat K_I$ the level of relative deviation $\delta \hat w$ depends on the fluid behaviour index, $n$. In general the lower the value of $n$ being analyzed, the bigger the reduction in $\hat K_I$ that is needed to achieve a certain level of discrepancy. Naturally, the biggest deviation between $\hat w_T$ and $\hat w_V$ is observed at the crack tip, where the respective asymptotics do not match. However, it is sufficient to take $\hat K_I=0.05$ to obtain a relative difference from the viscosity dominated solution of the order $10^{-3}$ for any $n$ on practically the whole crack length. One can conclude
that, when $\hat K_I<0.05$, it is acceptable to use the solution of the viscosity dominated variant of the problem in order to achieve the accuracy needed in any practical application.
These results provide an independent indirect verification of the computational accuracy of the algorithm. However, accurate computing in the limiting case of small toughness necessitates proper handling of the computational parameters (the regularization parameter $\varepsilon$ and mesh density).

\subsection{Comparison with other available results}
\label{sec:comparison}
Having identified the level of accuracy achievable by the presented numerical scheme we can now conduct computations for the reference data available in literature. In the following sections we shall present a comparative analysis for a few solutions delivered by other authors for PKN and KGD models.

\subsubsection{PKN model}
\label{sec:PKN_ss_2}

In \cite{linkov_2013} an analytical solution for the PKN model was delivered in terms of an infinite power series for the so-called "proper variable" $\hat w^{n+2}$ and the particle velocity $\hat v$. Due to the complexity of the problem the coefficients of the series are represented in the form of recurrent sums, which makes computing them a complicated task in and of itself. For this reason, the author presented explicit formulae for the first five coefficients in respective expansions. Moreover, the representation of the solution also contains an additional parameter, the \textit{self-similar crack length} (see \cite{linkov_2013}), $\xi_*$, which is to be determined from the influx boundary condition. Unfortunately, the slowest power series convergence over the whole range of $x$ takes place for $x=0.$ Thus one cannot expect good accuracy for the self-similar crack length approximation when using merely five terms. On the other hand, the author observed that the unknown $\xi_*$ lies between $\xi_*=1.04004$ (for $n=0$) and $\xi_*=1.00101$ (for $n=1$). This led him to conclude that the averaged value of the self-similar crack length $\xi_*=1.02$ is sufficient to produce, with 5-term approximation, a solution whose error does not exceed a few percent for any $n$.

In the following, with the accurate computational results obtained using the universal algorithm, we shall: i) recreate the respective numerical values of the self-similar crack length $\xi_*$ pertaining to the problem formulated in \cite{linkov_2013} -- here an approximate formula for $\xi_*(n)$ will be given; ii) estimate the error of the 5-term solution approximation, given in \cite{linkov_2013} for three variants of $\xi_*$: the actual (numerical) values, the averaged value $\xi_*=1.02$, and a linear approximation between the extreme values, $\xi_*(0)=1.04004$ and $\xi_*(1)=1.00101$, given in \cite{linkov_2013}.

All computations were carried out using a mesh composed of 200 points which, according to the accuracy analysis given in Section \ref{sec:PKN_ss_1}, provides a solution accuracy of up to order $10^{-9}$ for the crack opening and $10^{-8}$ for the particle velocity. Appropriate rescaling of the problem to the formulation used in \cite{linkov_2013} was carried out under the assumption, accepted there, that the self-similar crack propagation speed is a known function of $\xi_*$. A number of computations  for different values of the fluid behaviour index $n$ allowed us to recreate the relation $\xi_*(n)$ numerically. Based on the numerical simulations we have established an approximate formula for the self-similar crack length $\xi_*(n)$:
\begin{equation}
\label{xi_n}
\xi_*^{-\frac{2n+3}{n+2}}=(1-n)\left[\frac{2\sqrt{3}}{3}+n\frac{0.07221n-0.04222}{n^2+3.247n+2.458}\right]+0.99831387n,
\end{equation}
which imitates the accurate numerical results with an accuracy of order $10^{-7}$.

\begin{figure}[h!]
    \hspace{5mm}
    \includegraphics [scale=0.40]{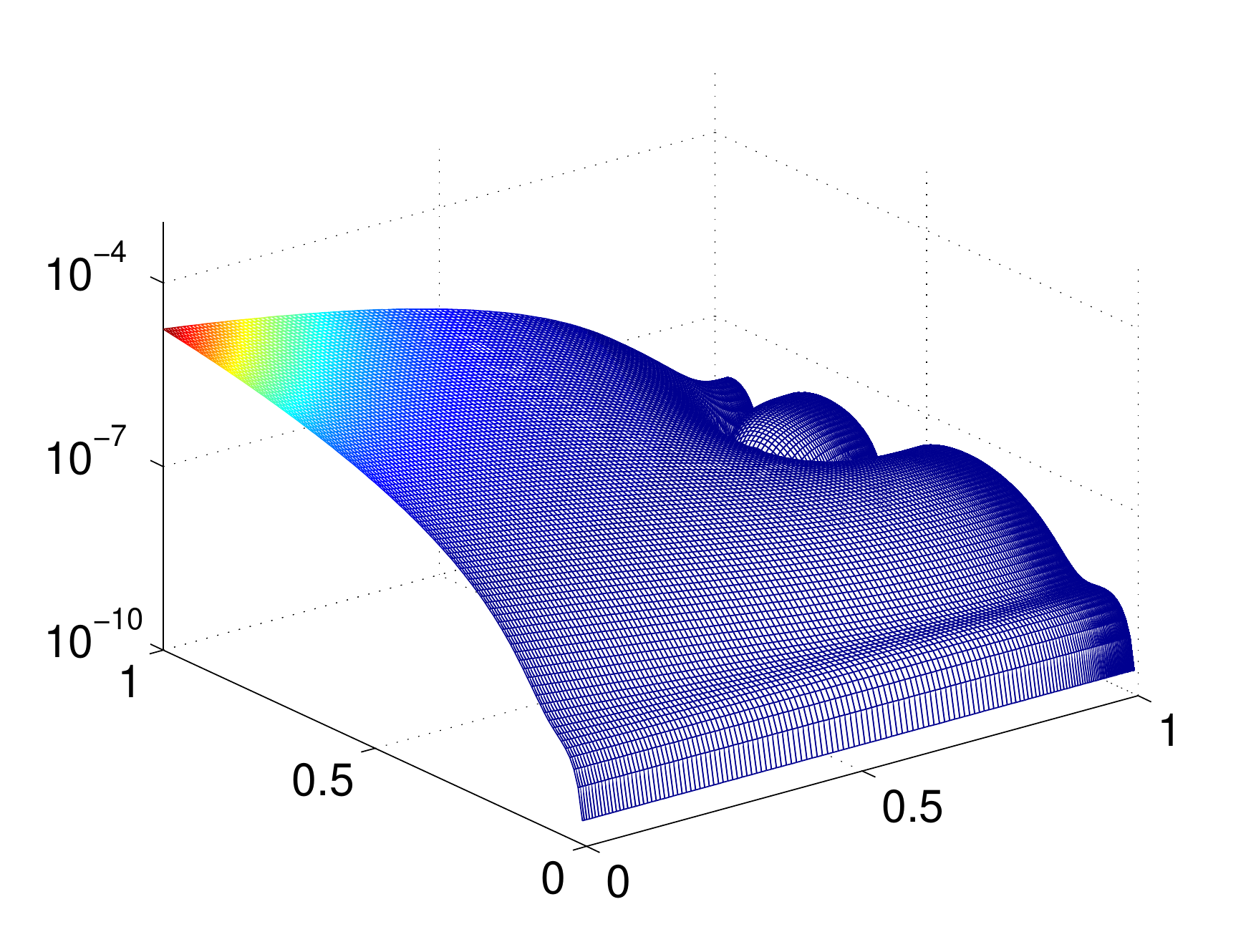}
		\put(-160,10){$n$}
		\put(-45,10){$x$}
    \put(-210,80){$\delta \hat w$}
    \put(-210,140){$\textbf{a)}$}
    \hspace{6mm}
    \includegraphics [scale=0.40]{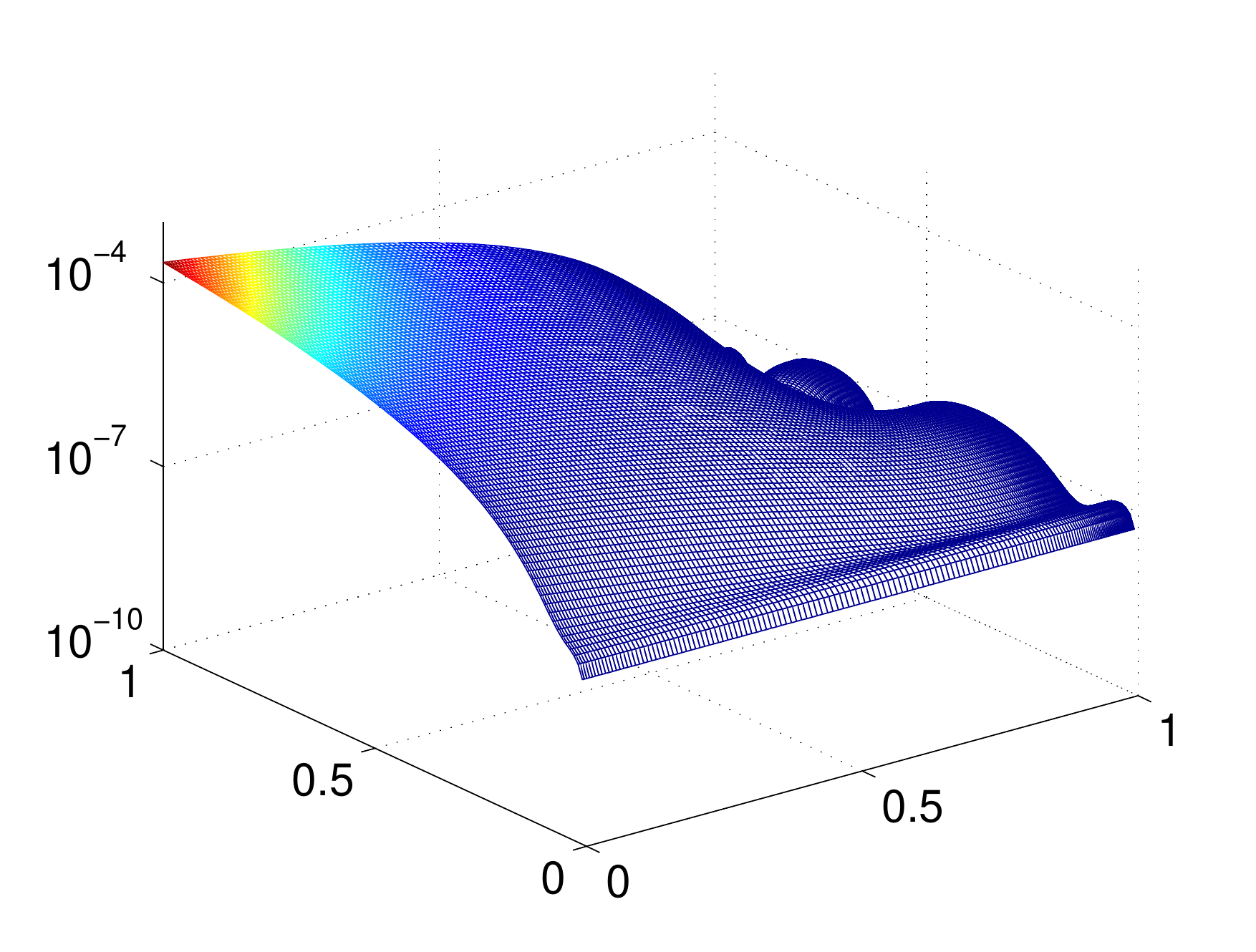}
		\put(-160,10){$n$}
		\put(-45,10){$x$}
    \put(-210,80){$\delta \hat v$}
    \put(-210,140){$\textbf{b)}$}

    \caption{PKN model -- comparison of the 5-term approximation solution presented in \cite{linkov_2013} with the accurate numerical solution for $N=200$: relative errors of: a) the crack opening, $\hat w(x)$, b) the particle velocity, $\hat v(x)$. Self-similar crack length was computed according to \eqref{xi_n}.}

\label{linkov_PKN_h_V_q0}
\end{figure}
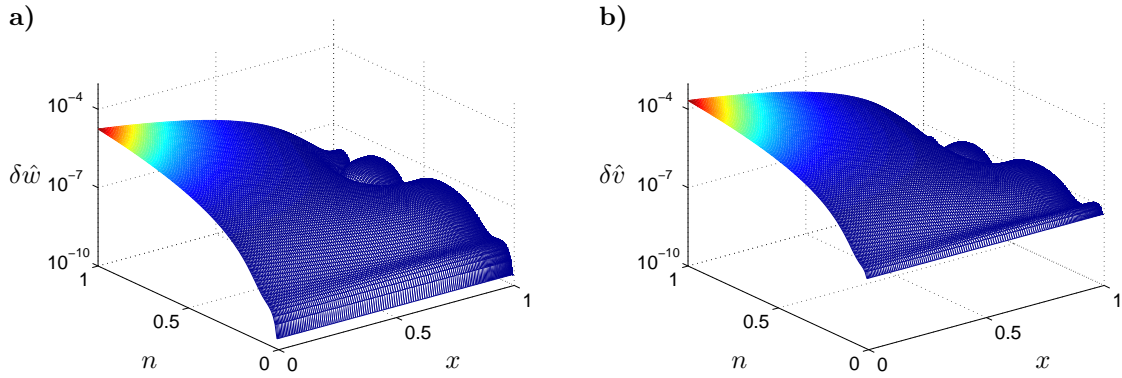

Having the values of $\xi_*(n)$ for an arbitrary fluid behaviour index enables us to
assess the accuracy of the 5-term solution approximation proposed in \cite{linkov_2013}.
We have checked that the error of the 5-term approximate solution remains exactly the same when recreated using relation \eqref{xi_n} or with the accurate numerical value of $\xi_*$.
The respective results are presented in Fig. \ref{linkov_PKN_h_V_q0}.

It shows that for both dependent variables such an approximation provides a maximal error for $\hat w$ of the level $10^{-5}$ and $10^{-4}$ for $\hat v$. The accuracy improves with decreasing $n$. The crack opening, $\hat w$, has greater potential for further error reduction.

\begin{figure}[h!]
    \hspace{5mm}
    \includegraphics [scale=0.40]{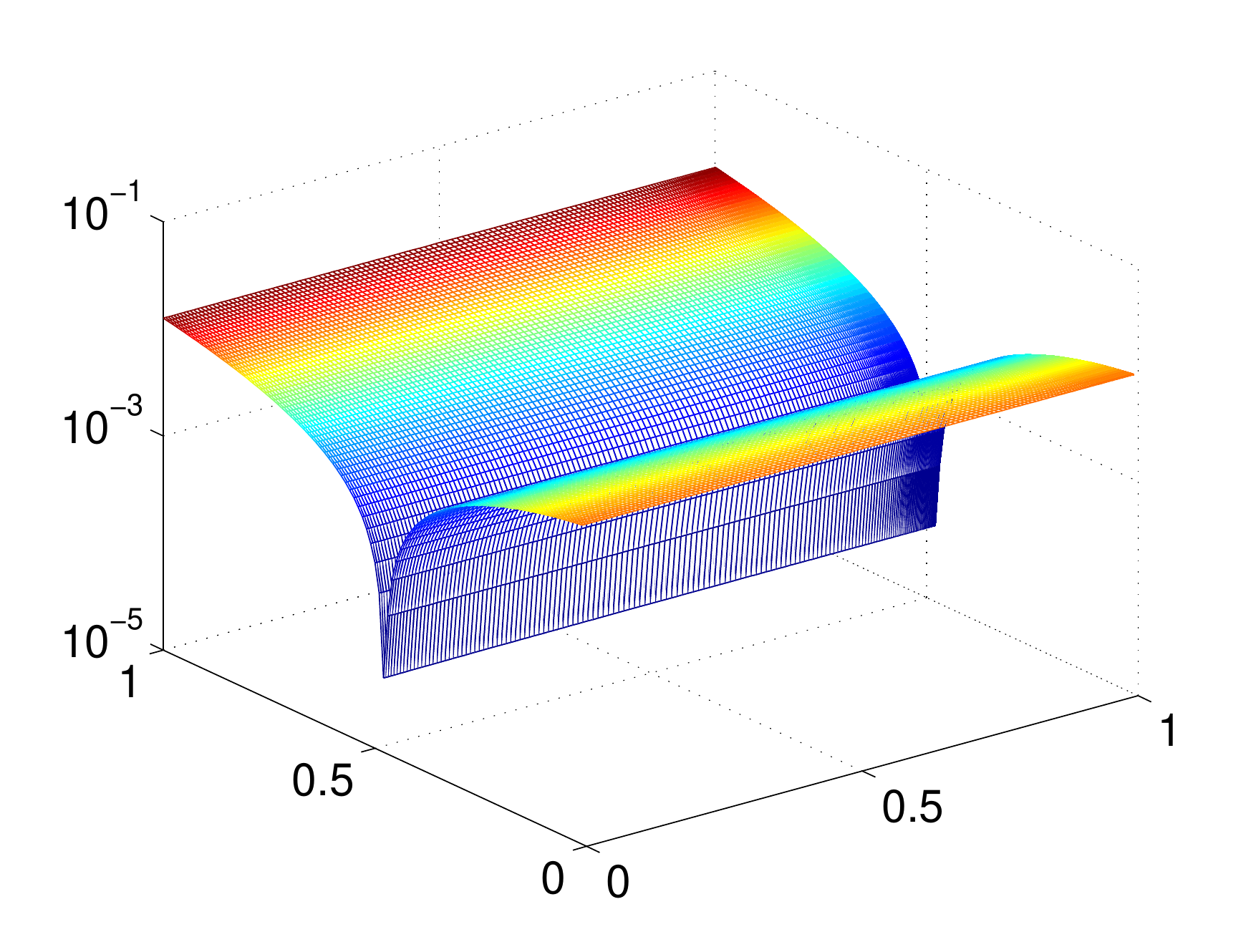}
		\put(-160,10){$n$}
		\put(-45,10){$x$}
    \put(-210,100){$\delta \hat w$}
    \put(-210,140){$\textbf{a)}$}
    \hspace{6mm}
    \includegraphics [scale=0.40]{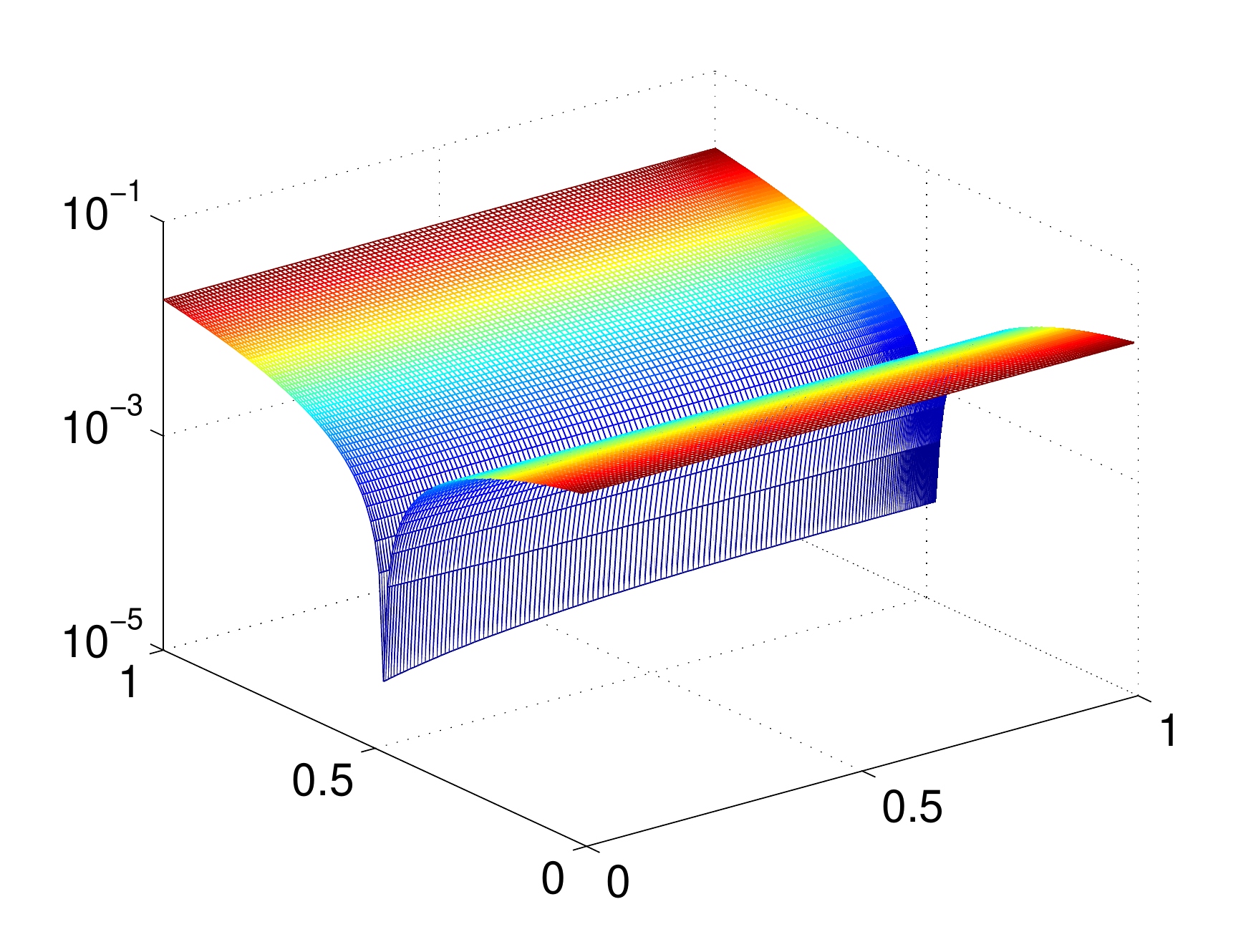}
		\put(-160,10){$n$}
		\put(-45,10){$x$}
    \put(-210,100){$\delta \hat v$}
    \put(-210,140){$\textbf{b)}$}

    \caption{PKN model -- comparison of the 5-term approximation solution presented in \cite{linkov_2013} with the accurate numerical solution for $N=200$: relative errors of: a) the crack opening, $\hat w(x)$, b) the particle velocity, $\hat v(x)$. The averaged (constant over $n$) self-similar crack length was taken: $\xi_*=1.02$ (see page 16 in \cite{linkov_2013}).}

\label{linkov_PKN_h_V_1_02}
\end{figure}

\begin{figure}[h!]
    \hspace{5mm}
    \includegraphics [scale=0.40]{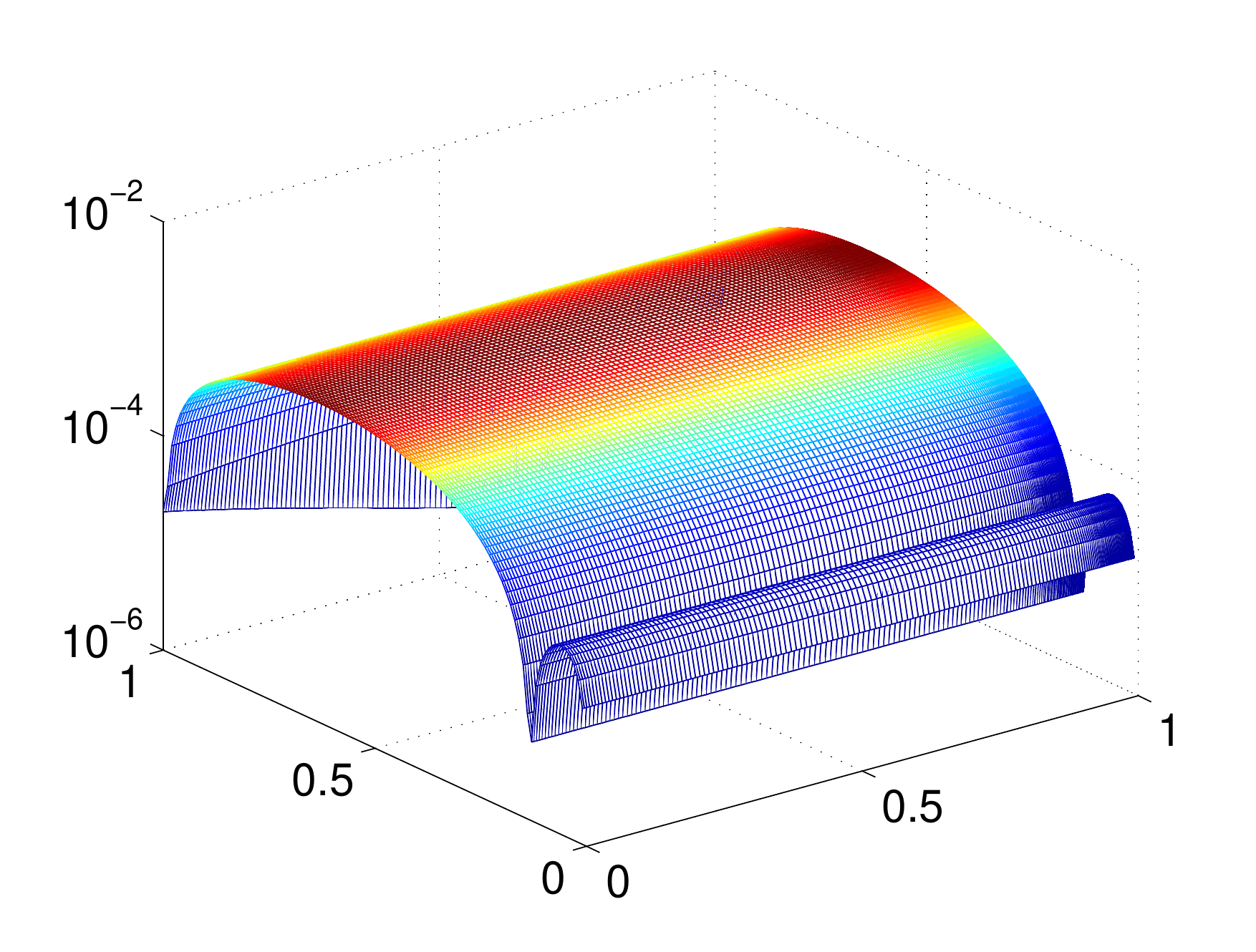}
		\put(-160,10){$n$}
		\put(-45,10){$x$}
    \put(-210,100){$\delta \hat w$}
    \put(-210,140){$\textbf{a)}$}
    \hspace{6mm}
    \includegraphics [scale=0.40]{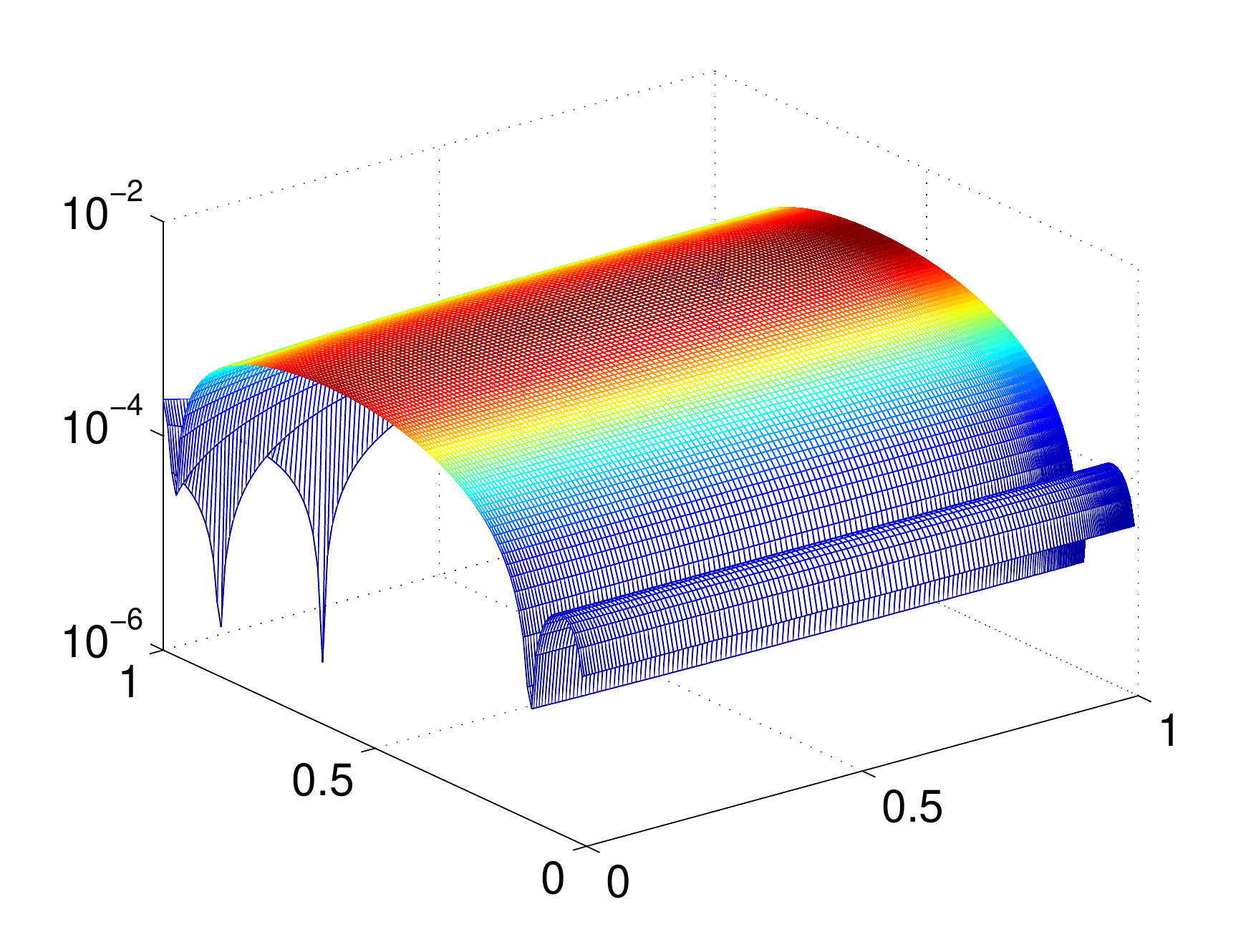}
		\put(-160,10){$n$}
		\put(-45,10){$x$}
    \put(-210,100){$\delta \hat v$}
    \put(-210,140){$\textbf{b)}$}

    \caption{PKN model -- comparison of the 5-term approximation solution presented in \cite{linkov_2013} with the accurate numerical solution for $N=200$: relative errors of: a) the crack opening, $\hat w(x)$, b) the particle velocity, $\hat v(x)$. The self-similar crack length, $\xi_*$, was taken as a linear function of $n$.}

\label{linkov_PKN_h_V_Lin}
\end{figure}

 For comparison, in Fig. \ref{linkov_PKN_h_V_1_02} we depict  the accuracy of the 5-term approximations for $\hat w$ and $\hat v$, obtained using the averaged value of the self-similar crack opening $\xi_*=1.02$ suggested in \cite{linkov_2013}. Now, $\delta \hat w$ and $\delta \hat v$ are of a similar level with maximal values (around 1 percent as predicted) located at $n=0$ and $n=1$. The best results are obtained in the vicinity of $n=1/2$, where the averaged $\xi_*$ is close to the real value. A natural suggestion emerging from this test is that a linear approximation of the self-similar crack length, based on the two extreme values (for $n=0$ and $n=1$), should produce better results than the constant averaged value. In Fig. \ref{linkov_PKN_h_V_Lin}  we depict the errors of $\hat w$ and $\hat v$ obtained for such a variant. The maximal errors have been reduced by one order of magnitude with respect to the previous case. The best coincidence with our numerical solution is obtained at the ends of the interval $n$.

\begin{figure}[h!]
\center
    \hspace{5mm}
    \includegraphics [scale=0.43]{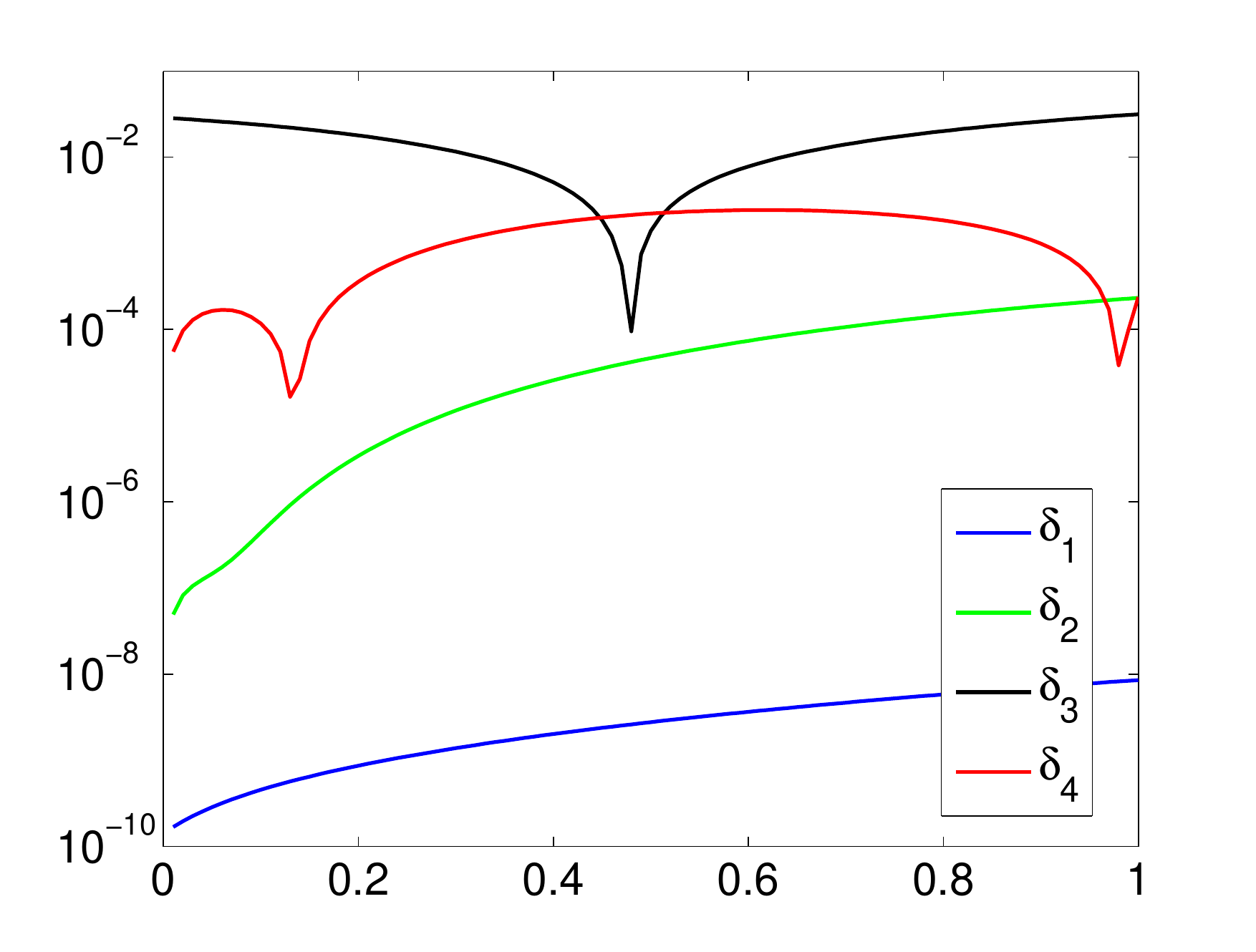}
		\put(-107,0){$n$}
    \put(-230,80){$\delta \hat q_0$}

    \caption{PKN model - fulfillment of the influx boundary condition by: $\delta_1$ -- the numerical solution, $\delta_2$ -- the 5-term approximation with $\xi_*$ computed from \eqref{xi_n}, $\delta_3$ -- the 5-term approximation with averaged $\xi_*=1.02$, $\delta_4$ -- the 5-term approximation with $\xi_*$ linearly dependent on $n$.}

\label{linkov_PKN_h_V_spl_q0}
\end{figure}

In order to cross-check the credibility of the respective results, let us investigate to what degree they satisfy the influx boundary condition. In the problem considered in \cite{linkov_2013} a predefined unit value of the influx was assumed (the normalization of the governing equations taken as in \cite{linkov_2013}): $\hat q_0=1$. In Fig. \ref{linkov_PKN_h_V_spl_q0} we show the relative errors of $\hat q_0$, recreated as a product of the crack opening and the particle velocity ($\hat q_0=\hat w(0) \hat v(0)$), for our numerical solution and different variants of the 5-term approximation. As can be seen in the figure, the numerical solution fulfills the influx boundary condition with a stable accuracy of the level $10^{-8}$ over the whole range of $n$. The 5-term approximation with $\xi_*$ computed from \eqref{xi_n} yields more accurate results for lower values of the fluid behaviour index, with a monotonous error increase up to a level of $10^{-4}$ for growing $n$. As anticipated, for $\xi_*$ linearly dependent on $n$ one obtains the best accuracy for the extreme values of the latter, while the averaged  $\xi_*=1.02$ produces optimal results for $n\approx 0.5$. To summarize, the 5-term approximation given in \cite{linkov_2013}, complemented with the approximated value of the self-similar crack opening $\xi_*(n)$ provided by formula \eqref{xi_n}, can be considered a very good benchmark for the PKN problem with non-Newtonian fluid.

Recently, two new analytical solutions of the PKN problem have been proposed in \cite{fareo_mason_2013}  for impermeable rock. One of them refers to the case when the fracture volume does not change with time for a zero injection flux rate (shut-in regime). We will not analyze this benchmark as it has no  relevance to the real (practical) situation. However, it demonstrates that there exists a set of nontrivial solutions to the problem if the initial crack opening is different from zero.

The second benchmark solution proposed in \cite{fareo_mason_2013} assumes that the particle velocity is constant along the entire fracture: $v(t,x)=v_0(t)=L'(t)$. Below we present a reproduction of this solution in terms of the self-similar formulation of type \eqref{psi_def}$_2$ for $\gamma=1/(n+2)$. It can be proved that the following set of equations constitutes a solution to the considered self-similar problem:
\begin{equation}
\label{w_n_bench}
\hat w(x)= \hat w_0 (1-x)^{1/(n+2)},\quad \hat w_0=(n+2)^{-\frac{1}{n(n+2)}}
\end{equation}
\begin{equation}
\label{v_n_bench}
\hat v \equiv \hat v_0 = (n+2)^{-\frac{n+1}{n^2}}.
\end{equation}
Note that this solution can be directly obtained from the analytical solution given in \cite{linkov_2013}.

\begin{figure}[h!]
	\begin{center}
		\hspace{5mm}
    \includegraphics [scale=0.4]{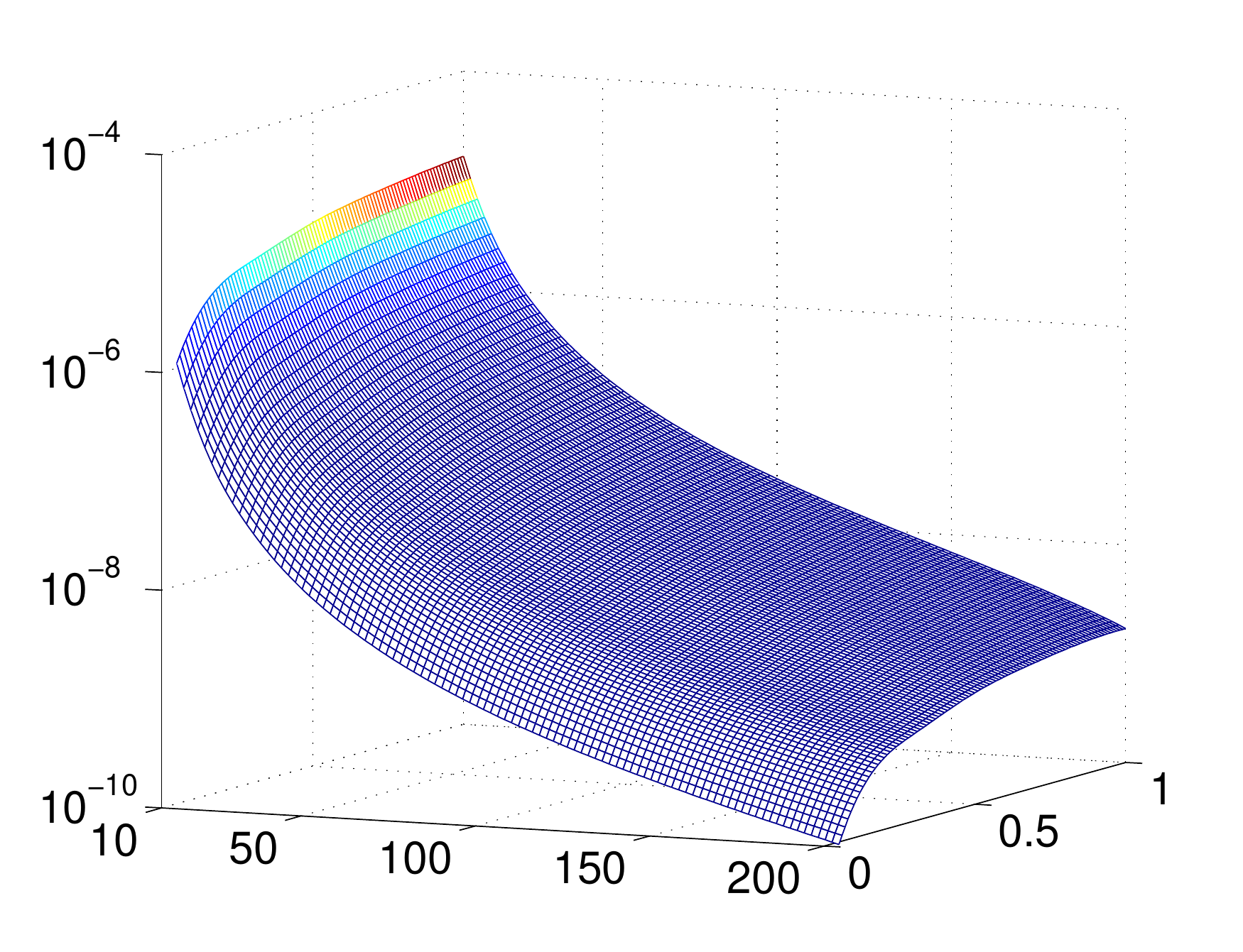}
		\put(-150,0){$N$}
		\put(-35,5){$n$}
    \put(-210,80){$\delta \hat w_{av}$}
    \put(-230,140){$\textbf{a)}$}
    \hspace{6mm}
    \includegraphics [scale=0.4]{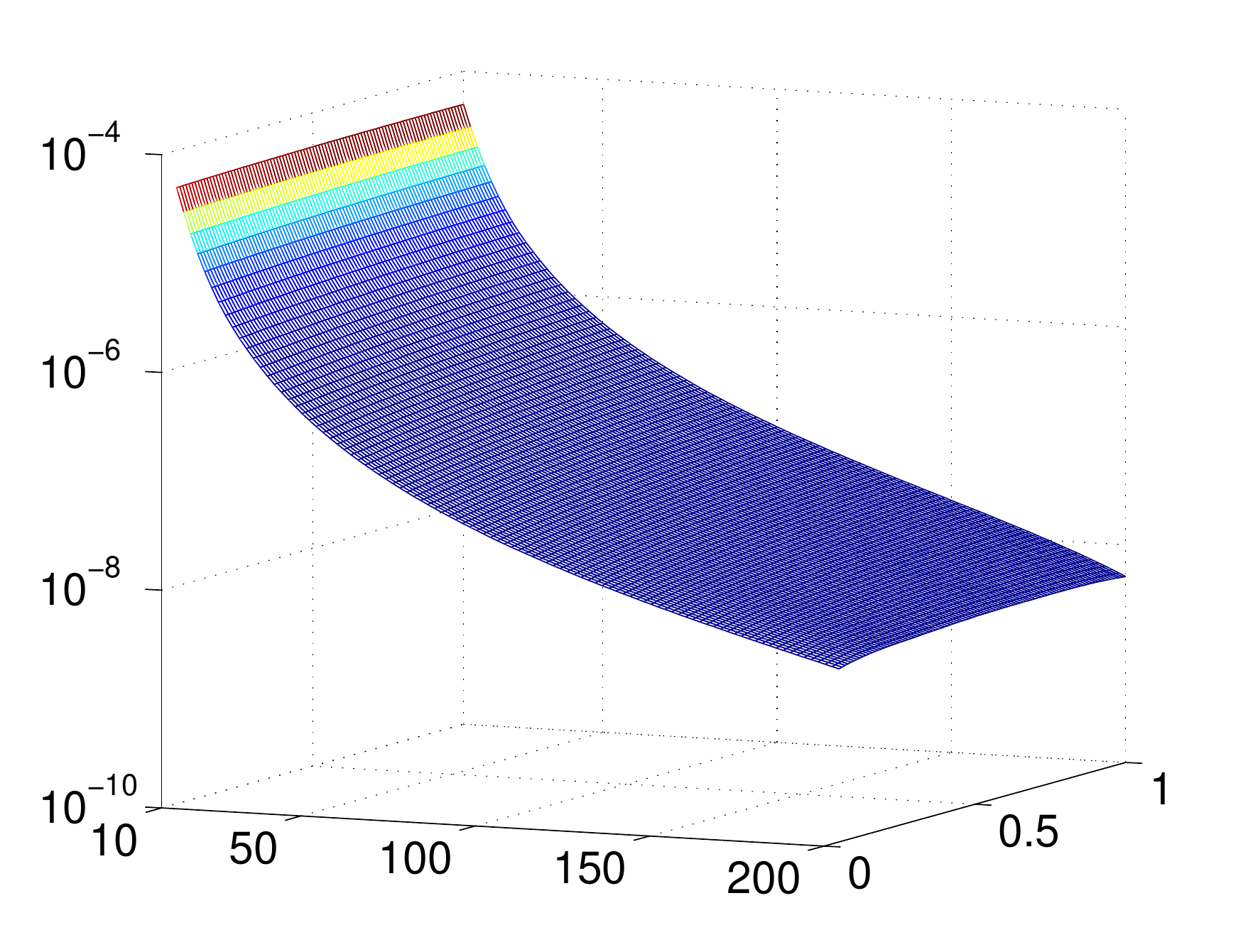}
		\put(-150,0){$N$}
		\put(-35,5){$n$}
		\put(-210,80){$\delta \hat v_{av}$}
    \put(-230,140){$\textbf{b)}$}
		\end{center}

    \caption{PKN model -- verification of accuracy of the computations by the universal algorithm against the analytical benchmark \eqref{w_n_bench} -- \eqref{v_n_bench}: average relative error of the numerical solution with different numbers of nodal points, $N$, and values of the fluid behaviour index, $n$: a) the crack opening $\hat w$, b) the particle velocity $\hat v$.  }

\label{av_rel_PKN_stale_V}
\end{figure}

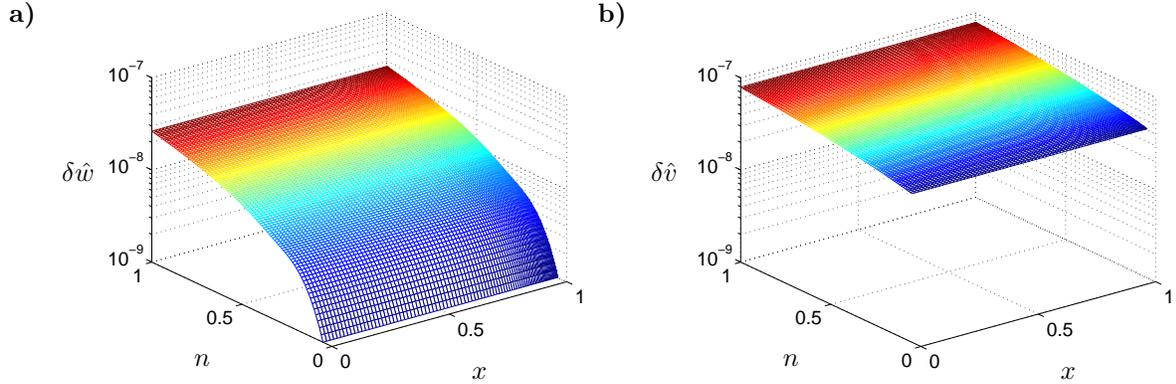
\begin{figure}[h!]
	\begin{center}
		\hspace{5mm}
    \includegraphics [scale=0.4]{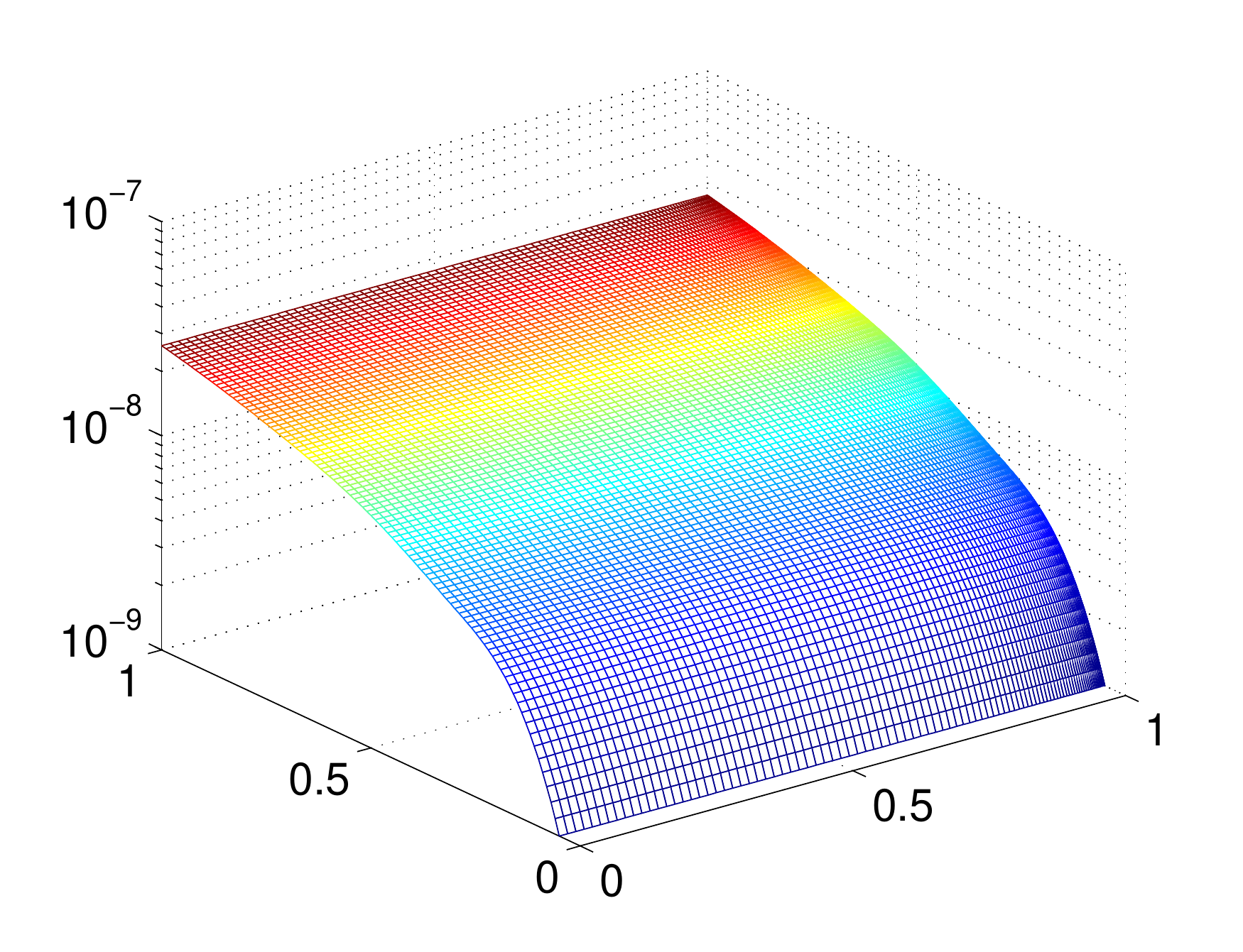}
		\put(-160,10){$n$}
		\put(-55,5){$x$}
    \put(-210,80){$\delta \hat w$}
    \put(-230,140){$\textbf{a)}$}
    \hspace{6mm}
    \includegraphics [scale=0.4]{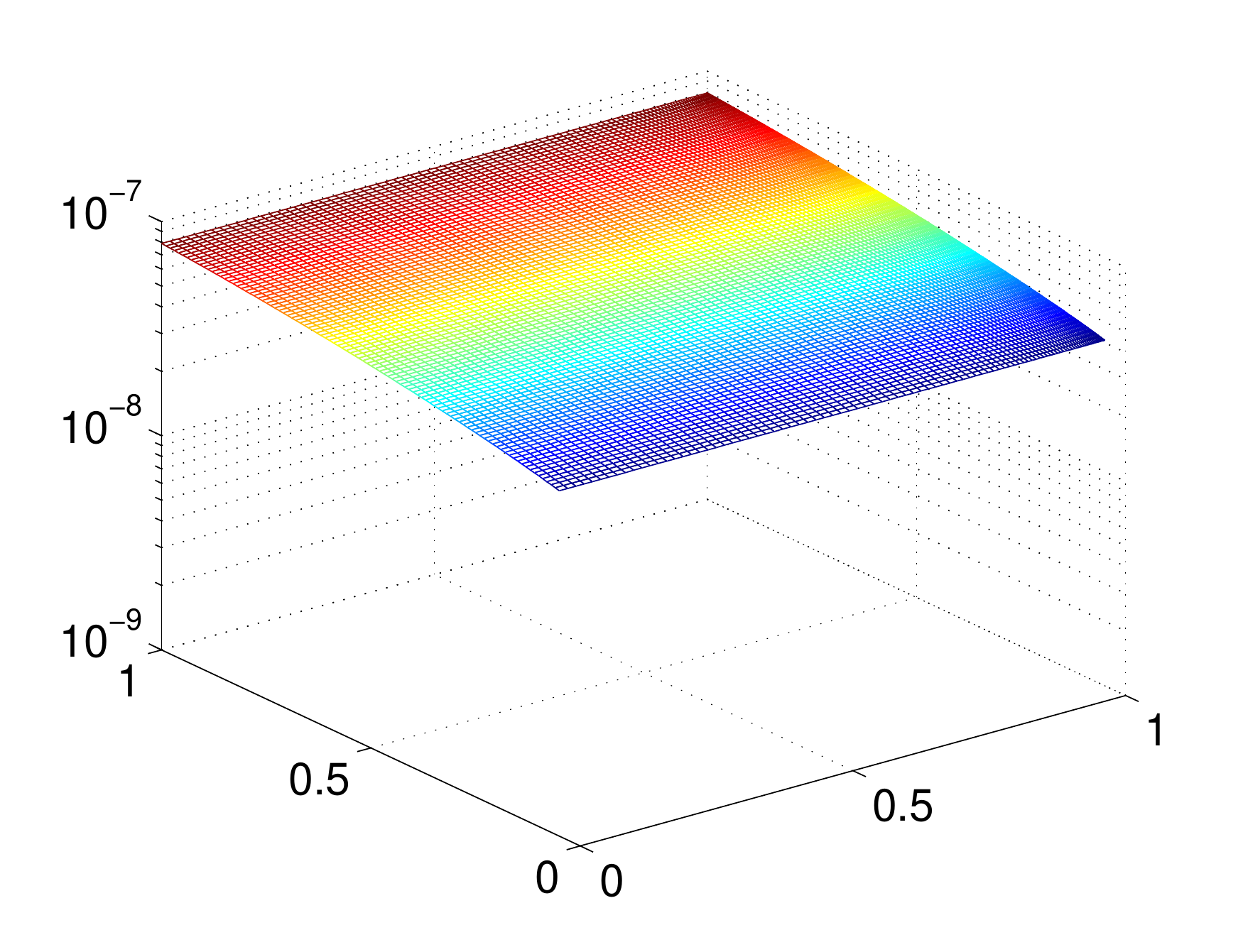}
		\put(-160,10){$n$}
		\put(-55,5){$x$}
		\put(-210,80){$\delta \hat v$}
    \put(-230,140){$\textbf{b)}$}
		\end{center}

    \caption{PKN model -- verification of accuracy of the computations by the universal algorithm against the analytical benchmark \eqref{w_n_bench} -- \eqref{v_n_bench} for $N=100$: average relative error of: a) the crack opening $\hat w$, b) the particle velocity $\hat v$. }

\label{rel_error_PKN_stale_V}
\end{figure}

Having this benchmark  we can  now conduct a similar accuracy analysis to that presented in the previous section. In Fig. \ref{av_rel_PKN_stale_V} -- \ref{rel_error_PKN_stale_V} we depict the average relative  error, over $x$, of the crack opening and the particle velocity for $10\leq N \leq 200$. It shows that the error level is hardly sensitive to the value of $n$. In the analyzed range of $N$ no accuracy saturation is observed. Although the general trends in error distribution are similar to those reported for the non-zero leak-off benchmark in Fig. \ref{av_rel_PKN}, the overall error level is slightly higher. This allows one to conclude that the accuracy of the computations by the universal algorithm is related to the type of problem being analyzed. We believe that this feature is not unique to the proposed scheme, as similar trends were reported in \cite{kus_mis_wr_2013,Lecampion_Brisbane,wr_mis_2013}.

Fig. \ref{rel_error_PKN_stale_V} illustrates the distribution of the relative error of $\hat w$ and $\hat v$ over the spatial interval for $N=100$ with different values of the fluid behaviour index $n$. For all analyzed $n$ we see an almost constant error over the whole range of $x$. No error magnification at the crack tip (as in Fig. \ref{rel_error_PKN}) is observed. Indeed, in this case $\delta \hat w_0$ and $\delta \hat v_0$ define the error level over $x$ (compare \eqref{w_n_bench} and \eqref{v_n_bench}). Compared to the results presented in Fig. \ref{rel_error_PKN}, the level of $\delta \hat v$ remains the same while $\delta \hat w$ is reduced by one order of magnitude. Interestingly, a pronounced reduction in $\delta \hat w$ takes place for $n$ approaching zero. We believe that this fact further confirms the good performance and stability of our scheme, as the analyzed benchmark degenerates for the perfectly plastic fluid.


\subsubsection{KGD model in viscosity dominated regime}
\label{sec:KGD_f_ss_2}

Let us now consider a reference solution given in \cite{ad_det_2002}. The authors analyzed the viscosity dominated regime of the KGD model for a number of shear-thinning fluids. They also assumed a constant influx and impermeability of the rock formation ($q_l=0$). Semi-analytical solutions to the problem were proposed for a variety of shear-thinning fluids ($n=0,0.1,0.3,0.5,0.7,0.9,1$).

In the following we compare our numerical results with those given in the recalled paper in terms of: i) self-similar crack opening, ii) self-similar fluid pressure, iii) self-similar fluid flow rate, iv) self-similar particle velocity. Although in \cite{ad_det_2002} there is no data for the particle velocity, it can be easily obtained through the fluid flow rate and the fracture opening ($\hat v=\hat q/ \hat w$). As for results obtained using the universal algorithm, the particle velocity is retrieved directly from the reduced particle velocity, while computing the fluid pressure requires additional postprocessing (integration). The predefined influx value was taken from the data given in \cite{ad_det_2002} (p.591 Table I).

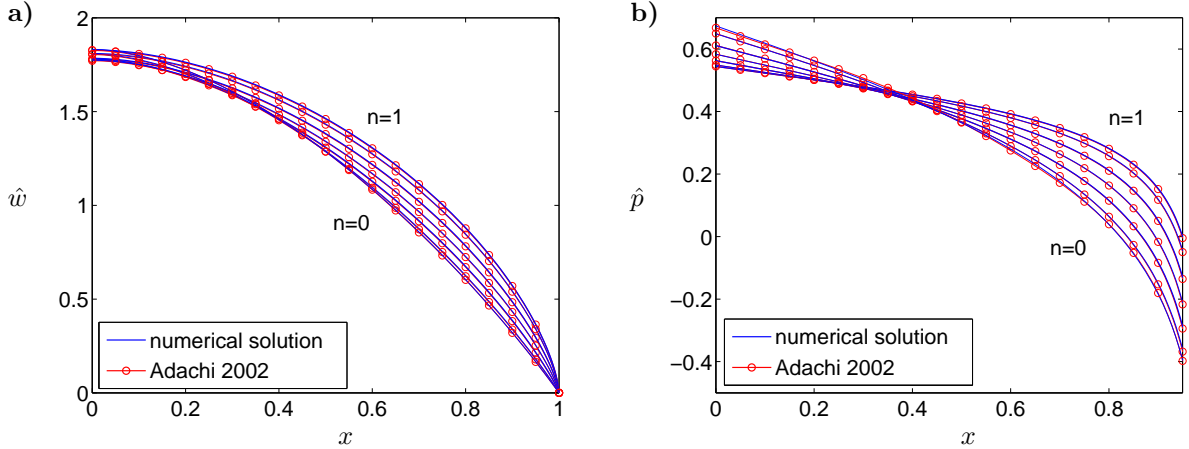
\begin{figure}[h!]

    \includegraphics [scale=0.45]{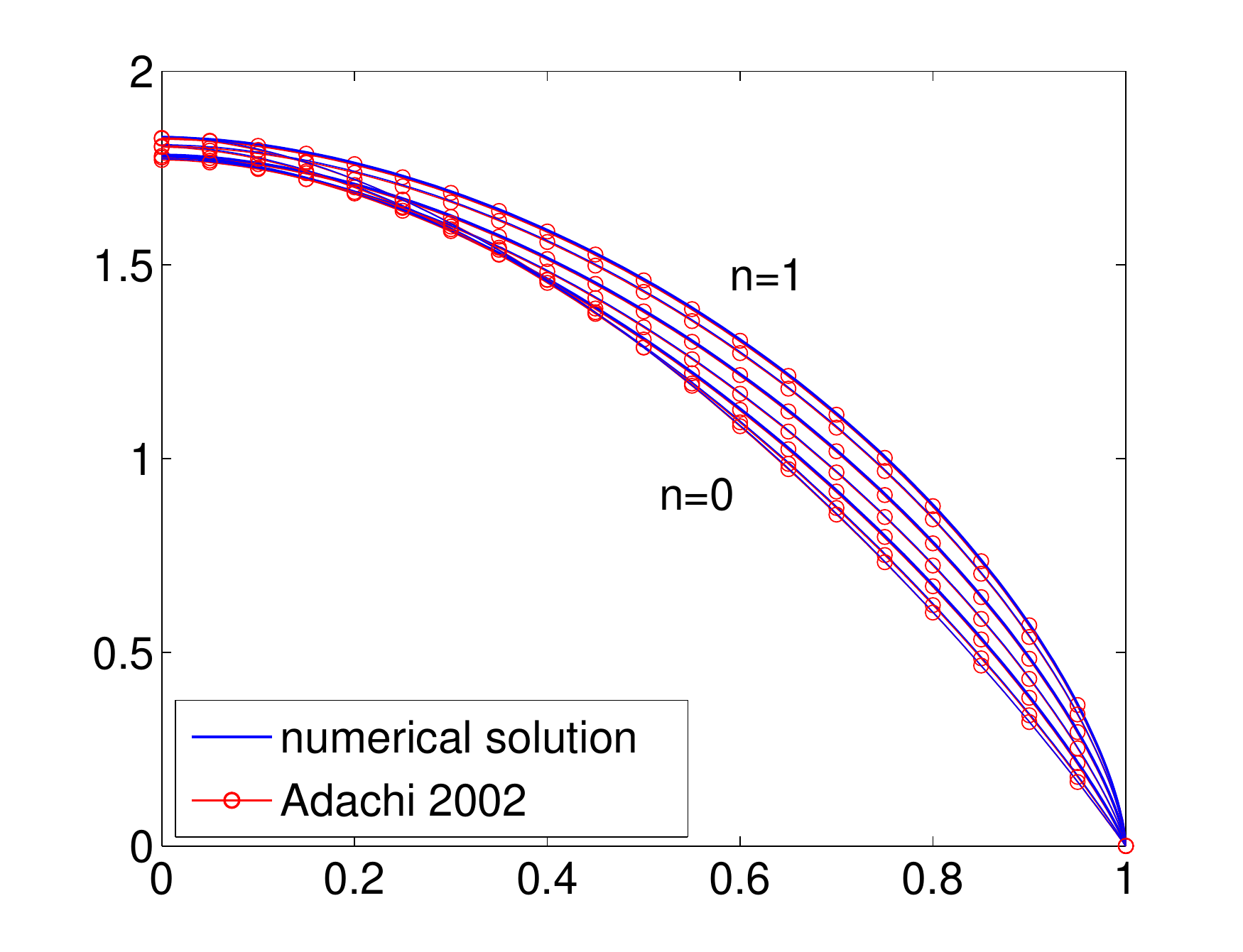}
    \put(-105,0){$x$}
    \put(-230,90){$ \hat w$}
    \put(-230,160){$\textbf{a)}$}
    \hspace{2mm}
    \includegraphics [scale=0.45]{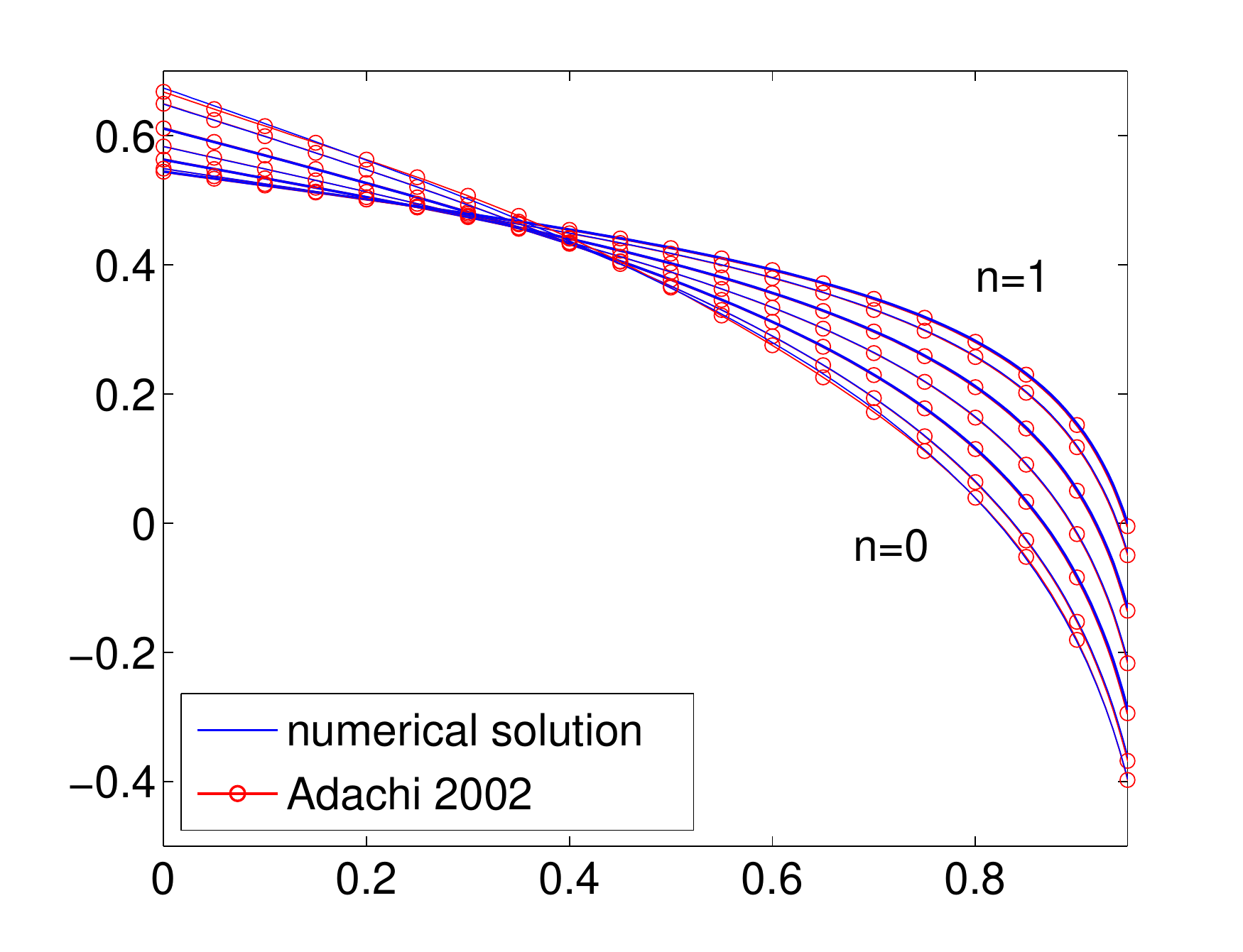}
    \put(-105,0){$x$}
    \put(-230,90){$\hat p$}
    \put(-230,160){$\textbf{b)}$}

    \caption{KGD model (viscosity dominated regime) -- comparison of the semi-analytical solution presented in \cite{ad_det_2002} with our numerical solution for $N=100$: a) the crack opening $\hat w$, b) the fluid pressure $\hat p$.}

\label{wyniki_Ad_wp}
\end{figure}

\begin{figure}[h!]

    \includegraphics [scale=0.45]{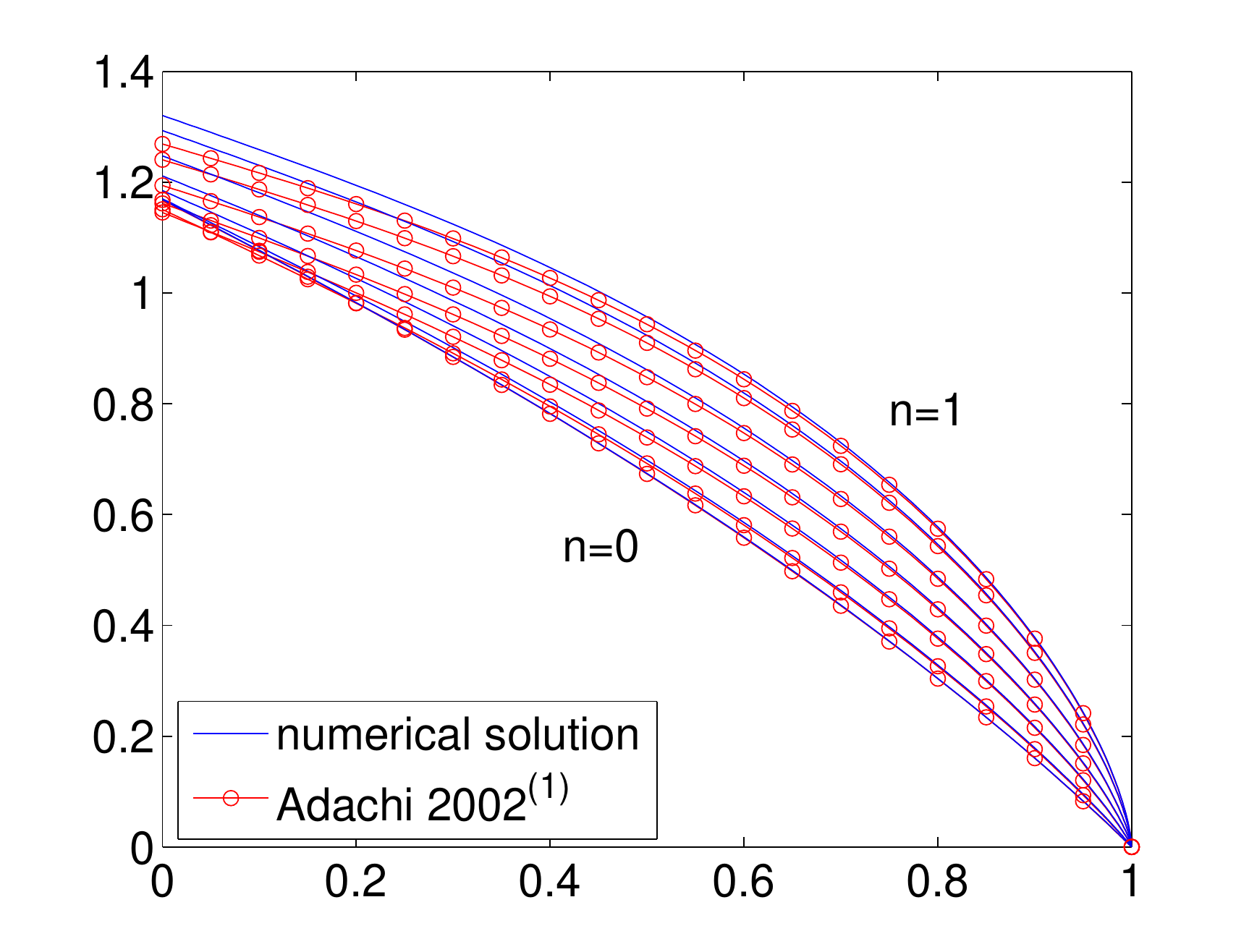}
    \put(-105,0){$x$}
    \put(-230,90){$ \hat q^{(1)}$}
    \put(-230,160){$\textbf{a)}$}
    \hspace{5mm}
		\includegraphics [scale=0.45]{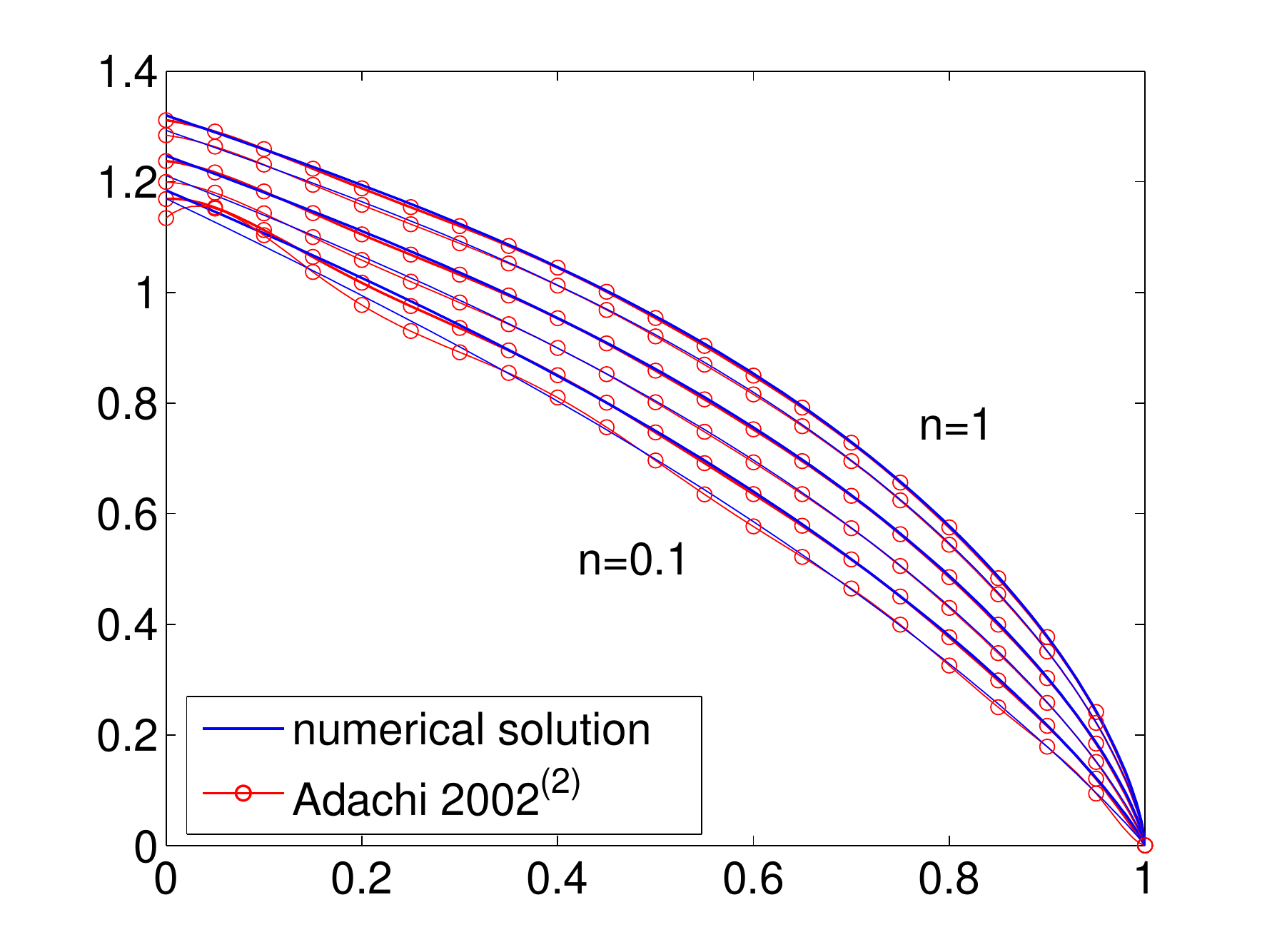}
		\put(-105,0){$x$}
    \put(-230,90){$\hat q^{(2)}$}
    \put(-230,160){$\textbf{b)}$}
		
    \caption{KGD model (viscosity dominated regime) -- comparison of the semi-analytical solution presented in \cite{ad_det_2002} with our numerical solution for $N=100$: the fluid flow rate $\hat q$.}

\label{wyniki_Ad_q}
\end{figure}

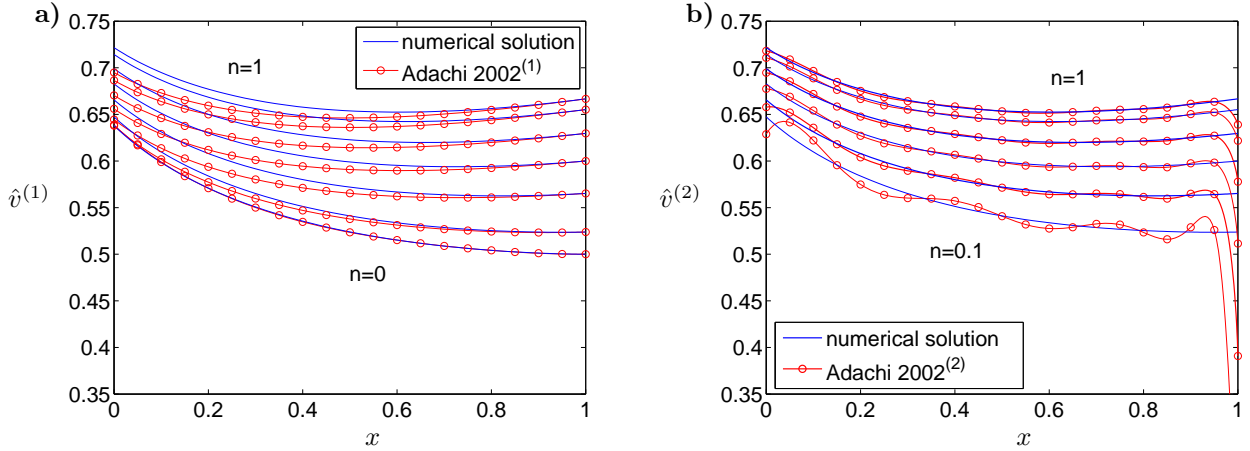
\begin{figure}[h!]

    \includegraphics [scale=0.45]{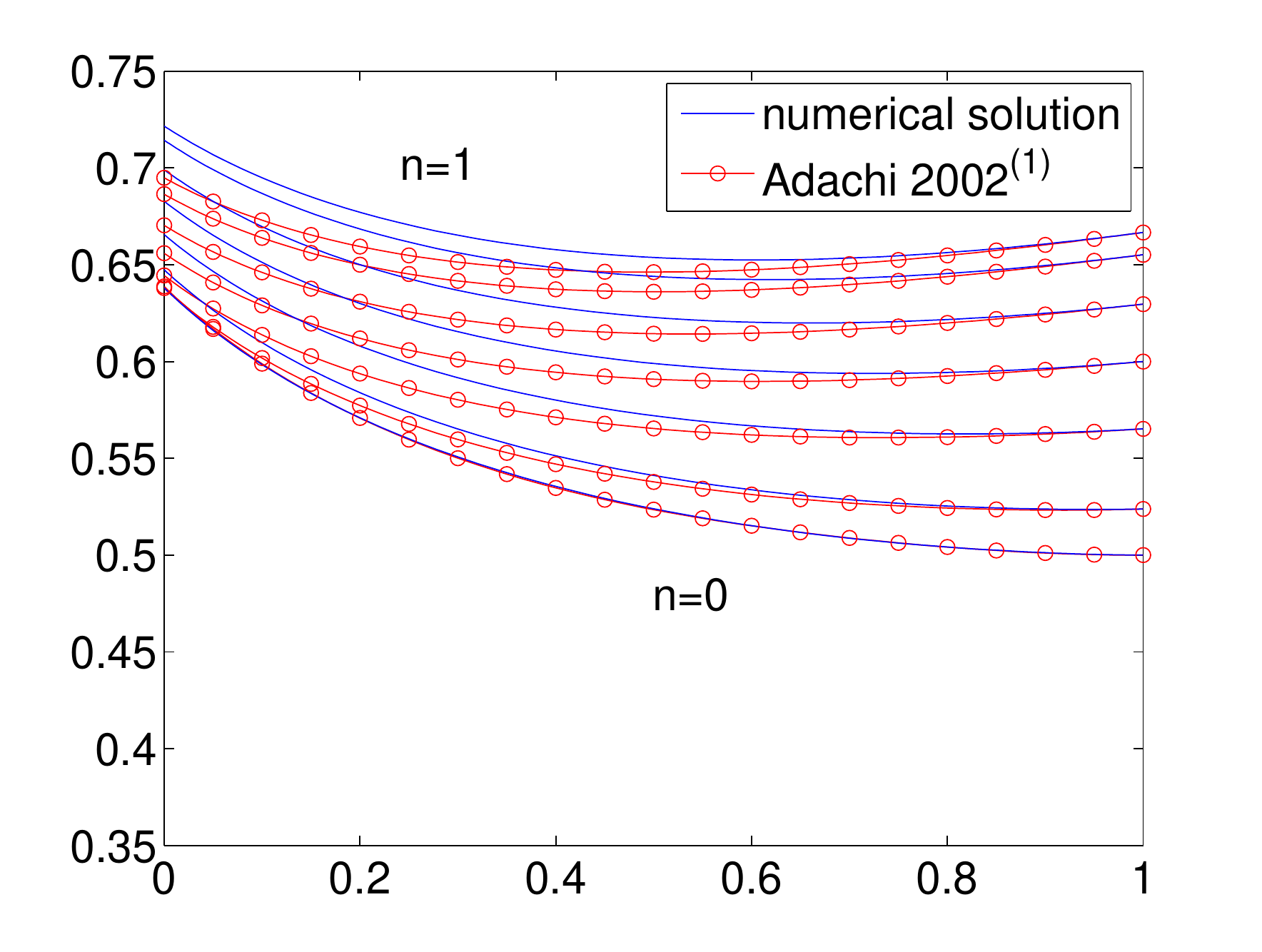}
    \put(-105,0){$x$}
    \put(-240,90){$ \hat v^{(1)}$}
    \put(-230,160){$\textbf{a)}$}
    \hspace{5mm}
    \includegraphics [scale=0.45]{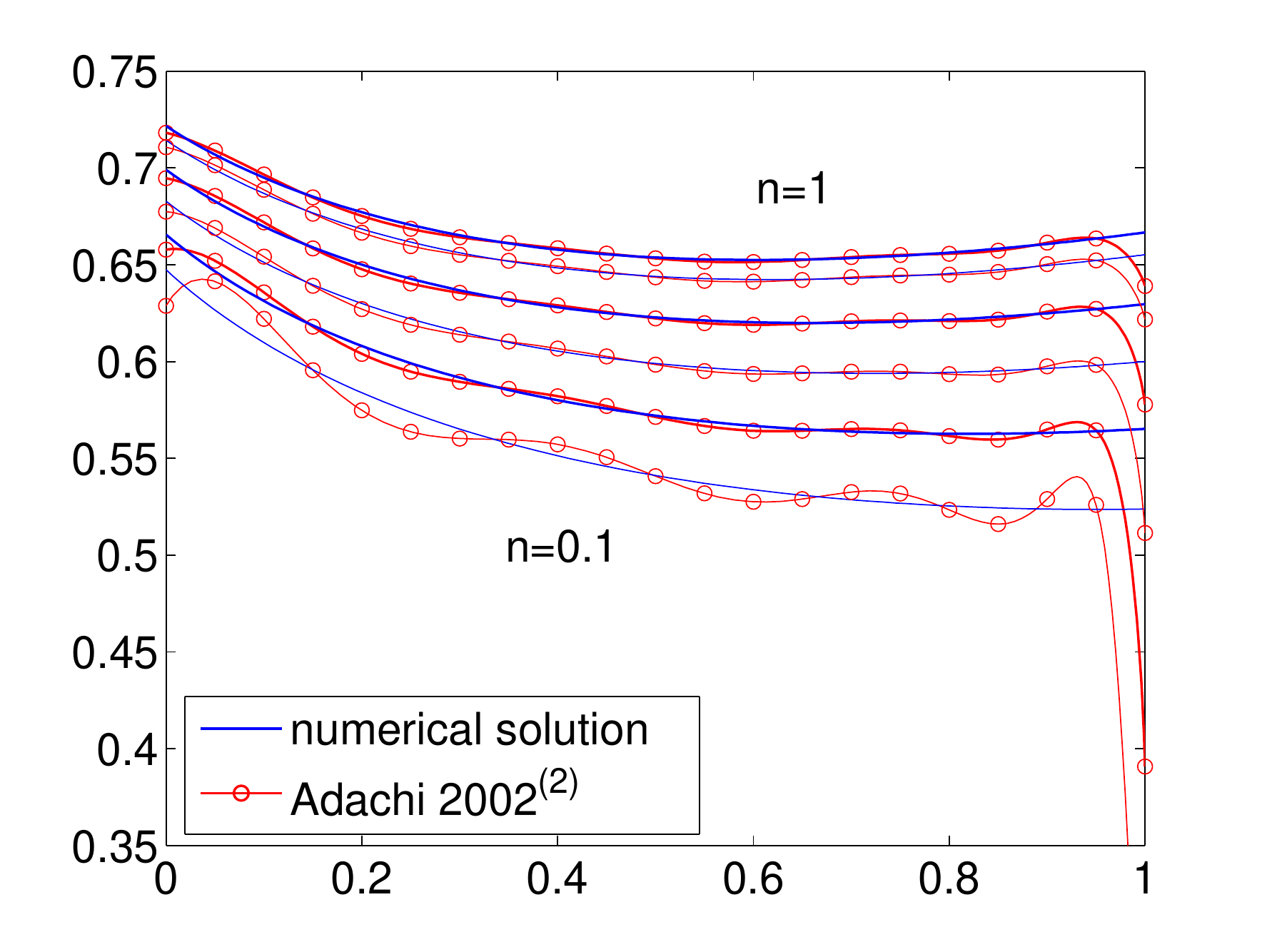}
    \put(-105,0){$x$}
    \put(-240,90){$\hat v^{(2)}$}
    \put(-230,160){$\textbf{b)}$}
		
    \caption{KGD model (viscosity dominated regime) -- comparison of the semi-analytical solution presented in \cite{ad_det_2002} with our numerical solution for $N=100$: the particle velocity $\hat v$.}

\label{wyniki_Ad_v}
\end{figure}

Our computations were carried out on an adaptive mesh composed of $200$ nodes, which according to the conducted accuracy analysis should produce errors of order $10^{-7}-10^{-8}$. Comparison of the numerical results with those from \cite{ad_det_2002} is shown in Fig. \ref{wyniki_Ad_wp} -- \ref{wyniki_Ad_v}.

As there are two ways of calculating the fluid flow rate $\hat q$ in \cite{ad_det_2002} -- either by  a sum of the base functions ($\bar \Psi_{\text{m} 0}$ in equation (41) in the aforementioned paper) or using $\hat q=\hat w \hat v$ (where $\hat v$ is computed from \eqref{v_ss}), we shall analyze both. From now on we will refer to these methods as $\hat q^{(1)}$ and $\hat q^{(2)}$, respectively. Pertaining formulae for the particle velocity are defined as: $\hat v^{(1)}=\hat q^{(1)}/\hat w$, while $\hat v^{(2)}$ is computed according to \eqref{v_ss}. Note that for the exact solution obtained using both computational methods should produce the same result. For the perfectly plastic fluid ($n=0$) neither $\hat q^{(2)}$ nor $\hat v^{(2)}$ can be obtained, as equation \eqref{v_ss} degenerates to yield \eqref{n0_special_case} and thus it can no longer be used to define the fluid flow rate $\hat q$ or the particle velocity $\hat v$ as it was explained in Section \ref{n_0_section}. Therefore the respective curves are not presented in Fig. \ref{wyniki_Ad_q}b and \ref{wyniki_Ad_v}b.

As can be seen for the crack opening and fluid pressure, the curves corresponding to the universal algorithm and to the solution obtained by \cite{ad_det_2002} are indistinguishable from one another in the presented scale (the $x$ interval in Fig. \ref{wyniki_Ad_wp}b was truncated because the pressure tends to infinity as $x\to 1$). When one analyzes the particle velocity it turns out that the values given by the universal algorithm are in good agreement with $\hat v^{(2)}$ over almost the entire interval. A distinct divergence in results can be observed in the near-tip region, where the quality of approximation $\hat v^{(2)}$ apparently deteriorates. On the other hand, when retrieving the particle velocity from the data given in \cite{ad_det_2002} as $\hat v^{(1)}$, we have very good coincidence in the vicinity of the fracture front, with the results diverging from each other as the crack inlet is approached. In general, better coincidence of $\hat v^{(1)}$ with our accurate numerical result for $\hat v$ takes place for lower values of fluid behaviour index $n$, while a reverse relation is observed for $\hat v^{(2)}$. A similar situation to that reported for $\hat v$ has been revealed for the fluid flow rate. Again, $\hat q^{(2)}$ gives better results away from the crack tip, while $\hat q^{(1)}$ produces better coincidence in the near-tip region.

The above analysis confirms the credibility of our solution, which together with the previous accuracy estimation allows us to now treat it as the numerical benchmark data. Following the idea from \cite{ad_det_2002} and \cite{linkov_2012}, we propose a new improved approximation of the dependent variables analyzed above. It provides a solution to the problem with far greater accuracy than any other known semi-analytical formulae and thus can be treated as the reference data itself when testing computational algorithms for hydraulic fractures.
The new solution approximation presented below is valid for any value of the fluid behaviour index. Let the fracture opening, $\hat w$, the fluid pressure, $\hat p$, and the particle velocity, $\hat v$, be expressed as:
\begin{equation}
\label{w_Ad_ap}
\hat w(x)=\hat w_0 (1-x^2)^{\frac{2}{n+2}}+\hat w_1 (1-x^2)^{4/3}+\hat w_2\left[\sqrt{1-x^2}-\frac{2}{3}(1-x^2)^{3/2}-x^2\ln \frac{1+\sqrt{1-x^2}}{x}\right],
\end{equation}
\begin{equation}
\label{p_Ad_ap}
\hat p(x)=\frac{\hat w_0 B\left(\frac{1}{2},\frac{2}{n+2}\right)}{(n+2)\pi} {_2}F_1\left(\frac{1}{2}-\frac{2}{n+2},1;1/2;x^2 \right)+\sum_{i=1}^6\hat p_i(1-x)^{(6-i)/2},
\end{equation}
\begin{equation}
\label{v_Ad_ap}
\hat v(x)=\frac{n+1}{n+2}+ \sum_{i=1}^5\hat v_i(1-x)^i,
\end{equation}
while the respective approximation of the fluid flow rate, $\hat q$, can be easily obtained from the product of the fracture opening \eqref{w_Ad_ap} and the
particle velocity \eqref{v_Ad_ap}.

In the equations above, $B$ is the beta function, $_2F_1$ denotes the Gauss hypergeometric function, $C_{{\cal A}}$ is as expressed in \eqref{LC_KGD_fluid} and
$\hat w_0$ is computed using the formula ($0\le n\leq 1$):
\begin{equation}
\label{w0_aproks}
\hat w_0=\left(\frac{n+1}{n+2}\right)^{n/(n+2)} C_{{\cal A}}^{-1/(n+2)}.
\end{equation}
Coefficients $\hat w_i(n)$, $\hat p_i(n)$ and $\hat v_i(n)$ are functions of the fluid behaviour index $n$. Their values are collected in Table \ref{T2} (Appendix B).

\begin{figure}[h!]

    \includegraphics [scale=0.45]{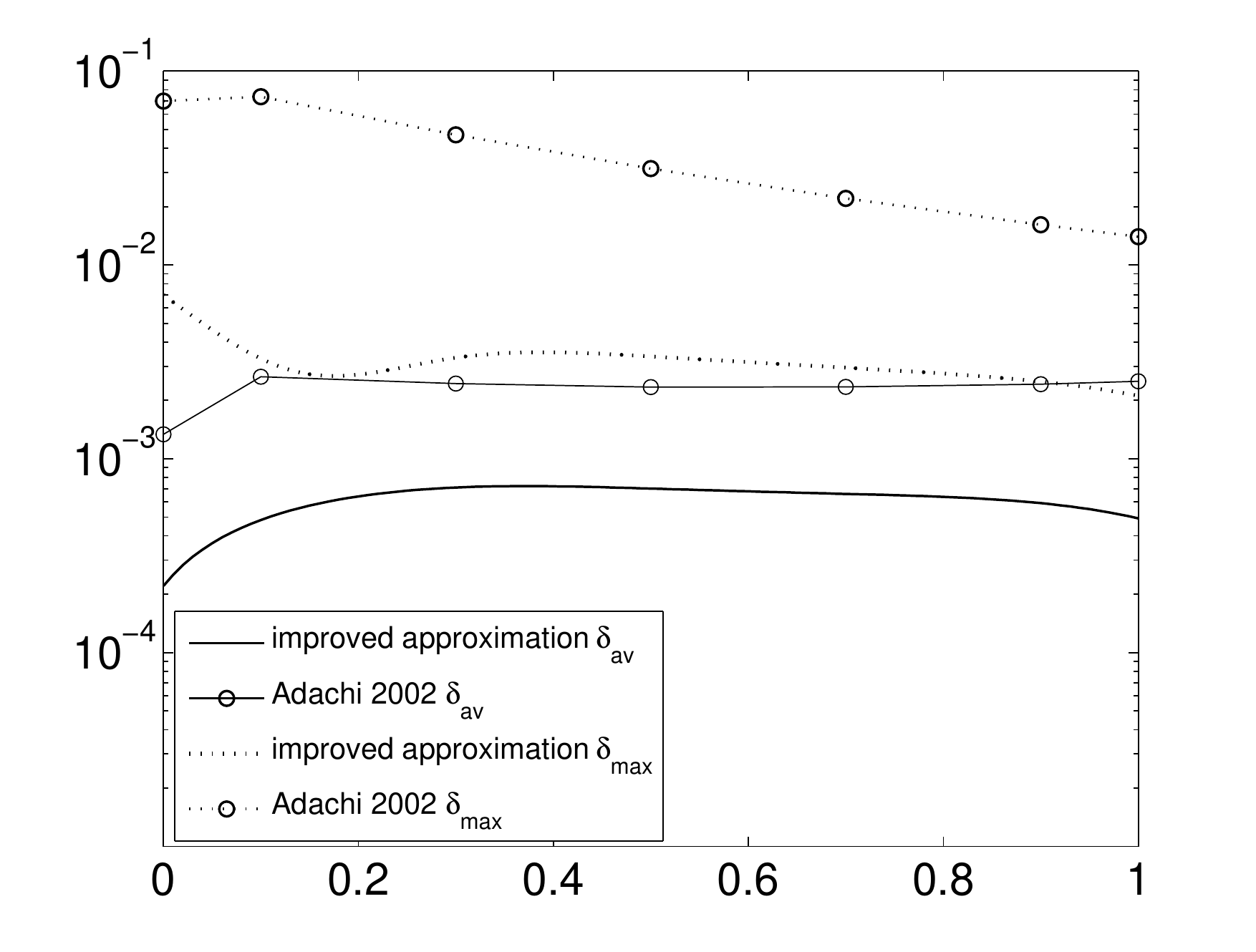}
    \put(-105,0){$n$}
    \put(-230,90){$ \delta \hat w$}
    \put(-230,160){$\textbf{a)}$}
    \hspace{5mm}
    \includegraphics [scale=0.45]{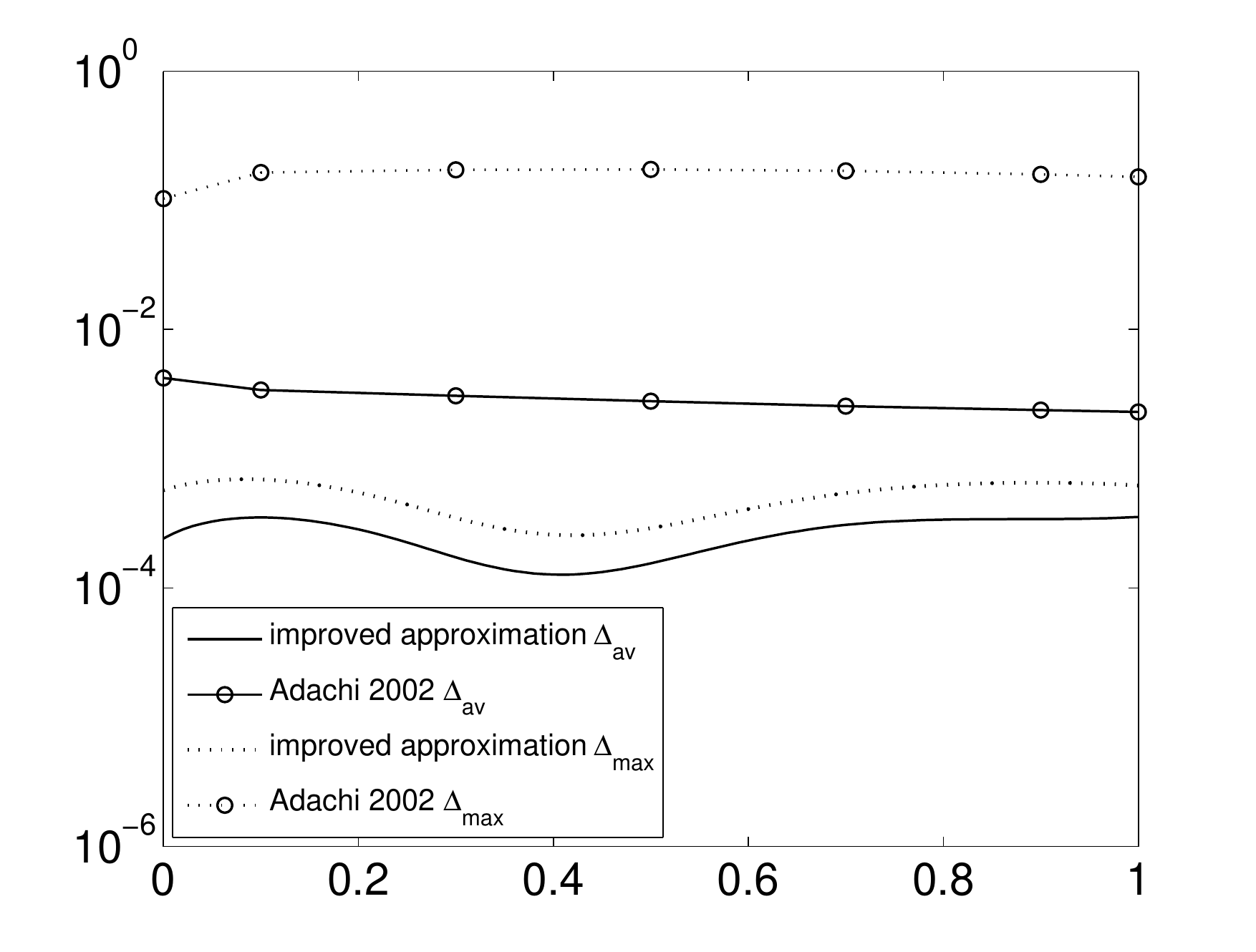}
    \put(-105,0){$n$}
    \put(-230,90){$\Delta \hat p$}
    \put(-230,160){$\textbf{b)}$}

    \caption{KGD model (viscosity dominated regime) -- comparison of the new approximate solution \eqref{w_Ad_ap} -- \eqref{v_Ad_ap} and the semi-analytical solution from \cite{ad_det_2002} with accurate numerical results for $N=100$: a) the average and maximal relative error of the formulas for the crack opening $\delta \hat w_{av}$ and $\delta \hat w_{max}$; b) average and maximal error of the fluid pressure $\Delta \hat p_{av}$ and $\Delta \hat p_{max}$. }

\label{error_av_max_adachi_imp_app_w_p}
\end{figure}
\begin{figure}[h!]

    \includegraphics [scale=0.45]{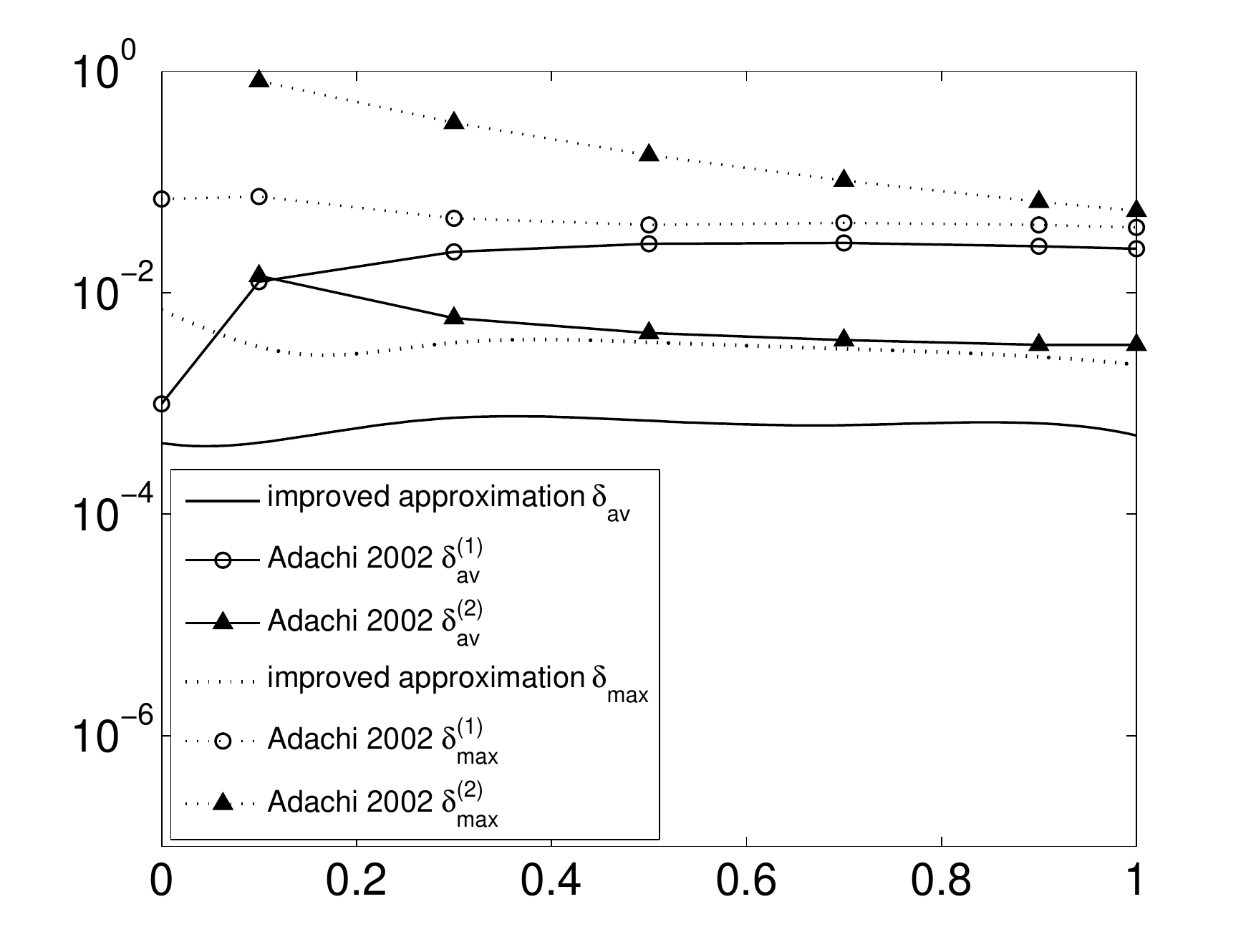}
    \put(-105,0){$n$}
    \put(-230,90){$ \delta \hat q$}
    \put(-230,160){$\textbf{a)}$}
    \hspace{5mm}
    \includegraphics [scale=0.45]{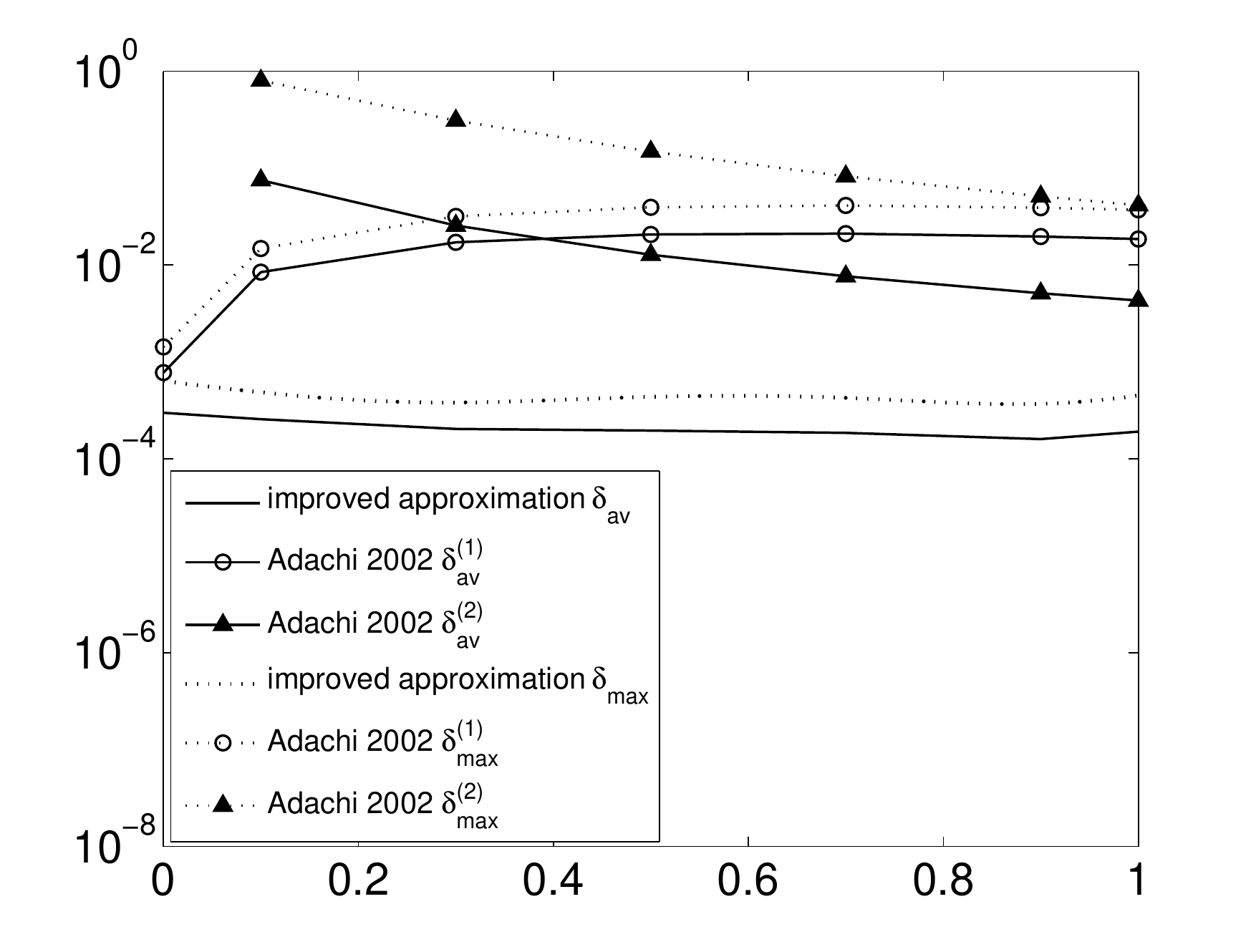}
    \put(-105,0){$n$}
    \put(-230,90){$\delta \hat v$}
    \put(-230,160){$\textbf{b)}$}

    \caption{KGD model (viscosity dominated regime) -- comparison of the new approximate solution \eqref{w_Ad_ap} -- \eqref{v_Ad_ap} and the semi-analytical solution from \cite{ad_det_2002} with accurate numerical results for $N=100$: a) the fluid flow rate $\hat q$, b) particle velocity $\hat v$. Superscript here defines which evaluation formula for the results from \cite{ad_det_2002} has been employed. Both average relative, $\delta f_{av}$,  and maximal relative, $\delta f_{max}$, errors are presented.}

\label{error_av_max_adachi_imp_app_q_v}
\end{figure}
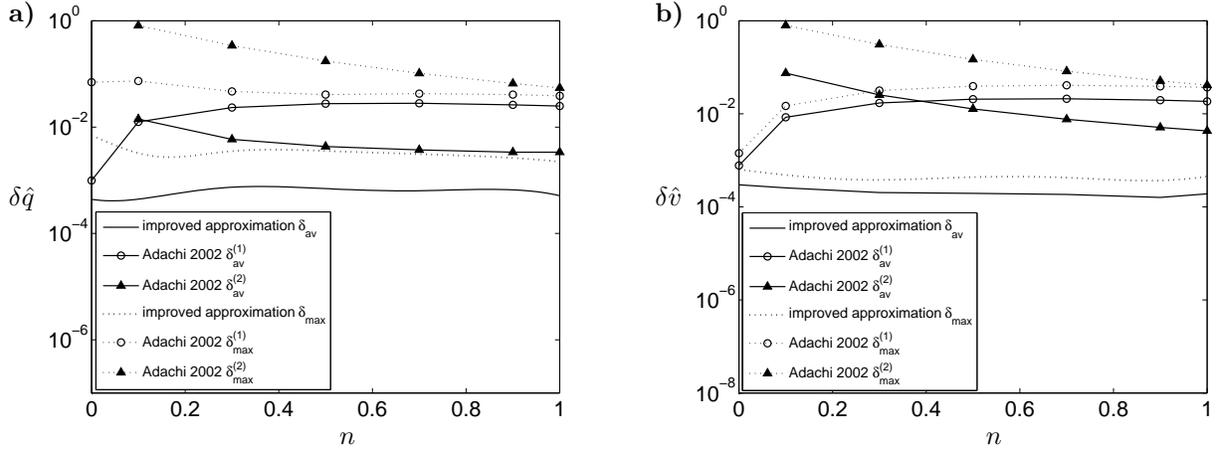

In Fig. \ref{error_av_max_adachi_imp_app_w_p} -- \ref{error_av_max_adachi_imp_app_q_v} we present a comparison of the new approximation \eqref{w_Ad_ap} -- \eqref{v_Ad_ap} and the semi-analytical formulae from \cite{ad_det_2002} with the exact numerical solution. Bearing in mind the accuracy level provided by the universal algorithm we claim that the depicted deviations describe, in fact, the error of the respective approximations. Fig. \ref{error_av_max_adachi_imp_app_w_p}a illustrates the relative deviations of the crack opening computed as the averaged, $\delta_{av}$, and maximal, $\delta_{max}$, values. For the net fluid pressure (Fig. \ref{error_av_max_adachi_imp_app_w_p}b) we show only the absolute deviations (again in the mean and maximal formulations), as the pressure graph intersects the $x$ axis. The deviation of the fluid flow rate and particle velocity are presented in Fig. \ref{error_av_max_adachi_imp_app_q_v}. Here, the data from \cite{ad_det_2002} was computed in two alternative ways: as $\hat q^{(1)}$, $\hat v^{(1)}$ and $\hat q^{(2)}$, $\hat v^{(2)}$.

The first straightforward conclusion that emerges from the presented results is that, regardless of the error measure and analyzed component of the solution, the new approximation gives at least one order of magnitude higher accuracy than the approximation from \cite{ad_det_2002}. Moreover, better stability of the results produced by \eqref{w_Ad_ap} -- \eqref{v_Ad_ap} is observed when varying $n$. When analyzing the graphs for $\delta \hat q$ and $\delta \hat v$ it shows that the quality of the solution from \cite{ad_det_2002} depends essentially on the way in which the fluid flow rate and the particle velocity are computed. In general, the method referred to as $\hat q^{(2)}$, $\hat v^{(2)}$ is better over almost the entire range of $n$, while both approaches are of similar accuracy for $n$ close to unity. However, this conclusion does not reflect the total complexity of the problem. The spatial distribution of respective errors reveals that $\hat q^{(2)}$ and $\hat v^{(2)}$ are much better than $\hat q^{(1)}$ and $\hat v^{(1)}$ over almost the whole crack length, except for the near-tip region where pronounced error magnification takes place. On the other hand, $\hat q^{(1)}$ and $\hat v^{(1)}$ prove to be very accurate when approaching the crack tip and diverge from accurate results with decreasing $x$. These trends can be observed in Fig. \ref{wyniki_Ad_q} -- Fig. \ref{wyniki_Ad_v}. Moreover, exactly the same trend was identified for Newtonian fluid in \cite{wr_mis_2015}. However, if one needs to utilize the approximation given in \cite{ad_det_2002} it seems that the optimal way to do so should be based on appropriate merging of both representations for $\hat q$ and $\hat v$.

\begin{figure}[h!]

    \includegraphics [scale=0.45]{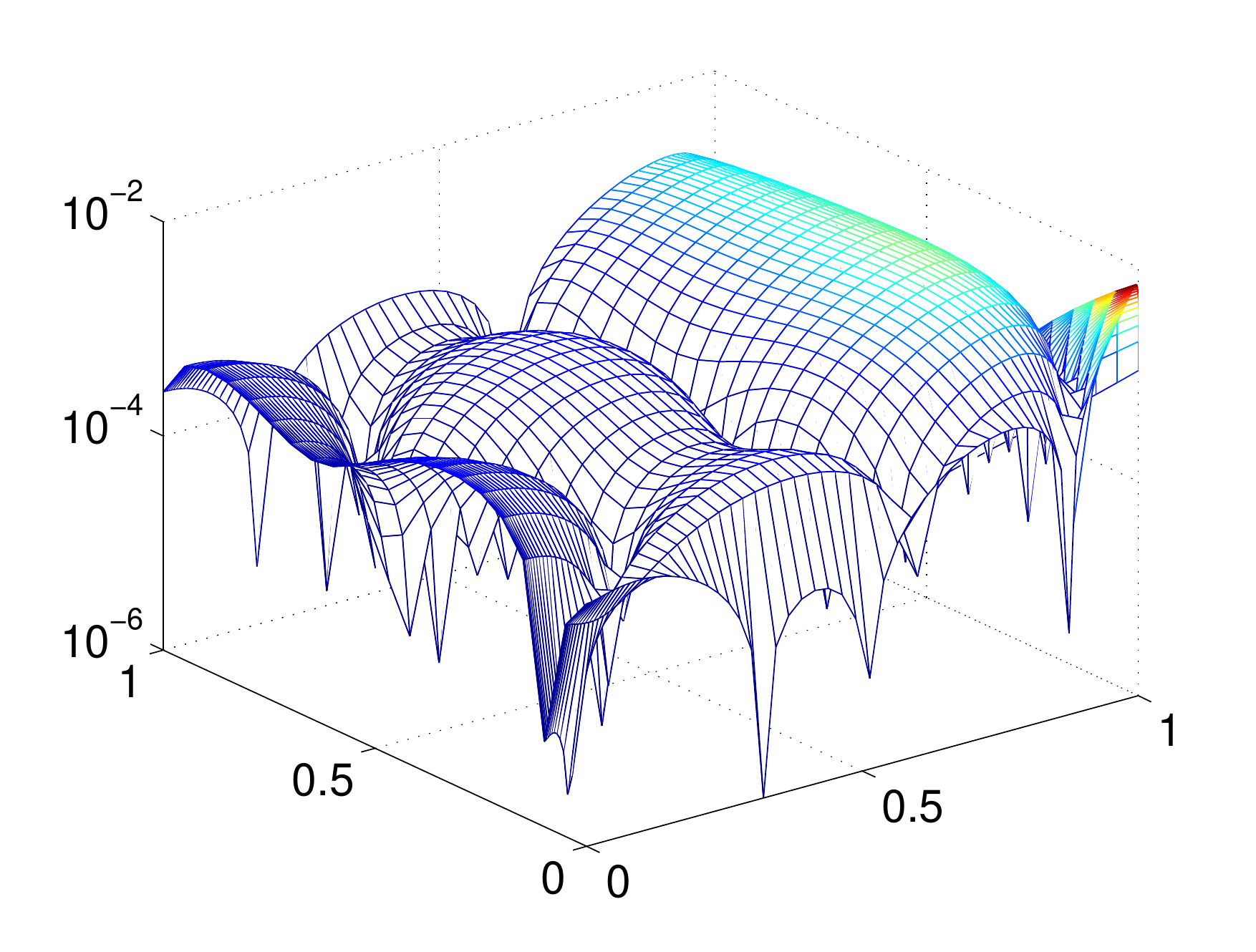}
    \put(-59,10){$x$}
    \put(-180,13){$n$}
    \put(-230,90){$ \delta \hat w$}
    \put(-230,160){$\textbf{a)}$}
    \hspace{2mm}
    \includegraphics [scale=0.45]{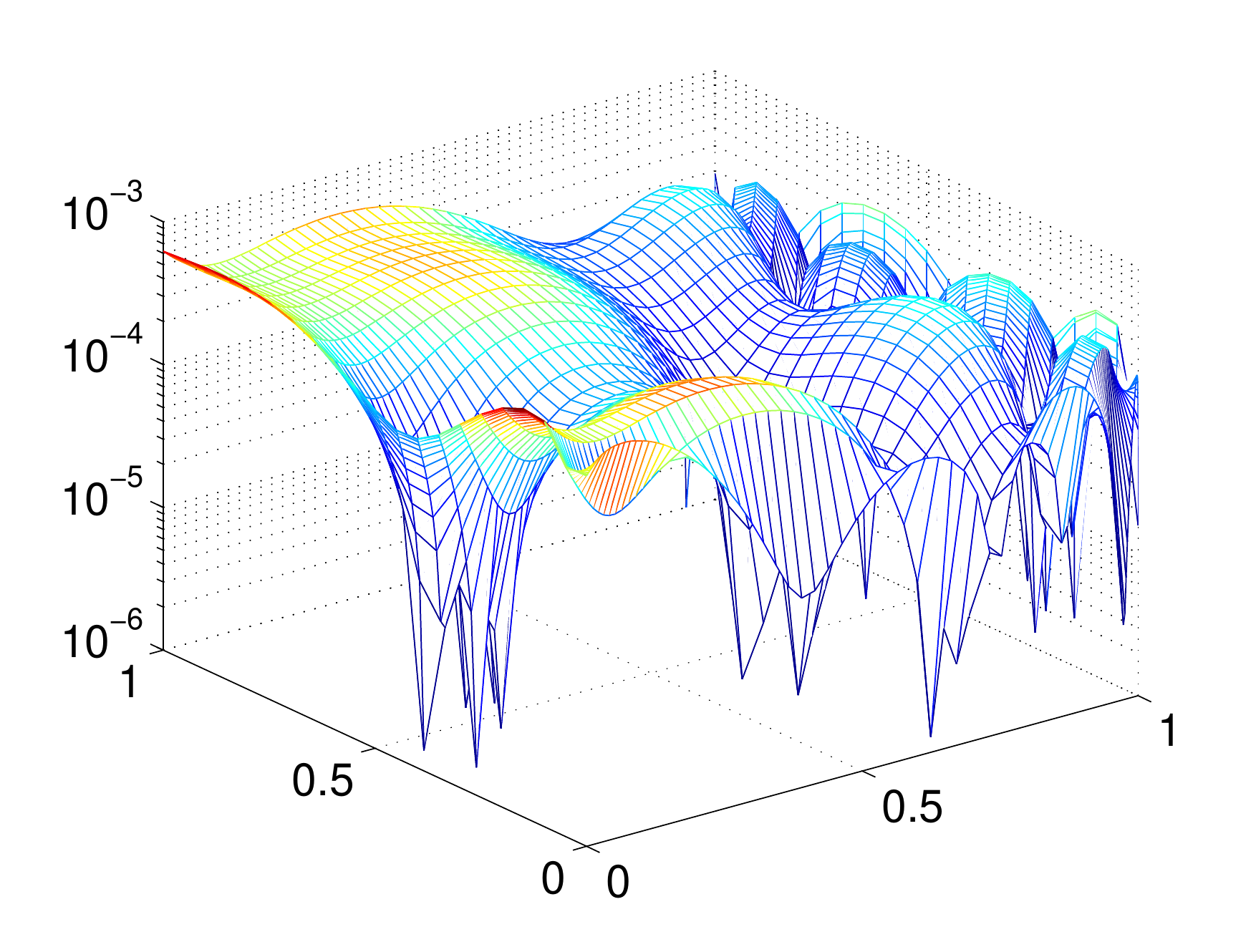}
    \put(-59,10){$x$}
    \put(-180,13){$n$}
    \put(-235,90){$\Delta \hat p$}
    \put(-230,160){$\textbf{b)}$}

    \caption{KGD model (viscosity dominated regime) -- comparison of the new approximate solution \eqref{w_Ad_ap} -- \eqref{v_Ad_ap} with accurate numerical results for $N=100$: a) relative deviation of crack width $\hat w$, b) absolute deviation of fluid pressure $\hat p$.}

\label{imp_app_3d_w_p}
\end{figure}

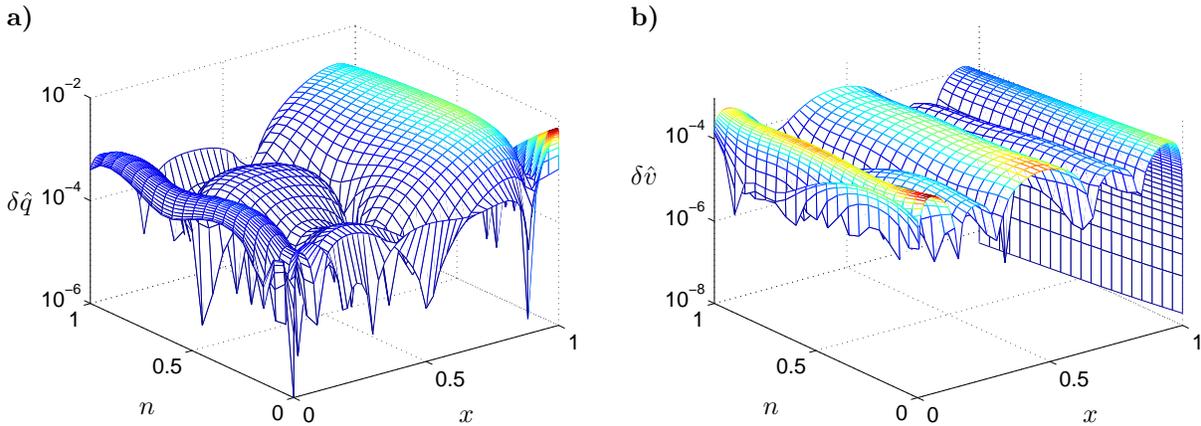
\begin{figure}[h!]

    \includegraphics [scale=0.45]{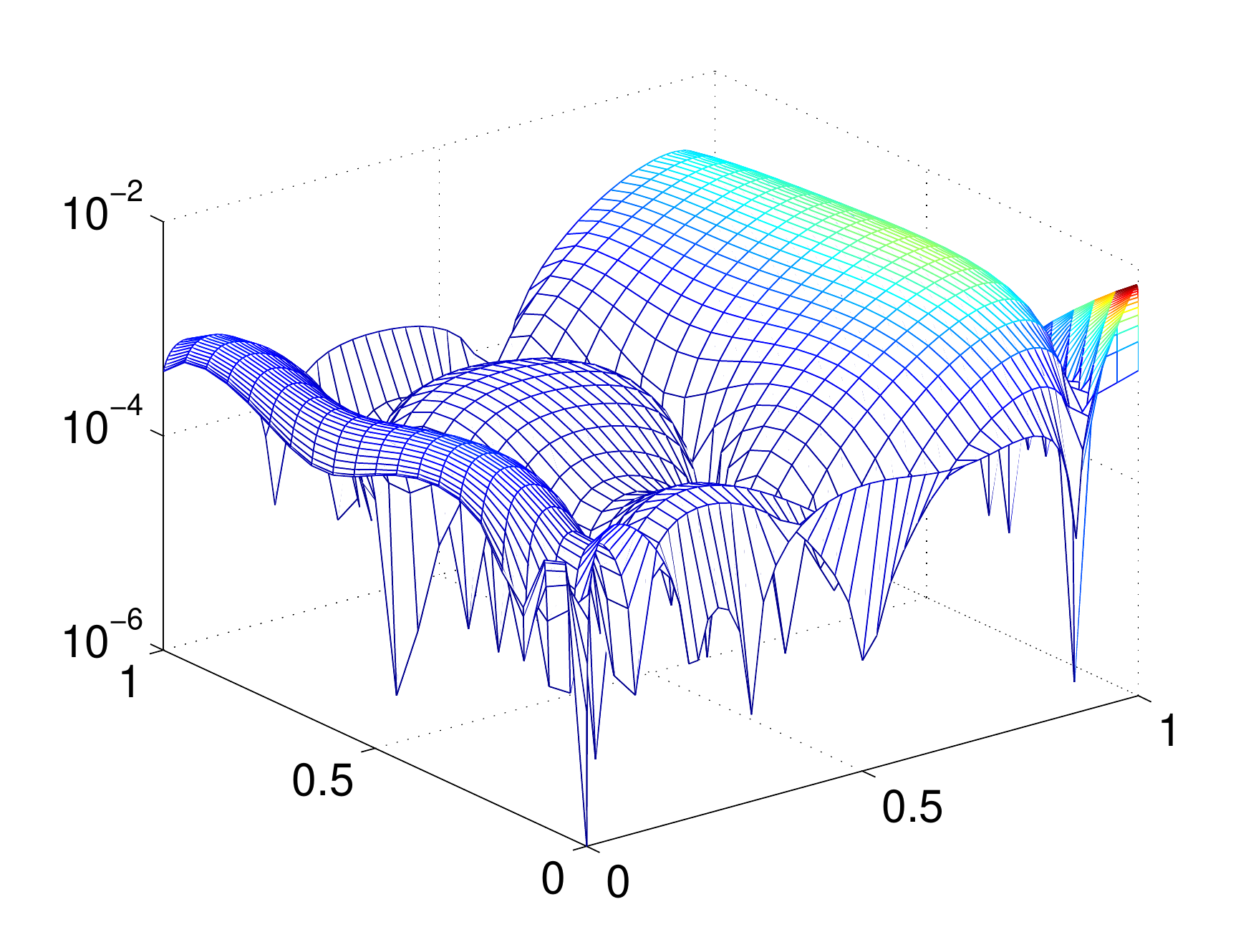}
    \put(-59,10){$x$}
    \put(-180,13){$n$}
    \put(-230,90){$ \delta \hat q$}
    \put(-230,160){$\textbf{a)}$}
    \hspace{2mm}
    \includegraphics [scale=0.45]{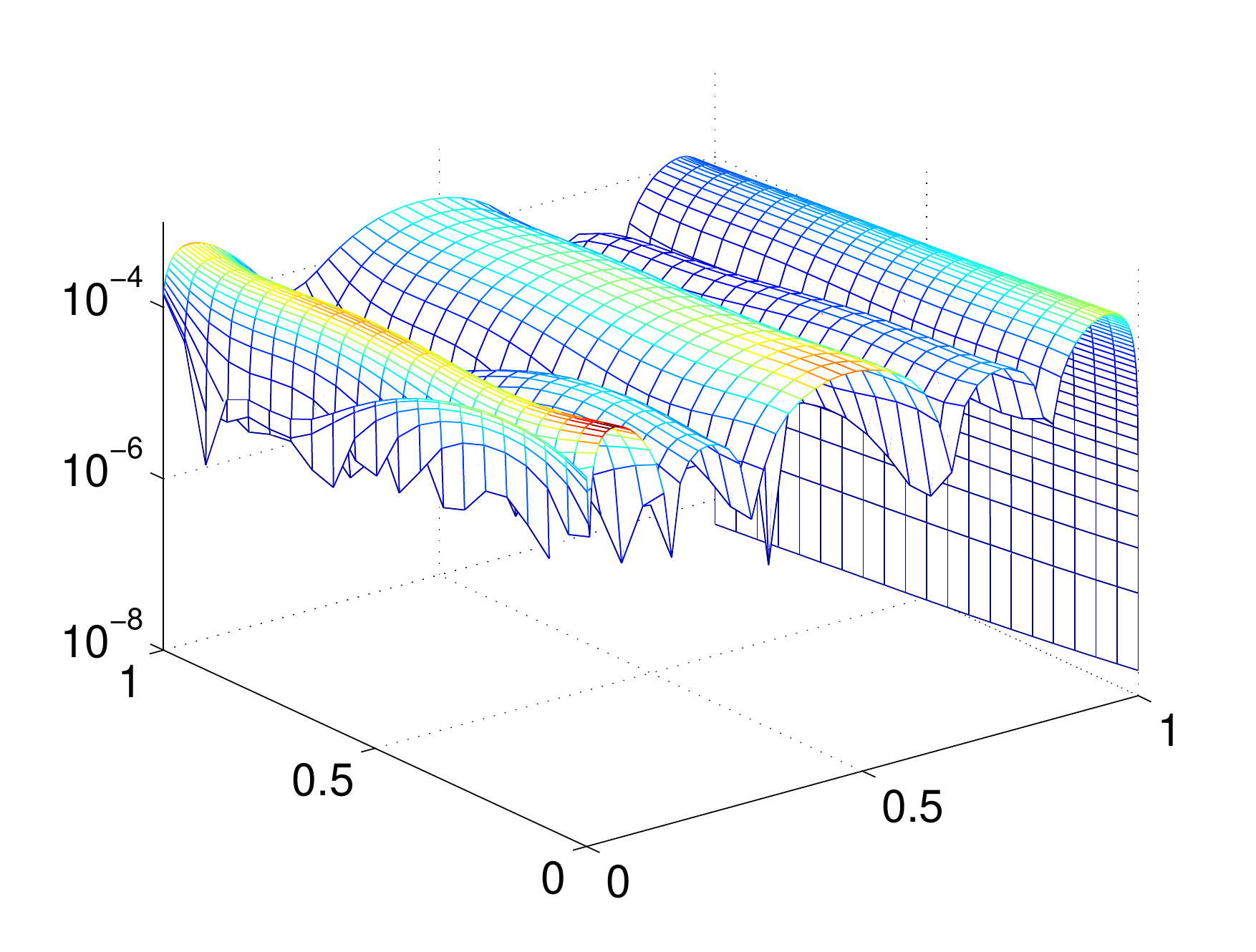}
    \put(-59,10){$x$}
    \put(-180,13){$n$}
    \put(-230,100){$\delta \hat v$}
    \put(-230,160){$\textbf{b)}$}

    \caption{KGD model (viscosity dominated regime) -- comparison of the new approximate solution \eqref{w_Ad_ap} -- \eqref{v_Ad_ap} with accurate numerical results for $N=100$: relative deviation of: a) fluid flow rate $\hat q$, b) particle velocity $\hat v$.}

\label{imp_app_3d_q_v}
\end{figure}

In order to supplement the foregoing analysis we shall now present a detailed description of the accuracy of approximation \eqref{w_Ad_ap} -- \eqref{v_Ad_ap}. In Fig. \ref{imp_app_3d_w_p} -- Fig. \ref{imp_app_3d_q_v} we depict the error distributions over $n$ and $x$ for the crack opening, $\delta \hat w$, the net fluid pressure, $\Delta \hat p$, the fluid flow rate, $\delta \hat q$, and the particle velocity, $\delta \hat v$. As previously, the relative errors are shown for $\hat w$, $\hat q$ and $\hat v$, while inaccuracy of $\hat p$ is described by its absolute measure. The first conclusion from the presented data is that the new improved approximation provides accurate and stable results regardless of the value of the fluid behaviour index, $n$. None of the presented errors exhibit sharp magnification at the crack tip, which means that the proposed formulae preserve the tip asymptotics very well. The maximal relative errors of the crack opening and the fluid flow rate are far below one percent (and far below the errors generated by the solution from \cite{ad_det_2002}), while the error of the particle velocity is even lower and does not exceed $2 \cdot 10^{-4}$. The net fluid pressure gives very accurate results even at the crack tip, where the singular behaviour holds. Furthermore, discrepancies between the improved approximation and the reference numerical data decrease as the fracture front is approached.

In the light of the foregoing analysis we can conclude that the new improved approximation \eqref{w_Ad_ap} -- \eqref{v_Ad_ap} constitutes a very good imitation of the reference solution for any $0\leq n \leq 1$ and thus can be itself used as a semi-analytical benchmark for testing other numerical algorithms.
In any case, it is appreciably more accurate than the solution proposed in \cite{ad_det_2002} and, unlike the former, avoids rapid error growth at the crack tip.

\begin{remark}
\label{remark_approx}
Respective formulae from \eqref{w_Ad_ap} -- \eqref{v_Ad_ap}, as defined on the basis of a pre-computed numerical solution, should be considered independent from each other. This means that if one wants to recreate e.g. the particle velocity by substituting \eqref{w_Ad_ap} and \eqref{p_Ad_ap} to \eqref{v_ss}, the resulting accuracy may be lower than that reported for \eqref{v_Ad_ap}.
\end{remark}

\clearpage

\subsubsection{KGD model in toughness dominated regime}
\label{sec:KGD_t_ss_2}

Let us now compare the numerical solution produced by the universal algorithm with the reference data for the toughness dominated regime given in \cite{gar_2006}. There the author presents a semi-analytical solution for the crack opening and the net fluid pressure in terms of a series expansion for three values of the fluid behaviour index: $n=\{0,0.5,1\}$. Results for $n=1$ have already been analyzed in \cite{wr_mis_2015}. Now we shall investigate the remaining variants ($n=0$ and $n=0.5$). Following the conclusions from Section \ref{sec:KGD_t_ss_1}, we conducted the computations using a spatial mesh consisting of $N=100$ nodes, which previous accuracy analysis showed provides a solution error of at most order $10^{-5}$.

\begin{figure}[h!]
    \hspace{5mm}
		\includegraphics [scale=0.43]{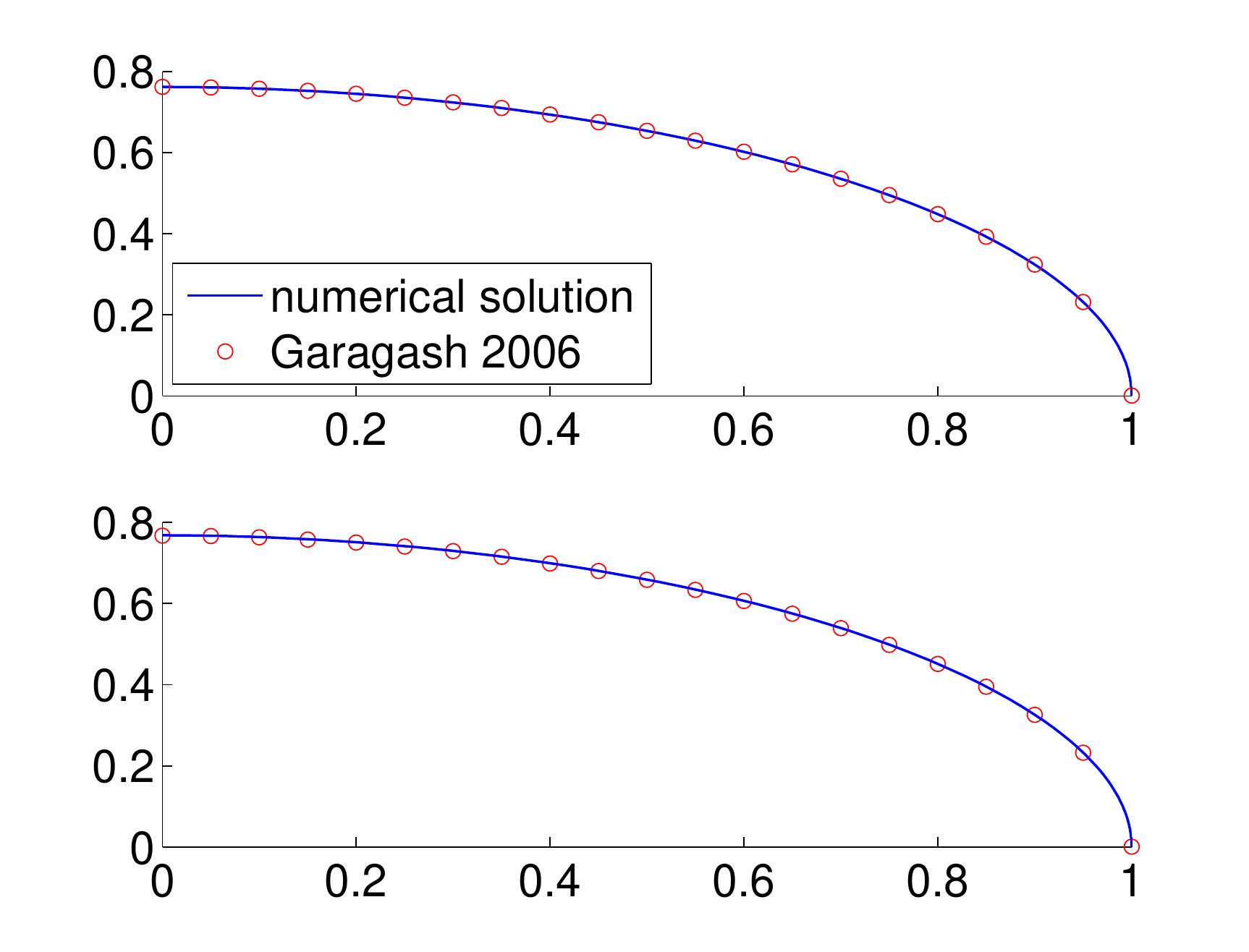}
		\put(-105,-5){$x$}
    \put(-230,145){$\textbf{a)}$}
		\put(-230,65){$\textbf{b)}$}
		\put(-210,120){$\hat w$}
    \put(-210,43){$\hat w$}
    \hspace{6mm}
    \includegraphics [scale=0.43]{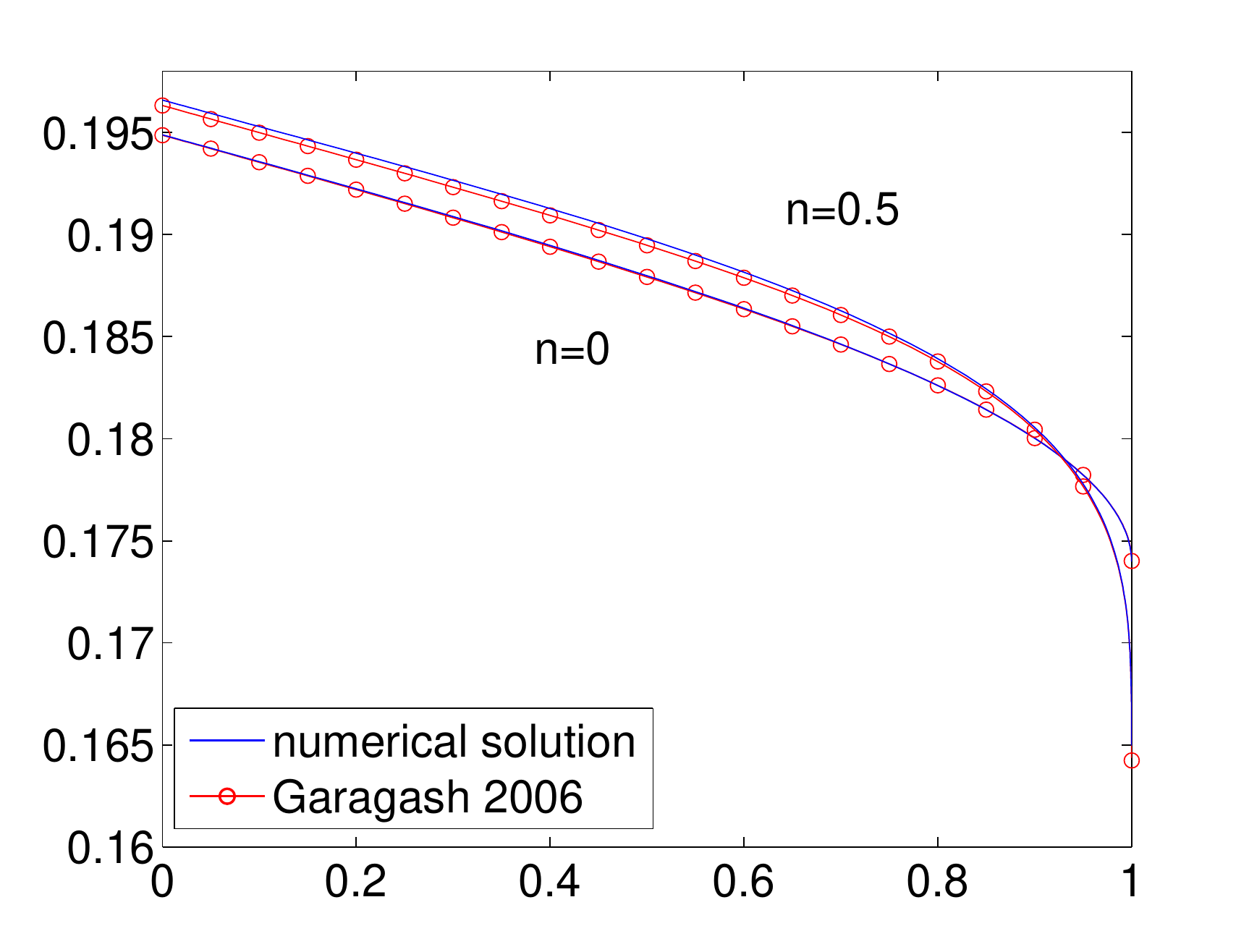}
    \put(-105,-5){$x$}
    \put(-220,80){$\hat p$}
    \put(-230,140){$\textbf{c)}$}

    \caption{KGD model (toughness dominated regime) -- comparison of our numerical solution for $N=100$ with semi-analytical solution presented in \cite{gar_2006}: a) the crack opening $\hat w$ for $n=0$, b) the crack opening $\hat w$ for $n=0.5$, c) the fluid pressure $\hat p$. }

\label{h_p_garagash}
\end{figure}

The plots of the crack opening, $\hat w$, and the net fluid pressure, $\hat p$, for $n=0$ and $n=0.5$, together with respective data from \cite{gar_2006}, are drawn in Fig. \ref{h_p_garagash}. Since the normalized values of the fracture apertures for these cases are very similar (even near the crack tip), they are presented in the separate graphs (Fig.~\ref{h_p_garagash}a and Fig.~\ref{h_p_garagash}b). Interestingly, the pertaining curves for the net fluid pressure are not that close to each other, especially in the near tip region. An explanation for this phenomenon is as follows. Although the leading terms of $\hat w$ are of the same order ($\alpha_0=1/2$, see Table \ref{T1}), the second terms depend on the fluid behavior index ($\alpha_1=(3-n)/2$). Consequently, on substitution of the asymptotic representation of $\hat w$ into the elasticity operator \eqref{p_ss_2}$_2$, the square root term produces a constant value, while the second term gives an expression of power law singularity that depends on the fluid behaviour index ($\hat p(x) \sim (1-x^2)^{(1-n)/2}$ as $x \to 1$). For this reason the tip behaviour of the net fluid pressure is essentially different for both cases. Moreover, it hinges on the value of the multiplier of the second term of asymptotic expansion of the fracture aperture, $\hat w_1$. That is why even very good coincidence of the crack widths cannot guarantee similar effect for the net fluid pressure. This also emphasizes importance of preserving further asymptotic terms of the solution for the quality of computations.

\begin{figure}[h!]
    \hspace{5mm}
		\includegraphics [scale=0.43]{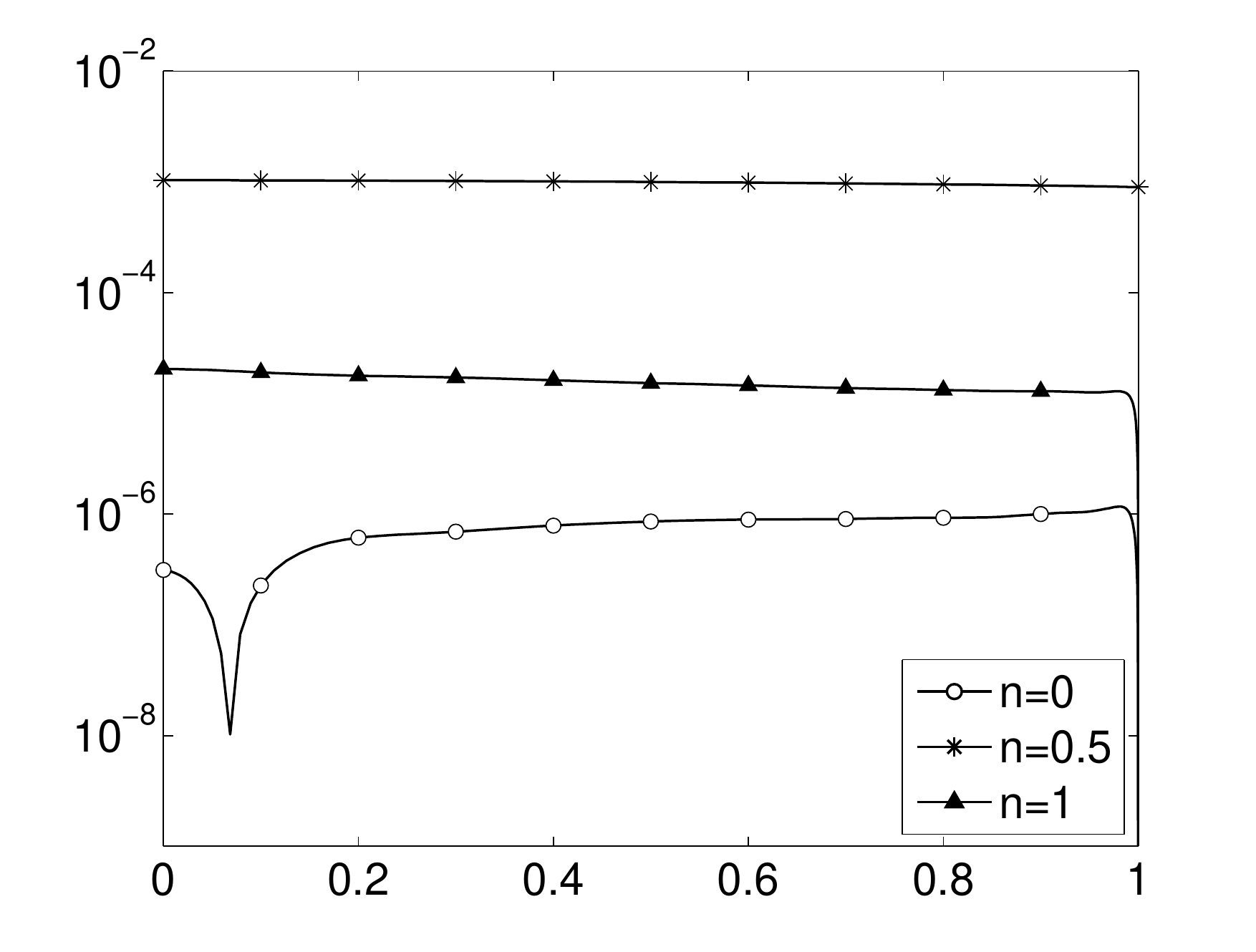}
		\put(-105,-5){$x$}
    \put(-230,80){$\delta \hat w$}
    \put(-230,140){$\textbf{a)}$}
    \hspace{6mm}
    \includegraphics [scale=0.43]{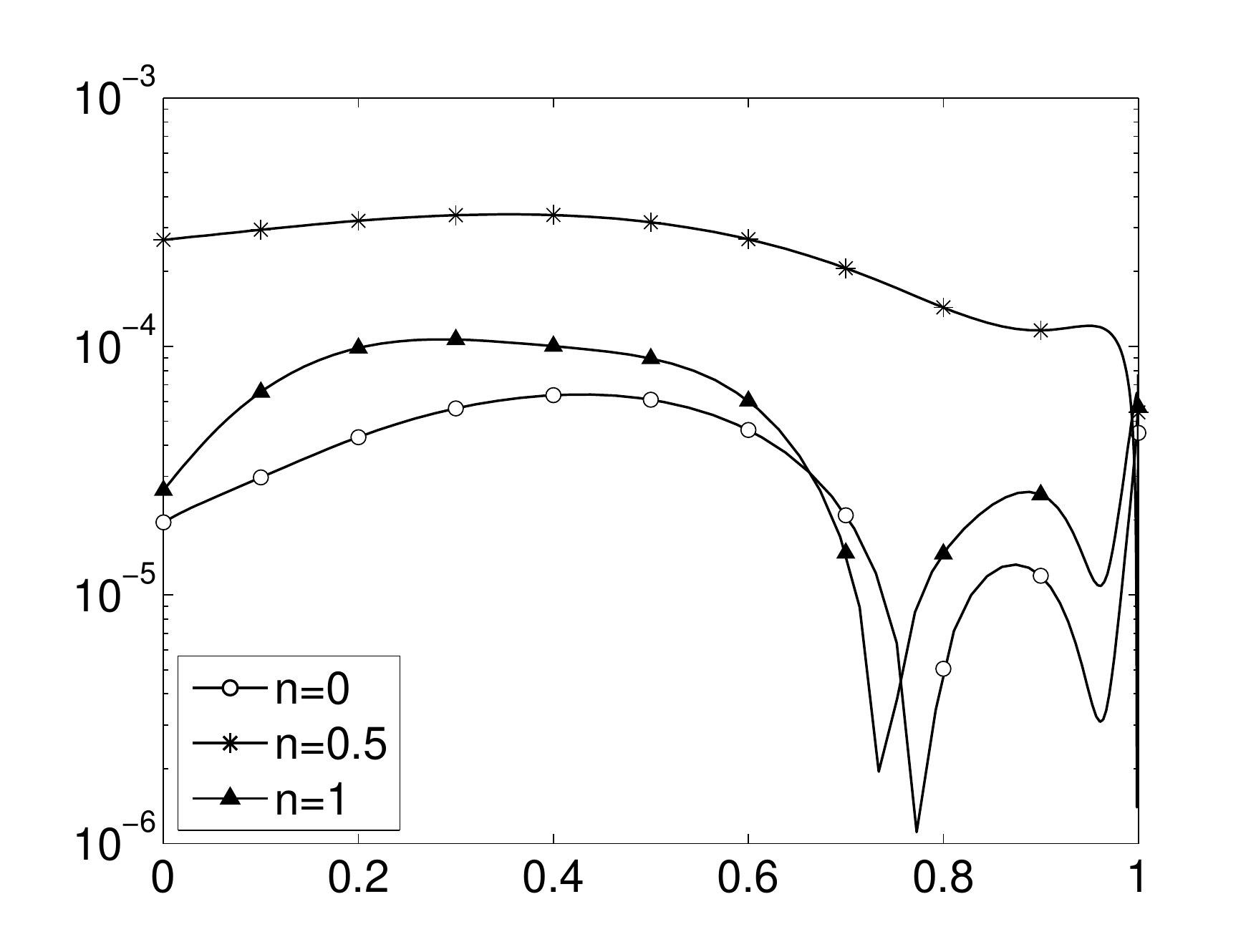}
    \put(-105,-5){$x$}
    \put(-230,80){$\Delta \hat p$}
    \put(-230,140){$\textbf{b)}$}

    \caption{KGD model (toughness dominated regime)  -- comparison of the semi-analytical solution presented in \cite{gar_2006} with our accurate numerical results for $N=100$: a) relative deviation of the crack opening, $\hat w$, b) absolute error of the fluid pressure, $\hat p$. }

\label{blad_garagash}
\end{figure}
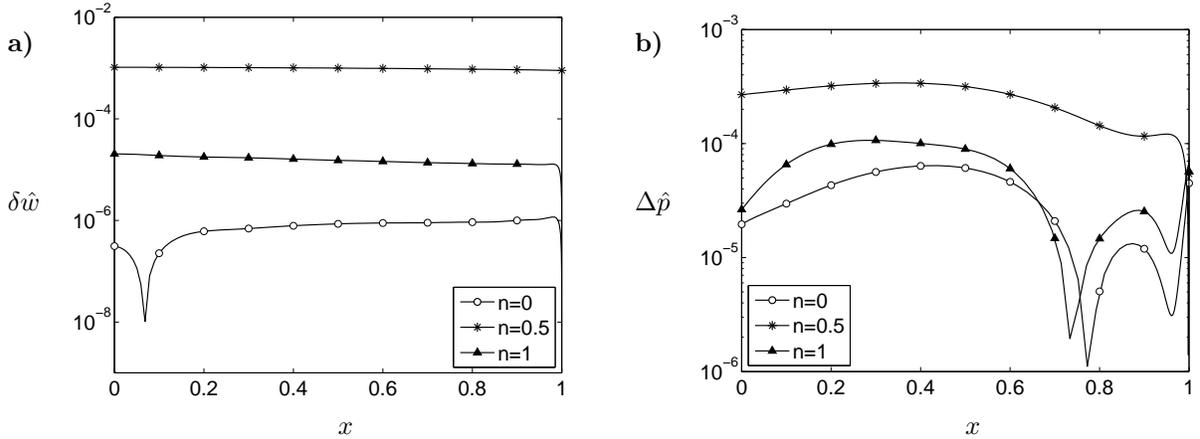

The relative deviations of the approximations given in \cite{gar_2006} from our numerical results are presented in Fig. \ref{blad_garagash}. To complement this picture, we also present  the case with $n=1$ (Newtonian fluid) discussed in  \cite{wr_mis_2015}.

It can be seen that we have very good coincidence of the analyzed results for $n=0$. The graphs of $\hat w$ and $\hat p$ are indistinguishable from each other. However, this is no longer the case for $n=0.5$, where the net fluid pressure curves diverge. The data for the case
$n=1$ is again very consistent with those obtained by the universal algorithm in \cite{wr_mis_2015}.

In order to explain this peculiarity, we should refer to the way in which the solution from \cite{gar_2006} is reconstructed, and how the scaling between the respective equations in our formulation and that  utilized {in \cite{gar_2006} is executed. In the recalled paper the author introduces a dimensionless crack half-length. This parameter is present in the leading terms of the power series representations for both the crack opening and the net fluid pressure, and thus in order to recreate the solution one needs to know its value. Moreover, the dimensionless crack half-length is employed in scaling when transforming the governing ODE from the formulation given in \cite{gar_2006} to the formula \eqref{ODE_gen}. In fact, the scaling criterion relates this value to the crack propagation speed $\hat v(1)$.

 In general, there are two alternative methods of obtaining the dimensionless crack half-length: one utilizing the integral of $\hat w$ over the spatial interval (the fracture volume), and the other  based on the fluid influx. Additionally, for $n=0$ and $n=1$ explicit formulae relating this parameter to one of the coefficients given in \cite{gar_2006} can be found. For an exact solution, all methods should produce the same results, but naturally one cannot expect such an effect when using an approximation. Unfortunately, the author does not provide  the numerical values of the dimensionless crack half-length.  In \cite{wr_mis_2015}, when analyzing the data for a Newtonian fluid, the aforementioned explicit formula was used in computations.  The same approach was implemented for $n=0$. It is evident that such a method produces far better coincidence of the tip asymptotes between our results and the data from  \cite{gar_2006}, which leads to appreciably better convergence between the corresponding solutions than in the generic case ($0<n<$1). Applying the alternative methods resulted in reduced solution accuracy for both, $n=0$ and $n=1$. For $n=0.5$ we computed the dimensionless crack half-length from the non-local relation (based on the fracture volume). We believe that it constitutes a more reasonable choice than the influx criterion, as the global condition is less sensitive to the point-wise deterioration of the accuracy. It is noteworthy that, when computing the sought for parameter from the influx, its value differs from the previous one by approximately ten percent.

\section{Solution of the problem in transient regime}
\label{sec:transient}

In this section we discuss an extension of the algorithm presented above to the time-dependent variant of the problem. The main assumptions and blocks of the algorithm remain the same.
The new features introduced here are subroutines for approximating the temporal derivative and the crack length. Presented numerical examples demonstrate the performance of the general variant of the algorithm.

\subsection{Problem formulation}

Analogously to that done for the self-similar variant of the problem, let us write the basic system of equations.

We use the following representation of the temporal derivative of the crack opening, as was proposed in \cite{wr_mis_2013}:
\begin{equation}\label{dwdt}
\frac{\partial w}{\partial t}\bigg|_{t_{i+1}}=2\frac{w(t_{i+1},x)-w(t_i,x)}{t_{i+1}-t_i}-\frac{\partial w}{\partial t}\bigg|_{t_i}.
\end{equation}
The basic advantage of the presented formula over the ordinary finite difference is that it provides higher order approximation $O(\Delta t^2))$ than the latter $(O(\Delta t))$. Note that the value of the derivative at the initial time step can be computed analytically from the continuity equation (see details in \cite{wr_mis_2015}).

The basic set of equations employed in the universal algorithm for the transient regime is listed below:
\begin{itemize}
\item equation \eqref{dwdt} to compute the temporal derivative,
\item equation defining the reduced velocity \eqref{phi_cont} in the following transformed version:
\begin{equation}\label{fi_trans}
\phi=\frac{L}{w}\int_x^1\left(\frac{\partial w}{\partial t}+\frac{\left({\cal{C}}_A {\cal{L}}(w)\right)^{1/n}}{L^{m/n+1}}w+q_l\right)d\eta,
\end{equation}
\item solvability condition resulting from \eqref{fi_trans} and \eqref{q0_phi} utilized to compute the parameter ${\cal L}(w)$:
\begin{equation}\label{solv_trans}
\int_0^1 \frac{\partial w(t,x)}{\partial t}dx-\frac{q_0(t)}{L(t)}+\frac{\left({\cal{C}}_A {\cal{L}}(w)\right)^{1/n}}{L^{m/n+1}(t)}\int_0^1w(t,x)dx+\int_0^1q_l(t,x)dx=0,
\end{equation}
\item equation to compute the crack opening, obtained by combining \eqref{p_prim} with the respective inverse elasticity operator \eqref{inv_norm_PKN} -- \eqref{inv_norm_KGD_2}:
\begin{equation}\label{w_trans}
\hat w ={\cal B} \left[L(t)\left(\hat\phi(\eta)+\eta \frac{\left({\cal{C}}_A {\cal{L}}(w)\right)^{1/n}}{L^{m/n}(t)}\right)^n\right]
\end{equation}
\item the influx boundary condition \eqref{q0_phi} and the regularized conditions at the crack tip,
\item initial conditions, which also allow computation of the temporal derivative at the initial time step,
\item relation to compute the crack length \eqref{L_int}.
\end{itemize}

\subsection{Computational algorithm for the transient regime}

The solution to the transient variant of the problem is computed at each time step by the iterative algorithm and is an extension of that presented for the self-similar formulation. By analogy to the description given in Section \ref{sec:num_alg_self_sim}, we can define the following stages of computations at time $t_{i+1}$:
\begin{itemize}

\item \emph{Preconditioning.} The first approximation of the crack opening $w^{(j-1)}(t_{i+1},x)$ is specified based on the temporal derivative from the previous time step $w'_t(t_i,x)$ and the initial condition $w(t_i,x)$.

\item \emph{First step.} The temporal derivative of the crack opening is computed using \eqref{dwdt}. Note that when obtaining the final solution $w(t_{i+1},x)$, one automatically has its temporal derivative as well. Then, we calculate the crack length $L^{(j-1)}(t_{i+1})$ using formula \eqref{L_int} with ${\cal{L}}(w^{(j-1)}(t_{i+1},x))$ (or ${\cal{L}}(w(t_{i},x))$ at the first iteration). Next equation \eqref{solv_trans} yields ${\cal{L}}^{(j)}(w(t_{i+1},x))$ which, substituted into \eqref{fi_trans}, serves to compute the reduced particle velocity $\phi ^{(j)}(t_{i+1},x)$. The $\varepsilon$-regularization technique is applied to carry out the integral in \eqref{fi_trans}. As a result, functions ${\cal{L}}^{(j)}(w(t_{i+1},x))$ and $\phi ^{(j)}(t_{i+1},x)$ computed at this stage satisfy: i) fluid balance equation \eqref{solv_trans}, ii) continuity equation \eqref{fi_trans}, iii) regularized boundary condition for $\phi$ which is an equivalent of \eqref{phi_tip}, iv) the influx boundary condition \eqref{q0_phi} indirectly through the fluid balance equation.

\item \emph{Second step.} Next the approximation of the crack opening $w^{(j)}(t_{i+1},x)$ is obtained from \eqref{w_trans}. The technique of numerically computing the operator ${\cal B}_w$ is exactly the same as for the self-similar variant of the problem.

\end{itemize}

The aforementioned two steps of the algorithm are repeated until respective components of the solution have converged within a prescribed tolerance.

\begin{remark}\label{modular_crack_length}
The modular algorithm architecture enables one to easily introduce the subroutine for crack length computation as an additional block. Naturally, this block was not present in the self-similar variant of the algorithm.
\end{remark}

\subsection{Algorithm performance in the transient regime}

In this part of the paper we present a brief investigation of the algorithm performance in the transient regime. The few examples shown below constitute a demonstration of the potential of the proposed numerical scheme. This analysis is not exhaustive, as such a task would massively increase the volume of the paper (it would require e.g. presenting a number of simulations for various spatial meshing variants and time stepping strategies conducted for various values of the fluid behaviour index). As the influence of the spatial meshing on computational accuracy was thoroughly analyzed for the self-similar formulation of the problem, we will not consider this issue here. All results presented below were obtained for an adaptive spatial mesh composed of $N=100$ nodal points, which for the self-similar variant produced a maximal solution error of at most order $10^{-6}$ for both the crack opening and particle velocity, regardless of the HF model. It is obvious that in the transient regime one cannot expect the same level of accuracy. Indeed, as was shown in \cite{wr_mis_2015}, the solution accuracy is naturally determined by the relation between the spatial and temporal meshing. For each of the HF models we present the results of computations for two values of the fluid behaviour index ($n=0.3$ and $n=0.7$) and two different temporal meshes. The number of time steps, denoted by $M$, was set to $M=20$ and $M=100$, while the same normalized target time $t=100$ was assumed for all computations. The time meshing strategy which provides an approximately parabolic time step distribution (with finer meshing for small times) was adopted from \cite{wr_mis_2013}.

The benchmark solutions utilized here are the time-dependent extensions of self-similar benchmarks used in Sections \ref{sec:PKN_ss_1} (PKN), \ref{sec:KGD_f_ss_1} (viscosity dominated regime of KGD) and \ref{sec:KGD_t_ss_1}  (toughness dominated regime of KGD). The analyzed error measures are: the relative error of crack opening, $\delta w(t,x)$, the relative error of the particle velocity, $\delta v(t,x)$, and the relative error of the crack length, $\delta L(t)$.

\subsubsection{Analysis of the algorithm - PKN model}

In Fig. \ref{PKN_dyn_n03_t20} -- \ref{PKN_dyn_n03_t100} the distributions of relative errors over space and time for the crack opening, $\delta w$, and the particle velocity, $\delta v$, are shown for $n=0.3$. Regardless of the temporal mesh density we see some general trends. Namely, for a fixed number of the time steps, $M$, the graphs for $\delta w$ and $\delta v$ are practically identical. Almost uniform distribution of the computational errors over the crack length, $x$, is observed. When analyzing the time characteristic it is evident that the greatest error growth takes place for small times before the level of relative error stabilizes. In the case under consideration, by taking $M=100$ instead of $M=20$, one reduces the error by over an order of magnitude. One should remember that $M=20$ produces a very coarse temporal mesh (only 20 steps from the initiation of fracture growth to the large time asymptote).

\begin{figure}[h!]

    \includegraphics [scale=0.45]{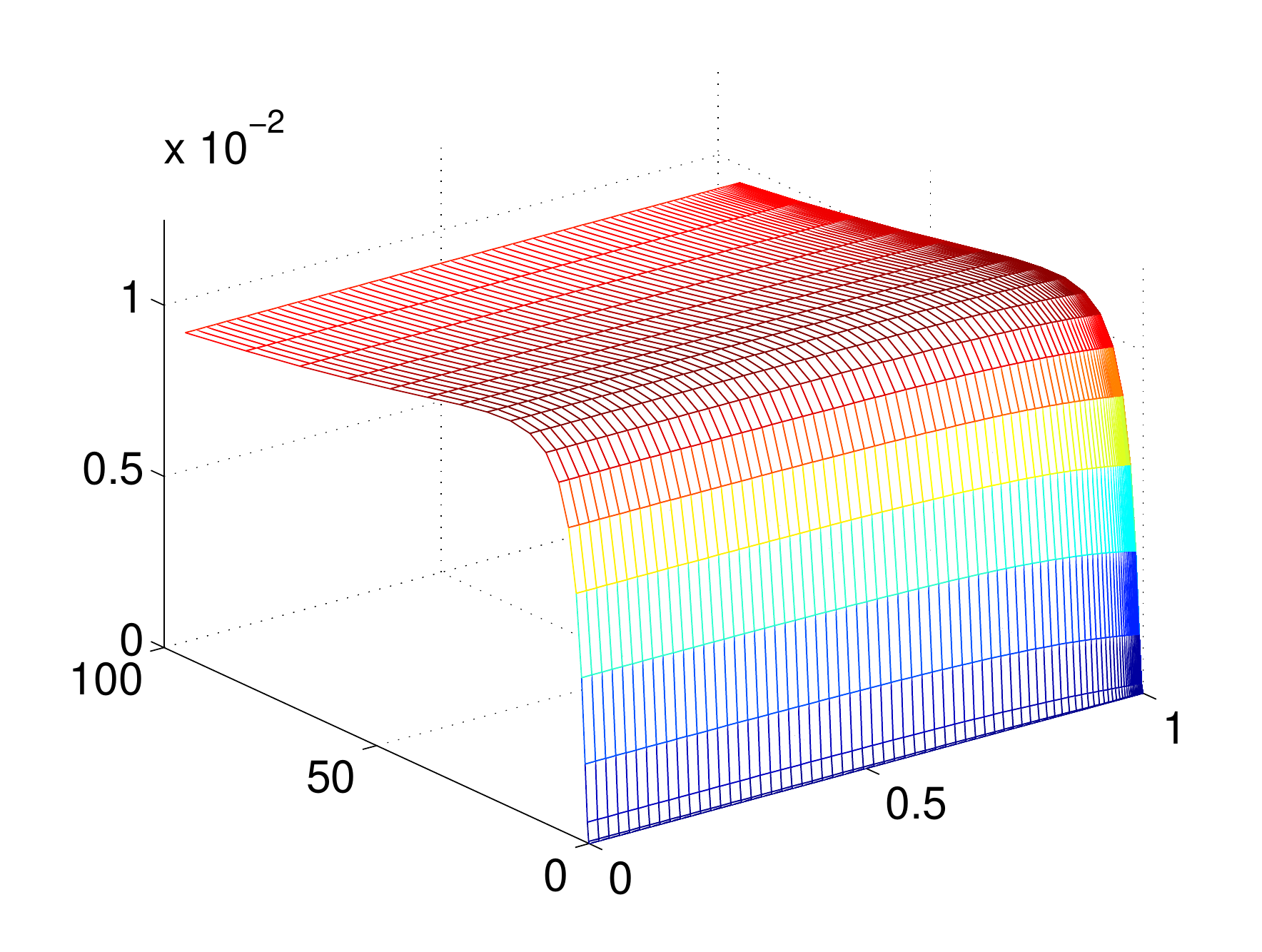}
    \put(-62,13){$x$}
    \put(-175,15){$t$}
		\put(-230,101){$ \delta w$}
    \put(-230,160){$\textbf{a)}$}
    \hspace{2mm}
    \includegraphics [scale=0.45]{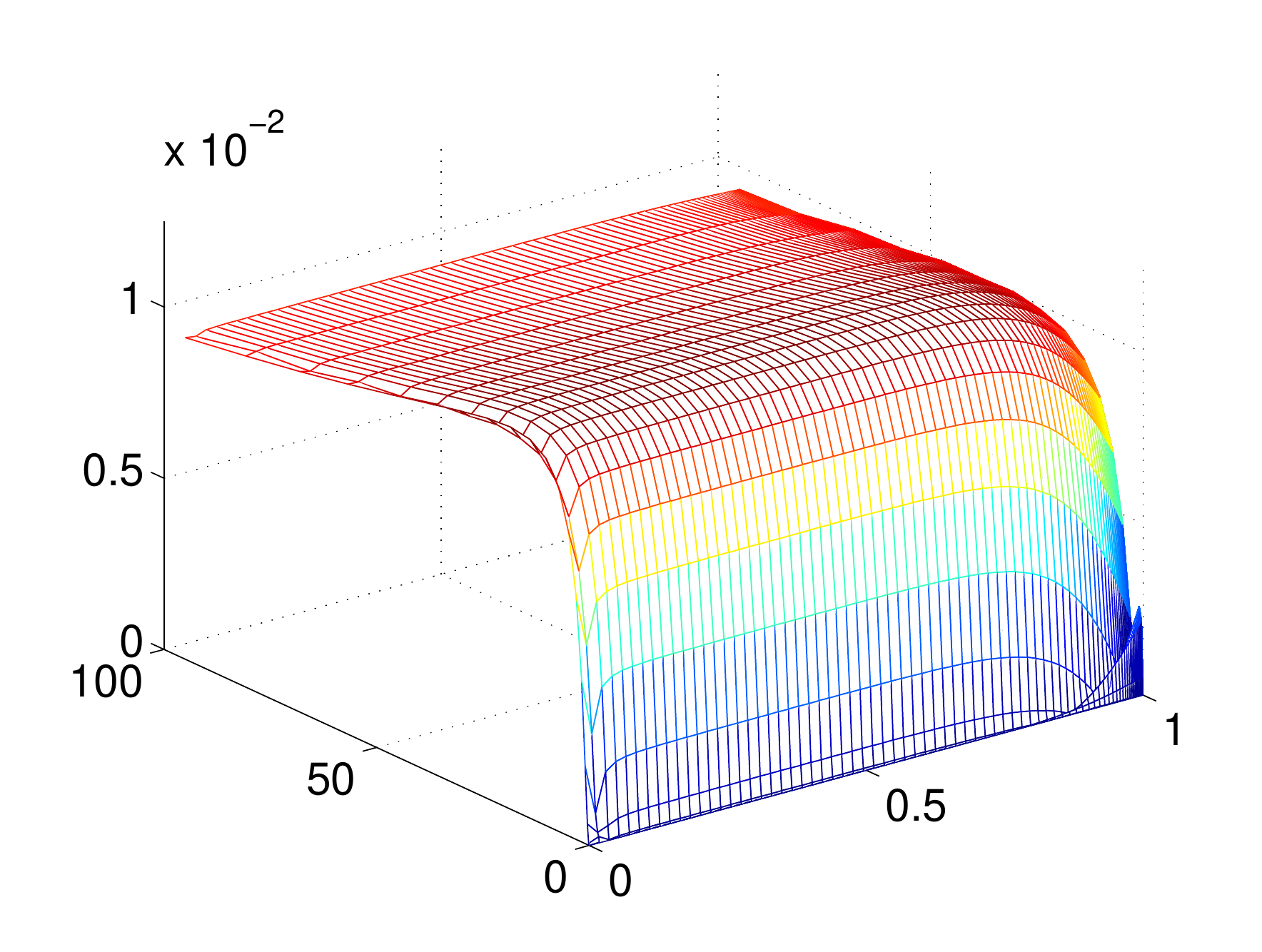}
    \put(-62,13){$x$}
    \put(-175,15){$t$}
		\put(-230,107){$\delta v$}
    \put(-230,160){$\textbf{b)}$}

    \caption{\textit{PKN model}. Relative solution error for $n=0.3$ with $N=100$ (spatial mesh), $M=20$ (temporal
mesh): a) the crack opening $\delta w$, b) the particle velocity $\delta v$.}

\label{PKN_dyn_n03_t20}
\end{figure}

\begin{figure}[h!]

    \includegraphics [scale=0.45]{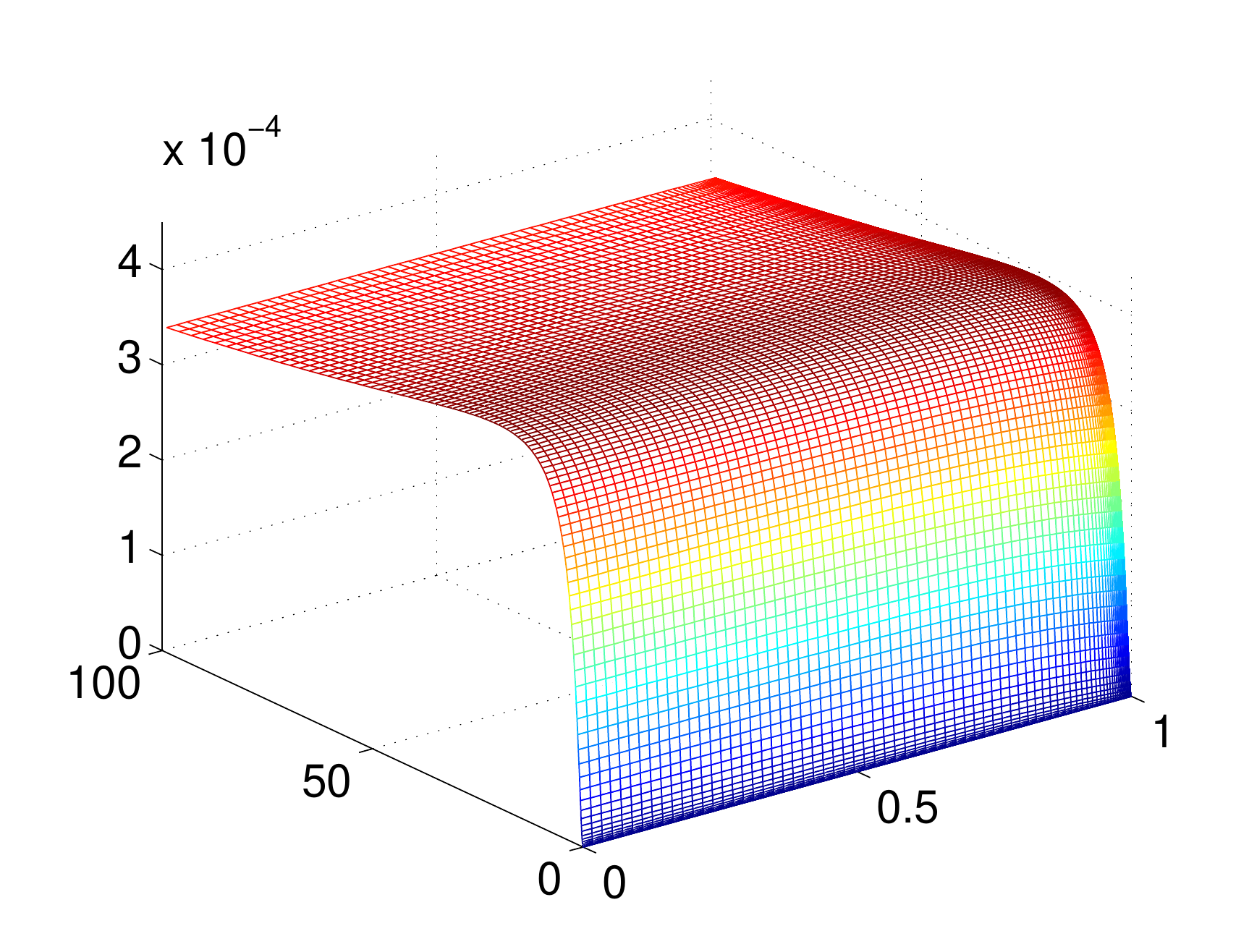}
    \put(-62,13){$x$}
    \put(-175,15){$t$}
    \put(-230,100){$ \delta w$}
    \put(-230,160){$\textbf{a)}$}
    \hspace{2mm}
    \includegraphics [scale=0.45]{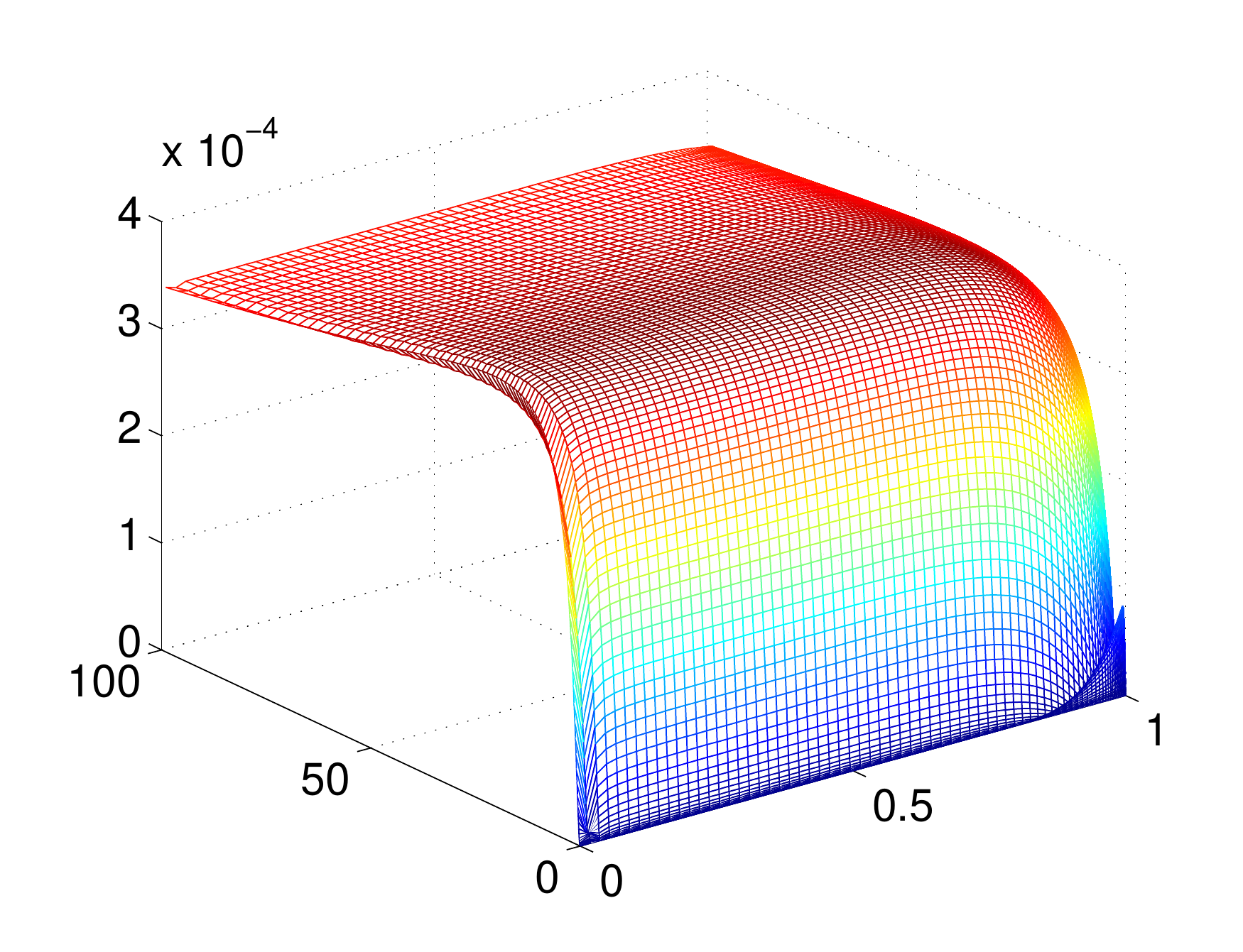}
    \put(-62,13){$x$}
    \put(-175,15){$t$}
    \put(-230,100){$\delta v$}
    \put(-230,160){$\textbf{b)}$}

    \caption{\textit{PKN model}. Relative solution error for $n=0.3$ with $N=100$ (spatial mesh), $M=100$ (temporal
mesh): a) the crack opening $\delta w$, b) the particle velocity $\delta v$.}

\label{PKN_dyn_n03_t100}
\end{figure}
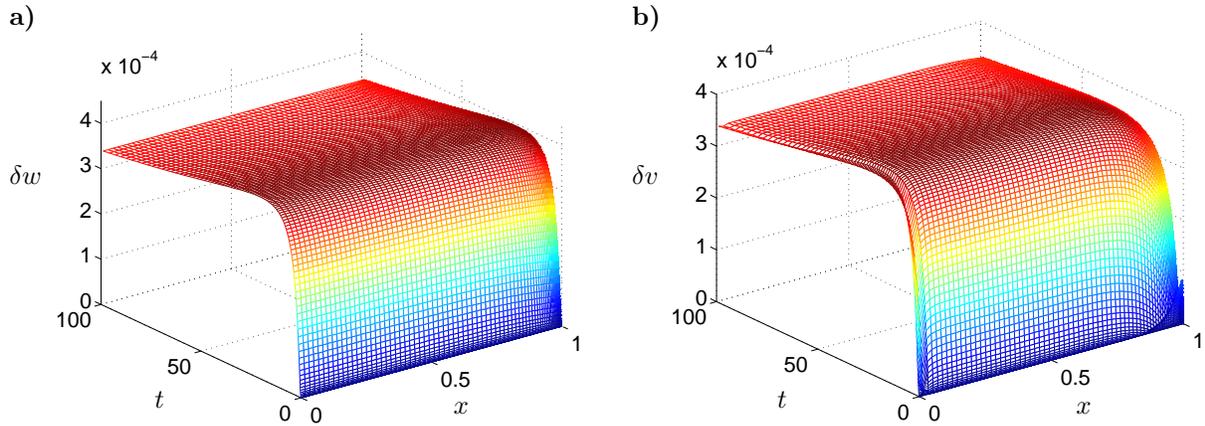

A bit different trend is observed in Fig. \ref{PKN_dyn_n07_t20} -- \ref{PKN_dyn_n07_t100}, where the relative errors for $n=0.7$ are presented. First, the overall accuracy has improved by approximately one order of magnitude with respect to the case of $n=0.3$. The maxima of $\delta w$ are now about two times smaller than the maxima of $\delta v$ for the same $M$. Also, the potential for error reduction when increasing $M$ is larger.

When analyzing the crack aperture it  shows that the transition from $M=20$ to $M=100$ gives two orders of magnitude better accuracy. In the case of particle velocity this reduction is smaller, but still distinctly greater than it was for $n=0.3$. Nonetheless, it is difficult to speculate whether the reason for this change in general trends can be entirely attributed to changing the value of $n$. Note that when expanding the self-similar benchmarks to their time-dependent versions they produce quantitatively different  dynamic behaviour (temporal derivatives, crack propagation speed etc.).

\begin{figure}[h!]

    \includegraphics [scale=0.45]{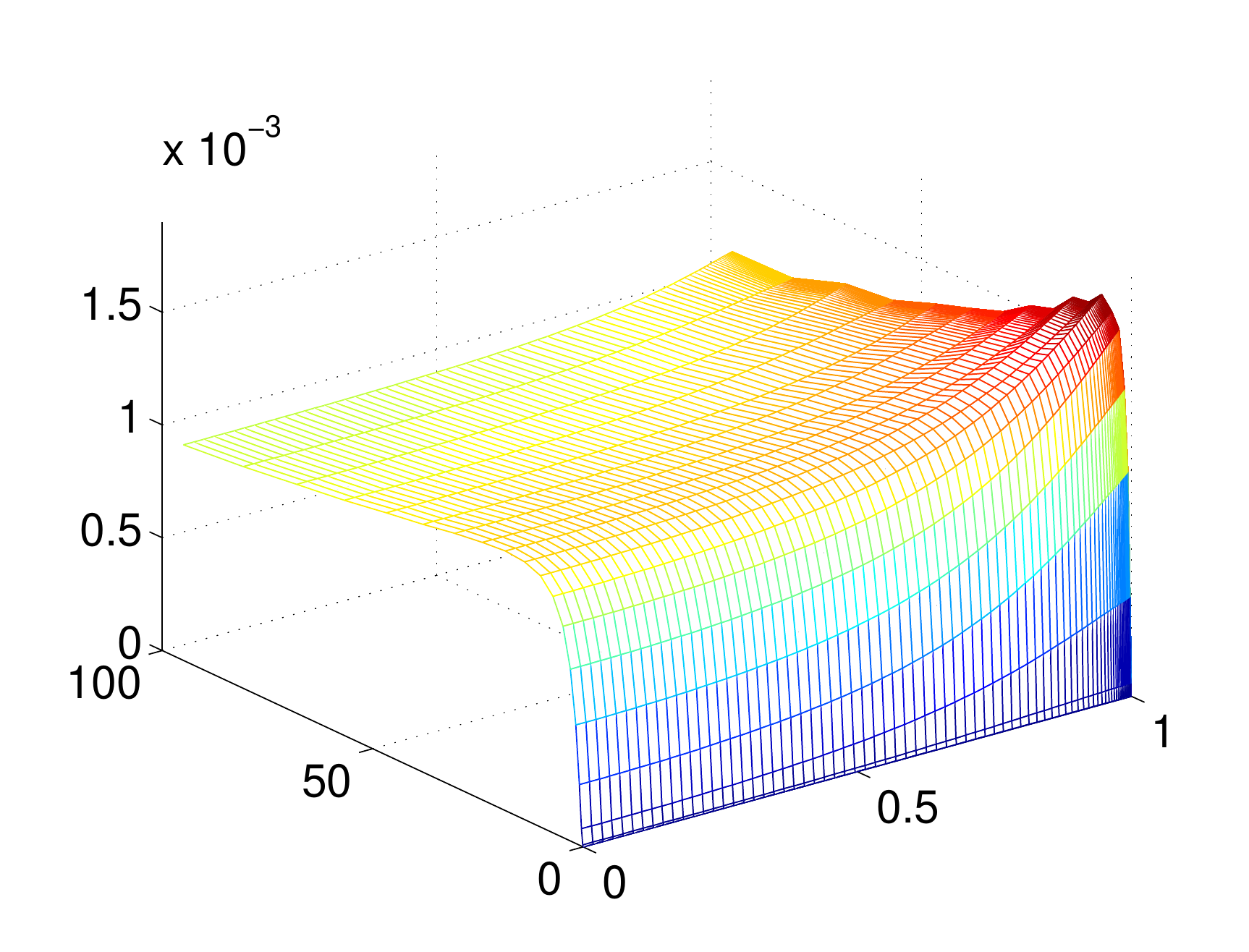}
    \put(-62,13){$x$}
    \put(-175,15){$t$}
    \put(-230,90){$ \delta w$}
    \put(-230,160){$\textbf{a)}$}
    \hspace{5mm}
    \includegraphics [scale=0.45]{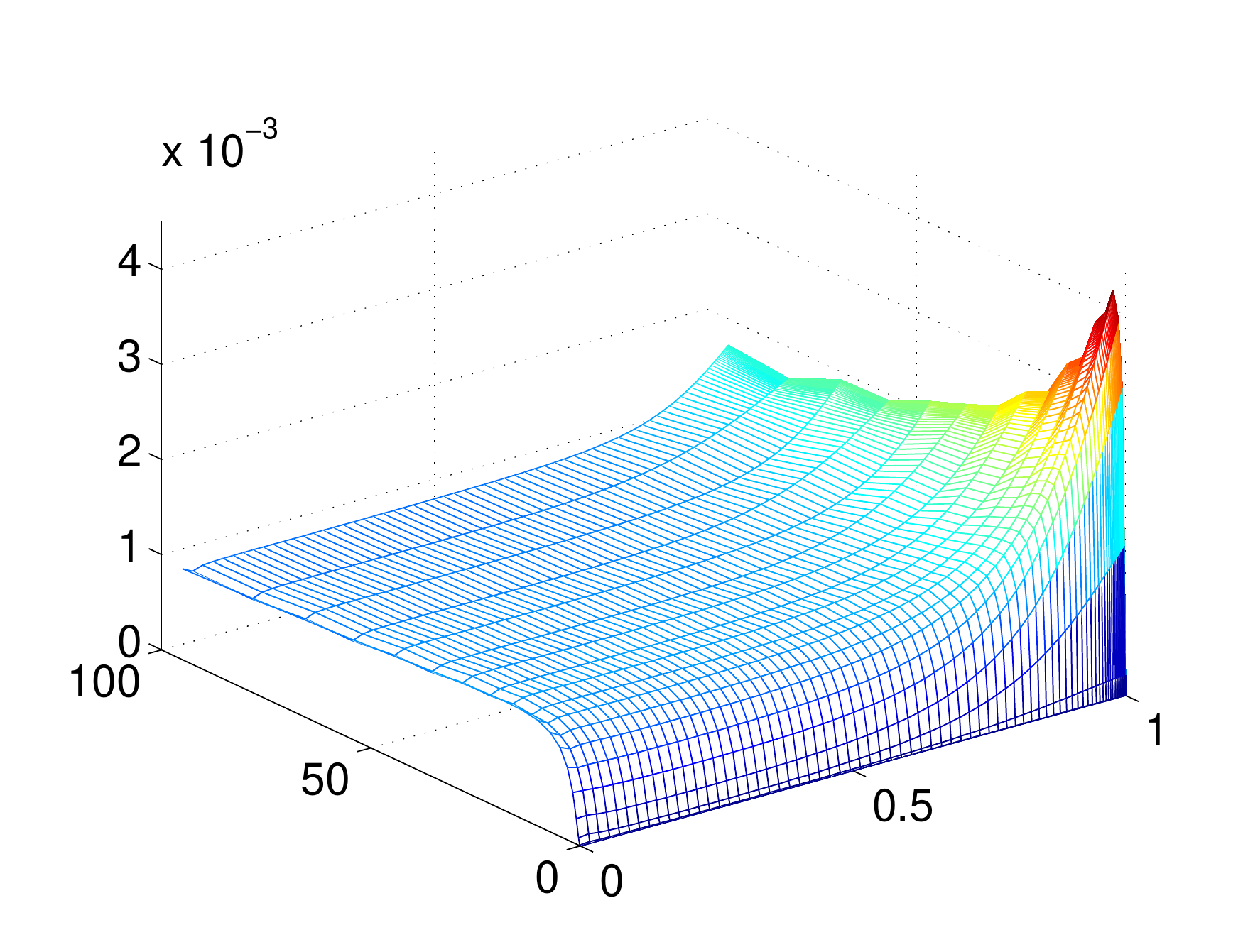}
    \put(-62,13){$x$}
    \put(-175,15){$t$}
    \put(-230,90){$\delta v$}
    \put(-230,160){$\textbf{b)}$}

    \caption{\textit{PKN model}. Relative solution error for $n=0.7$ with $N=100$ (spatial mesh), $M=20$ (temporal
mesh): a) the crack opening $\delta w$, b) the particle velocity $\delta v$.}

\label{PKN_dyn_n07_t20}
\end{figure}

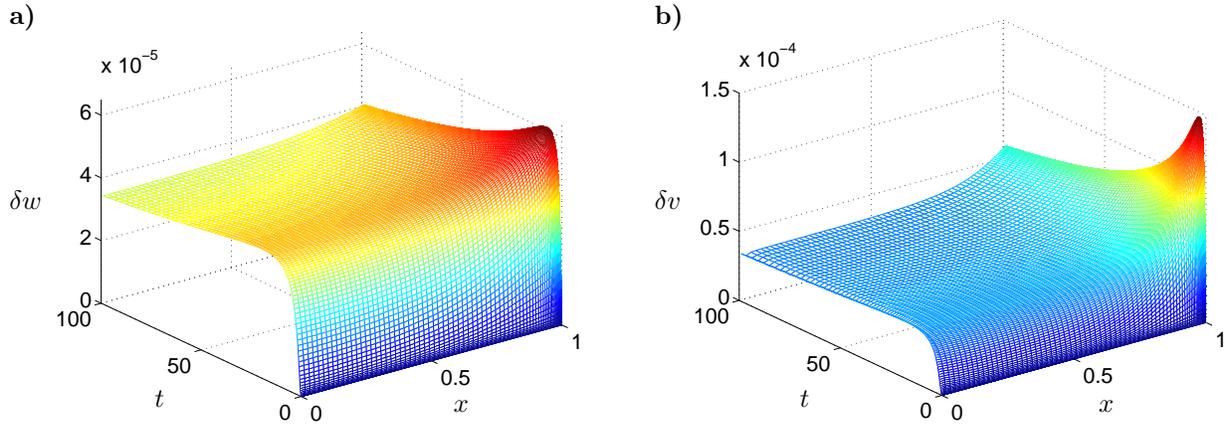
\begin{figure}[h!]

    \includegraphics [scale=0.45]{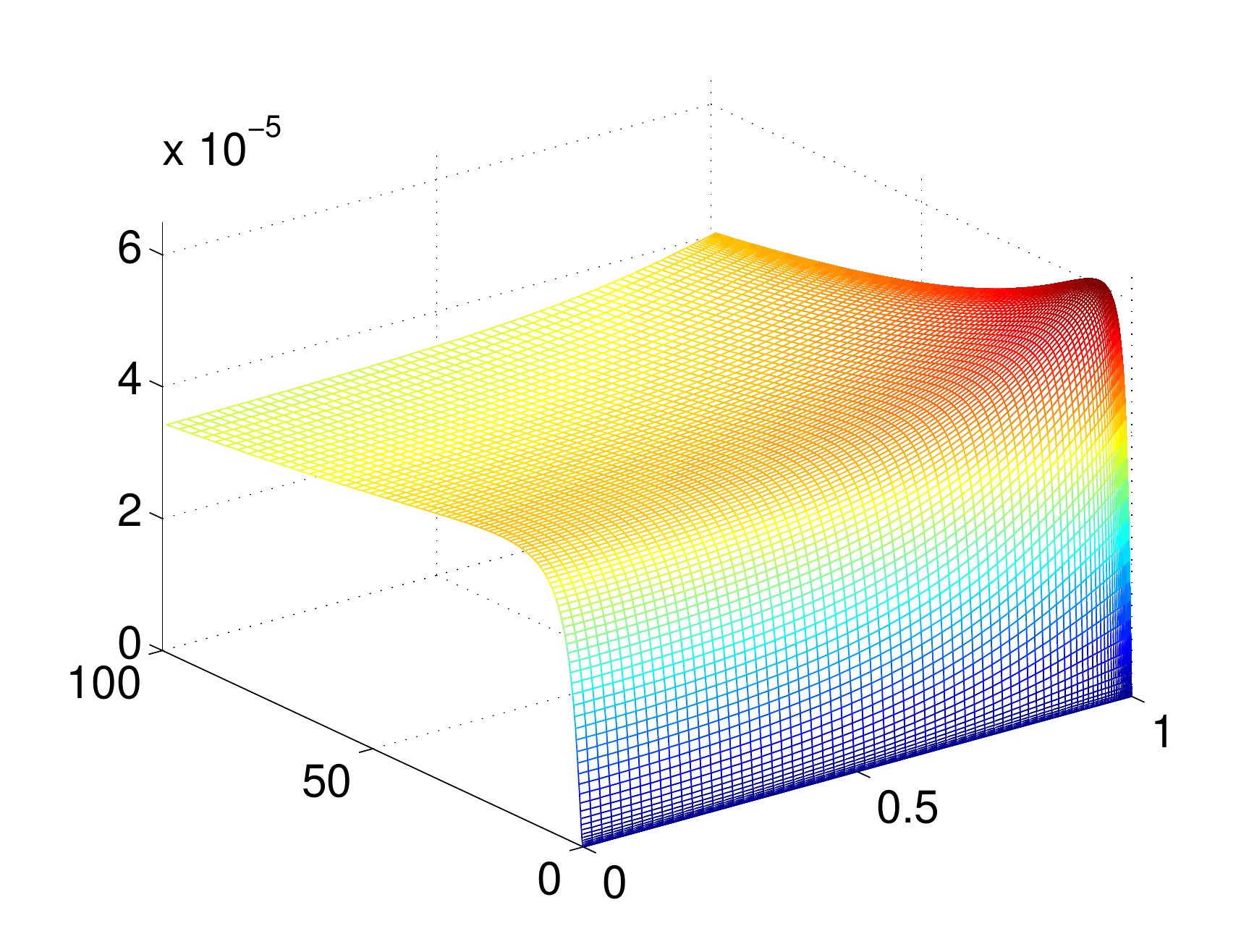}
    \put(-62,13){$x$}
    \put(-175,15){$t$}
    \put(-230,90){$ \delta w$}
    \put(-230,160){$\textbf{a)}$}
    \hspace{5mm}
    \includegraphics [scale=0.45]{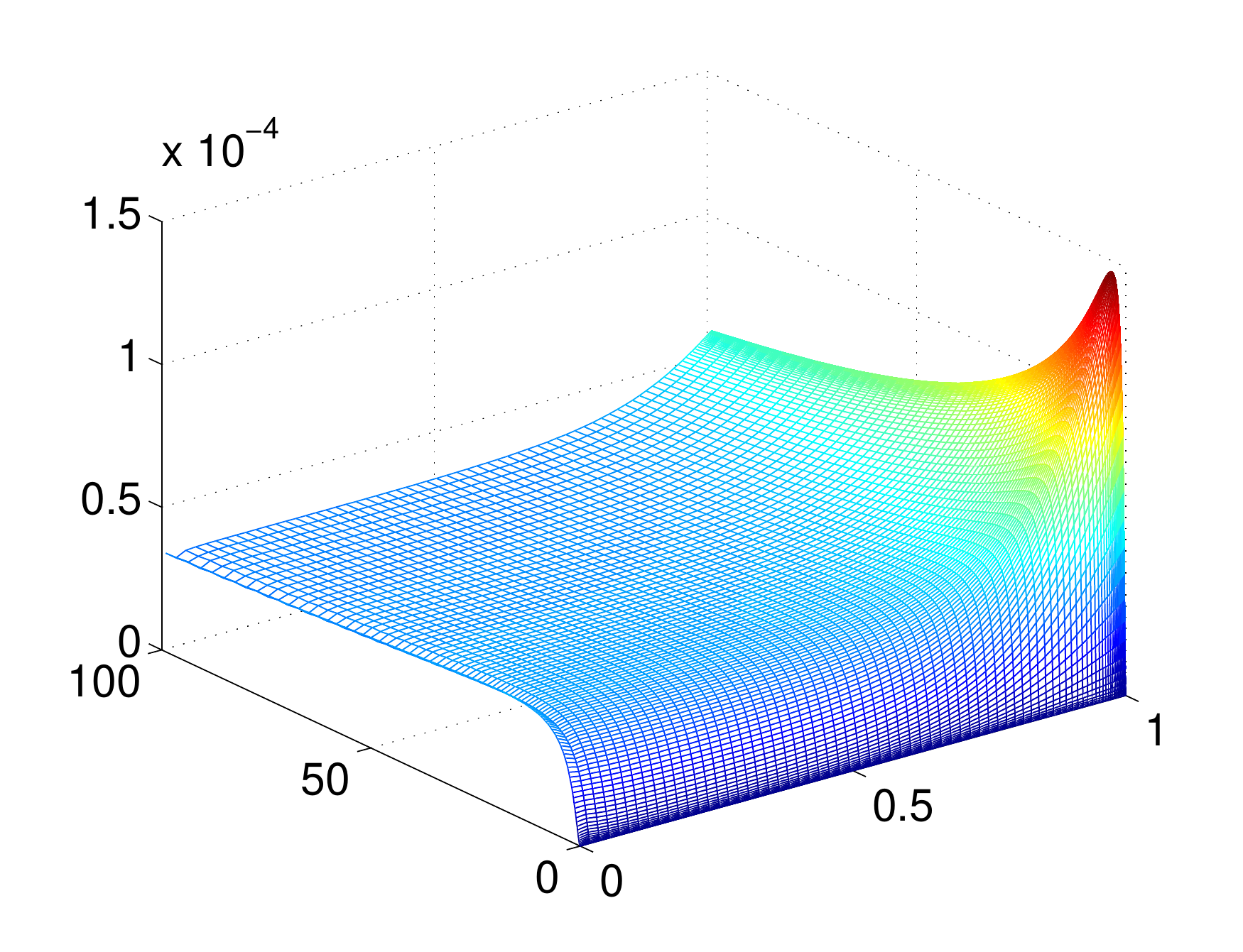}
    \put(-62,13){$x$}
    \put(-175,15){$t$}
    \put(-230,90){$\delta v$}
    \put(-230,160){$\textbf{b)}$}

    \caption{\textit{PKN model}. Relative solution error for $n=0.7$ with $N=100$ (spatial mesh), $M=100$ (temporal
mesh): a) the crack opening $\delta w$, b) the particle velocity $\delta v$.}

\label{PKN_dyn_n07_t100}
\end{figure}

To complement this part of the analysis let us look at Fig. \ref{dyn_blad_L}a, where the crack length error, $\delta L$, for the respective cases is depicted. It shows that, regardless of the value of fluid behaviour index $n$ and the number of time steps $M$, the error level stabilizes after the initial growth, however its magnitude depends on the considered variant. In general, a better solution for $w$ and $v$ produces better results for $L$, which is an intuitive trend.

\subsubsection{Analysis of the algorithm - KGD model in viscosity dominated regime}

Let us now analyze the accuracy of computations  for the viscosity dominated regime of the KGD model. In Fig. \ref{KGD_f_dyn_n03_t20} -- \ref{KGD_f_dyn_n03_t100} the relative errors of $w(t,x)$ and $v(t,x)$ for $n=0.3$ are presented. This time (unlike in the PKN variant) the maximal values of $\delta v$ are approximately two times greater than the maxima of $\delta w$. Moreover, the distribution of $\delta w$ exhibits different shapes than that of $\delta v$. Increasing the number of time steps from 20 to 100 entails a two order improvement in accuracy for the crack opening, and one order for the particle velocity. In comparison with the PKN case (Fig. \ref{PKN_dyn_n03_t20} -- Fig. \ref{PKN_dyn_n03_t100}), the level of $\delta w$ increased approximately two times, while the magnification of $\delta v$ was even greater. The latter effect is caused by the error increase at the crack tip for the small time range.

\begin{figure}[h!]

    \includegraphics [scale=0.45]{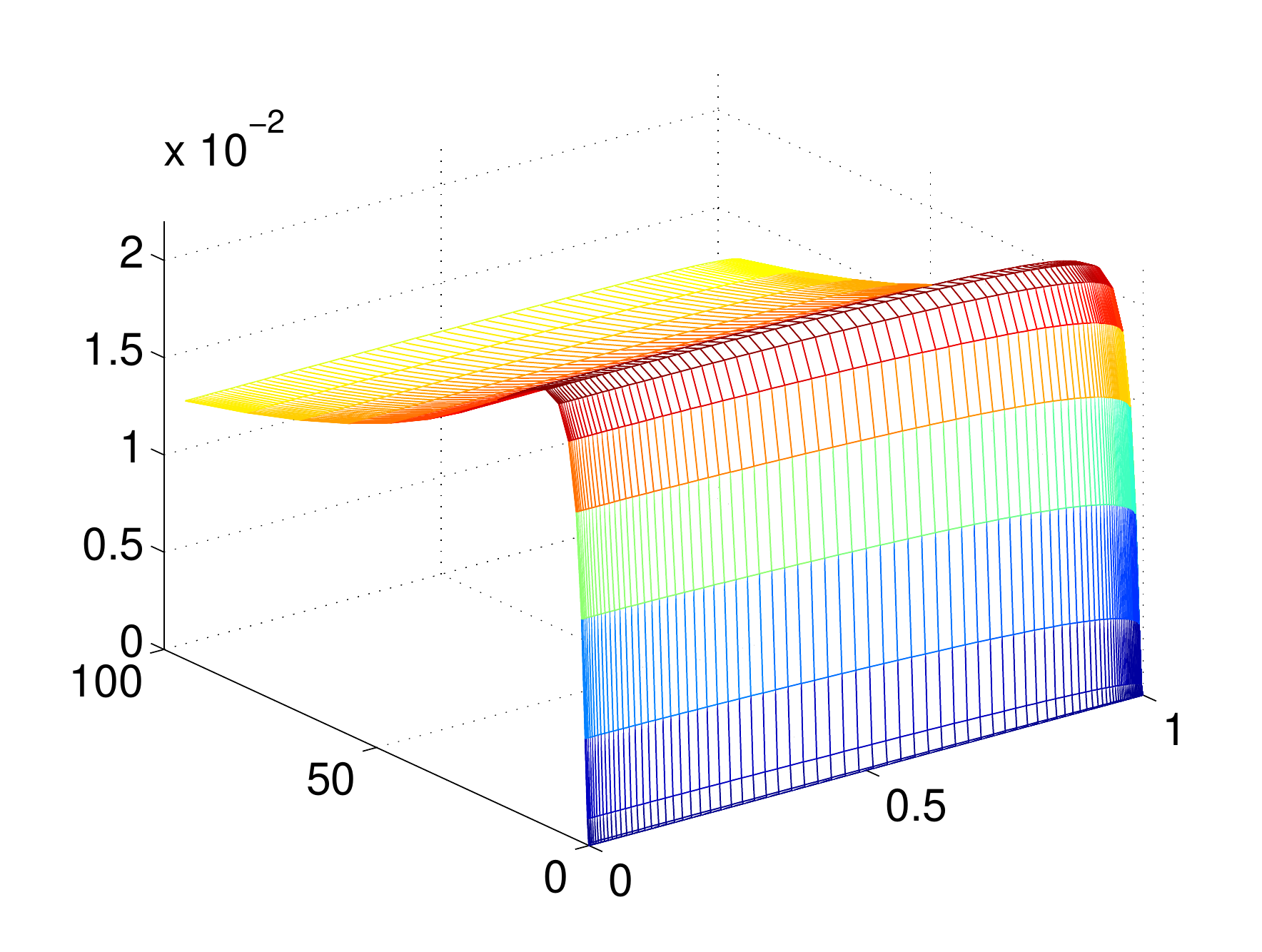}
    \put(-62,13){$x$}
    \put(-175,15){$t$}
    \put(-235,90){$ \delta w$}
    \put(-230,160){$\textbf{a)}$}
    \hspace{1mm}
    \includegraphics [scale=0.45]{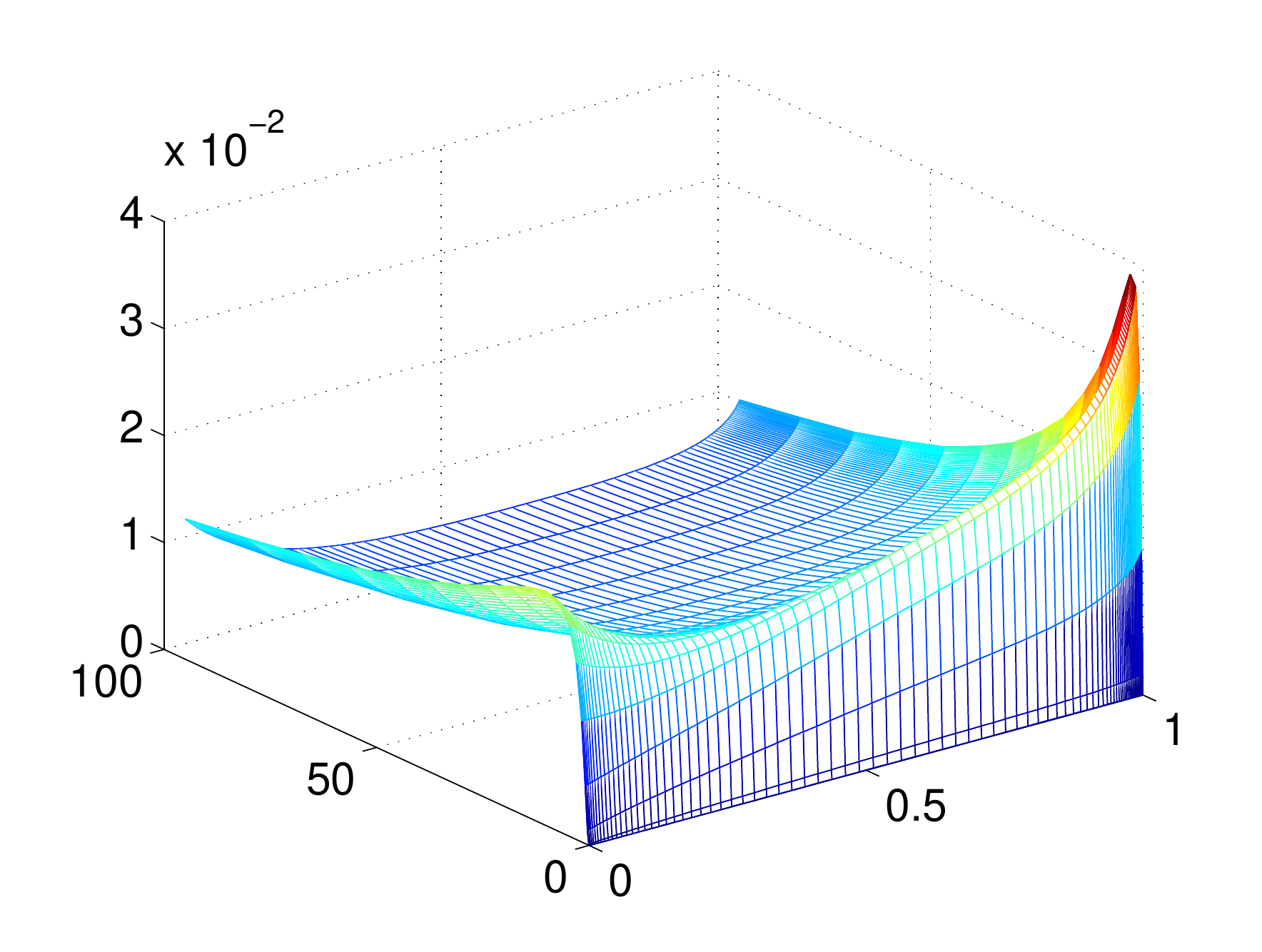}
    \put(-62,13){$x$}
    \put(-175,15){$t$}
    \put(-235,90){$\delta v$}
    \put(-230,160){$\textbf{b)}$}

    \caption{\textit{KGD model (viscosity dominated regime)}. Relative solution error for $n=0.3$ with $N=100$ (spatial mesh), $M=20$ (temporal
mesh): a) the crack opening $\delta w$, b) the particle velocity $\delta v$.}

\label{KGD_f_dyn_n03_t20}
\end{figure}

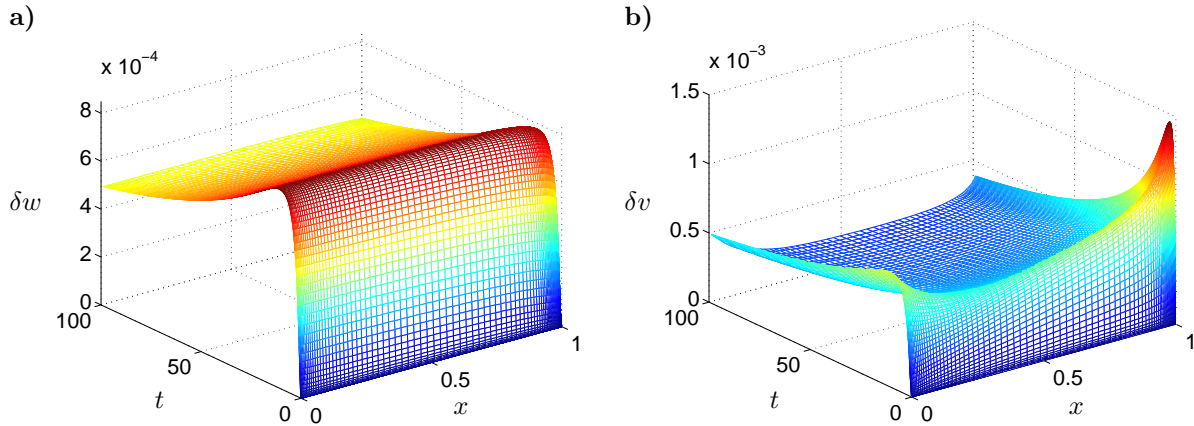
\begin{figure}[h!]

    \includegraphics [scale=0.45]{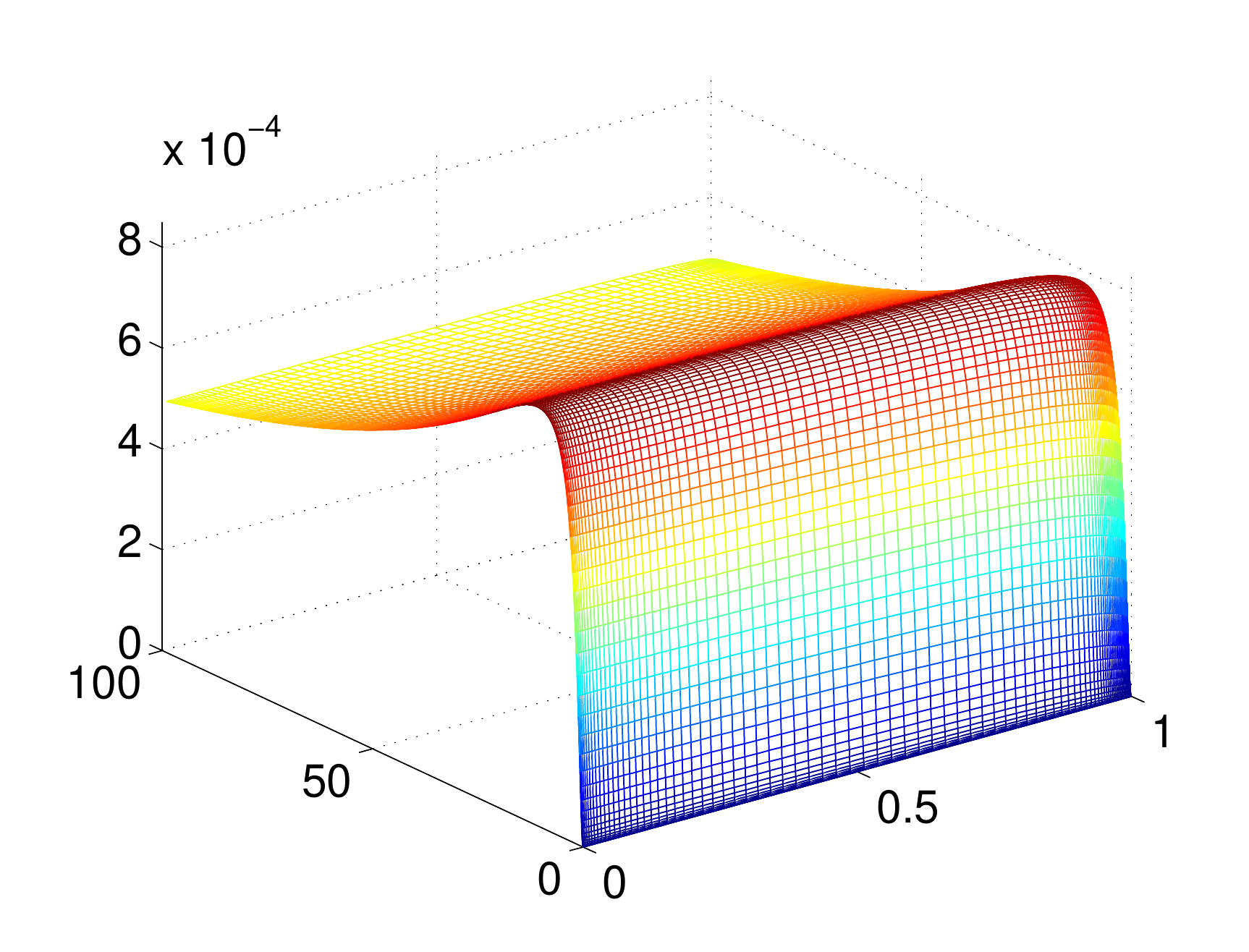}
    \put(-62,13){$x$}
    \put(-175,15){$t$}
    \put(-230,90){$ \delta w$}
    \put(-230,160){$\textbf{a)}$}
    \hspace{1mm}
    \includegraphics [scale=0.45]{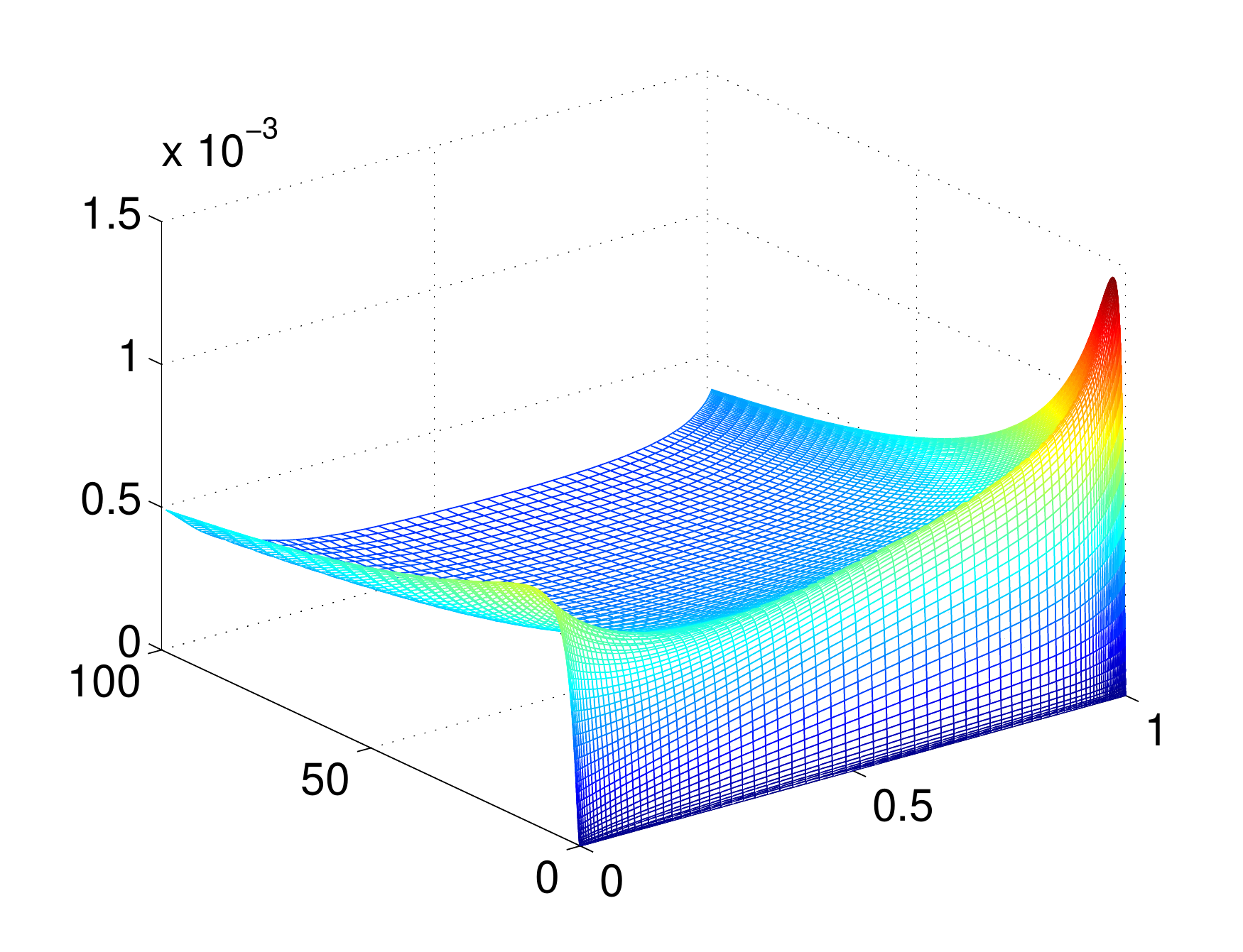}
    \put(-62,13){$x$}
    \put(-175,15){$t$}
    \put(-230,90){$\delta v$}
    \put(-230,160){$\textbf{b)}$}

    \caption{\textit{KGD model (viscosity dominated regime)}. Relative solution error for $n=0.3$ with $N=100$ (spatial mesh), $M=100$ (temporal
mesh): a) the crack opening $\delta w$, b) the particle velocity $\delta v$.}

\label{KGD_f_dyn_n03_t100}
\end{figure}

Fig. \ref{KGD_f_dyn_n07_t20} -- \ref{KGD_f_dyn_n07_t100} depict the solution error for $n=0.7$. This time, for fixed $M$, we obtain the same level of inaccuracy for both the crack opening and the particle velocity. In this case the magnification of $\delta v$ at the crack tip is not observed. Both errors remain stable over the whole range of $x$. In comparison with the case when $n=0.3$, the accuracy is of at least one order of magnitude better.

\begin{figure}[h!]

    \includegraphics [scale=0.45]{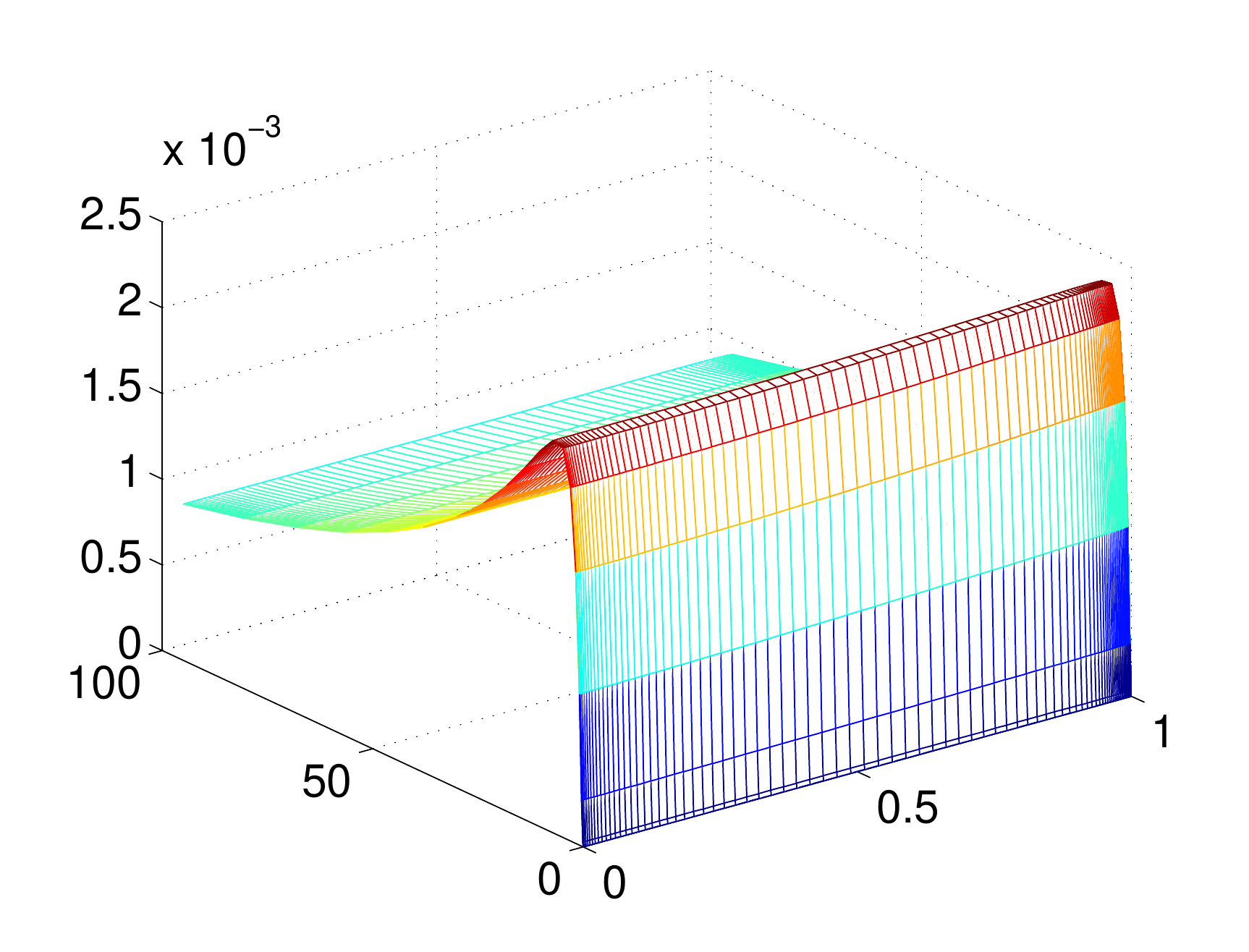}
    \put(-62,13){$x$}
    \put(-175,15){$t$}
    \put(-230,90){$ \delta w$}
    \put(-230,160){$\textbf{a)}$}
    \hspace{1mm}
    \includegraphics [scale=0.45]{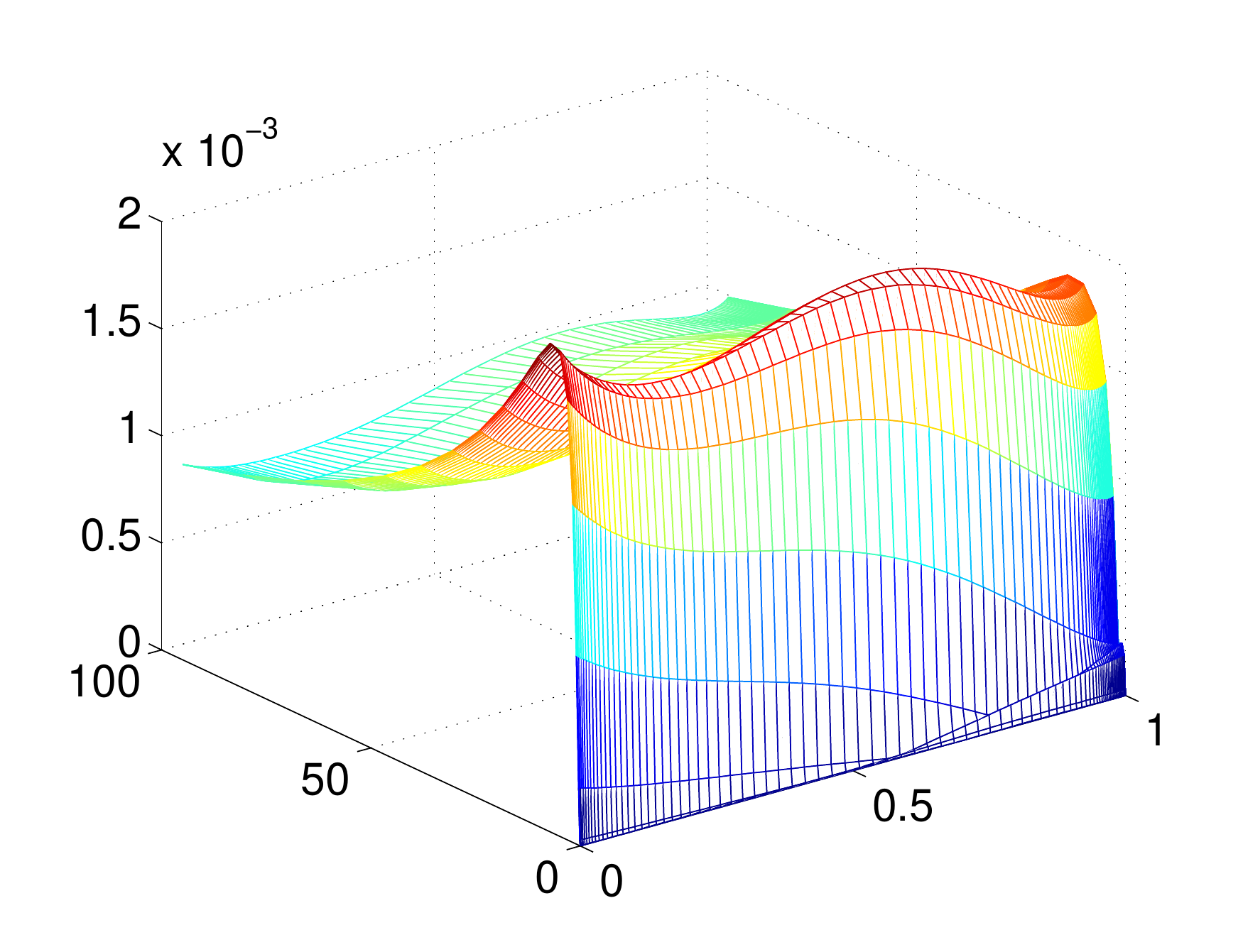}
    \put(-62,13){$x$}
    \put(-175,15){$t$}
    \put(-230,90){$\delta v$}
    \put(-230,160){$\textbf{b)}$}

    \caption{\textit{KGD model (viscosity dominated regime)}. Relative solution error for $n=0.7$ with $N=100$ (spatial mesh), $M=20$ (temporal
mesh): a) the crack opening $\delta w$, b) the particle velocity $\delta v$.}

\label{KGD_f_dyn_n07_t20}
\end{figure}

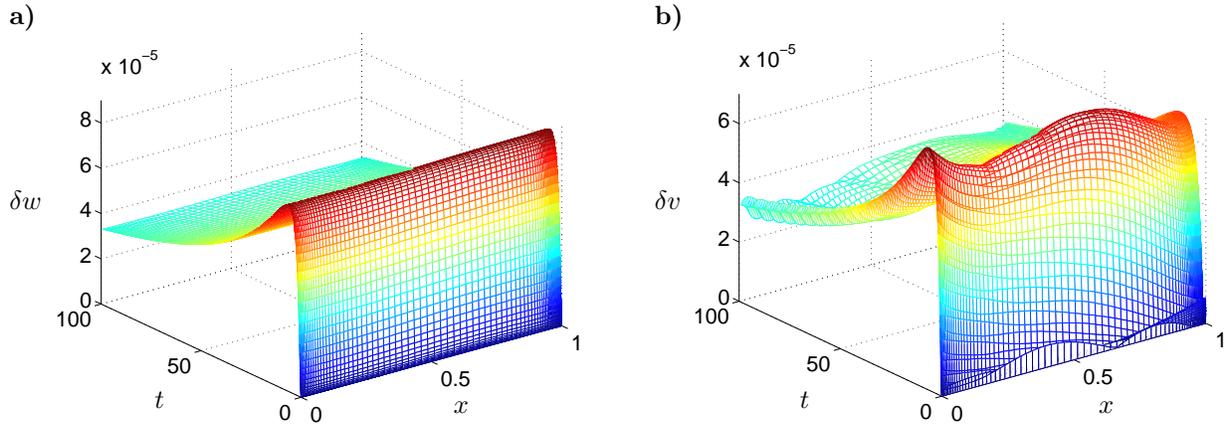
\begin{figure}[h!]

    \includegraphics [scale=0.45]{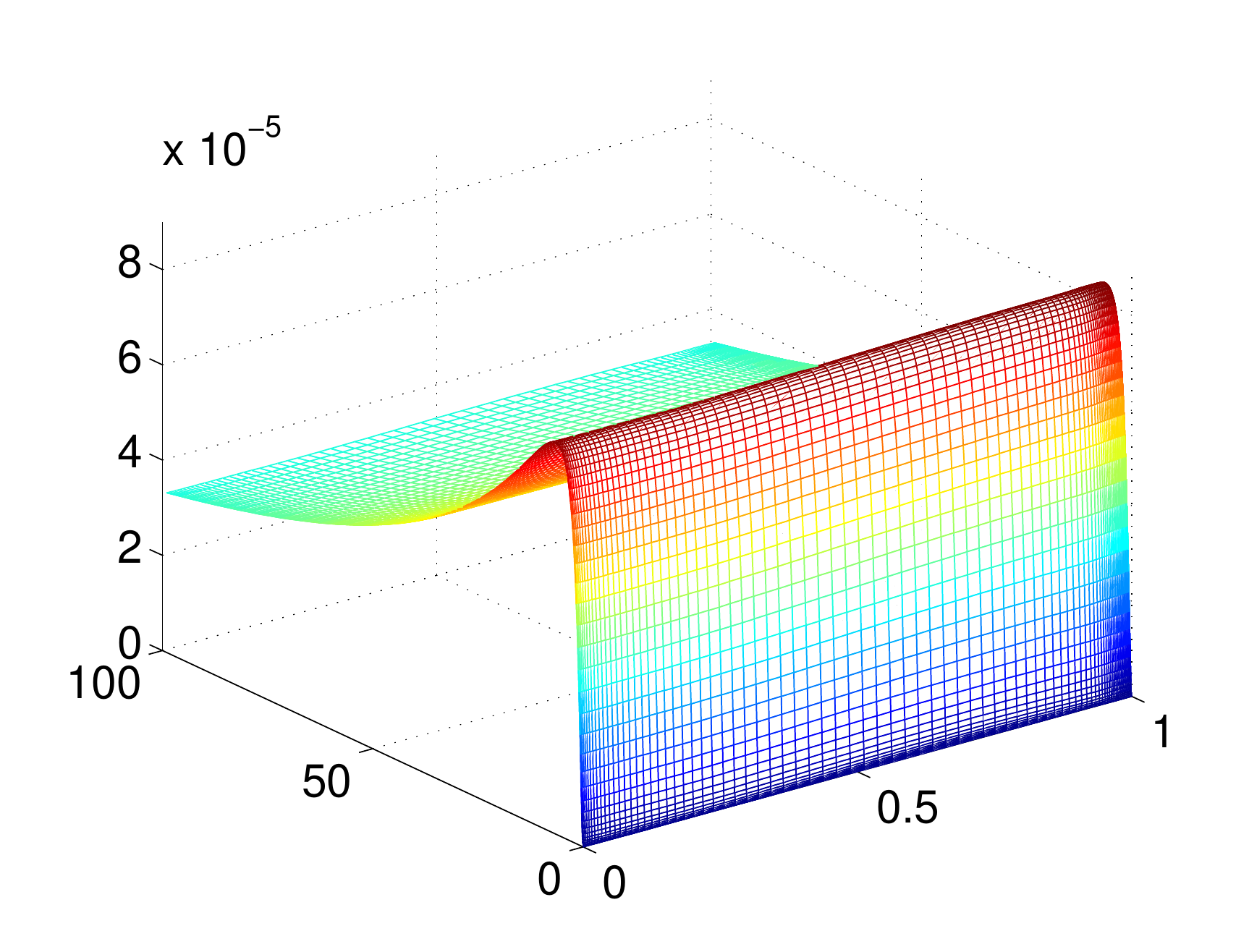}
    \put(-62,13){$x$}
    \put(-175,15){$t$}
    \put(-230,90){$ \delta w$}
    \put(-230,160){$\textbf{a)}$}
    \hspace{5mm}
    \includegraphics [scale=0.45]{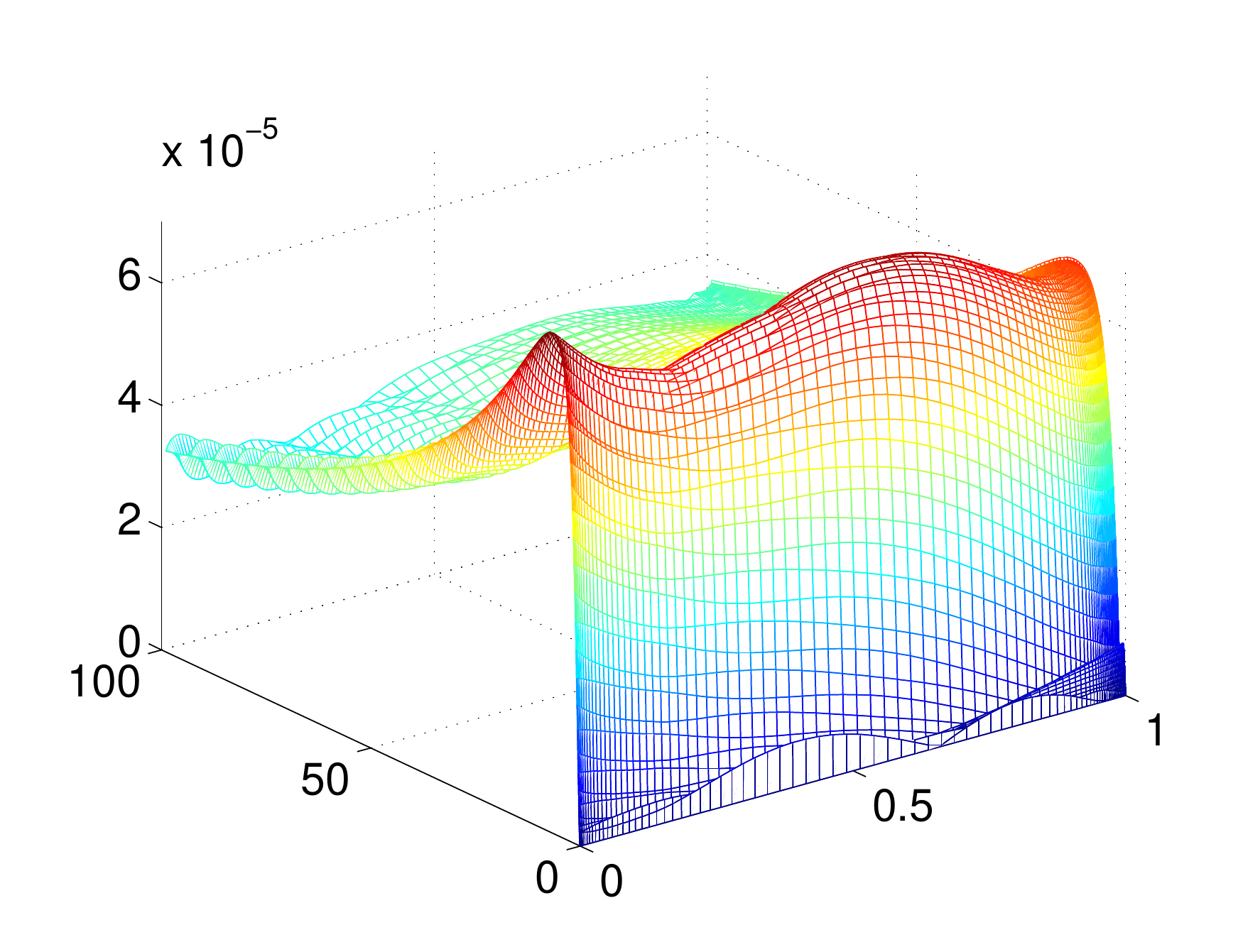}
    \put(-62,13){$x$}
    \put(-175,15){$t$}
    \put(-230,90){$\delta v$}
    \put(-230,160){$\textbf{b)}$}

    \caption{\textit{KGD model (viscosity dominated regime)}. Relative solution error for $n=0.7$ with $N=100$ (spatial mesh), $M=100$ (temporal
mesh): a) the crack opening $\delta w$, b) the particle velocity $\delta v$.}

\label{KGD_f_dyn_n07_t100}
\end{figure}

The relative errors for the crack length $L(t)$ are presented in Fig. \ref{dyn_blad_L}b. The character of growth for $\delta L$, as well as its magnitude,
are very similar to those reported previously for the PKN model (Fig. \ref{dyn_blad_L}a).

\clearpage

\subsubsection{Analysis of the algorithm - KGD model in toughness dominated regime}

The results of computations for $n=0.3$, illustrated by the relative solution error, are presented in Fig. \ref{KGD_t_dyn_n03_t20} -- \ref{KGD_t_dyn_n03_t100}. The level of $\delta w$ for different $M$ is now one order of magnitude lower than that for the viscosity dominated benchmark, which we believe is not a feature of the algorithm itself but rather of the analyzed benchmark (as was shown in the self-similar formulation, the solution accuracy depends on the benchmark type). In the case of particle velocity, the magnitude of maximal errors remains comparable to those obtained previously in the viscosity dominated regime, but the spacing of $\delta v$ is completely different. This time a distinct error magnification is observed at the crack tip over the whole time interval. It can be explained by the fact that the crack propagation speed is now computed based on the two leading asymptotic terms of $w$ (instead of one for the PKN model and KGD model in viscosity dominated regime, compare \eqref{LC_PKN} -- \eqref{LC_KGD_fluid}). The multiplier of the second term is approximated, by its very nature, in a less accurate way than the coefficient of the leading term, which may affect the overall accuracy of $v(t,1)$.

\begin{figure}[h!]

    \includegraphics [scale=0.45]{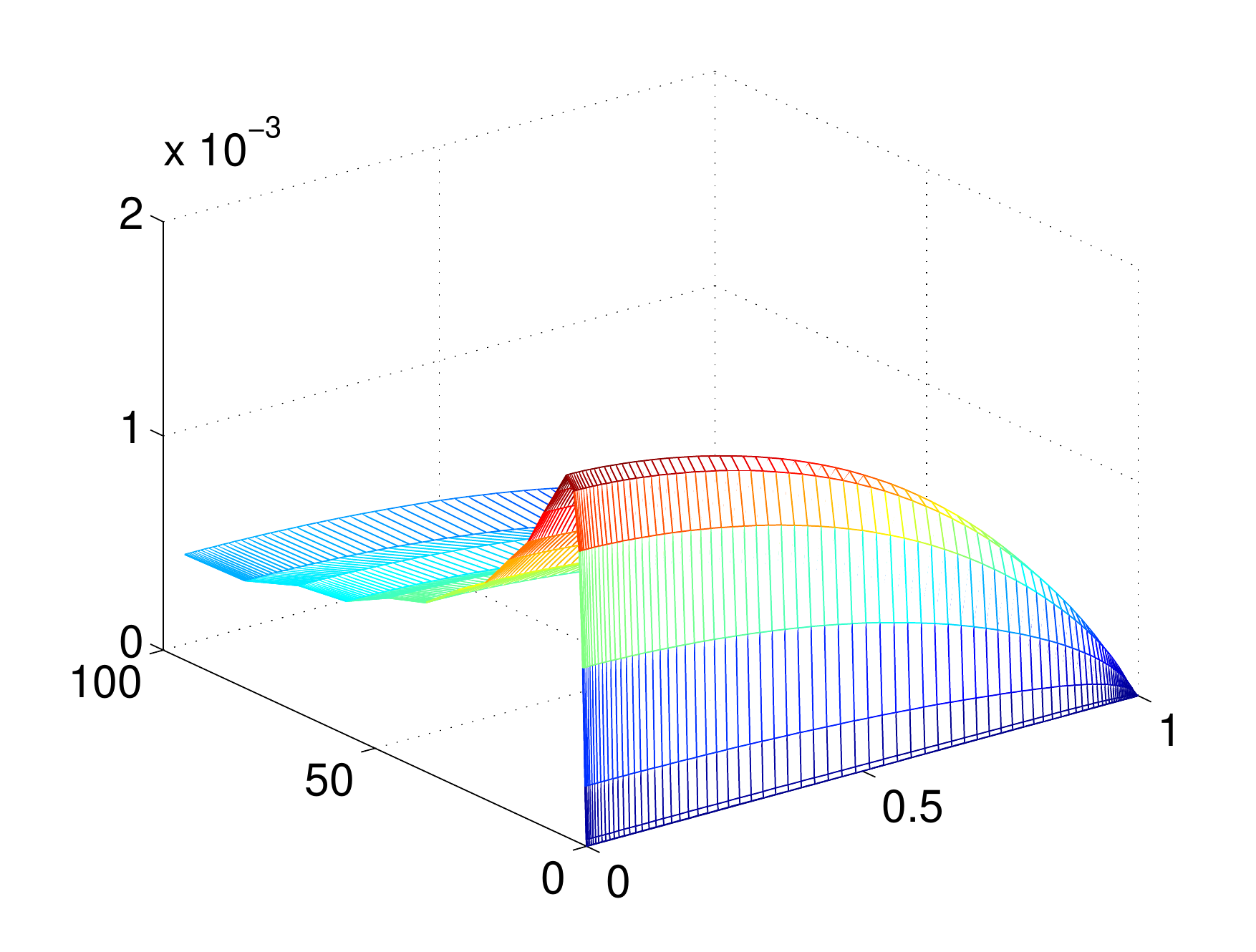}
    \put(-62,13){$x$}
    \put(-175,15){$t$}
    \put(-230,110){$ \delta w$}
    \put(-230,160){$\textbf{a)}$}
    \hspace{1mm}
    \includegraphics [scale=0.45]{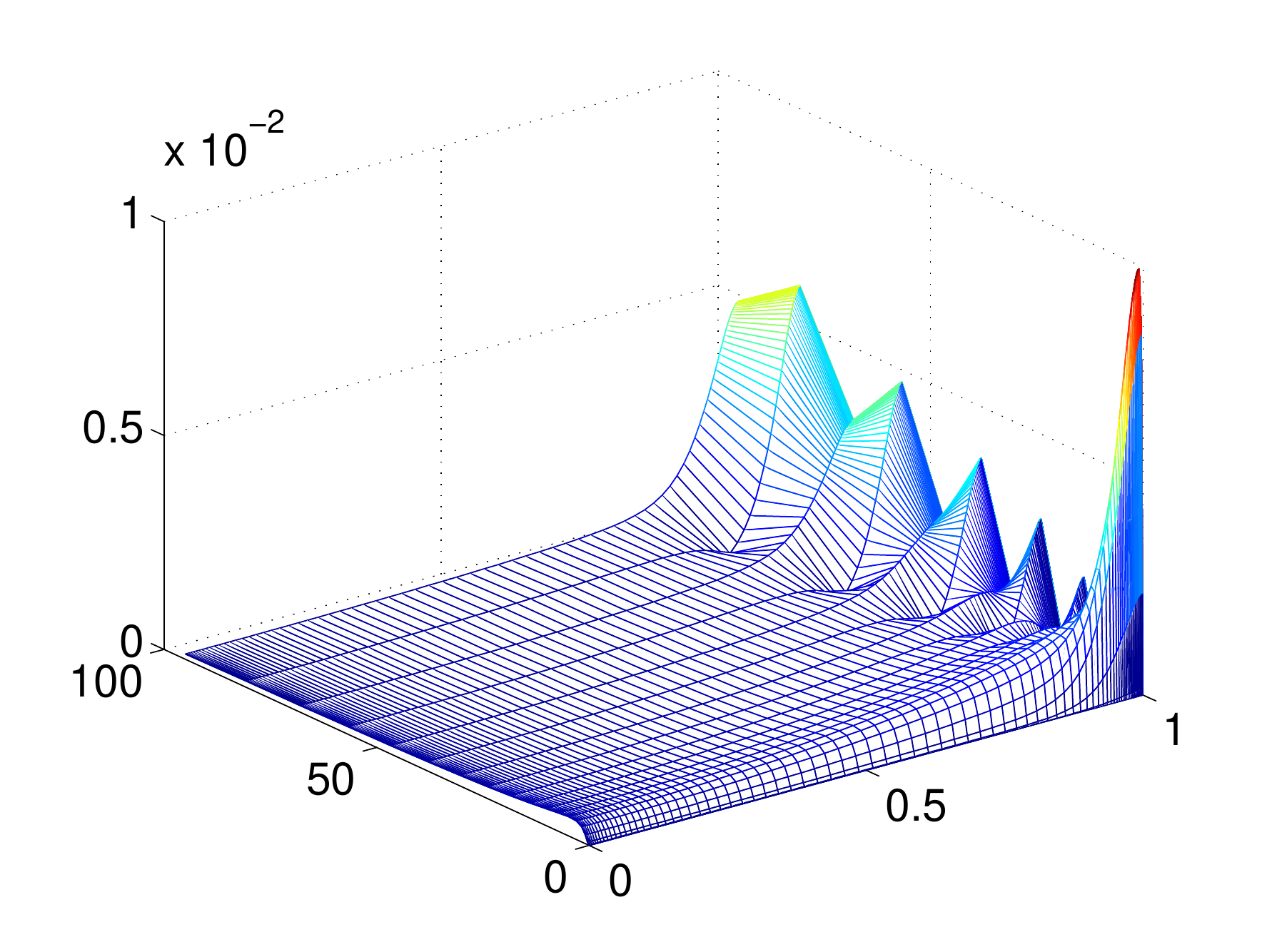}
    \put(-62,13){$x$}
    \put(-175,15){$t$}
    \put(-230,110){$\delta v$}
    \put(-230,160){$\textbf{b)}$}

    \caption{\textit{KGD model (toughness dominated regime)}. Relative solution error for $n=0.3$ with $N=100$ (spatial mesh), $M=20$ (temporal
mesh): a) the crack opening $\delta w$, b) the particle velocity $\delta v$.}

\label{KGD_t_dyn_n03_t20}
\end{figure}

\begin{figure}[h!]

    \includegraphics [scale=0.45]{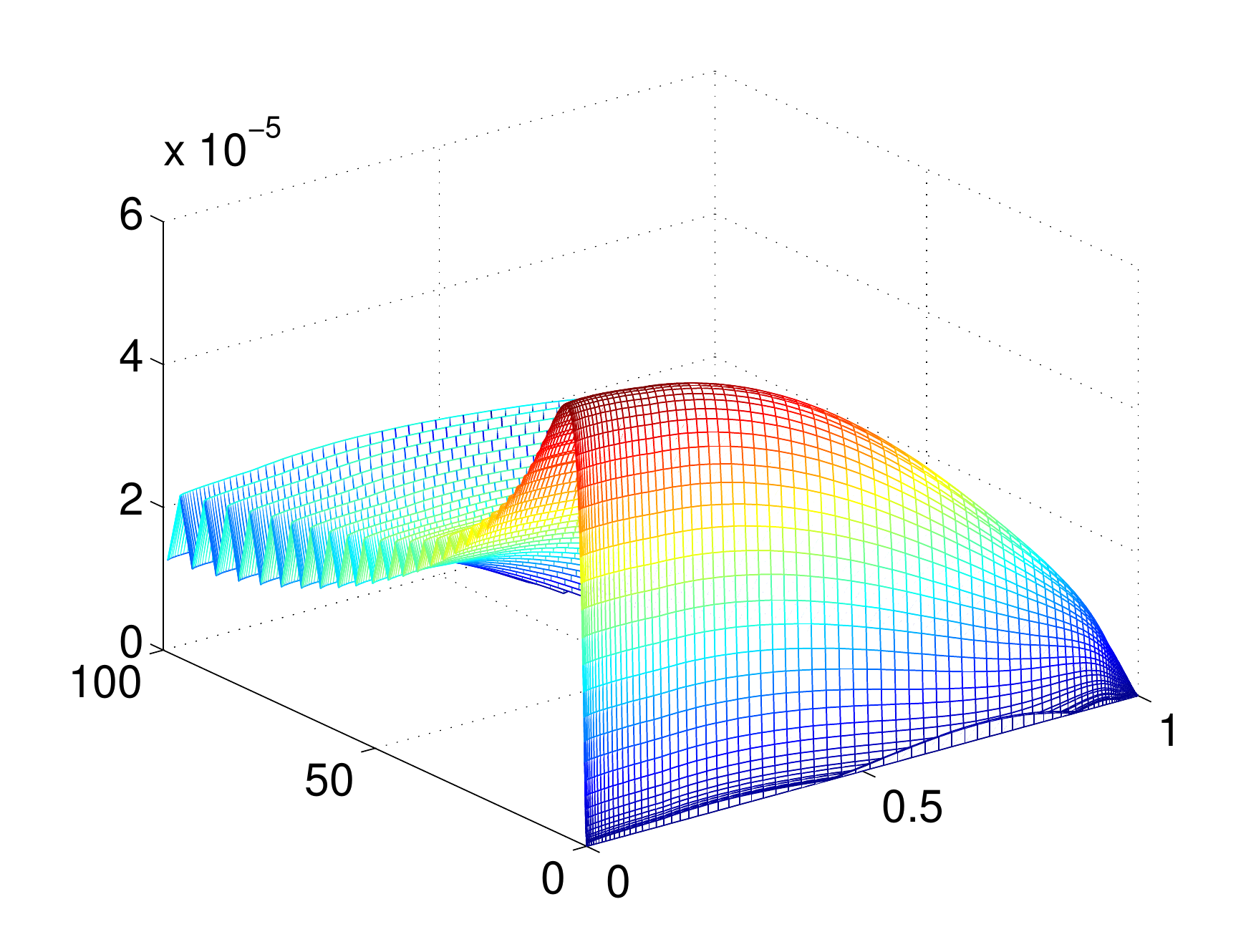}
    \put(-62,13){$x$}
    \put(-175,15){$t$}
    \put(-230,90){$ \delta w$}
    \put(-230,160){$\textbf{a)}$}
    \hspace{1mm}
    \includegraphics [scale=0.45]{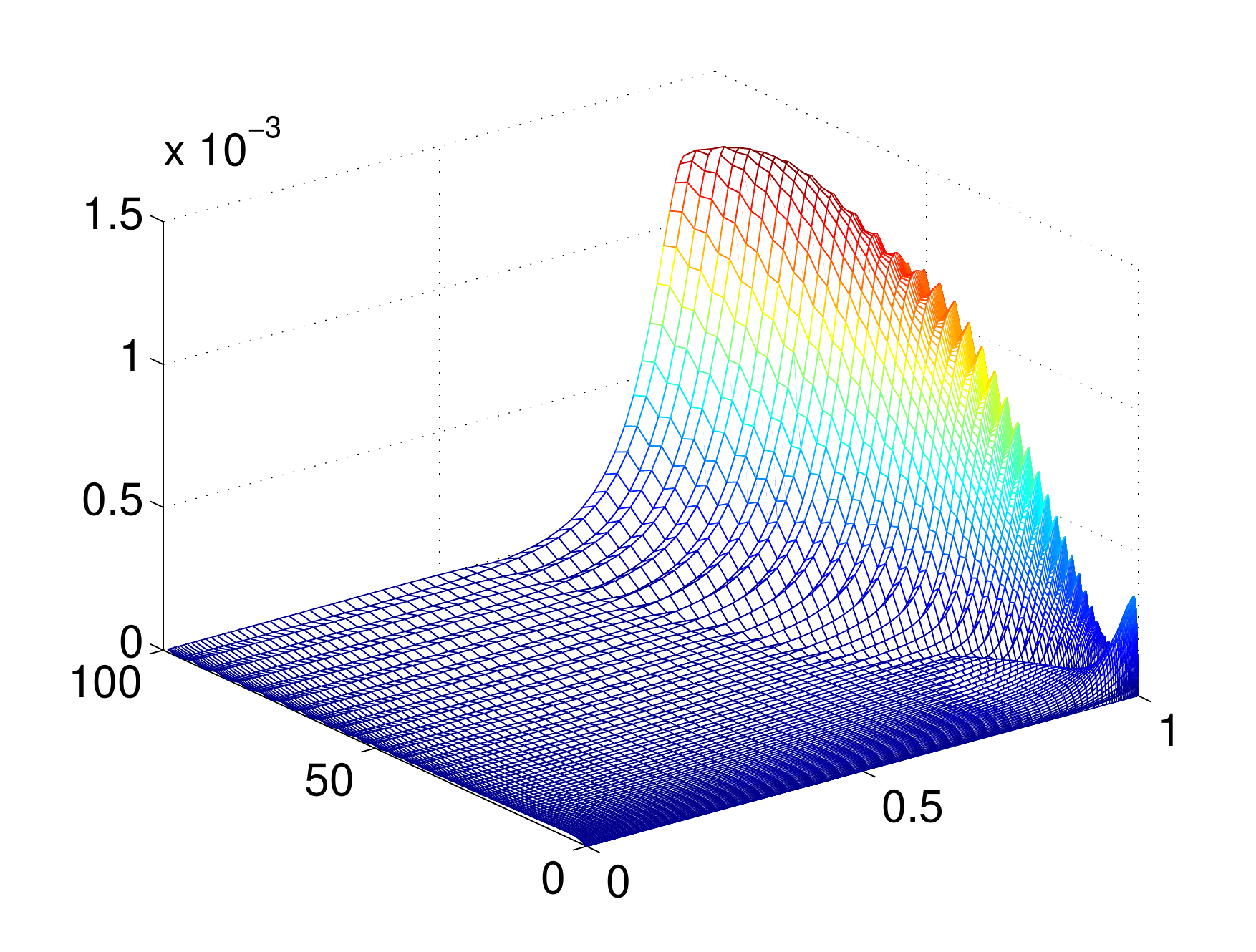}
    \put(-62,13){$x$}
    \put(-175,15){$t$}
    \put(-230,90){$\delta v$}
    \put(-230,160){$\textbf{b)}$}

    \caption{\textit{KGD model (toughness dominated regime)}. Relative solution error for $n=0.3$ with $N=100$ (spatial mesh), $M=100$ (temporal
mesh): a) the crack opening $\delta w$, b) the particle velocity $\delta v$.}

\label{KGD_t_dyn_n03_t100}
\end{figure}
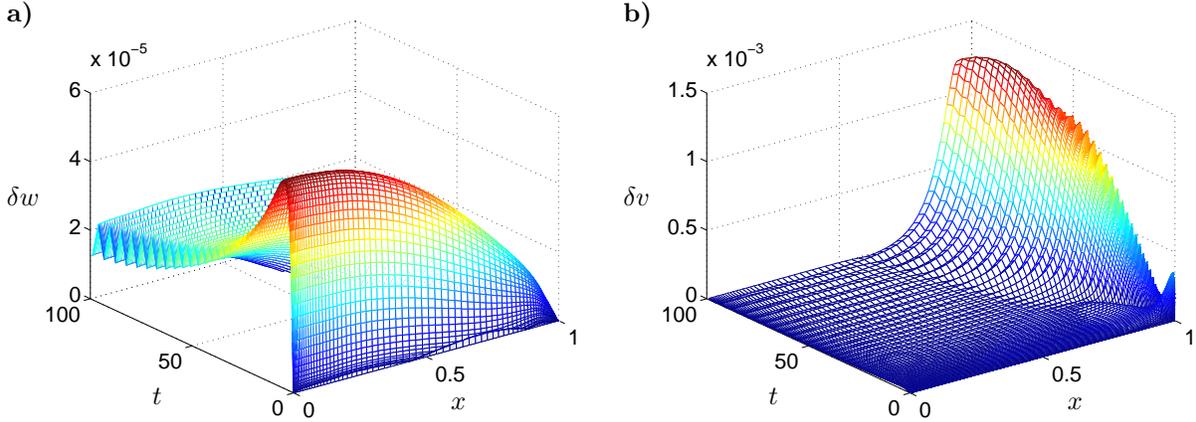

The results for $n=0.7$ are shown in Fig. \ref{KGD_t_dyn_n07_t20} -- Fig. \ref{KGD_t_dyn_n07_t100}. The graphs for $\delta w$ resemble those obtained for $n=0.3$ in both the level and distribution of errors. By taking $M=100$ instead of $M=20$ one increases the accuracy by two orders of magnitude. In the case of particle velocity we do not have a characteristic error increase in the near-tip region, which was present for $n=0.3$. However, a sharp magnification of $\delta v(t,1)$ for small time is still in place. Also, some minor fluctuations in the tip velocity error for larger times are observed. Compared to the variant with $n=0.3$ the maximal level of $\delta v$ remained the same for $M=20$, but was reduced by one order of magnitude for the refined temporal meshing ($M=100$).

\begin{figure}[h!]

    \includegraphics [scale=0.45]{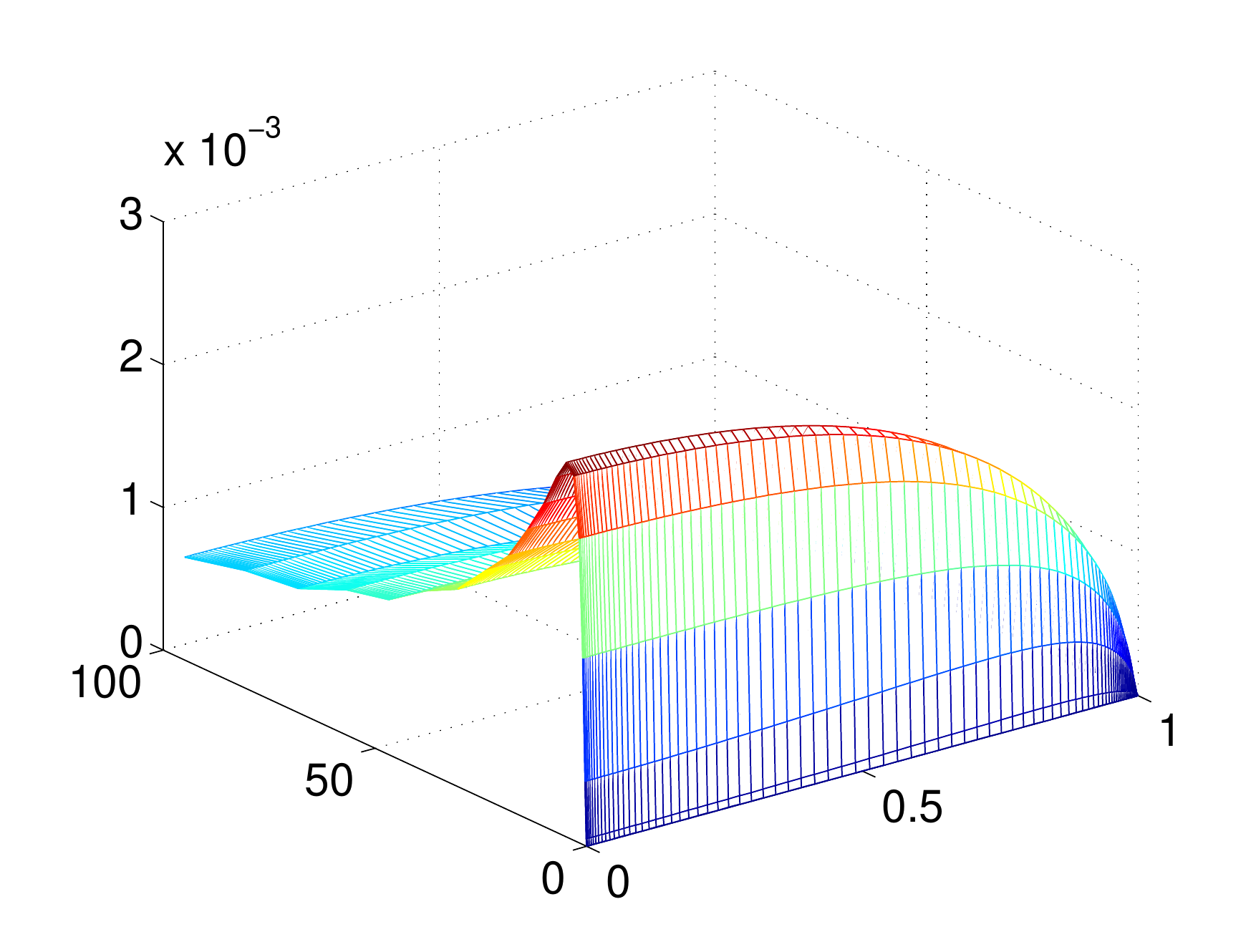}
    \put(-62,13){$x$}
    \put(-175,15){$t$}
    \put(-230,110){$ \delta w$}
    \put(-230,160){$\textbf{a)}$}
    \hspace{5mm}
    \includegraphics [scale=0.45]{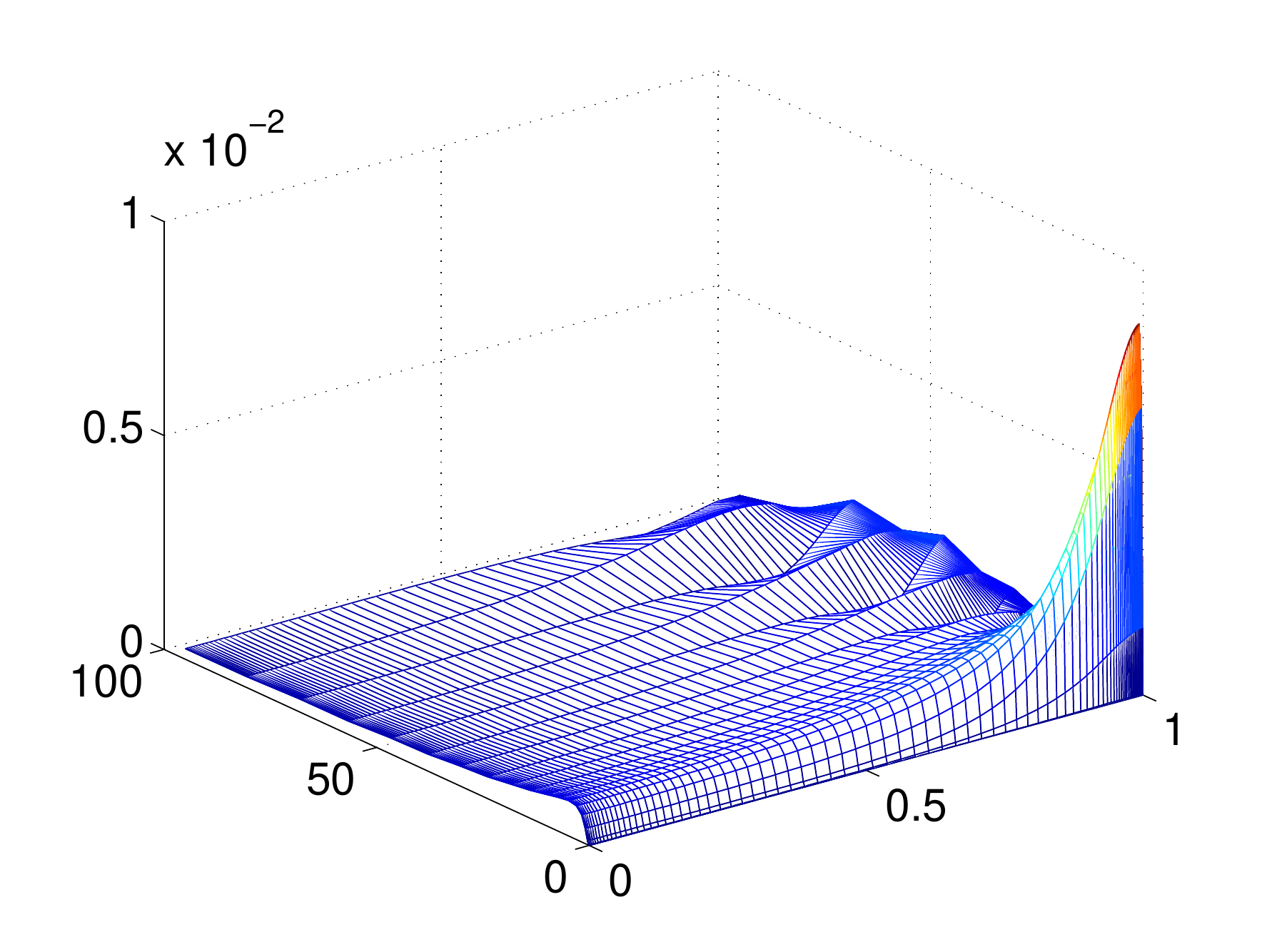}
    \put(-62,13){$x$}
    \put(-175,15){$t$}
    \put(-230,110){$\delta v$}
    \put(-230,160){$\textbf{b)}$}

    \caption{\textit{KGD model (toughness dominated regime)}. Relative solution error for $n=0.7$ with $N=100$ (spatial mesh), $M=20$ (temporal
mesh): a) the crack opening $\delta w$, b) the particle velocity $\delta v$.}

\label{KGD_t_dyn_n07_t20}
\end{figure}

\begin{figure}[h!]

    \includegraphics [scale=0.45]{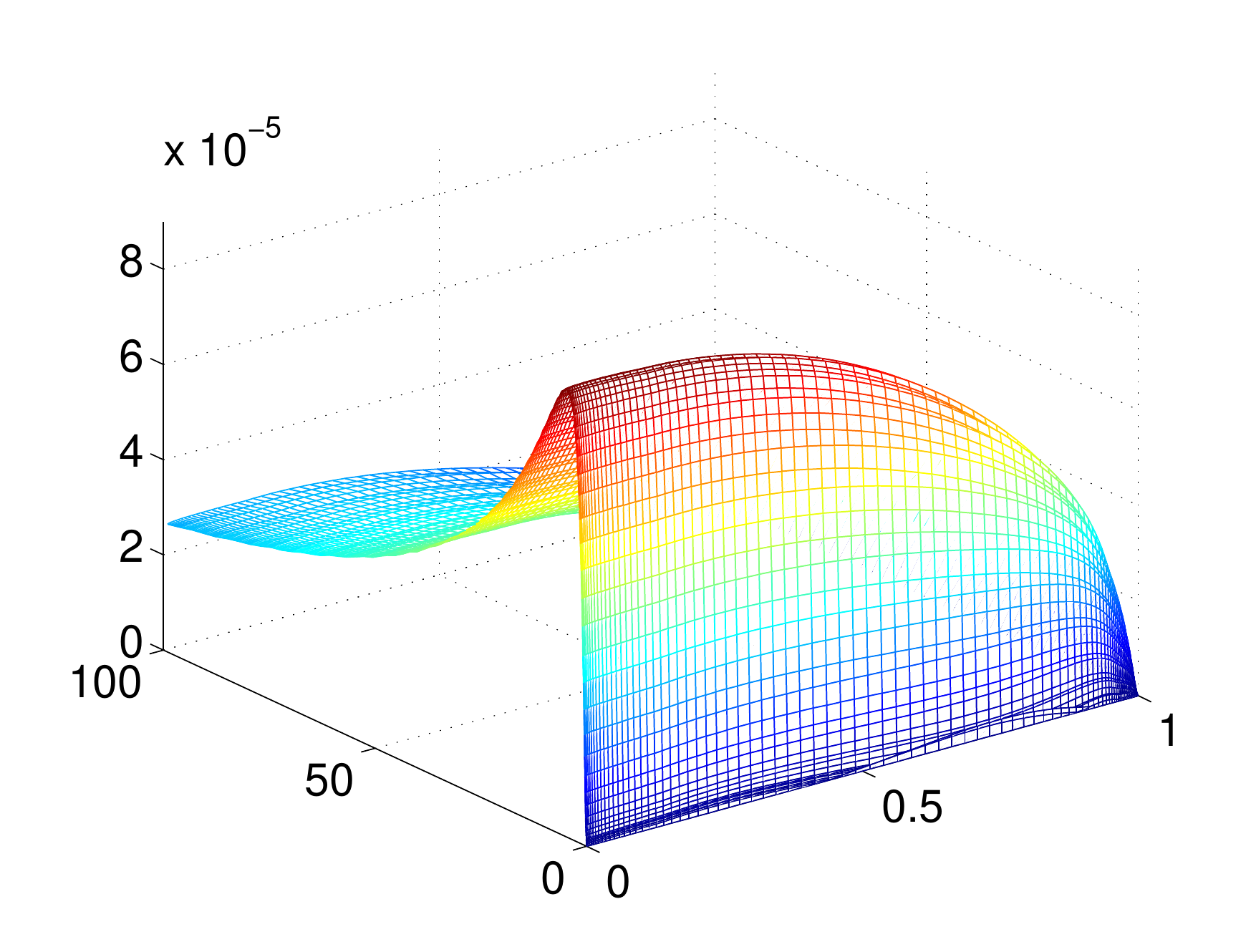}
    \put(-62,13){$x$}
    \put(-175,15){$t$}
    \put(-230,90){$ \delta w$}
    \put(-230,160){$\textbf{a)}$}
    \hspace{5mm}
    \includegraphics [scale=0.45]{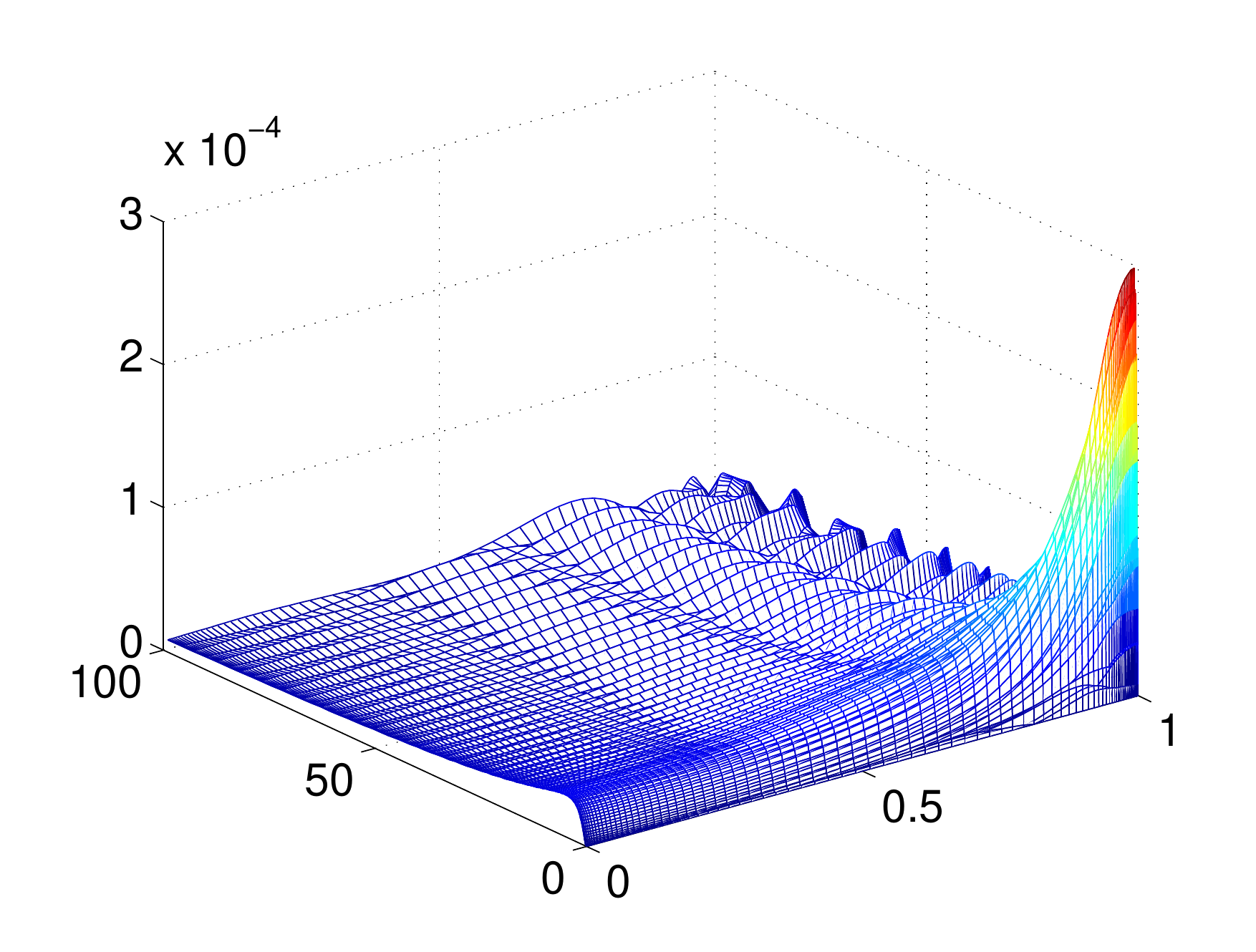}
    \put(-62,13){$x$}
    \put(-175,15){$t$}
    \put(-230,90){$\delta v$}
    \put(-230,160){$\textbf{b)}$}

    \caption{\textit{KGD model (toughness dominated regime)}. Relative solution error for $n=0.7$ with $N=100$ (spatial mesh), $M=100$ (temporal
mesh): a) the crack opening $\delta w$, b) the particle velocity $\delta v$.}

\label{KGD_t_dyn_n07_t100}
\end{figure}

Finally, the graphs for the crack length error are shown in Fig. \ref{dyn_blad_L}c. This time the differences in the value of $\delta L$ are much smaller than that for other HF models. Moreover, unlike the previous cases, now it is $n=0.3$ which gives slightly better results. However the general trend of error increase in the early time, and its stabilization after the initial growth, is still present.

\begin{figure}[h!]
		\hspace{-6mm}
		\includegraphics [scale=0.32]{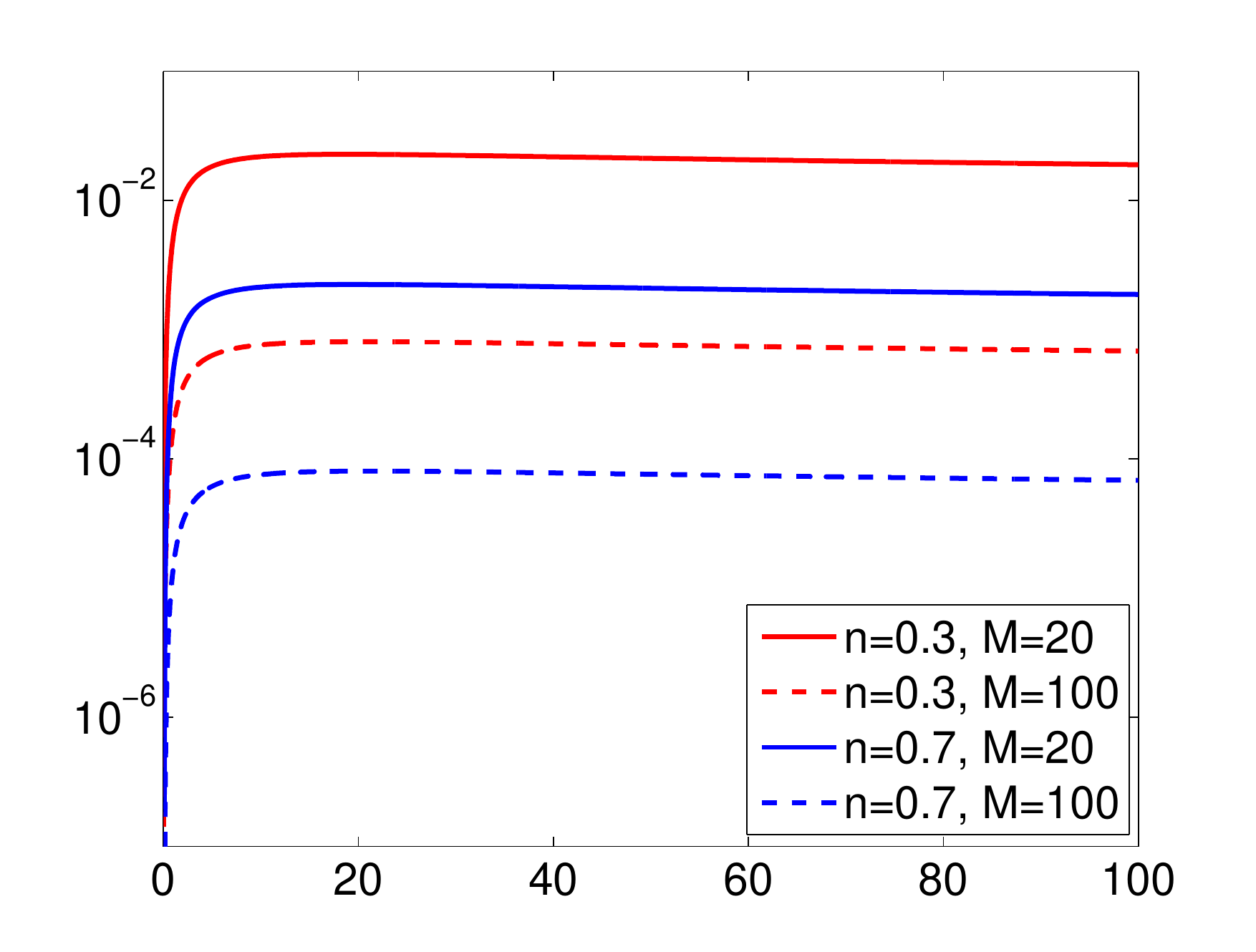}
		\put(-155,105){$\delta L$}
		\put(-85,-5){$t$}
    \hspace{-6mm}
    \includegraphics [scale=0.32]{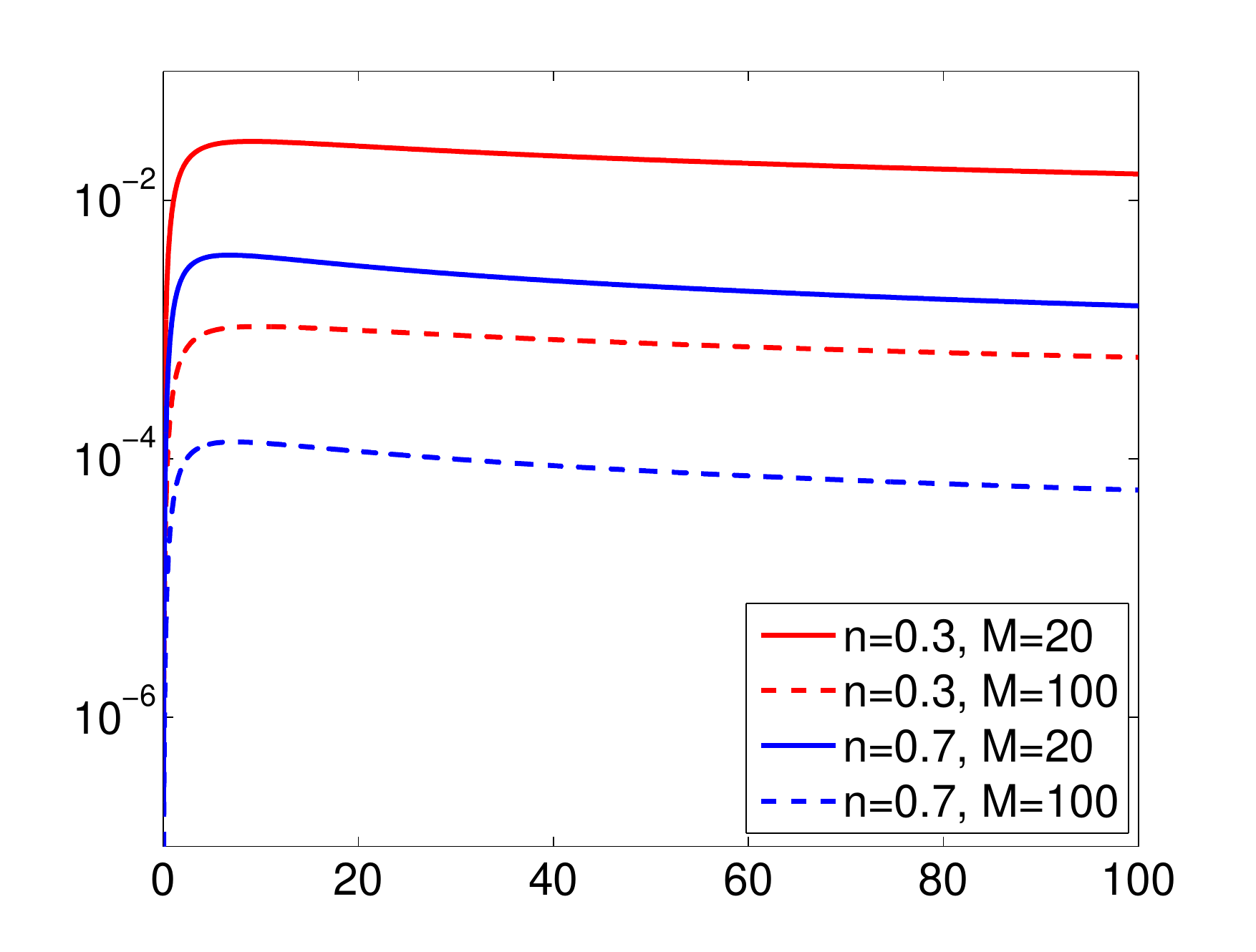}
		\put(-155,105){$\delta L$}
		\put(-85,-5){$t$}
    \hspace{-6mm}
    \includegraphics [scale=0.32]{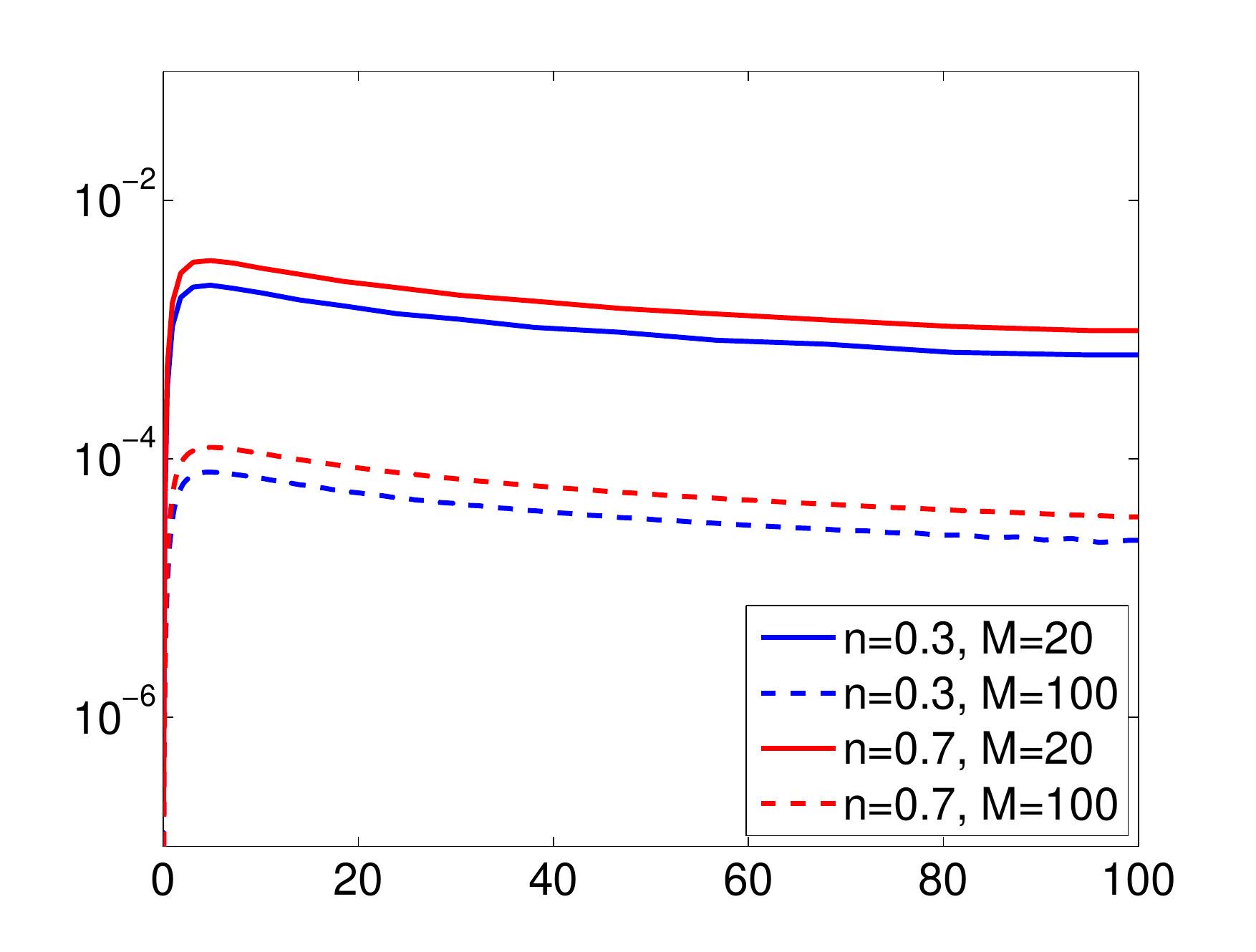}
		\put(-155,105){$\delta L$}
  	\put(-85,-5){$t$}

    \caption{Relative error of the fracture length: a) PKN model, b) KGD model in viscosity dominated regime , c) KGD model in toughness dominated regime. Spatial mesh: $N=100$.}

\label{dyn_blad_L}
\end{figure}

\section{Discussion and Conclusions}
\label{sec:conclusions}

In the paper a problem of 1D hydraulic fracture has been analyzed. The power law rheological model of the fracturing fluid has been employed for the classical PKN and KGD (in viscosity and toughness dominated regimes) models. Following the idea from \cite{wr_mis_2015} the problem has been reformulated in terms of a new pair of dependent variables: the crack opening, $w$, and the reduced particle velocity, $\phi$. A unified equation \eqref{v_0_univ} defining the interrelation between the crack propagation speed and the solution tip asymptotics has been derived for various elasticity operators and crack propagation regimes. Self-similar formulations of the problem for different HF models have been delivered.

A universal algorithm for solving the problem, being an extension of that proposed in \cite{wr_mis_2015}, has been introduced. It enables us to tackle various elasticity operators, fluid flow and fracture propagation regimes in the framework of a unified numerical scheme. The computational algorithm is comprised of two basic modules:
\begin{itemize}
\item{Universal module to compute the reduced particle velocity -- this block remains the same regardless of the HF model and crack propagation regime. }
\item{Module to compute the crack opening -- this block is adjusted depending on the elasticity relation in use. The asymptotic parameters embedded here facilitate adaptation of the module to various crack propagation regimes.}
\end{itemize}

Crucial elements of the proposed algorithm are:
\begin{itemize}
\item{The mechanism of fracture front tracing based on the local condition of Stefan type (speed equation). This device utilizes an explicit formula \eqref{v_0_univ} relating the crack propagation speed to the crack opening tip asymptotics. This in turn allows us to compute in an explicit form the crack length from the equation \eqref{L_int}.}
\item{Appropriate handling of the independent (spatial and temporal meshing) and dependent variables.}
\item{Improved approximation of the temporal derivative. }
\item{Dedicated regularization techniques (the so-called $\varepsilon$-regularization and operator regularization of the governing equations)}.
\end{itemize}

The algorithm performance has been thoroughly investigated. To this end a set of analytical benchmarks, introduced in Appendix A, have been utilized. The accuracy of computations has also been verified against other solutions available in the literature.
The following conclusions can be drawn about the performance of the proposed numerical scheme:
\begin{itemize}
\item{The algorithm is efficient and stable regardless of the hydraulic fracture model and crack propagation regime. }
\item{High solution accuracy can be obtained at low computational cost. Proper utilization of the basic devices embedded in the algorithm allows good results to be obtained for even coarse meshing.}
\item{Even in the extreme regimes of crack propagation (e.g the small toughness case) high solution quality can be retained, however it necessitates appropriate measures such as skilled handling of the $\varepsilon$-regularization technique and spatial meshing. As is to be expected, the computational efficiency decreases for such limiting variants of the problem. }
\item{The numerical results provided by the computational scheme are much more accurate than any other data available in the literature (e.g. for 100 spatial mesh points at least three orders better accuracy can be obtained regardless of the analyzed HF model).}
\end{itemize}

The developed numerical algorithm also allowed us to improve two important classical benchmark solutions. For the PKN model the  formula \eqref{xi_n} for the self-similar crack half-length, $\xi_*$, when combined with the
solution given in \cite{linkov_2013} provides an improved semi-analytical benchmark for an arbitrary value of the fluid behaviour index.
Additionally, a new improved solution approximation of the viscosity dominated regime of the KGD model has been proposed. The approximation guarantees  much better accuracy than any other semi-analytical solution available in the literature. Its quality is equally good for the whole class of shear-thinning fluids. The simple form of the proposed formulae facilitates their application in testing other numerical algorithms.

To summarize, the proposed algorithm has been proven to posses all of the advantages of the basic version originally introduced in  \cite{wr_mis_2015} for the Newtonian fluids. It constitutes a flexible and reliable tool for numerically modeling hydraulic fractures driven by shear-thinning fluids.

\vspace{7mm}
\textbf{Acknowledgments.}

The authors acknowledge support from FP7 Marie Curie IAPP projects PIAP-GA-2009-251475. MW is also grateful for the support of PIAP-GA-2011-286110.

\section*{Appendix A: Benchmark solutions}

Let us describe a set of benchmark solutions used to analyze the accuracy of computations. The main concept for their construction was already presented and employed in \cite{wr_mis_2015}. The basic assumption is that, if we can provide an analytical solution for the self-similar problem defined in Section \ref{sec:self_sim_formulation}, it can be immediately extended to the time-dependent form by relations \eqref{self_sim}.

The general idea of finding the benchmark solution is as follows. First, let us assume that the crack opening function can be expressed as a weighted sum of appropriately chosen base functions:
\begin{equation}\label{w_rep}
\hat w(x)=\sum_{i=0}^R \lambda_i \hat \Omega_i(x).
\end{equation}
The functions $\hat \Omega_i(x)$ are selected in a way which enables one to: i) comply with  the respective asymptotic representation \eqref{w_asym}, ii) compute the pressure operators \eqref{p_ss_2} analytically, iii) satisfy the boundary conditions \eqref{exp_ss_BC} (and \eqref{sym_ss} for KGD). Provided that ii) is fulfilled, the pressure function can be calculated in a closed form from \eqref{p_ss_2}:
\begin{equation}\label{p_rep}
\hat p(x)=\sum_{i=0}^R \lambda_i \hat \Pi_i(x),
\end{equation}
where each of the functions $\hat \Pi_i(x)$ corresponds to respective function $\hat \Omega_i(x)$.
Then, according to \eqref{v_ss}, the particle velocity function gives:
\begin{equation}
\label{v_rep}
\hat v(x)=\left[-\left(\sum_{i=0}^R \lambda_i \hat \Omega_i(x)\right)^{n+1}\sum_{i=0}^R \lambda_i \frac{d \hat \Pi_i}{dx}\right]^{\frac{1}{n}}.
\end{equation}
The reduced velocity can then be determined by using \eqref{v_rep} in \eqref{fi_sep}:
\begin{equation}
\label{fi_rep}
\hat \phi(x)=\left[-\left(\sum_{i=0}^R \lambda_i \hat \Omega_i(x)\right)^{n+1}\sum_{i=0}^R \lambda_i \frac{d \hat \Pi_i}{dx}\right]^{\frac{1}{n}}-x\Big(C_{\cal A}{\cal L}(\hat w)\Big)^{\frac{1}{n}}.
\end{equation}

Next, by applying \eqref{w_rep} and \eqref{fi_rep} in \eqref{ODE_gen}, the benchmark leak-off function $\hat q_l(x)$ can be defined.
The benchmark value of $\hat q_0$ is determined by substituting \eqref{w_rep} and \eqref{fi_rep} into the boundary condition \eqref{exp_ss_BC}$_3$:
\begin{equation}
\label{q0_rep}
\hat q_0=\left[-\left(\sum_{i=0}^R \lambda_i \hat \Omega_i(0)\right)^{n+2}\sum_{i=0}^R \lambda_i \frac{d \hat \Pi_i}{dx} \big|_{x=0}\right]^{\frac{1}{n}} .
\end{equation}
In this way, the analytical benchmark solution is fully defined by \eqref{w_rep},
\eqref{fi_rep}, the corresponding leak-off function and the corresponding influx value \eqref{q0_rep}. The respective fluid balance equation \eqref{balance_ss} is satisfied automatically.

\subsection*{A1: PKN model}
For the PKN model let us adopt the following base functions $\hat w_i(x)$:
\begin{equation}
\label{h_PKN}
\hat \Omega_i(x)=(1-x)^{i+\frac{1}{n+2}}, \quad i=0,..,R-1, \quad \hat \Omega_R(x)=e^x(1-x)^{R+\frac{1}{n+2}}.
\end{equation}
The function $\hat \Omega_R(x)$ was taken in this manner in order to introduce an additional non-linear effect to the benchmark, without
violating the asymptotic behaviour.

By applying representation \eqref{h_PKN} in \eqref{w_rep} and elasticity operator \eqref{p_ss_inv_PKN}, one obtains the formula for the pressure function, $\hat p(x)$,
which after differentiation yields:
\begin{equation}
\label{pp_PKN}
\frac{d}{dx}\hat p(x)=-\left[\sum_{i=0}^{R-1}\lambda_i \left(i+\frac{1}{n+2}\right)(1-x)^{i-\frac{n+1}{n+2}}+\lambda_R e^x(1-x)^{R+\frac{1}{n+2}}\left(\frac{R+\frac{1}{n+2}}{1-x} -1\right)\right].
\end{equation}
Then by formulae \eqref{v_rep} -- \eqref{q0_rep} one can construct the benchmark solution, taking ${\cal L}(\hat w)=\lambda_0^{n+2}$.

\subsection*{A2: KGD model}

For this model we assume that the self-similar crack aperture is defined by the following base functions:
\begin{equation}\label{h_KGD_fluid}
\hat \Omega_i(x)=(1-x^2)^{\alpha_i} C^{\alpha_i-1/2}_{2(i+1)-2}(x), \quad i=0,..,R-1, \quad \hat \Omega_R(x)=\sqrt{1-x^2}-\frac{2}{3}(1-x^2)^{3/2}-x^2\ln \left|{\frac{1+\sqrt{1-x^2}}{x}}\right|,
\end{equation}
where $C^{\alpha_i-1/2}_{2(i+1)-2}(x)$ is the ultraspherical polynomial. Respective powers $\alpha_i$ are to be taken in accordance with the asymptotic behaviour of the crack opening (see e.g. Table \ref{T1}). The term $\Omega_R(x)$ was introduced to obtain a non-zero pressure gradient for $x=0$. Note that:
\begin{equation}\label{w_N fluid}
\hat \Omega_R(x)=O\left((1-x^2)^{5/2}\right), \quad x\to 1.
\end{equation}
By applying \eqref{h_KGD_fluid} in \eqref{p_ss_2}$_2$ and subsequent differentiation one obtains an analytical formula for the pressure gradient:
\begin{equation}\label{pp_KGD}
\frac{d}{dx}\hat p(x)=-\sum_{i=0}^{R-1}\lambda_i H_i(x)-\lambda_R \left(\frac{\pi}{4} - x\right),
\end{equation}
where:
\begin{equation}\label{H0_KGD_1}
H_0(x)=\frac{1}{\pi}\alpha_0(2\alpha_0-1)B(1/2,\alpha_0)x\cdot_2F_1(3/2-\alpha_0,2;3/2;x^2),
\end{equation}
\begin{equation}\label{H0_KGD_2}
H_i(\alpha_i,x)=\frac{(1-2\alpha_i)(2i+1)}{4\pi(i+\alpha)}\,xB\big(-\frac{1}{2}-i,\alpha+i+1\big) \cdot \Big\{
\big[2(i+1)(i+1+\alpha)-2-5i\big]\times
\end{equation}
\[
{}_2F_1\big(\frac{3}{2}-i-\alpha,i+1;\frac{3}{2};x^2\big)-\frac{2\alpha}{3}(i+1)(2i-3+2\alpha)x^2\cdot{}_2F_1\big(\frac{5}{2}-i-\alpha,i+2;\frac{5}{2},x^2\big)  \Big\},
\quad
i \geq 1.
\]
Here $B$ is the beta function and $_2F_1$ is the Gauss hypergeometric function. Then from \eqref{v_rep} -- \eqref{q0_rep} we have the complete benchmark solution, where ${\cal L}(\hat w)=\lambda_0^{n+2}$ (viscosity dominated regime) or ${\cal L}(\hat w)=\lambda_0^{n+1}\lambda_1$ (toughness dominated regime). Moreover, by properly handling the values of respective coefficients $\lambda_i$, one can control the asymptotic behaviour of the particle velocity and leak-off function, and in the case of the toughness dominated mode the value of the self-similar stress intensity factor $\hat K_I$.

\begin{remark}
For the toughness dominated regime of the KGD model we take:
\begin{equation}
\label{w_0_toughness}
\hat \Omega_0(x)=\sqrt{1-x^2}.
\end{equation}
As a result the corresponding pressure component is constant along the entire fracture length which implies that:
\begin{equation}
\label{H0_tough}
H_0(x)=0.
\end{equation}

\end{remark}

\begin{remark}
If need be one can also utilize a simplified version of the benchmark solution. Namely, when functions $\Omega_i(x)$ from the basis \eqref{h_KGD_fluid} are taken as the power terms $\hat \Omega_i(x)=(1-x^2)^{\alpha_i}$, respective entries $H_i(x)$ from \eqref{pp_KGD} are then computed according to \eqref{H0_KGD_1}. All other benchmark components are determined in the same way as previously.
\end{remark}

\subsubsection*{A2.1: Perfectly plastic fluid (n=0)}
In the case of a perfectly plastic fluid ($n=0$) the Poiseuille equation degenerates to the form  \eqref{n0_special_case}.  For this reason it can no longer be used to compute the particle velocity, and as such the method of constructing benchmark examples described above fails. In order to circumvent this deficiency we shall modify the problem slightly and propose a set of analytical benchmark solutions for such a formulation.

As before we take some representation of the crack opening \eqref{w_rep} which accounts for the proper tip asymptotics. When substituting it into the elasticity relation \eqref{p_ss_2}$_2$ one arrives at the expression for the net fluid pressure in the form  \eqref{p_rep}. Next we impose the following constraints on the values of the first coefficients $\lambda_i$:
\begin{itemize}
\item{the viscosity dominated regime of the KGD model}
\begin{equation}
\label{lam_0_fluid}
\lambda_0=\sqrt{\pi},
\end{equation}
\item{the toughness dominated regime of the KGD model}
\begin{equation}
\label{lam_0_tough}
\lambda_0\lambda_1=-\frac{4}{3\pi}.
\end{equation}
\end{itemize}

As a result of these assumptions the equivalent of \eqref{n0_special_case} yields:
\begin{equation}
\label{n0_cond_new}
-\hat w \frac{d \hat p}{dx}=g(x), \quad g(1)=1,
\end{equation}
where $g(x)$ is a known function depending on the chosen $\hat w$ and $\hat p$.

In order to construct the set of benchmark solutions we assume that \eqref{n0_cond_new} holds instead of \eqref{n0_special_case}. Consequently, the computational formula \eqref{w_operator_n0} needs to be modified to:

\begin{equation}\label{w_operator_n0_mod}
\hat w(x)= \frac{4}{\pi}\int_0^1 \frac{g(x)}{\hat w(\eta)}
K(\eta,x) d\eta+\frac{4}{\sqrt{\pi}}\hat K_I\sqrt{1- x^2}.
\end{equation}

In this way, by making slight modifications to the problem (introduction of relations \eqref{n0_cond_new} and \eqref{w_operator_n0_mod} instead of \eqref{n0_special_case} and \eqref{w_operator_n0}) we arrive at the formulation which allows one to find a class of analytical benchmark solutions that can be used to test the numerical algorithm.

The benchmark value of the self-similar crack propagation speed $\hat v_0$ is computed according to \eqref{v0_n0_ss}, with the leak-off function and influx magnitude chosen for convenience. Finally, the reduced particle velocity is obtained in an analogical way from \eqref{ss_1}.

Throughout this paper we use the following variants of the benchmark solutions for $n=0$.
\begin{itemize}
\item{The viscosity dominated regime.}

Three base functions $\hat \Omega_i$ are employed:
\[ \hat \Omega_0(x)=1-x^2, \quad \hat \Omega_1(x)=(1-x^2)^2, \quad \hat \Omega_2(x)=\hat \Omega_R(x), \]
where $\hat \Omega_R$ is defined in \eqref{h_KGD_fluid}. The corresponding components of the pressure derivative are computed according to \eqref{pp_KGD} -- \eqref{H0_KGD_1}.

\item{The toughness dominated regime.}

The base functions $\hat \Omega_i$ are taken as:
\[ \hat \Omega_0(x)=\sqrt{1-x^2}, \quad \hat \Omega_1(x)=(1-x^2)^{3/2}\log(1-x^2), \quad \hat \Omega_2(x)=\hat \Omega_R(x), \]
where $\hat \Omega_R$ is defined in \eqref{h_KGD_fluid}. As usual, the square root term produces a constant pressure component (zero gradient). The remaining components of the pressure derivative are:
\[\hat p_1'(x)=\left(3\log 2-\frac{5}{2}\right)x-\frac{3}{2}\arcsin(x)\frac{1-2x^2}{\sqrt{1-x^2}},\quad \hat p_2'(x)=x-\frac{\pi}{4}.\]
\end{itemize}

Note that for the foregoing representation of the crack opening, the integrals needed to define the benchmark values of particle velocity and reduced particle velocity are computed analytically.

\section*{Appendix B: Coefficients of the formulas \eqref{w_Ad_ap} -- \eqref{v_Ad_ap}}

Coefficients $\hat w_i(n)$, $\hat p_i(n)$ and $\hat v_i(n)$ in formulas \eqref{w_Ad_ap} -- \eqref{v_Ad_ap} are all functions of the fluid behaviour index $n$ and are approximated by the following expression:
\begin{equation}
\label{approx_values}
\vartheta(n)=\sum_{k=0}^4 r_k^\vartheta n^k,
\end{equation}
with values of $r_k^\vartheta$ given in Table \ref{T2}.

\begin{table}[h]
\vspace{3mm}
\renewcommand{\arraystretch}{1.5}
\begin{center}
\begin{tabular}{|c|c|c|c|c|c|c|c|c|c|c|}
  \hline
    $\vartheta(n)$ & $r_0^\vartheta$ & $r_1^\vartheta$ &$r_2^\vartheta$& $r_3^\vartheta$ &$r_4^\vartheta$\\
  \hline\hline
  $\hat w_1$&-0.165 &0.529&-0.7587&0.6006&-0.1985 \\
  \hline
  $\hat w_2$&0.673&-0.4886&0.00747&0.1486&-0.0692\\
  \hline
  $\hat p_1$&0.6228&-0.1334&-0.5781&0.5167&-0.1536\\
  \hline
  $\hat p_2$&-1.388&0.1244&1.977&-1.873&0.5961\\
	\hline
	$\hat p_3$&1.911&-1.521&-0.4046&0.8542&-0.2925\\
  \hline
  $\hat p_4$&-1.129&1.952&-1.674&0.9952&-0.3012\\
  \hline
	$\hat p_5$&0.04245&-0.1925&0.2071&-0.1114&0.02931\\
  \hline
	$\hat p_6$&0.05075&-0.1895&0.2868&-0.2337&0.07845\\
  \hline
	$\hat v_1$&0.00987&-0.1373&0.1017&-0.0497&0.0122\\
  \hline
  $\hat v_2$&0.01939&-0.00148&0.02551&-0.01331&0.00323\\
	\hline
	$\hat v_3$&0.2264&-0.06635&-0.04062&0.03081&-0.00845\\
  \hline
  $\hat v_4$&-0.3275&0.1403&0.04508&-0.04654&0.01473\\
  \hline
	$\hat v_5$&0.2098&-0.09169&-0.01172&0.01446&-0.00373\\
  \hline
\end{tabular}
\end{center}

\vspace{-2mm}

\caption{The values of coefficients $r_k^\vartheta$ in equation \eqref{approx_values}. }
\label{T2}
\end{table}

\section*{Nomenclature}

\begin{tabular}{ll}
$K_I$ & stress intensity factor\\
$K_{IC}$ & material toughness \\
$l(t)$ & crack length ($L(t)$ after normalization) \\
$n$ & fluid behaviour index\\
$p(t,x)$ & net fluid pressure \\
$q(t,x)$ & fluid flow rate \\
$q_l(t,x)$ & leak-off\\
$q_0(t)$ & influx \\
$v(t,x)$ & particle velocity\\
$w(t,x)$ & crack opening \\
$\alpha_0$, $\alpha_1$, $\alpha_2$ & exponents of leading terms in asymptotic expansion of the crack opening\\
$\beta_1$, $\beta_2$ & exponents of leading terms in asymptotic expansion of the particle velocity\\
$\zeta_0$, $\zeta_1$, $\zeta_2$ & exponents of leading terms in asymptotic expansion of the reduced particle velocity\\
$\phi(t,x)$ & reduced particle velocity \\
\end{tabular}

\vspace{3mm}

All variables with "$\sim$" symbol refer to the normalized formulation, while "$\wedge$" symbolizes self-similar formulation.

\end{document}